\documentclass[prd,twocolumn,superscriptaddress,preprintnumbers,showpacs,nofootinbib,notitlepage,floatfix]{revtex4-2}
\usepackage{hyperref}
\usepackage{graphicx}
\usepackage{subfigure}
\usepackage{amsmath}
\usepackage{amssymb}
\usepackage{slashed}
\usepackage{amsfonts}
\usepackage{xcolor}
\usepackage{multirow}
\usepackage{array}
\usepackage{mwe}
\usepackage{mathrsfs}

\usepackage[utf8]{inputenc}
\usepackage[normalem]{ulem}

\DeclareRobustCommand{\eq}[1]{Eq.~\eqref{eq:#1}}

\DeclareRobustCommand{\eqs}[2]{Eqs.~\eqref{eq:#1} and \eqref{eq:#2}}

\DeclareRobustCommand{\fig}[1]{Fig.~\ref{fig:#1}}
\DeclareRobustCommand{\figs}[2]{Figs.~\ref{fig:#1} and \ref{fig:#2}}
\DeclareRobustCommand{\app}[1]{App.~\ref{app:#1}}
\DeclareRobustCommand{\sec}[1]{Sec.~\ref{sec:#1}}

\DeclareRobustCommand{\eq}[1]{Eq.~(\ref{eq:#1})}
\DeclareRobustCommand{\eqs}[2]{Eqs.~(\ref{eq:#1}) and (\ref{eq:#2})}

\newcommand{\nn}{\nonumber}

\newcommand\bets{\begin{table*}}
\newcommand\eets[1]{\label{tb:#1}\end{table*}}

\begin{document}


\title{LaMET's Asymptotic Extrapolation vs. Inverse Problem}

\author{Jiunn-Wei Chen}
\affiliation{Department of Physics and Center for Theoretical Physics,
National Taiwan University, Taipei, Taiwan 106}
\affiliation{Physics Division, National Center for Theoretical Sciences, Taipei 10617, Taiwan}

\author{Xiang Gao}
\affiliation{Physics Department, Brookhaven National Laboratory, Upton, New York 11973, USA}

\author{Jinchen He}
\affiliation{Department of Physics, University of Maryland, College Park, MD 20850, USA}
\affiliation{Physics Division, Argonne National Laboratory, Lemont, IL 60439, USA}

\author{Jun Hua}
\affiliation{Key Laboratory of Atomic and Subatomic Structure and Quantum Control (MOE), Guangdong Basic Research Center of Excellence for Structure and Fundamental Interactions of Matter, Institute of Quantum Matter, South China Normal University, Guangzhou 510006, China}
\affiliation{Guangdong-Hong Kong Joint Laboratory of Quantum Matter, Guangdong Provincial Key Laboratory of Nuclear Science, Southern Nuclear Science Computing Center, South China Normal University, Guangzhou 510006, China}

\author{Xiangdong Ji}
\affiliation{Department of Physics, University of Maryland, College Park, MD 20850, USA}

\author{Andreas Sch\"afer}
\affiliation{Institut f\"ur Theoretische Physik, Universit\"at Regensburg, D-93040 Regensburg, Germany}
\affiliation{Department of Physics, National Taiwan University, Taipei, Taiwan 106}

\author{Yushan Su}
\affiliation{Department of Physics, University of Maryland, College Park, MD 20850, USA}

\author{Wei Wang}
\affiliation{State Key Laboratory of Dark Matter Physics, School of Physics and Astronomy, Shanghai Jiao Tong University,  Shanghai 200240, China}

\author{Yi-Bo Yang}
\affiliation{CAS Key Laboratory of Theoretical Physics, Institute of Theoretical Physics, Chinese Academy of Sciences, Beijing 100190, China}
\affiliation{University of Chinese Academy of Sciences, School of Physical Sciences, Beijing 100049, China}
\affiliation{School of Fundamental Physics and Mathematical Sciences, Hangzhou Institute for Advanced Study, UCAS, Hangzhou 310024, China}
\affiliation{International Centre for Theoretical Physics Asia-Pacific, Beijing/Hangzhou, China}

\author{Jian-Hui Zhang}
\affiliation{School of Science and Engineering, The Chinese University of Hong Kong, Shenzhen 518172, China}

\author{Qi-An Zhang}
\affiliation{School of Physics, Beihang University, Beijing 102206, China}

\author{Rui Zhang}
\affiliation{Physics Division, Argonne National Laboratory, Lemont, IL 60439, USA}

\author{Yong Zhao}
\affiliation{Physics Division, Argonne National Laboratory, Lemont, IL 60439, USA}

\begin{abstract}

Large-Momentum Effective Theory (LaMET) is a physics-guided systematic expansion to calculate light-cone parton distributions, including collinear (PDFs) and transverse-momentum-dependent ones, at any fixed momentum fraction $x$ within a range of $[x_{\rm min}, x_{\rm max}]$. It theoretically solves the ill-posed inverse problem that afflicts other theoretical approaches to collinear PDFs, such as short-distance factorizations. 
Recently, arXiv:2504.17706~\cite{Dutrieux:2025jed} raised practical concerns about whether current or even future lattice data will have sufficient precision in the sub-asymptotic correlation region to support an error-controlled extrapolation---and if not, whether it becomes an inverse problem where the relevant uncertainties cannot be properly quantified. While we agree that not all current lattice data have the desired precision to qualify for an asymptotic extrapolation, some calculations do, and more are expected in the future. We comment on the analysis and results in Ref.~\cite{Dutrieux:2025jed} and argue that a physics-based systematic extrapolation still provides the most reliable error estimates, even when the data quality is not ideal. In contrast, re-framing the long-distance asymptotic extrapolation as a data-driven-only inverse problem with {\it ad hoc} mathematical conditioning could lead to unnecessarily conservative errors.

\end{abstract}

\maketitle

\section{The Problem}

Over the past decade, there has been significant progress in the lattice QCD calculation of parton physics, including parton distribution functions (PDFs), generalized parton distributions (GPDs), and transverse-momentum-dependent distributions (TMDs), etc. During this period, quite a few methods~\cite{Liu:1993cv,Detmold:2005gg,Braun:2007wv,Davoudi:2012ya,Ji:2013dva,Radyushkin:2017cyf,Ma:2017pxb,Chambers:2017dov,Shindler:2023xpd} have been developed to complement the traditional approaches of computing the lowest few moments of the parton distributions, aiming at either higher moments or the $x$-dependence of parton distributions. Among them, Large-Momentum Effective Theory (LaMET)~\cite{Ji:2013dva,Ji:2014gla,Ji:2020ect}, also known as the quasi-PDF method in the calculation of collinear PDFs, provides the most general framework for computing PDFs, GPDs, and TMDs, as well as light-cone wave functions. It has so far been the most widely adopted approach, achieving significant successes across all frontiers of parton physics~\cite{Xiong:2013bka,Lin:2014zya,Ji:2014hxa,Ji:2014lra,Sufian:2014jma,Ji:2015qla,Xiong:2015nua,Alexandrou:2016eyt,Chen:2016utp,Lin:2016qia,Yang:2016nfc,Yang:2016plb,Alexandrou:2017dzj,Alexandrou:2017huk,Alexandrou:2017qpu,Chen:2017mie,Chen:2017mzz,Constantinou:2017sej,Ishikawa:2017iym,Ji:2017oey,Ji:2017rah,Lin:2017ani,Stewart:2017tvs,Xiong:2017jtn,Zhang:2017bzy,Zhang:2017zfe,Alexandrou:2018eet,Alexandrou:2018pbm,Alexandrou:2018yuy,Chen:2018xof,Ebert:2018gzl,Fan:2018dxu,Izubuchi:2018srq,Ji:2018hvs,LatticeParton:2018gjr,Lin:2018pvv,Liu:2018hxv,Liu:2018tox,Zhang:2018diq,Zhang:2018nsy,Zhang:2018rls,Zhao:2018fyu,Alexandrou:2019dax,Alexandrou:2019lfo,Chai:2019rer,Chen:2019lcm,Constantinou:2019vyb,Ebert:2019tvc,Liu:2019urm,Shanahan:2019zcq,Wang:2019msf,Zhang:2019qiq,Alexandrou:2020qtt,Alexandrou:2020uyt,Alexandrou:2020zbe,Bhattacharya:2020cen,Bhattacharya:2020jfj,Chai:2020nxw,Chen:2020arf,Chen:2020iqi,Chen:2020ody,Ebert:2020gxr,Fan:2020nzz,Gao:2020ito,Hua:2020gnw,Ji:2020brr,Ji:2020ect,Ji:2020jeb,LatticeParton:2020uhz,Lin:2020fsj,Lin:2020rxa,Lin:2020ssv,Shanahan:2020zxr,Shugert:2020tgq,Vladimirov:2020ofp,Zhang:2020dkn,Zhang:2020gaj,Zhang:2020rsx,Alexandrou:2021bbo,Alexandrou:2021oih,Bhattacharya:2021moj,Bhattacharya:2021rua,Constantinou:2021nbn,Dodson:2021rdq,Gao:2021hxl,Gao:2021dbh,LatticePartonLPC:2021gpi,Li:2021wvl,Lin:2021brq,Lin:2021ukf,Scapellato:2021uke,Schlemmer:2021aij,Shanahan:2021tst,Bhattacharya:2022aob,Constantinou:2022fqt,Deng:2022gzi,Ebert:2022fmh,Gao:2022iex,Gao:2022uhg,LatticeParton:2022xsd,LatticeParton:2022zqc,LatticePartonCollaborationLPC:2022myp,LatticePartonLPC:2022eev,Scapellato:2022mai,Schindler:2022eva,Zhang:2022xuw,Alexandrou:2023ucc,Avkhadiev:2023poz,Bhattacharya:2023jsc,Bhattacharya:2023nmv,Bhattacharya:2023tik,Cichy:2023dgk,Deng:2023csv,Gao:2023ktu,Gao:2023lny,Holligan:2023jqh,Holligan:2023rex,Ji:2023pba,LatticeParton:2023xdl,LatticePartonLPC:2023pdv,Lin:2023gxz,Zhao:2023ptv,Liu:2023onm,Avkhadiev:2024mgd,Baker:2024zcd,Bollweg:2024zet,Chen:2024rgi,Cloet:2024vbv,Ding:2024saz,Gao:2024fbh,Good:2024iur,Han:2024min,Holligan:2024umc,Holligan:2024wpv,Ji:2024hit,LatticeParton:2024mxp,LatticeParton:2024vck,LatticeParton:2024zko,Miller:2024yfw,Mukherjee:2024xie,Spanoudes:2024kpb,Zhang:2024omt,Bollweg:2025ecn,Bollweg:2025iol,Han:2025odf,Holligan:2025ydm,LPC:2025spt,Zhang:2025hvf,Ji:2025mvk,Ji:2015jwa,Chen:2016fxx,Zhang:2018ggy,Braun:2018brg,Wang:2019tgg,Ji:2021uvr,Su:2022fiu,Zhu:2022bja,Yao:2022vtp,Pang:2024sdl,Han:2024cht}.

LaMET is a physics-guided effective field theory (EFT) expansion of light-cone parton distributions, including the PDFs, GPDs, and TMDs, at any fixed momentum fraction $x$ within a moderate range of $[x_{\rm min}, x_{\rm max}]$~\cite{Ji:2024oka}, in terms of the quark and gluon momentum distributions in a fast moving hadron, aka the quasi distributions~\cite{Ji:2013dva}. As an important advantage, it bypasses the inverse problem (IP) of reconstructing the PDF $x$-dependence from correlations of limited range, which afflicts other theoretical approaches such as the short-distance factorization (SDF), including the pseudo-PDF~\cite{Radyushkin:2017cyf,Orginos:2017kos} and ``good lattice cross section''~\cite{Ma:2017pxb}. The SDF approach has also gained notable traction and led to substantial progress in extracting the lowest Mellin moments and phenomenological fitting of collinear PDFs and GPDs~\cite{Karpie:2017bzm,Orginos:2017kos,Radyushkin:2017lvu,Radyushkin:2017sfi,Karpie:2018zaz,Radyushkin:2018cvn,Radyushkin:2018nbf,Balitsky:2019krf,Joo:2019bzr,Joo:2019jct,Karpie:2019eiq,Radyushkin:2019mye,Radyushkin:2019owq,Sufian:2019bol,Bhat:2020ktg,Bringewatt:2020ixn,DelDebbio:2020rgv,Fan:2020cpa,Joo:2020spy,Li:2020xml,Shugert:2020tgq,Sufian:2020vzb,Zhao:2020bsx,Balitsky:2021bds,Balitsky:2021cwr,Balitsky:2021qsr,Egerer:2021ymv,Fan:2021bcr,HadStruc:2021qdf,HadStruc:2021wmh,Karpie:2021pap,Salas-Chavira:2021wui,Bhat:2022mjv,Bhat:2022zrw,Delmar:2022plq,Fan:2022kcb,Gao:2022uhg,Gao:2022vyh,HadStruc:2022nay,HadStruc:2022yaw,JeffersonLabAngularMomentumJAM:2022aix,Bhattacharya:2023ays,Delmar:2023agv,Dutrieux:2023zpy,Good:2023gai,Karpie:2023nyg,Nurminen:2023qok,Radyushkin:2023ref,Bhattacharya:2024qpp,Bhattacharya:2024wtg,Blossier:2024wyx,Cheng:2024wyu,Cichy:2024afd,Dutrieux:2024rem,HadStruc:2024rix,Karpie:2024bof,Kovner:2024pwl,SanJosePerez:2024axh,SanJosePerez:2024fad,Pang:2024kza,Ji:2025mvk}, despite challenges in controlling the model uncertainty in the latter and its limited applicability to a broader range of parton physics.
 
\begin{figure}
    \centering
    \includegraphics[width=0.9\linewidth]{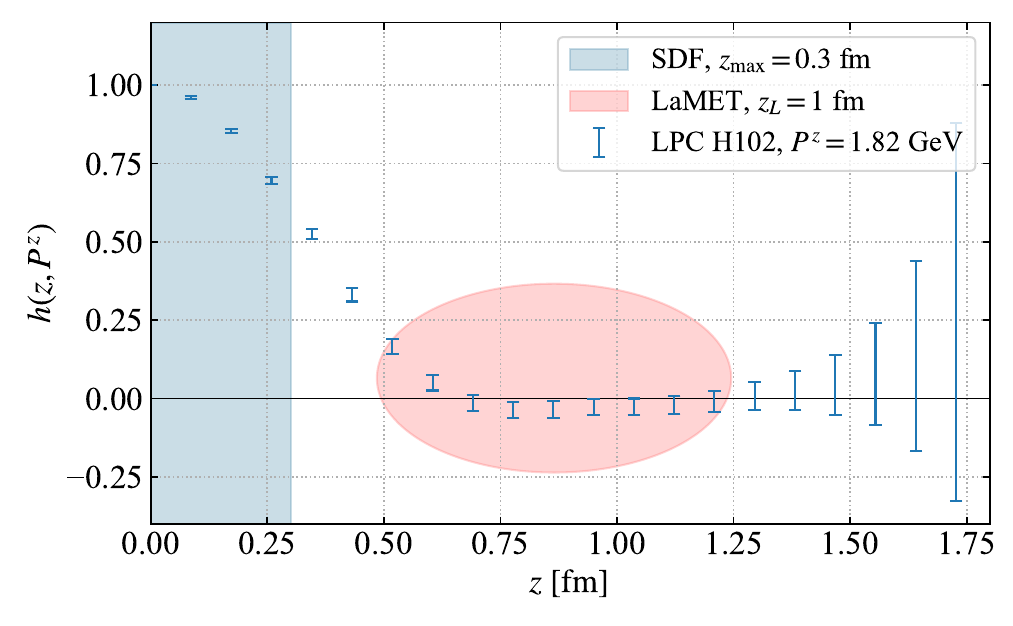}
    \caption{An illustration of the difference between LaMET and SDF: LaMET in principle utilizes data at all $z$ to sharply resolve the momentum of a parton, though in practice they are usually truncated at $z_L\sim 1.0$~fm due to noisy data and extrapolated using physics-motivated asymptotic forms to $z=\infty$~\cite{Ji:2020brr}. Meanwhile, the SDF can only use data up to about $z_{\rm max}\sim 0.2-0.3$~fm (shaded blue region) to fit a model of the PDF, which constitutes an IP. Plotted are the hybrid-scheme nucleon transversity quasi-PDF matrix elements computed in Ref.~\cite{LatticeParton:2022xsd}. The shaded-red ellipse approximately indicates the sub-asymptotic region where the exponential decay starts to dominate. The concern raised in ~\cite{Dutrieux:2025jed} is whether the data precision in this region allows for a reliable error estimate of the extrapolation; otherwise, it must be treated as an IP.}
    \label{fig:hz}
\end{figure}

We illustrate the difference between LaMET and SDF in \fig{hz}, both of which start from the lattice matrix element~\cite{Ji:2013dva,Radyushkin:2017cyf},
\begin{align}
    h(z,P) &= {1\over 2P^t}\langle P| \bar{\psi}(z) W(z,0) \gamma^t \psi(0)|P\rangle\,,
\end{align}
where $|P\rangle$ is a plane-wave hadron state with momentum $P^\mu=(P^t,0,0,P^z)$, and $W(z,0)$ with $z^\mu=(0,0,0,z)$ is a spatial Wilson line that guarantees the gauge invariance of the correlator. In SDF, the data contain leading-twist information up to about $z=0.2-0.3$~fm, beyond which the framework breaks down. For instance, the average size of the instanton, a non-perturbative gluon configuration, is about 0.3~fm~\cite{Nowak:1996aj,Schafer:1996wv}. At finite momentum $P^z$, only a limited range of leading-twist correlations can be accessed, which is insufficient for a reliable forward determination of the PDFs. Instead, this segment of correlations can be used for phenomenological fittings of the PDFs.
In contrast, LaMET legitimately utilizes matrix elements at all $z$ to compute the quasi-PDF, thereby providing a sharply-defined momentum fraction $x$ for quarks and gluons. This explains why LaMET can calculate the PDFs at a fixed momentum fraction $x$ through a systematic expansion. This distinction between LaMET and SDF has been extensively discussed in the literature~\cite{Ji:2020ect,Ji:2020brr,Ji:2020byp,Ji:2022ezo,Ji:2024oka}, where it is argued that, given the same lattice data at a large $P^z$, LaMET offers the most accurate determination of a PDF within a moderate range of $x$.
 
Recently, Ref.~\cite{Dutrieux:2025jed} raised practical concerns about whether lattice data in the sub-asymptotic region, i.e., $0.7~\text{fm} \lesssim z \lesssim 1.0~\text{fm}$ presented there, are precise enough for a controlled physical extrapolation to larger $z$. If not, this could adversely impact the Fourier transform (FT) used to obtain quasi-PDFs for the LaMET expansion, rendering the error estimation in the literature possibly unreliable. Various data-driven IP methods~\cite{rietsch1976maximum,burnier2013bayesian,liang2020towards,backus1968resolving,DelDebbio2025Bayesian}
were proposed to treat the FT, yielding large uncertainties of the reconstructed quasi-PDF, dubbed as ``inverse problem in the LaMET framework''.

In this paper, we directly respond to these concerns. In \sec{theory}, we first briefly review the fundamental difference in the predictive capabilities of LaMET and SDF, based on the extensive discussions in Refs.~\cite{Ji:2020brr,Ji:2020ect,Ji:2020byp,Ji:2022ezo,Ji:2024oka}, emphasizing LaMET as a forward-problem (FP) formalism of calculating, not fitting, parton distributions. Then in \sec{asymptotic} we point out that the alleged IP in LaMET application to lattice data is a standard asymptotic extrapolation problem in lattice QCD, which can be systematically treated with reliable error estimates. In \sec{response}, we examine in detail the criticisms and data analysis in Ref.~\cite{Dutrieux:2025jed}, and conclude that the observed FT errors can be bounded by a theoretical estimate based on asymptotic extrapolation, and the data-driven IP methods used are less reliable as they do not satisfy the physical constraints. We conclude in \sec{end}. Finally, since the authors of Ref.~\cite{Dutrieux:2025jed} recently commented~\cite{Dutrieux:2025axb} on our current paper, we provide a response in \app{reply}.

\section{LaMET As a Forward-Problem Formalism}
\label{sec:theory}

LaMET is an application of the EFT methods developed by Weinberg and Wilson, among others~\cite{Ji:2024oka}.
This EFT allows for a systematic expansion of the light-cone observables in terms of similar ones in a large-momentum hadron. Although motivated by lattice QCD, LaMET is applicable to any other theories that can generate
observables in a large-momentum hadron state, for example, the instanton liquid formalism for hadrons~\cite{Nowak:1996aj,Schafer:1996wv}.

Here is a brief summary of the main idea of LaMET. Light-cone physics corresponds to observables in a hadron with infinite momentum. If the equivalent observables in a large-momentum hadron are known, then the light-cone observables can be obtained through a large-momentum expansion. For example, the light-cone PDF $f(x, \mu)$ can be expanded from the momentum distribution (quasi-PDF) in a hadron of large momentum, $\tilde f(y,P^z)$, through the EFT formula
\begin{align}\label{eq:lamet2}
    f(x,\mu) &=  \int_{-\infty}^\infty {dy\over |y|} C\Big({x\over y}, {\mu \over y P^z}\Big) \tilde f(y,P^z) \nn\\
    &\qquad + {\cal O}\Big( {\Lambda_{\mathrm{QCD}}^2\over (xP^z)^2}, {\Lambda_{\mathrm{QCD}}^2\over ((1-x)P^z)^2}\Big)\,,
\end{align}
where $C$ is a perturbative matching kernel, given as a series in the strong coupling $\alpha_s(\mu)$, and the power corrections are controlled by the external parton and spectator momenta $xP^z$ and $(1-x)P^z$, respectively. After taking into account evolution and higher-order perturbative corrections~\cite{Holligan:2023rex,Ji:2023pba,Ji:2024hit}, it is found that the matching at the leading power is reliable within a moderate range of $x\in [x_{\rm min}, x_{\rm max}]$ with $x_{\rm min} \approx 1- x_{\rm max} \sim {\Lambda_{\rm QCD}/(2P^z)}$. Another example of such an EFT expansion in terms of the quasi-PDF 
can be found in the deep-inelastic scattering cross section in Ref.~\cite{Ji:2024oka}.

The momentum distribution $\tilde f(y,P^z)$ is the starting point of the LaMET expansion for PDFs, which is obtained from the coordinate-space correlation $h(z,P)$ computed on the lattice. The latter is a spatial correlation function that decays exponentially in the \textbf{asymptotic} region $z>z_A$. In contemporary lattice simulations, high-precision data are reachable 
in the \textbf{sub-asymptotic} region $z_{SA} \le z \le z_A$ before the decay is fully dominant. 
For the nucleon, one may expect $z_{SA} \sim 0.5$~fm and $z_A \sim 1.2$~fm, which can be more precisely identified with high-quality lattice data. 
Augmented with a physics-driven systematic extrapolation from the information in the sub-asymptotic region, one can obtain the full coordinate-space
correlations. The results provide the needed input to calculate the quasi-distributions, 
\begin{align}\label{eq:qpdf}
    \tilde f(y, P^z) \equiv P^z\int_{-\infty}^\infty {d z \over 2\pi}e^{iy P^z z} h(z, P)\,.
\end{align}
As presented in \eq{qpdf}, the alleged IP in Ref.~\cite{Dutrieux:2025jed} is essentially a matter of convergence and precision of the numerical Fourier integral, which are influenced by the interpolation and extrapolation of lattice data points in $z$-space. The smoothness of correlation functions in a bounded hadron ensures that the interpolation errors are small, while their asymptotic behavior at large distances controls the extrapolation uncertainty.

Since perturbative matching is already encoded in an EFT, it does not introduce an additional IP in the region where the framework is valid. In the literature, the coefficient $C$ in \eq{lamet2} has been derived by inverting the kernel that matches the PDF onto the quasi-PDF\cite{Xiong:2013bka,Izubuchi:2018srq}, which raised the question of whether this procedure constitutes an IP. To address this concern, we note that the leading-power matching kernel can be inverted order by order in $\alpha_s$, since the calculation is performed in the asymptotic limit $P^z \to \infty$ where perturbation theory is valid across the entire kinematic range. However, at finite $P^z$, both perturbation theory and the power expansion break down in certain kinematic regions. As a result, \eq{lamet2} and the expansion of the quasi-PDF in terms of the PDF correspond to two distinct EFT expansions, which cannot be transformed into one another. This comparison is analogous to that between the quasi- and pseudo-PDFs~\cite{Ji:2020brr,Ji:2020ect,Ji:2020byp,Ji:2022ezo,Ji:2024oka}.

In practice, there may still be concern about whether the FT uncertainty at small-$x$ in the quasi-PDF will propagate to moderate $x$ after its convolution with the matching kernel in \eq{lamet2}. First, matching is performed between the collinear modes in the quasi-PDF and those in the PDF~\cite{Ji:2020ect,Ji:2020byp,Ji:2022ezo,Ji:2024oka}, with a hard momentum $xP^z$. In contrast, contributions from soft modes with $x\to0$ are power suppressed. 

This suppression can also be derived from the structure of the matching kernel $C(x/y)$. For example, up to next-to-leading order (NLO),
\begin{align}
    C\Big({x\over y}\Big) &= \delta \Big({x\over y}-1\Big) + {\alpha_s \over 2\pi} C^{(1)}\Big({x\over y}\Big)\,,\\
    C^{(1)}\Big({x\over y}\Big) &= C^{(1)}_r\Big({x\over y}\Big) - \delta\Big({x\over y}-1\Big) \int_{-\infty}^\infty dy' C^{(1)}_r(y')\,.
\end{align}
At leading order, matching does not change the quasi-PDF. At NLO, in the limit $|x/y|\gg1$,
\begin{align}
    \lim_{|x/y|\gg1}C^{(1)}_r\Big({x\over y}\Big) \propto \Big({y\over x}\Big)^2\,.
\end{align}
For finite $x$, the dominant contribution to the matching integral arises from the region $|y| \gtrsim |x|$, while contributions from $|y| \ll x$ are suppressed quadratically. 
In practice, the range of $x$ targeted in LaMET analyses lies well above the small-$x$ region where FT uncertainties become significant, rendering any contamination from the latter negligible.

Therefore, according to \eqs{lamet2}{qpdf}, LaMET is an FP formalism in the lattice calculation of PDFs at moderate $x$, which is systematically improvable without an adverse IP.

It is important to emphasize that in this formulation, the shapes and mathematical properties of the PDFs are not assumed a \textit{priori}, in contrast to the IP formulation for SDF~\cite{Dutrieux:2025jed}.
In fact, PDFs are calculated one $x$ at a time, and the $x$-space functions come out as a prediction. Since no connections with partons are made before the large-momentum expansion, and the terms in SDF for the coordinate-space correlations---such as ``higher-twist contamination at large $z$''~\cite{Dutrieux:2025jed}---are irrelevant. 
It is also unnecessary to invoke a notion of so-called smearing of order $O(\Lambda_{\rm QCD}/P_z)$ in $x$-space~\cite{Dutrieux:2025jed} to understand how LaMET works; in the standard EFT framework, the relevant systematics are accurately described by the power accuracy, which begins at $O(\Lambda_{\rm QCD}^2/P_z^2)$ in LaMET~\cite{Zhang:2023bxs}.
Moreover, any variation in the PDF---as hypothesized in Ref.~\cite{Dutrieux:2025jed}---must either be predicted by LaMET or lie within its overall uncertainty band. If an undetected large, sharp variation exists in the quasi-PDF, it would correspond to a persistent long-range mode in coordinate space, which is ruled out by color confinement. If such a feature were to appear only in the PDF at moderate $x$, it would require an equally large, oppositely signed variation in the power correction. This would contradict the convergence of the EFT expansion in the region where it is expected to hold. Importantly, this scenario can be readily tested by performing calculations at multiple values of $P^z$, yet no such behavior has been observed so far.

On the other hand, reconstructing the PDFs from a limited segment of the leading-twist correlation in SDF is an IP that has uncontrolled uncertainties without additional physics input. For example, the pseudo-PDF method is based on the SDF of the same $h(z,P)$~\cite{Ji:2017rah,Radyushkin:2017lvu,Izubuchi:2018srq},
\begin{align}\label{eq:sdf1}
    h(z,P) &= \int_0^1 d\alpha\ {\cal C}(\alpha,z^2\mu^2) h_{\rm LC}(\alpha\lambda,\mu) + {\cal O}(z^2\Lambda_{\rm QCD}^2)\,,
\end{align}
where the quasi light-cone distance $\lambda=zP^z$, and the light-cone correlation function $h_{\rm LC}(\lambda,\mu)$ is the inverse FT of the PDF,
\begin{align}\label{eq:pdf}
    f(x,\mu) &\equiv \int_{-\infty}^\infty {d\lambda \over 2\pi}e^{i x\lambda} h_{\rm LC}(\lambda,\mu)\,.
\end{align}
Here ${\cal C}$ is the matching kernel, and the power corrections are suppressed by the short distance $z$. 
After taking into account higher-order perturbative corrections~\cite{Gao:2021hxl,Holligan:2023rex,Zhang:2023bxs}, the SDF is found to be reliable up to $ z=0.2-0.3$~fm, which corresponds to the strong coupling constant $\alpha_s(1/z) \sim 1$ and the breakdown of perturbation theory. As a result, the largest $\lambda$ that can be reached in contemporary lattice simulations is limited to $\lambda_{\rm max}\sim 3.0-4.0$, which is far from the asymptotic region where the light-cone correlation decays algebraically. Therefore, the $x$-dependence of PDFs cannot be calculated directly and must instead be fitted through parametrizations and Bayesian or neural-network approaches with assorted numerical and mathematical assumptions~\cite{Orginos:2017kos,Karpie:2019eiq,DelDebbio:2020rgv,Dutrieux:2024rem}.

\section{Asymptotic Extrapolation for Long-Range Correlations vs. Inverse Problem}
\label{sec:asymptotic}

To facilitate the LaMET analysis, one needs
to have coordinate-space correlation functions in a large-momentum hadron at all distances $z\in [0,\infty]$, as information at negative $z$ can be obtained from discrete symmetries. It is well known that lattice QCD cannot produce quality data for $z\gg 1$~fm due to the exponentially worsening signal-to-noise ratio (SNR). Therefore, we cannot have all the data that LaMET requires. Fortunately, physical principles offer important guidance. Even without doing 
a very challenging calculation, QCD predicts that the correlation function must effectively vanish at sufficiently large $z$ due to color confinement. In fact, more precise theoretical predictions are available. According to dispersion theory, the correlation function decays exponentially at large $z$ with the decay length controlled by the residual mass of a heavy-light meson~\cite{Gao:2021dbh}. Moreover, this mass is an intrinsic property of the correlation function in the sense that it is independent of the external hadron momentum. Through a more detailed analysis~\cite{Ji:2026vir}, one can derive subleading behaviors of the correlation function as well. Armed with these known rigid physics constraints, the lattice data can be reliably extrapolated to larger $z$ in the asymptotic region.

\subsection{Asymptotic Extrapolation}

Asymptotic expansion is essentially a generalization of the Taylor expansion in small parameters, which is widely used in theoretical physics, such as EFTs. It is also one of the most frequently used techniques in lattice QCD analyses, where the physical conditions are imposed as a prior. The precision of calculations is often assessed based on this procedure.

Asymptotic fitting has been employed when extracting the ground-state energy from two-point correlation functions, where the SNR typically decreases exponentially at large Euclidean time separations. A stable extraction relies on whether the correlation functions at finite time have entered the asymptotic region dominated by the ground state, which is typically indicated by a plateau in the standard logarithmic ratio plot. To better describe data at finite time separations, sub-asymptotic contributions from excited states are often included, leading to more sophisticated fits with additional parameters. In more elaborate fits, contributions from two-body scattering states may also be included~\cite{Mondal:2020cmt,Mondal:2020ela,Mondal:2021oot}.

Other examples include continuum and chiral extrapolations~\cite{Chen:2001eg,Arndt:2001ye,Chang:2018uxx}, where subleading effects from finite lattice spacing and unphysical pion masses are derived using EFTs and incorporated into fits to lattice data. In the same spirit, since dispersion relation analysis is a rigorous and model-independent method in theoretical physics, it supports the use of asymptotic extrapolation for spatial correlation functions as a standard and justified procedure.

\subsection{Inverse Problem}
\label{sec:def-inv}

When data in the sub-asymptotic region is noisy and the number of fit parameters exceeds the number of reliable data points, the fits are unstable. This problem happens frequently in physics, and certain parameters are not well fitted or over-fitted, and the $\chi^2$ will have flat directions in the parameter space. In this case, one should either generate more and better data or limit the number of parameters in the fitting and estimate the errors based on the relevant physics conditions. Nevertheless, it is not usually regarded as an IP in the literature. 

On the other hand, it has been suggested in Ref.~\cite{Dutrieux:2025jed} that in LaMET analysis, if the lattice data have large errors in the sub-asymptotic domain, one should treat the extrapolation as an IP, just like in the case of SDF where $\lambda_{\rm max} \sim 3.0-4.0$ is far from the asymptotic region of the leading-twist correlation function. While the IP method may initially appear to be an attractive option for tackling such problems, it is not suitable for LaMET analysis in its current form in Ref.~\cite{Dutrieux:2025jed}.

IPs arise when there is a limited set of data points, yet there is an extraordinarily large number of parameters or degrees of freedom due to the absence of a coherent physical guidance. To proceed, one typically introduces specific parametrizations or models to reduce the degrees of freedom and make the fit tractable. However, if there are a large number of models possible, it is very difficult to assess the systematic errors, because it is hard to argue, based on physics, which model is better. IPs are generally ill-posed, in contrast to well-posed problems. Mathematically, the ill-posedness of an IP is characterized by three criteria proposed by Jacques Hadamard: ``existence, uniqueness, and stability of the solution with respect to the input data'' \cite{Hadamard:1923,Kirsch:2011}. A problem is considered ill-posed if it fails to meet any of these conditions. The ill-posedness typically manifests as an instability of the solution, which makes error estimation very difficult. 

The global analysis of PDF from experimental data is a standard IP. As a continuous function of the momentum fraction $x$, PDFs are difficult to model. In fact, one might go so far as to say that {\it a PDF is a collection of an infinite number of parameters}. Of course, we believe that it is a continuous function and probably differentiable to a high degree except for $x=0$. A typical modelling of the PDF involves incorporating physics constraints like Regge behavior as $x\to0$ and power counting in the $x\to1$ limit. However, in the middle-$x$ region there is no physics-guided model, one has to interpolate the two end-point behaviors. Thus, unless there is a large number of experimental data points with wide kinematic coverage---which we fortunately have---and the smoothness assumptions are imposed, IP algorithms tend to generate many solutions with hard-to-estimate systematic errors. Indeed, at small and large $x$ where the data are sparse, the uncertainty of the PDF can become extremely large. It is challenging for IP methods to provide a reliable small-$x$ PDF in the absence of data constraints from high-energy experiments.

\begin{figure}
    \centering
     \subfigure[~Reconstruction of the correlation function and its tail in $\lambda$-space.]{\includegraphics[width=0.9\linewidth]{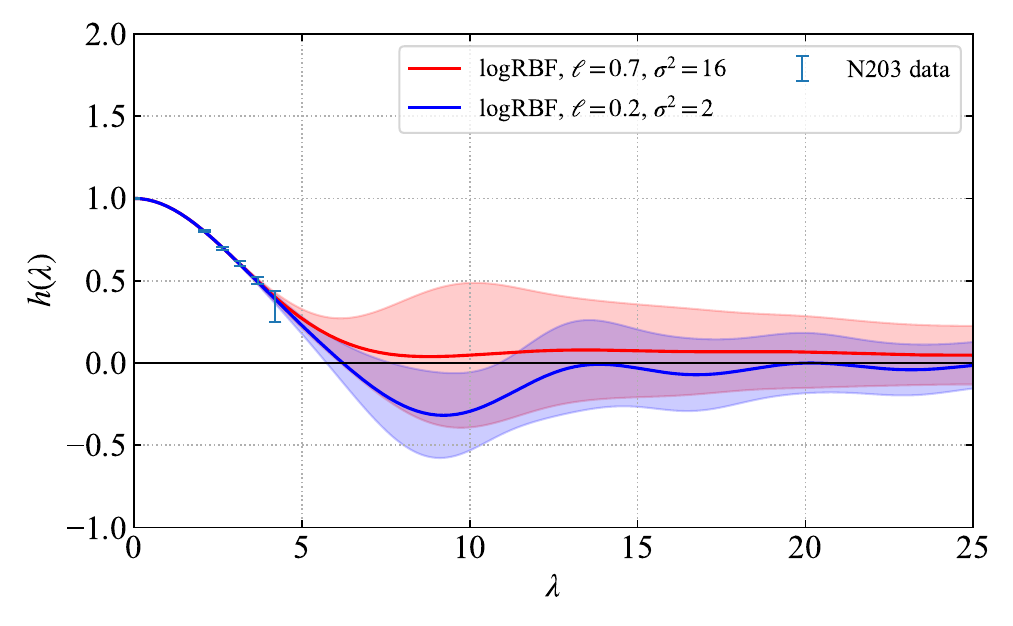}\label{fig:fx_conn_a}}
     \subfigure[~Reconstruction of the transversity PDF in $x$-space.]{\includegraphics[width=0.9\linewidth]{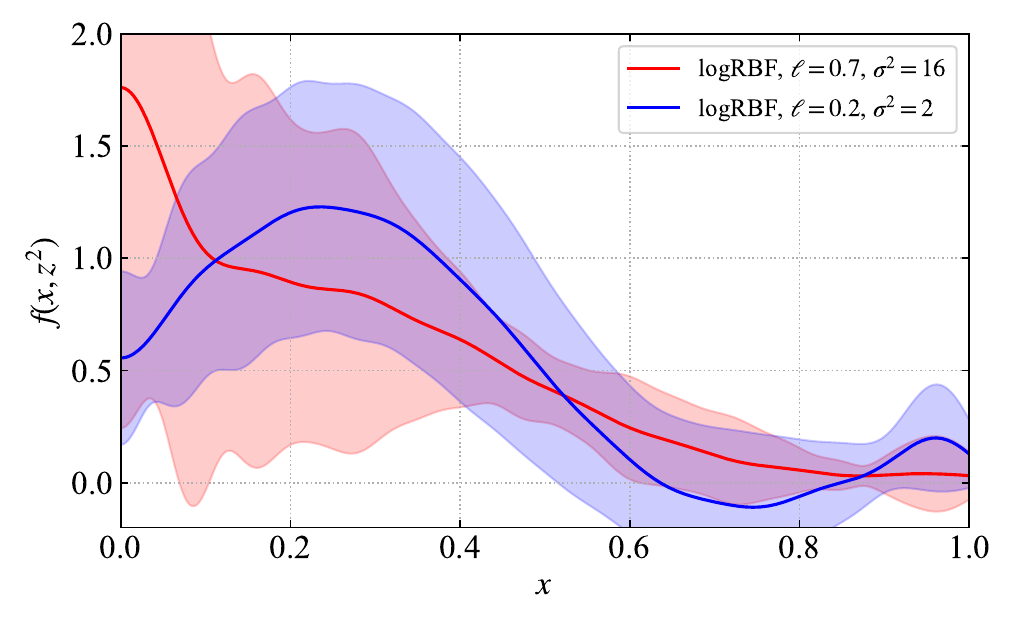}\label{fig:fx_conn_b}}
   \caption{An example of an IP in the SDF framework: reconstruction of the PDF from short-distance correlation functions at $z=0.26$ fm with proton momenta $P^z = 1.6,2.0,2.4,2.8,3.2$ GeV on the N203 ensemble, as reported in Ref.~\cite{LatticeParton:2022xsd}.
   The reconstruction is performed using the GPR method with the logRBF kernel proposed in Ref.~\cite{Dutrieux:2024rem}. 
   To illustrate the model uncertainty arising from different hyperparameter choices, two sets of priors are compared: the red band corresponds to $l = 0.7 \approx \ln (2)$, $\sigma^2=16$ and prior mean $f(x) = 4$; the blue band corresponds to $l = 0.2 \approx \ln(1.25) $, $\sigma^2=2$ and prior mean $f(x) = 1$. 
   }
    \label{fig:fx_con}
\end{figure}

The lattice QCD data analysis in the SDF framework is of a similar nature where one attempts to reconstruct a continuous PDF from a limited segment of leading-twist correlation. The model uncertainty in fitting the PDFs is difficult to quantify and has been shown to be significantly underestimated when using rigid parameterizations of their functional form---for example, see Refs.~\cite{Khan:2022vot,Chowdhury:2024ymm}. A particular IP method within the SDF framework was introduced in Ref.~\cite{Dutrieux:2024rem}, using the Gaussian Process Regression (GPR) with a logRBF kernel, defined as
\begin{align}
    K\left(x, x^{\prime}\right)=\sigma^2 \exp \left(-\frac{\left(\ln (x)-\ln \left(x^{\prime}\right)\right)^2}{2 \ell^2}\right) ~,
\end{align}
where the endpoint behaviors at $x = 0$ and $x = 1$ are not constrained by any fixed power-law exponents, as advocated in Ref.~\cite{Dutrieux:2024rem}. While the method is purported to be guided by physical motivations, there remains ambiguity not only in the choice of the kernel form and its hyperparameters but also in the motivations themselves. For example, the decision to fix the prior mean to ensure $100\%$ uncertainty at $x=0$, or to set the threshold of uncertainty in the extrapolation region to be $1.3$ times the error of $h(\lambda_{\rm max},P)$ or $0.3$ times its central value, reflects arbitrary prescriptions rather than principles derived from physical theory. In addition, $l=\ln(2)$ is not the only option to maintain a reasonable flexibility of $x$-dependence, a motivation for the GPR method. To illustrate the model uncertainty behind, an example in the SDF framework is presented in \fig{fx_con}. The transversity PDF is reconstructed from correlation functions at $z=0.26$~fm with proton momenta $P^z = 1.6, 2.0, 2.4, 2.8, 3.2$ GeV on the N203 ensemble, as reported in Ref.~\cite{LatticeParton:2022xsd}. Two sets of hyperparameters are employed for comparison, both of which preserve the smoothness and decorrelation between small and large $x$. As shown in \fig{fx_con}, these two settings lead to distinct posterior reconstructions, revealing a strong sensitivity to the choice of hyperparameters in the GPR method.

With limited lattice data, IP methods that rely primarily on mathematical constraints and minimal physics input often lead to uncontrolled and unrealistic uncertainties. Unlike asymptotic fits, which use rigorous, physically motivated functional forms with a small number of parameters, the IP method presented in Ref.~\cite{Dutrieux:2024rem} allows a broad space of solutions constrained only by abstract assumptions. While such flexibility is valuable for general problems, it limits reliability in extrapolating lattice data to the asymptotic region, where physical insight is essential. Until all the physical constraints in LaMET are fully integrated into IP frameworks, asymptotic analysis remains the most reliable approach for extrapolation.

\subsection{Asymptotic Extrapolation in LaMET}
\label{sec:phy-con} 

The asymptotic extrapolation in LaMET begins with the physics constraint that spatial correlators in the hadrons decay exponentially, independent of hadron momentum. The asymptotic exponential behavior can be argued on many grounds, as well as derived from a dispersion-relation analysis, such as the one in the appendix of Ref.~\cite{Gao:2021dbh}. One can use the same strategy for the spectral analysis of two-point functions to see when the correlation enters the exponential decay region. While not all lattice calculations have passed this test successfully, certain data sets clearly do.

\begin{figure}
    \centering
    \subfigure[~Bare matrix elements of the nucleon transversity quasi-PDF at zero momentum from four different lattice ensembles H102, N203, N302, and X650~\cite{LatticeParton:2022xsd}.]{\includegraphics[width=0.9\linewidth]{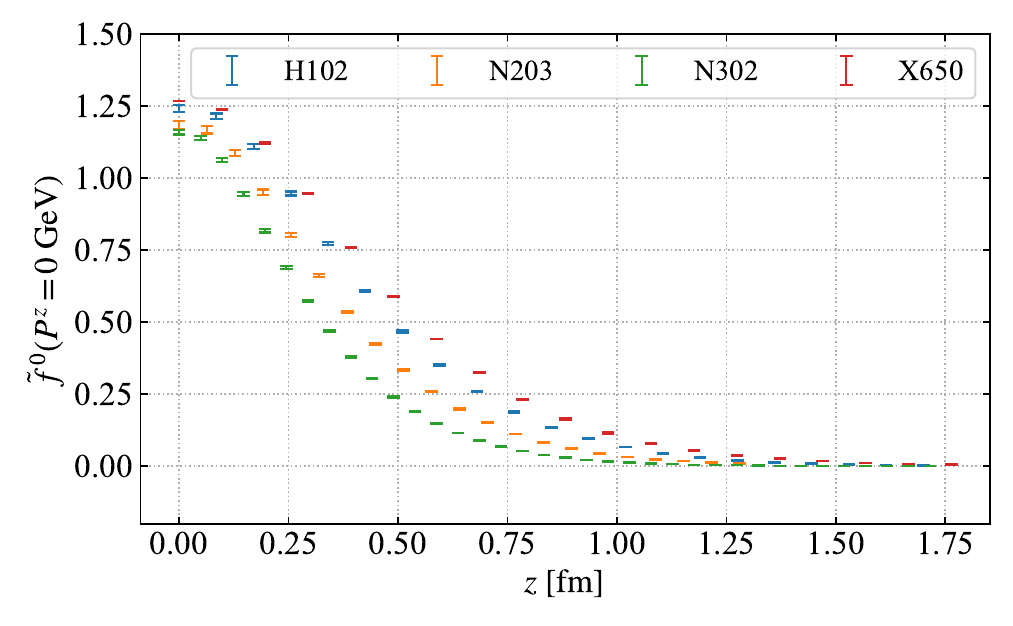}\label{fig:lpc_me_a}}
    \subfigure[~Effective mass ($m_{\mathrm{eff}} (z) = \ln (\tilde{f}^0 (z) / \tilde{f}^0 (z+a))$) extracted from the bare matrix elements, with colored bands indicating the outcomes of constant fits.]{\includegraphics[width=0.9\linewidth]{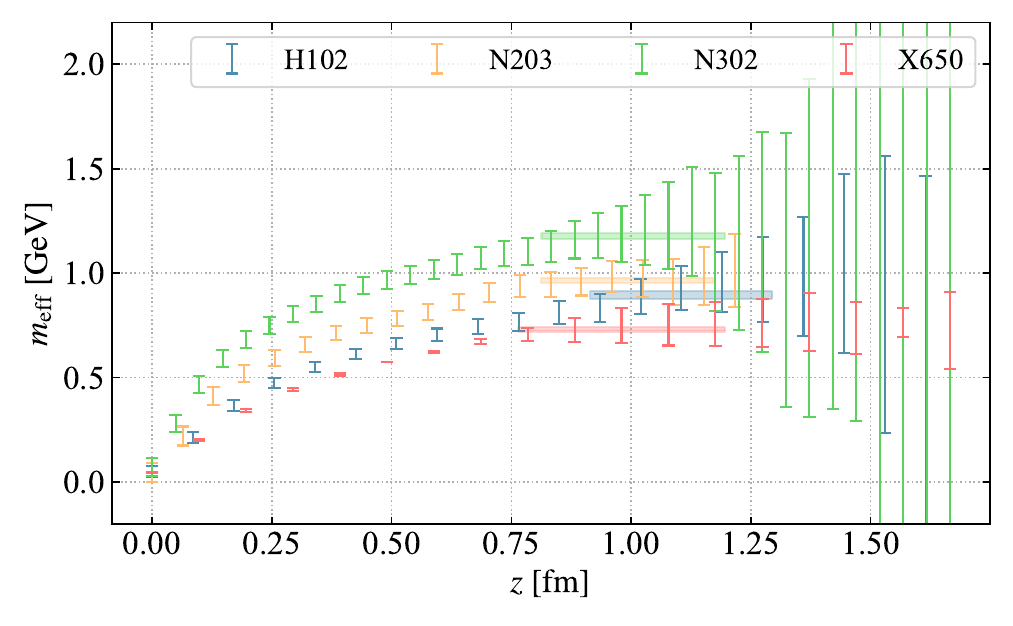}\label{fig:lpc_me_b}}
    \caption{An example of high-precision data in which we see clearly asymptotic exponential decay around $z=1$ fm.}
    \label{fig:lpc_me}
\end{figure}

An example is the zero-momentum lattice correlation function for the nucleon transversity quasi-PDF shown in \fig{lpc_me} from Ref.~\cite{LatticeParton:2022xsd}\footnote{Since Ref.~\cite{LatticeParton:2022xsd} is a state-of-the-art LaMET analysis for nucleon PDFs, we will use it as our example throughout this paper.}, where the exponential decay is clearly visible in the data. We use the standard ground-state fitting to find the decay constant $m_{\rm eff}$ in \fig{lpc_me_b}. The plateau or asymptotic region is reached after $z\sim 0.75$~fm for all lattice ensembles, where the data still have relatively high precision. The mass obtained here corresponds to a ``heavy-light'' meson, which has a linear divergence from the ``heavy quark''~\cite{Gao:2021dbh}. The linear divergence can be renormalized through a particular mass subtraction scheme (a constant shift in \fig{lpc_me_b}), and the residual mass in the renormalized correlator can be chosen to be on the order of $\Lambda_{\rm QCD}$, say, $0.2-0.3$ GeV, which depends on the renormalon regularization scheme in the matching coefficient~\cite{Zhang:2023bxs}. Since the light-cone PDF is renormalon free, this scheme dependence is canceled by the corresponding matching coefficient. Therefore, one can choose a proper scheme to shift the Wilson-line mass subtraction within $O(\Lambda_{\rm QCD})$ while maintaining the perturbative convergence and numerical stability of matching, so that the renormalized correlation function decays faster. This does not change the final uncertainty in the PDF but makes the FT easier to control. Note that this physical mechanism contradicts the interpretation in Ref.~\cite{Dutrieux:2025jed}, which attributes the exponential decay to the pion mass.

Since the leading exponential decay arises from gluon contributions that are independent of the hadron momentum, we expect the same matrix element to exhibit identical exponential decay behavior in a hadron state with large $P^z$. Therefore, the decay constant fitted from the zero-momentum matrix element can be used to constrain the exponential decay in large-$P^z$ matrix elements. 

The extrapolation can be made more accurate by incorporating the asymptotic behavior of the correlation function, which can be derived from the dispersion relation~\cite{Gao:2021dbh}. For example, at large $z$ the pion matrix element behaves as
\begin{align}\label{eq:asym}
    \lim_{z\to\infty}h(z, P^z) &= A(P^z) {e^{-m_{\rm eff} z}\over z^d} e^{-iP^zz}\,,
\end{align}
where $A(P^z)$ denotes the product of pion-to-heavy-light meson transition form factors, which decrease with increasing $P^z$, reflecting a Lorentz contraction effect. Therefore, at nonzero $P^z$, the phase factor $e^{-iP^z z}$ predicts that the matrix element will change sign in the asymptotic region, a feature observed in many calculations, such as Refs.~\cite{Dutrieux:2025jed,LatticeParton:2022xsd}, which can be incorporated into extrapolation models. 

While asymptotic extrapolation might be perceived as a model, it is in fact grounded in a rigorous derivation from the dispersion relation, effectively functioning as an EFT expansion. In the asymptotic limit, one typically begins with the leading behavior characterized by a finite number of parameters, and can systematically improve accuracy by including subleading terms. A more accurate and general asymptotic form for extrapolation will be provided in~\cite{Ji:2026vir}. 

Of course, understanding the long tail of correlation functions—such as excited-state contamination—remains a challenging problem in many scenarios. For example, in the hadronic vacuum polarization (HVP) contribution to the muon’s anomalous magnetic moment, control over the tail is achieved through spectral decomposition~\cite{DellaMorte:2017dyu}; by exploiting the positivity of the correlator to impose rigorous upper and lower bounds~\cite{Borsanyi:2016lpl,Budapest-Marseille-Wuppertal:2017okr,RBC:2018dos,Gerardin:2019rua,Djukanovic:2024cmq}; or by using Bayesian model averaging to systematically combine multiple fit models and account for model-selection uncertainty~\cite{Jay:2020jkz,MILC:2024ryz,FermilabLatticeHPQCD:2024ppc}. The asymptotic extrapolation in LaMET faces similar issues and could benefit from techniques developed in such settings with EFT constraints. However, unlike the HVP case---which involves integrating the correlation function with a kernel that enhances the long-range contribution---the FT kernel $\exp(ixP^z z)$ in LaMET naturally suppresses this contribution at finite $x$. Even if uncertainty remains in reconstructing the long tail, the FT kernel still ensures an upper bound on it, provided that the exponential decay is encoded. We demonstrate this in the example in \sec{ft-error} below.

If the lattice data within the sub-asymptotic region $[z_{SA}, z_A]$ (for the nucleon, $z_{SA} \sim 0.5$~fm and $z_A \sim 1.2$~fm) have sufficiently small errors, then the asymptotic extrapolation can be tightly constrained. This level of precision has been demonstrated for the renormalized transversity quasi-PDF matrix elements in Ref.~\cite{LatticeParton:2022xsd}, as shown in \fig{lpc_a}. A reliable extrapolation can be obtained by fitting the asymptotic form to data within a chosen window inside $[z_{SA}, z_A]$. The truncation point $z_L$---beyond which data are replaced by the extrapolated form with its associated error band---may lie either below or above $z_A$, depending on the data precision.

In practical analyses, the asymptotic extrapolations should satisfy the following physical conditions. 
\begin{enumerate}
    \item The lattice data should be truncated at a point within the sub-asymptotic and asymptotic regions where they have decreased smoothly to near zero.
    \item The extrapolation model must preserve the correct asymptotic behavior of the correlation function without introducing unphysical distortions.
    \item The extrapolated result---including its error---should continue to decrease with increasing $z$, consistent with the asymptotic decay. In cases where oscillations or sign changes are present, the amplitude of oscillation should decay exponentially.
\end{enumerate}
Since the exact asymptotic behavior of a correlation function depends on the operator and external state, the parameters used in the extrapolation must be determined on a case-by-case basis. After performing the extrapolation, it remains necessary to estimate the model uncertainty in the FT. While most LaMET analyses have imposed the physical conditions for extrapolation, not all have provided an estimate of the associated uncertainty, as criticized by Ref.~\cite{Dutrieux:2025jed}. In Ref.~\cite{Gao:2021dbh}, such an estimate was obtained by varying the truncation point~\footnote{In Ref.~\cite{Dutrieux:2025axb}, it was pointed out that the pion data points beyond $z_L \sim 0.9$ fm should not be discarded, as they are statistically nonzero. However, since their magnitude continues to grow to unphysically large values, they were ruled out in Ref.~\cite{Gao:2021dbh} based on the pion's size, such as its charge radius of about $0.6$ fm~\cite{Gao:2021xsm}. After all, this is a special case of LaMET analyses. In the future, such a restriction will be relaxed with more dedicated lattice calculations of the long-range correlations.}, the prior for the decay constant, and the extrapolation model. Notably, unphysical power-law decay models were also included to emphasize the differences. It was found that the variations among extrapolations are negligible at moderate $x$, so they are not included in the final error budget.

Nevertheless, when the precision of data in the sub-asymptotic region is not sufficiently high, the FT uncertainty can become non-negligible. In such cases, the method in Ref.~\cite{Gao:2021dbh} may be less reliable, as it does not exhaust the model space. For a more accurate quantification of this error, one can first determine $m_{\rm eff}$, $z_{SA}$ and $z_A$ from high-precision zero-momentum matrix elements and then use them as priors at nonzero momenta. Subsequently, techniques analogous to those used in HVP calculations can be applied to constrain the long-distance tail and yield a more reliable uncertainty estimate~\cite{Ji:2026vir}. Since the asymptotic decay arises from an EFT expansion in which subleading terms fall off more rapidly than the leading contribution, this uncertainty is controllable. A more accurate derivation of the asymptotic form and its corrections~\cite{Ji:2026vir} will refine this estimate.

\section{Comments on Data and Analysis in arXiv:2504.17706}
\label{sec:response}

The claim about the IP in LaMET in Ref.~\cite{Dutrieux:2025jed} focuses on the practical concern that noisy lattice data in the sub-asymptotic region may prevent a reliable asymptotic extrapolation and, therefore, should be analyzed by IP methods with minimal physics constraints. The authors also provided numerical evidence through tests on a set of nucleon matrix elements, demonstrating that variations among different IP methods with exponential decay still result in considerable differences from moderate to large $x$. This is consistent with our expectation that IP methods with minimal physics input will lead to unnecessarily conservative errors. 

However, we still expect that the asymptotic extrapolation in LaMET is a better option for providing a reliable error estimate even when the data quality is not ideal. Besides, the arguments and data analysis presented in Ref.~\cite{Dutrieux:2025jed} are problematic. First, data precision in the asymptotic region can be improved with realistic resources and new lattice techniques~\cite{Zhang:2025hyo,Wagman:2024rid,Hackett:2024nbe,Gao:2023lny,Zhao:2023ptv}, which the authors described as exponentially hard. In addition, the numerical tests in Ref.~\cite{Dutrieux:2025jed} suffer from several problems, including large errors in the sub-asymptotic region, an incorrect renormalization scheme, and analysis methods that do not satisfy physical constraints. Ultimately, the uncertainty of the FT within the models remains bounded by our estimate from asymptotic extrapolation, which can serve as a realistic quantification of the error given the data quality. Therefore, FT in fact remains under control in the standard LaMET analysis, contrary to the conclusion drawn in~\cite{Dutrieux:2025jed}.

\subsection{Good Precision Achieved in the Asymptotic Region}
\label{sec:exp-decay}

One major argument of Ref.~\cite{Dutrieux:2025jed} is that it is unlikely for lattice data to exhibit the asymptotic exponential decay behavior before the error becomes too big, due to the exponential loss in the signal-to-noise ratio (SNR). This concern is misplaced for two reasons. First, identifying exponential decay in the data does not necessarily require an unrealistically high SNR. It is not an exponentially hard problem to reduce the errors of the matrix elements at $z\lesssim 1.0$ fm, which can be achieved with higher statistics and new lattice techniques such as the kinematically enhanced hadron interpolation operators~\cite{Zhang:2025hyo}. Second, the extrapolation error is not governed by the SNR itself, but by the absolute sizes of the signal and noise, which set the upper bound on the uncertainty~\cite{Gao:2021dbh}, as explained in \sec{ft-error} below. In other words, even when the data quality is not ideal,
the LaMET analysis with asymptotic extrapolation can still provide reliable error estimates. 

In the past, high-level precision was not always achieved in LaMET calculations, primarily because the lattice setups were not designed to prioritize higher statistics for large-$z$ matrix elements.
However, some calculations have already produced sufficiently precise data in the asymptotic region, such as nucleon transversity quasi-PDF~\cite{LatticeParton:2022xsd}, the nucleon unpolarized quasi-GPD~\cite{Holligan:2023jqh}, the pion and kaon quasi distribution amplitudes~\cite{LatticeParton:2022zqc}, and the pion quasi-TMD wave function~\cite{Avkhadiev:2024mgd}.
In the future, lattice groups can tailor their data production strategies to better meet the requirements of LaMET, such as in the case of spectroscopy calculations.

\subsection{Data Quality and Renormalization Problems}
\label{sec:ill-method}

Before commenting on the data quality, we would like to point out that using the ratio scheme~\cite{Orginos:2017kos} at large $z$ to obtain the quasi-PDFs is incorrect. First, it introduces uncontrolled non-perturbative effects at intermediate and large $z$, which subsequently contaminate all $x$ after the FT. Second, there is no QCD factorization for a ratio-scheme quasi-PDF in $x$-space despite the studies in early literature~\cite{Radyushkin:2017lvu,Zhang:2018ggy,Izubuchi:2018srq}. In fact, $x$-space factorization has been rigorously established for the $\overline{\mathrm{MS}}$ scheme only, implying that any scheme not perturbatively convertible to it at all $z$---such as the ratio scheme---is not factorizable.

Besides, since Ref.~\cite{Dutrieux:2025jed} aims to demonstrate a problem in LaMET analysis, it should at least adhere to the same procedure when making this claim. As discussed in \sec{asymptotic}, we expect the non-zero momentum matrix element to decay exponentially at the same rate as that in the rest frame. Taking a ratio of matrix elements at different momenta removes this key feature entirely, so no definitive conclusion can be drawn from such an example. Hence, the ratio scheme is logically inconsistent for the purpose of Ref.~\cite{Dutrieux:2025jed}. In order to proceed, we will treat the ratio-scheme matrix elements as ``pseudo data'' for LaMET analysis.

As shown in Figure 3 of Ref.~\cite{Dutrieux:2025jed}, the ratio of matrix elements changes sign and is consistent with zero with increasingly large error bars in the range $8<\lambda<11$ (or $0.75~\text{fm}<z<1.03~\text{fm}$). 
This ratio after the first zero is small in amplitude due to the Lorentz contraction effect in \eq{asym}. Most of the conclusions in Ref.~\cite{Dutrieux:2025jed} were drawn by analyzing this imprecise ratio, which should be insignificant to the FT when exponential decay is preserved with proper renormalization such as the hybrid scheme~\cite{Ji:2020brr,Holligan:2023rex,Zhang:2023bxs}. However, the sizable errors of data can lead to significant variation in the fits and extrapolations, contributing to the uncertainty of the FT.

This situation, if interpreted as pseudo data within the LaMET framework, can be significantly improved by reducing the errors in the region $0.75~\text{fm} < z <1.03~\text{fm}$ by a factor of two or more, allowing the truncation point to be set at $z_L \sim 1.03$~fm to better satisfy the physical conditions for asymptotic extrapolation. This precision is achievable with a four-fold increase in statistics or by using the kinematically enhanced interpolation operator~\cite{Zhang:2025hyo}, both of which are feasible nowadays.

\subsection{Data-Driven IP Methods May Conflict with Physical Constraints}
\label{sec:dd}

When data are consistent with zero but have large errors, asymptotic analysis remains a valid method to estimate errors, as discussed in \sec{asymptotic}. Nevertheless, the large uncertainties may prompt some to treat this as an IP and resort to data-driven approaches. However, two fundamental issues arise: first, data-driven fitting alone cannot distinguish whether a model is physical, since many models can fit the data equally well; second, without proper constraints, these approaches risk producing unphysical results in the extrapolation region.

If one treats the ratio-scheme matrix elements as pseudo data for LaMET, then the first issue is exemplified by the Gaussian-decay model used in Ref.~\cite{Dutrieux:2025jed}. Although it is known to contradict the exponential decay of nucleon correlation functions at long range~\footnote{In fact, the exponential decay cancels out in the ratio scheme, but it is even more difficult to justify why the Gaussian-decay would survive.}, the punishment here is not significant. 

The second issue is manifested by the comparison between extrapolations using the exact exponential model (Exact exp) and two Radial-Basis Function (RBF) models with exponential decay (RBF-exp and logRBF-exp) under the Gaussian process regression (GPR) framework. The RBF kernel is defined as~\cite{Dutrieux:2024rem}
\begin{align}
    K\left(x, x^{\prime}\right)=\sigma^2 \exp \left(-\frac{\left( x -x^{\prime}\right)^2}{2 \ell^2}\right)\,.
\end{align}
Unlike the Exact exp approach, which follows the LaMET procedure of truncation and extrapolation, the GPR methods infer posterior distributions based on all data from $z = 0$ to $1.13$~fm, despite the fact that exponential decay does not apply at smaller $z$. Moreover, despite being motivated by somewhat vague arguments listed in Ref.~\cite{Dutrieux:2024rem}, the choice of RBF kernels and their associated hyperparameters in the GPR framework remains \textit{ad hoc} and subject to model dependence~\cite{Dutrieux:2025jed,Karpie:2019eiq,Dutrieux:2024rem}, and there is no clear mechanism to ensure that they do not distort the physical conditions in the process. As we show below, this can result in unphysical extrapolations---even with high-precision data. Finally, the indiscriminate use of all noisy data also unnecessarily amplifies the fitting errors.

Therefore, the GPR methods in Ref.~\cite{Dutrieux:2025jed} do not qualify as asymptotic extrapolations, and it is difficult to assess their systematic uncertainties. From a theoretical standpoint, comparing them to an asymptotic extrapolation is an apples-to-oranges mismatch.

After all, to further analyze the issue with the GPR fits using RBF-based kernels, we will temporarily treat them as comparable to the asymptotic extrapolation. As it turns out, none of them satisfies all the physical conditions described in \sec{phy-con}. As shown in Figures 3 and 4 in Ref.~\cite{Dutrieux:2025jed}, the RBF-exp and logRBF-exp fits match the data up to $\lambda \sim 8$ (or $z \sim 0.75$~fm), beyond which they begin to deviate, giving the appearance of extrapolations starting at $\lambda_L \sim 8$. However, the matrix elements and their errors in the extrapolation region continue to grow in magnitude before slowly decaying as $\lambda \to \infty$, clearly violating the third physical condition. In particular, even the Exact exp extrapolation slightly violates the first physical condition, since it truncates at $z \sim 0.75$~fm, which is early for the amplitude of $h(z, P)$---including its slope---to decay sufficiently close to zero.

Due to the violations of physical conditions for extrapolation, the FTs are not reliable, so the resulting discrepancies among different quasi distributions in Figure 3 of Ref.~\cite{Dutrieux:2025jed} are in principle overstated. Nevertheless, given the data quality, these violations are acceptable mainly because the error bars of the GPR fits are also large.

\begin{figure}[htbp]
    \centering
    \subfigure[{~Nucleon transversity quasi-PDF matrix elements in the hybrid scheme at $P^z=1.82$ GeV from a lattice of spacing $a=0.085$ fm~\cite{LatticeParton:2022xsd}. The Exact exp, RBF-exp, and logRBF-exp models with different correlation lengths $r$~\cite{Dutrieux:2025jed} are used to extrapolate from $z_L=14a=1.19$ fm. The red data points are fitted exclusively by the Exact exp model.}]
    {\includegraphics[width=0.9\linewidth]{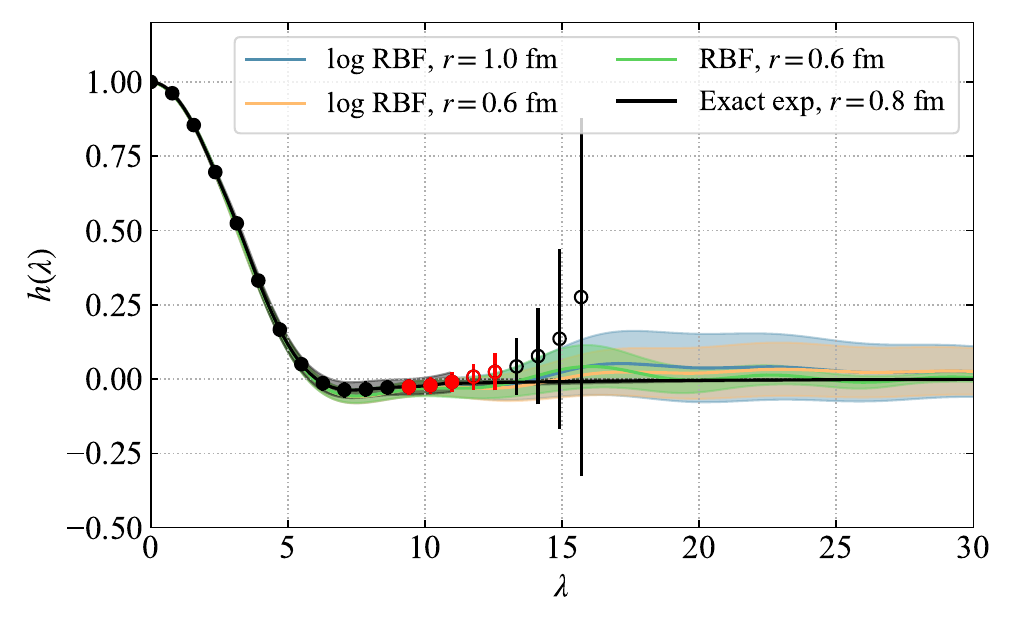}\label{fig:lpc_a}
    }
    \subfigure[{~Quasi-PDFs from different extrapolation models.}]
    {\includegraphics[width=0.9\linewidth]{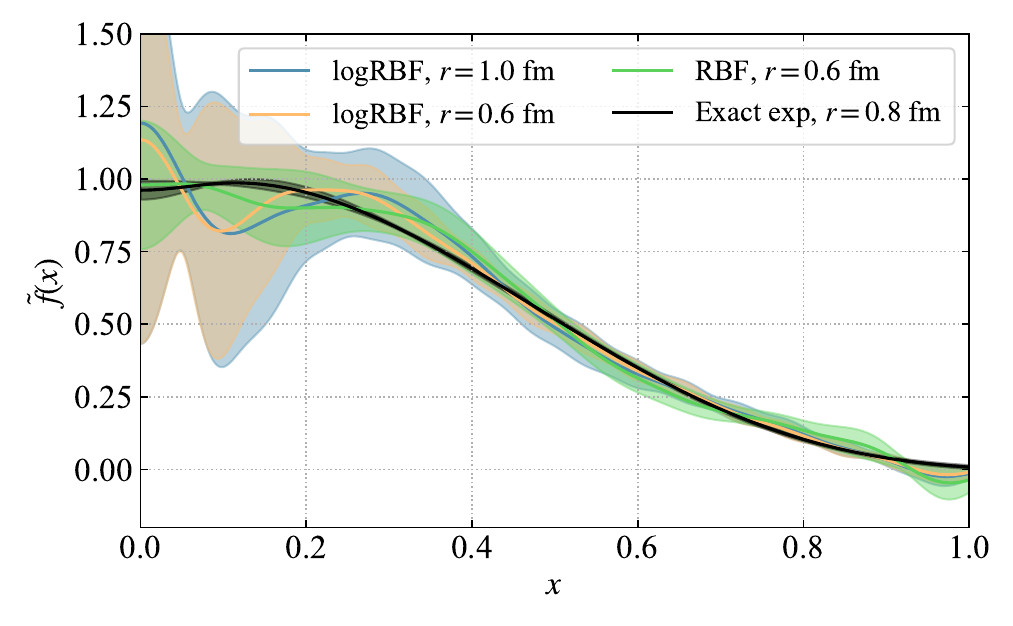}\label{fig:lpc_b}
    }
    \subfigure[{~Variation of different FTs around the mean of Exact exp. At moderate $x$ their central values are very close, despite the larger uncertainties of the RBF extrapolations due to violation of physical conditions.}]
    {\includegraphics[width=0.9\linewidth]{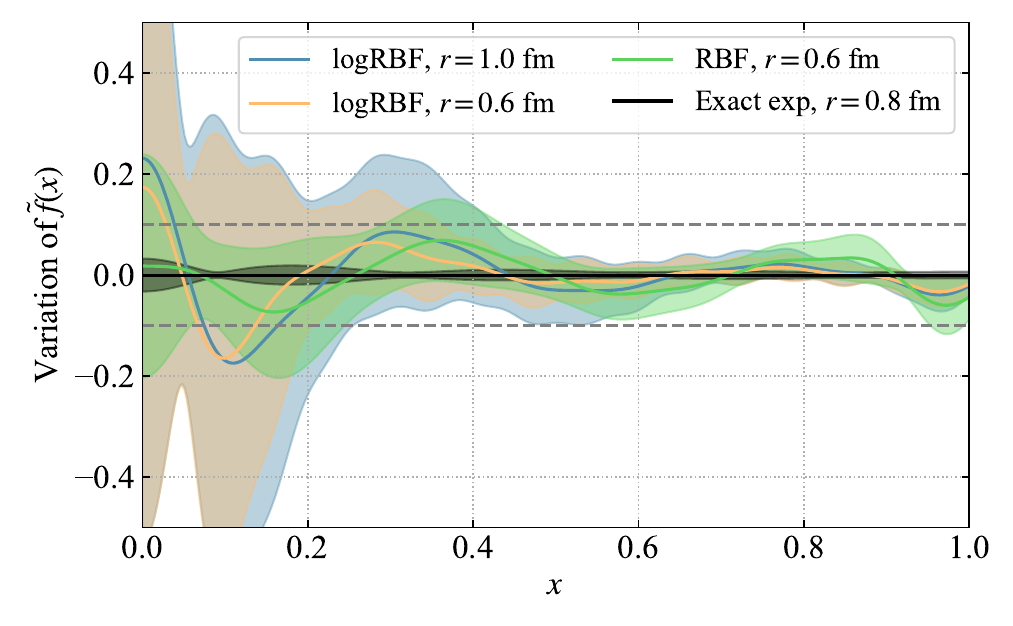}\label{fig:lpc_c}
    }
    \caption{FT tests with the nucleon transversity quasi-PDF matrix elements from Ref.~\cite{LatticeParton:2022xsd}.}
    \label{fig:lpc}
\end{figure}

For the purpose of comparison, we conducted analogous tests employing identical hyperparameters~\footnote{Unlike Ref.~\cite{Dutrieux:2025jed}, we do not impose a threshold on the uncertainty of the reconstructed tail when determining the hyperparameters, because such a choice is arbitrary. Our purpose is to show that GPR without physical constraints could lead to unphysical extrapolations. While adopting a threshold might help the GPR reconstruction satisfy the conditions in \sec{phy-con}, its physical basis is unclear, so the associated model uncertainty cannot be reliably quantified.}, as indicated in Reference 52 of Ref.~\cite{Dutrieux:2025jed}, using the hybrid-scheme nucleon transversity quasi-PDF matrix elements from Ref.~\cite{LatticeParton:2022xsd}. As shown in \fig{lpc_a}, these matrix elements are much more precise than those in Ref.~\cite{Dutrieux:2025jed} in the region $8<\lambda< 11$ (or $0.85~\text{fm}<z<1.19~\text{fm}$), which allows for truncation at $z_L=1.19$~fm for asymptotic extrapolation, thus significantly reducing the uncertainty of the FT. However, despite using the same range of precise data in $z \le z_L$, the GPR methods produce posteriors whose amplitudes and errors far exceed those of all data in the range $6 < \lambda < 11$, clearly violating the third physical condition. This suggests that the GPR methods used in Ref.~\cite{Dutrieux:2025jed} did not incorporate the exponential decay behavior in their prior settings, particularly when improper correlation lengths are chosen, causing the long-range behavior to be overly influenced by the smaller-$z$ region.

Therefore, the best approach to extrapolating the pseudo data in Ref.~\cite{Dutrieux:2025jed} is through the asymptotic analysis. Its uncertainty can be better controlled by using the decay constant $m_{\rm eff}$---fitted from the zero-momentum matrix element---as input, truncating at a larger $z_L$, and adopting a more accurate asymptotic form for both extrapolation and error estimation.

\subsection{Asymptotic Analysis Provides Realistic Error Estimate Regardless of Data Quality}
\label{sec:ft-error}

The lattice correlation functions include statistical errors that are correlated and propagate in the FT. Since FT is a linear operation, there are mature statistical methods to handle the error propagation~\cite{Young:2012kg}. In addition to statistical errors, the FT also includes model uncertainty arising from the extrapolation. Another main point advocated in Ref.~\cite{Dutrieux:2025jed} is that when the uncertainty from extrapolation is non-negligible, it cannot be reliably quantified. We disagree. As we show here, the uncertainty can be estimated within asymptotic extrapolation and does not constitute the IP encountered in SDF.

As explained above, a non-negligible model uncertainty is the outcome of insufficient data quality in the sub-asymptotic region, which can be reduced by improving the data precision. To quantify this uncertainty, it suffices to show that it is both bounded from above and systematically reducible. The upper bound can be estimated by evaluating the contribution of the extrapolated region to the Fourier integral. One bound has been derived in the appendix of Ref.~\cite{Gao:2021dbh} as
\begin{align}\label{eq:bd}
    \delta f(x, P, \lambda_L) < \frac{4N_x |h(z,P;\lambda_L)|_{\rm max}}{\pi x}\,,
\end{align}
where $|h(z,P;\lambda_L)|_{\rm max}$ is the maximal value of $|h(z,P)|$ for $\lambda_L \le \lambda < \infty$~\footnote{We have slightly adapted the derivation in Ref.~\cite{Gao:2021dbh}, which assumed that $|h(\lambda, P)|$ decays monotonically for $\lambda > \lambda_L$. In our version, the derivation also allows for oscillations in this region.}, and $N_x$ is an integer which satisfies $|h'(\lambda_L+2\pi N_x/x)|\ll |x|$ so that the contribution to the Fourier integral from $\lambda > \lambda_L + 2\pi N_x/x$ is negligible.

The upper bound in \eq{bd} applies to both exponential and algebraic decays, typically requiring $N_x \sim \mathcal{O}(1)$ in the former and $N_x \gg 1$ in the latter to saturate the Fourier integral. This contrast is expected: for the same $\lambda_L$, the extrapolated region contributes much less to the Fourier integral in the case of exponential decay, leading to a correspondingly smaller model uncertainty. For exponential decays, this upper bound is usually comparable to or smaller than the other uncertainties in the LaMET analyses,  making it a reasonable estimate of the FT model uncertainty. Moreover, it excessively estimates the true uncertainty, as it assumes that $|h(z, P)|$ can vary arbitrarily between 0 and its maximum value $|h(z, P; \lambda_L)|_{\rm max}$ at each $z$. With a more accurate derivation of the asymptotic behavior of spatial correlations~\cite{Ji:2026vir}, a significantly tighter bound on the model uncertainty can be established for the error budget.

Since the RBF extrapolations in Ref.~\cite{Dutrieux:2025jed} include an exponential-decay component, we assume that they satisfy the conditions used to derive \eq{bd}. Therefore, we may apply the same upper bound to the RBF extrapolations as to the Exact exp extrapolation. The FT model uncertainty can be estimated by the variation of central values across different models, which should persist even when the statistical errors approach zero. Because the extrapolations using the RBF kernel do not satisfy the third physical condition in \sec{phy-con}, we set $|h(z,P;\lambda_L)|_{\rm max} = 0.1$, reflecting the amplitude of the matrix elements in the range $0.75\,\text{fm} < z < 1.03\,\text{fm}$. At $x = 0.5$, taking $N_x = 1$ is sufficient to saturate the Fourier integral, yielding an upper bound of $\sim 0.25$ on the model uncertainty. Meanwhile, according to Figure 3 of Ref.~\cite{Dutrieux:2025jed}, the quasi-PDFs obtained from different models are consistent within errors at finite $x$, despite that their central values differ by an amount that generally decreases with increasing $x$, which is predicted by the upper bound in \eq{bd}. See Figure 5 of Ref.~\cite{Dutrieux:2025jed}. At $x = 0.5$, the spread among central values is around 0.2, implying an uncertainty of about 0.1, which is still bounded by the estimate from \eq{bd}.

Moreover, with the precise data in \fig{lpc_a}, the quasi-PDFs from different extrapolations show significantly better agreement, as illustrated in \figs{lpc_b}{lpc_c}. The variation around the mean of the Exact exp model is less than 0.05 at $x = 0.5$, which is again bounded by the estimated error of $\sim 0.06$, obtained using $|h(z,P;\lambda_L)|_{\rm max} = 0.025$ and $N_x = 1$. Notably, the uncertainties of the GPR fits are much larger, reflecting the uncontrolled model dependence of the GPR method, as the latter can violate the physical conditions for asymptotic extrapolation. See the discussion in \sec{dd}.

Therefore, the FTs in both examples behave just as theoretically expected, with model uncertainties at finite $x$ well bounded by the estimates derived from asymptotic extrapolation. Note that if an extrapolation satisfies all the physical conditions, then $|h(z,P;\lambda_L)|_{\rm max}$ should be suppressed by at least an additional factor of $\exp[-(m_{\rm eff}/P^z)\cdot 2\pi/x ]\sim 0.3$ compared to the estimates we provided above. This corresponds to an upper bound of $\sim 0.075$ for data in Ref.~\cite{Dutrieux:2025jed} and $\sim 0.02$ for the data in Ref.~\cite{LatticeParton:2022xsd}, which are small and can serve as a reasonable estimate of the FT model uncertainty. This estimate can be further tightened by using more accurate sub-asymptotic form and reaching larger $z_L$ to reduce both $|h(z,P;\lambda_L)|_{\rm max}$ and $N_x$, which in turn corresponds to better satisfying the physical conditions for extrapolation.

\subsection{Exponential Decay is the Key to Controlling FT Errors in LaMET}
\label{sec:misc}

Ref.~\cite{Dutrieux:2025jed} also made the claim that exponential decay plays a minimal role in controlling the FT at moderate to large $x$, so that the IP in LaMET and SDF is essentially the same. This claim is based on the observation that different $z$- and $x$-space reconstruction methods yield similar results for the quasi-PDF, regardless of whether exponential decay is imposed. 

However, the logic behind that conclusion is in fact reversed. It is the asymptotic decay of the pseudo data---not the extrapolation model---that ensures control of the FT. This is why similar results can still be obtained even when the decay behavior is not explicitly built into the model, exactly the first issue with the data-driven methods discussed in \sec{dd}. To match the exponentially decaying data, the extrapolation must also satisfy the corresponding physical behavior, otherwise the systematic errors cannot be reliably estimated.

In contrast, the matrix elements in SDF are limited to $\lambda \lesssim 4.0$---far from the asymptotic region---and decay much more slowly as a power law, which inevitably leads to significantly larger variations across different extrapolation models, as shown in \fig{fx_con}. Imposing any model, even those incorporating exponential decay, does not resolve the IP in this case.

\section{Conclusion}
\label{sec:end}

In this work, we have explained why the asymptotic analysis in LaMET enables a reliable extrapolation of the long-range lattice matrix elements and controlled FT at finite $x$. This ensures that LaMET can provide a point-by-point calculation---rather than fitting---of the $x$-dependence of parton distributions, which resolves the IP faced by the SDF due to limited range of accessible leading-twist correlations. In addition, we commented on the arguments and data analysis in Ref.~\cite{Dutrieux:2025jed}, and showed that the considerable FT model uncertainty is mainly caused by the insufficient data precision in the sub-asymptotic region, which is not due to the LaMET approach itself and can be systematically improved. Moreover, the observed discrepancies among different extrapolation models are still well bounded by the theoretical estimate from asymptotic extrapolations, thus showing that the FT is under control. This demonstrates that even when the data quality is not ideal, the asymptotic extrapolation in LaMET remains the most reliable approach for completing the FT and providing realistic error estimates. Meanwhile, re-framing the extrapolation as an IP and adopting a data-driven approach with minimal physical constraint is likely to produce unnecessarily conservative errors.\\

\begin{acknowledgments}
We thank Daniel Hackett, William Good, Yizhuang Liu, Swagato Mukherjee, Fei Yao, and Ismail Zahed for valuable discussions. We also thank Fei Yao for sharing the data of nucleon transversity quasi-PDF.
This material is based upon work supported by the U.S. Department of Energy, Office of Science, Office of Nuclear Physics through Contract No.~DE-SC0012704 and Contract No.~DE-AC02-06CH11357, the Scientific Discovery through Advanced Computing (SciDAC) award \textit{Fundamental Nuclear Physics at the Exascale and Beyond}, and the \textit{Quark-Gluon Tomography (QGT) Topical Collaboration} with Award DE-SC0023646, and the National Science and Technology Council, Taiwan, under Grants No. 113-2112-M-002-012. This work is supported in part by Natural Science Foundation of China under grant No. 12125503,  12335003, 12375069, 12205106. 
QAZ is also supported by the Fundamental Research Funds for the Central Universities. JH is supported by Guangdong Major Project of Basic and Applied Basic Research No. 2020B0301030008. YS is also supported by the 2024-2025 JSA/Jefferson Lab graduate fellowship.

\end{acknowledgments}

\appendix

\section{Response to arXiv:2506.24037}
\label{app:reply}

After the appearance of this paper on arXiv, the same authors of Ref.~\cite{Dutrieux:2025jed} published a comment in arXiv:2506.24037~\cite{Dutrieux:2025axb}, where they clarified that they ``\textit{did not conclude in Ref.~\cite{Dutrieux:2025jed} that it is not possible to quantify or control the uncertainties in the LaMET inverse problem}''. Taken literally, this clarification acknowledges that the FT uncertainty can indeed be controlled and quantified in LaMET. To us, this is the central issue that distinguishes LaMET from SDF, since the latter does not have control over the uncertainty in $x$-space with the same contemporary lattice setup. With this consensus, the quantification of the FT uncertainty in LaMET becomes a practical issue.

However, the authors of Ref.~\cite{Dutrieux:2025jed} adopted a definition of the IP broad enough to encompass LaMET as well as most standard lattice analyses, even though these are not commonly regarded as such by the lattice community. After all, the categorical statements made in Ref.~\cite{Dutrieux:2025axb} are that GPR is better than the asymptotic extrapolation in quantifying the FT uncertainty in LaMET, and that exponential decay has a negligible impact on the GPR reconstruction of the $x$-dependence. Moreover, the authors promoted GPR as a reliable approach for SDF as well, which underscores their statement in Ref.~\cite{Dutrieux:2025jed} that LaMET and SDF share a similar IP, contrary to the main point of this paper.

First of all, as explained in \sec{theory}, the uncertainty in SDF cannot be controlled due to the intrinsic limitation of effective data with the contemporary lattice setup. Even if the GPR reconstruction yields results consistent with the true value in LaMET, this does not imply that the method is similarly reliable in SDF.

Second, we clarify that our proposal is not merely to perform an extrapolation using a rigid model, as claimed in Ref.~\cite{Dutrieux:2025axb}, but to also provide an estimate of the associated uncertainty. While it is true that not all LaMET analyses in the literature have done so, in many cases this is because the uncertainty is either negligible or not dominant compared to other sources of systematics. Nevertheless, as we argued in \sec{phy-con}, a reliable estimate of the FT uncertainty can be obtained by incorporating knowledge of the asymptotic behavior of the correlation function. Key quantities such as the decay constant and the extent of the sub-asymptotic region can be extracted from high-precision zero-momentum matrix elements, which are frame-independent and can serve as priors at nonzero momentum. This approach removes the arbitrariness in implementing the first physical condition, which was criticized in Ref.~\cite{Dutrieux:2025axb}. Furthermore, since the correlation function admits a controlled asymptotic expansion with subleading terms that decay more rapidly, one can estimate the uncertainty from the missing contributions in the same spirit as in EFT calculations~\cite{Ji:2026vir}.

Third, the GPR reconstruction, at least in its current form in Refs.~\cite{Dutrieux:2024rem,Dutrieux:2025jed,Dutrieux:2025axb}, does not allow for a reliable quantification of the FT uncertainty in LaMET. Before elaborating on why, we note that a meaningful uncertainty quantification entails an upper bound that can be systematically improved. This upper bound can be rigorously derived from the asymptotic expansion of the correlation functions~\footnote{The conservative estimate in \eq{bd} is obtained by encoding only the leading exponential decay while leaving the remaining $z$-dependence unspecified, so $N_x$ remains indefinite, though it is expected to be close to 1. However, with more precise knowledge of the asymptotic behavior~\cite{Ji:2026vir}, this upper bound can be tightened without arbitrariness.}, which was acknowledged in Ref.~\cite{Dutrieux:2025axb} as the authors ``\textit{have never doubted that one can produce a reasonable estimate on the maximal uncertainty of the $x$-reconstruction}''. Meanwhile, GPR uses smoothness of the quasi-PDF as the sole physical regulator for solving the FT problem. Unlike the asymptotic decay of correlation functions, this notion of ``smoothness'' is vague and lacks a clear foundation in quantum field theory. In practice, it is primarily governed by the hyperparameters that must be chosen on a case-by-case basis~\cite{Dutrieux:2025axb}, yet the associated model uncertainty still needs to be estimated. Most importantly, smoothness alone does not enable a rigorous derivation of an upper bound on this uncertainty, so its reliable quantification cannot be made.

Finally, the authors continued to use reversed logic in Ref.~\cite{Dutrieux:2025axb} in discussing the role of exponential decay in the extrapolation. It is the exponential decay of the matrix elements themselves, not the reconstruction method, that ensures the FT uncertainty admits a systematically improvable upper bound and is therefore under control~\footnote{Note that exponential decay is a sufficient but not necessary property for the FT to be under control, as other rapidly decaying features would also suffice. However, in LaMET, it is the only relevant physical property of the correlation function.}. The small difference observed between GPR models with and without exponential decay is a consequence of this property, which relaxes the model space for extrapolation. A similar observation was also made in Ref.~\cite{Gao:2021dbh}, where both exponential and power-law decay models produced consistent results. However, when it comes to quantifying the FT uncertainty, a reliable estimate is only possible with knowledge of the asymptotic behavior of the correlation functions.

The authors of Ref.~\cite{Dutrieux:2025axb} also cited the deep-inelastic scattering structure function as a counterexample to our statement that the FT in LaMET is an FP. By expressing the PDF as a convolution of the matching coefficient with the structure function---i.e., as an FP---they argued that this contradicts the paradigm that PDF fitting is an IP. However, the key difference lies in the premise: so far one has not obtained the structure functions at a fixed $Q^2$ over a sufficiently wide range of the Bjorken variable $x_B$ for a reliable interpolation. In contrast, the lattice correlation functions can be computed at fixed $P^z$ within a wide range of $z$ and reliably extrapolated to $z \to \infty$.

Regarding ``point-by-point'' calculation of the PDF in LaMET, we are not referring to point-like features. Rather, our emphasis is on LaMET's capability to calculate, rather than fit, the PDFs. As explained in \sec{theory}, any sharp features of the quasi-PDF or PDF must either be detected or lie within the overall uncertainty band; otherwise, their presence would violate the EFT power expansion or asymptotic decay of the correlation functions. In the example shown in Figure 4 of Ref.~\cite{Dutrieux:2025axb}, the hypothetical bump at $x = 0.5$ corresponds to an oscillatory mode that decays too slowly in coordinate space, which can be ruled out by physical constraints.


\bibliography{Refs}

\begin{thebibliography}{266}%
\makeatletter
\providecommand \@ifxundefined [1]{%
 \@ifx{#1\undefined}
}%
\providecommand \@ifnum [1]{%
 \ifnum #1\expandafter \@firstoftwo
 \else \expandafter \@secondoftwo
 \fi
}%
\providecommand \@ifx [1]{%
 \ifx #1\expandafter \@firstoftwo
 \else \expandafter \@secondoftwo
 \fi
}%
\providecommand \natexlab [1]{#1}%
\providecommand \enquote  [1]{``#1''}%
\providecommand \bibnamefont  [1]{#1}%
\providecommand \bibfnamefont [1]{#1}%
\providecommand \citenamefont [1]{#1}%
\providecommand \href@noop [0]{\@secondoftwo}%
\providecommand \href [0]{\begingroup \@sanitize@url \@href}%
\providecommand \@href[1]{\@@startlink{#1}\@@href}%
\providecommand \@@href[1]{\endgroup#1\@@endlink}%
\providecommand \@sanitize@url [0]{\catcode `\\12\catcode `\$12\catcode
  `\&12\catcode `\#12\catcode `\^12\catcode `\_12\catcode `\%12\relax}%
\providecommand \@@startlink[1]{}%
\providecommand \@@endlink[0]{}%
\providecommand \url  [0]{\begingroup\@sanitize@url \@url }%
\providecommand \@url [1]{\endgroup\@href {#1}{\urlprefix }}%
\providecommand \urlprefix  [0]{URL }%
\providecommand \Eprint [0]{\href }%
\providecommand \doibase [0]{http://dx.doi.org/}%
\providecommand \selectlanguage [0]{\@gobble}%
\providecommand \bibinfo  [0]{\@secondoftwo}%
\providecommand \bibfield  [0]{\@secondoftwo}%
\providecommand \translation [1]{[#1]}%
\providecommand \BibitemOpen [0]{}%
\providecommand \bibitemStop [0]{}%
\providecommand \bibitemNoStop [0]{.\EOS\space}%
\providecommand \EOS [0]{\spacefactor3000\relax}%
\providecommand \BibitemShut  [1]{\csname bibitem#1\endcsname}%
\let\auto@bib@innerbib\@empty
\bibitem [{\citenamefont {Dutrieux}\ \emph
  {et~al.}(2025{\natexlab{a}})\citenamefont {Dutrieux}, \citenamefont {Karpie},
  \citenamefont {Monahan}, \citenamefont {Orginos}, \citenamefont {Radyushkin},
  \citenamefont {Richards},\ and\ \citenamefont
  {Zafeiropoulos}}]{Dutrieux:2025jed}%
  \BibitemOpen
  \bibfield  {author} {\bibinfo {author} {\bibfnamefont {H.}~\bibnamefont
  {Dutrieux}}, \bibinfo {author} {\bibfnamefont {J.}~\bibnamefont {Karpie}},
  \bibinfo {author} {\bibfnamefont {C.~J.}\ \bibnamefont {Monahan}}, \bibinfo
  {author} {\bibfnamefont {K.}~\bibnamefont {Orginos}}, \bibinfo {author}
  {\bibfnamefont {A.}~\bibnamefont {Radyushkin}}, \bibinfo {author}
  {\bibfnamefont {D.}~\bibnamefont {Richards}}, \ and\ \bibinfo {author}
  {\bibfnamefont {S.}~\bibnamefont {Zafeiropoulos}},\ }\href@noop {} {\
  (\bibinfo {year} {2025}{\natexlab{a}})},\ \Eprint
  {http://arxiv.org/abs/2504.17706} {arXiv:2504.17706 [hep-lat]} \BibitemShut
  {NoStop}%
\bibitem [{\citenamefont {Liu}\ and\ \citenamefont {Dong}(1994)}]{Liu:1993cv}%
  \BibitemOpen
  \bibfield  {author} {\bibinfo {author} {\bibfnamefont {K.-F.}\ \bibnamefont
  {Liu}}\ and\ \bibinfo {author} {\bibfnamefont {S.-J.}\ \bibnamefont {Dong}},\
  }\href {\doibase 10.1103/PhysRevLett.72.1790} {\bibfield  {journal} {\bibinfo
   {journal} {Phys. Rev. Lett.}\ }\textbf {\bibinfo {volume} {72}},\ \bibinfo
  {pages} {1790} (\bibinfo {year} {1994})},\ \Eprint
  {http://arxiv.org/abs/hep-ph/9306299} {arXiv:hep-ph/9306299} \BibitemShut
  {NoStop}%
\bibitem [{\citenamefont {Detmold}\ and\ \citenamefont
  {Lin}(2006)}]{Detmold:2005gg}%
  \BibitemOpen
  \bibfield  {author} {\bibinfo {author} {\bibfnamefont {W.}~\bibnamefont
  {Detmold}}\ and\ \bibinfo {author} {\bibfnamefont {C.~J.~D.}\ \bibnamefont
  {Lin}},\ }\href {\doibase 10.1103/PhysRevD.73.014501} {\bibfield  {journal}
  {\bibinfo  {journal} {Phys. Rev. D}\ }\textbf {\bibinfo {volume} {73}},\
  \bibinfo {pages} {014501} (\bibinfo {year} {2006})},\ \Eprint
  {http://arxiv.org/abs/hep-lat/0507007} {arXiv:hep-lat/0507007} \BibitemShut
  {NoStop}%
\bibitem [{\citenamefont {Braun}\ and\ \citenamefont
  {M\"uller}(2008)}]{Braun:2007wv}%
  \BibitemOpen
  \bibfield  {author} {\bibinfo {author} {\bibfnamefont {V.}~\bibnamefont
  {Braun}}\ and\ \bibinfo {author} {\bibfnamefont {D.}~\bibnamefont
  {M\"uller}},\ }\href {\doibase 10.1140/epjc/s10052-008-0608-4} {\bibfield
  {journal} {\bibinfo  {journal} {Eur. Phys. J. C}\ }\textbf {\bibinfo {volume}
  {55}},\ \bibinfo {pages} {349} (\bibinfo {year} {2008})},\ \Eprint
  {http://arxiv.org/abs/0709.1348} {arXiv:0709.1348 [hep-ph]} \BibitemShut
  {NoStop}%
\bibitem [{\citenamefont {Davoudi}\ and\ \citenamefont
  {Savage}(2012)}]{Davoudi:2012ya}%
  \BibitemOpen
  \bibfield  {author} {\bibinfo {author} {\bibfnamefont {Z.}~\bibnamefont
  {Davoudi}}\ and\ \bibinfo {author} {\bibfnamefont {M.~J.}\ \bibnamefont
  {Savage}},\ }\href {\doibase 10.1103/PhysRevD.86.054505} {\bibfield
  {journal} {\bibinfo  {journal} {Phys. Rev. D}\ }\textbf {\bibinfo {volume}
  {86}},\ \bibinfo {pages} {054505} (\bibinfo {year} {2012})},\ \Eprint
  {http://arxiv.org/abs/1204.4146} {arXiv:1204.4146 [hep-lat]} \BibitemShut
  {NoStop}%
\bibitem [{\citenamefont {Ji}(2013)}]{Ji:2013dva}%
  \BibitemOpen
  \bibfield  {author} {\bibinfo {author} {\bibfnamefont {X.}~\bibnamefont
  {Ji}},\ }\href {\doibase 10.1103/PhysRevLett.110.262002} {\bibfield
  {journal} {\bibinfo  {journal} {Phys. Rev. Lett.}\ }\textbf {\bibinfo
  {volume} {110}},\ \bibinfo {pages} {262002} (\bibinfo {year} {2013})},\
  \Eprint {http://arxiv.org/abs/1305.1539} {arXiv:1305.1539 [hep-ph]}
  \BibitemShut {NoStop}%
\bibitem [{\citenamefont
  {Radyushkin}(2017{\natexlab{a}})}]{Radyushkin:2017cyf}%
  \BibitemOpen
  \bibfield  {author} {\bibinfo {author} {\bibfnamefont {A.~V.}\ \bibnamefont
  {Radyushkin}},\ }\href {\doibase 10.1103/PhysRevD.96.034025} {\bibfield
  {journal} {\bibinfo  {journal} {Phys. Rev. D}\ }\textbf {\bibinfo {volume}
  {96}},\ \bibinfo {pages} {034025} (\bibinfo {year} {2017}{\natexlab{a}})},\
  \Eprint {http://arxiv.org/abs/1705.01488} {arXiv:1705.01488 [hep-ph]}
  \BibitemShut {NoStop}%
\bibitem [{\citenamefont {Ma}\ and\ \citenamefont {Qiu}(2018)}]{Ma:2017pxb}%
  \BibitemOpen
  \bibfield  {author} {\bibinfo {author} {\bibfnamefont {Y.-Q.}\ \bibnamefont
  {Ma}}\ and\ \bibinfo {author} {\bibfnamefont {J.-W.}\ \bibnamefont {Qiu}},\
  }\href {\doibase 10.1103/PhysRevLett.120.022003} {\bibfield  {journal}
  {\bibinfo  {journal} {Phys. Rev. Lett.}\ }\textbf {\bibinfo {volume} {120}},\
  \bibinfo {pages} {022003} (\bibinfo {year} {2018})},\ \Eprint
  {http://arxiv.org/abs/1709.03018} {arXiv:1709.03018 [hep-ph]} \BibitemShut
  {NoStop}%
\bibitem [{\citenamefont {Chambers}\ \emph {et~al.}(2017)\citenamefont
  {Chambers}, \citenamefont {Horsley}, \citenamefont {Nakamura}, \citenamefont
  {Perlt}, \citenamefont {Rakow}, \citenamefont {Schierholz}, \citenamefont
  {Schiller}, \citenamefont {Somfleth}, \citenamefont {Young},\ and\
  \citenamefont {Zanotti}}]{Chambers:2017dov}%
  \BibitemOpen
  \bibfield  {author} {\bibinfo {author} {\bibfnamefont {A.~J.}\ \bibnamefont
  {Chambers}}, \bibinfo {author} {\bibfnamefont {R.}~\bibnamefont {Horsley}},
  \bibinfo {author} {\bibfnamefont {Y.}~\bibnamefont {Nakamura}}, \bibinfo
  {author} {\bibfnamefont {H.}~\bibnamefont {Perlt}}, \bibinfo {author}
  {\bibfnamefont {P.~E.~L.}\ \bibnamefont {Rakow}}, \bibinfo {author}
  {\bibfnamefont {G.}~\bibnamefont {Schierholz}}, \bibinfo {author}
  {\bibfnamefont {A.}~\bibnamefont {Schiller}}, \bibinfo {author}
  {\bibfnamefont {K.}~\bibnamefont {Somfleth}}, \bibinfo {author}
  {\bibfnamefont {R.~D.}\ \bibnamefont {Young}}, \ and\ \bibinfo {author}
  {\bibfnamefont {J.~M.}\ \bibnamefont {Zanotti}},\ }\href {\doibase
  10.1103/PhysRevLett.118.242001} {\bibfield  {journal} {\bibinfo  {journal}
  {Phys. Rev. Lett.}\ }\textbf {\bibinfo {volume} {118}},\ \bibinfo {pages}
  {242001} (\bibinfo {year} {2017})},\ \Eprint
  {http://arxiv.org/abs/1703.01153} {arXiv:1703.01153 [hep-lat]} \BibitemShut
  {NoStop}%
\bibitem [{\citenamefont {Shindler}(2024)}]{Shindler:2023xpd}%
  \BibitemOpen
  \bibfield  {author} {\bibinfo {author} {\bibfnamefont {A.}~\bibnamefont
  {Shindler}},\ }\href {\doibase 10.1103/PhysRevD.110.L051503} {\bibfield
  {journal} {\bibinfo  {journal} {Phys. Rev. D}\ }\textbf {\bibinfo {volume}
  {110}},\ \bibinfo {pages} {L051503} (\bibinfo {year} {2024})},\ \Eprint
  {http://arxiv.org/abs/2311.18704} {arXiv:2311.18704 [hep-lat]} \BibitemShut
  {NoStop}%
\bibitem [{\citenamefont {Ji}(2014)}]{Ji:2014gla}%
  \BibitemOpen
  \bibfield  {author} {\bibinfo {author} {\bibfnamefont {X.}~\bibnamefont
  {Ji}},\ }\href {\doibase 10.1007/s11433-014-5492-3} {\bibfield  {journal}
  {\bibinfo  {journal} {Sci. China Phys. Mech. Astron.}\ }\textbf {\bibinfo
  {volume} {57}},\ \bibinfo {pages} {1407} (\bibinfo {year} {2014})},\ \Eprint
  {http://arxiv.org/abs/1404.6680} {arXiv:1404.6680 [hep-ph]} \BibitemShut
  {NoStop}%
\bibitem [{\citenamefont {Ji}\ \emph {et~al.}(2021{\natexlab{a}})\citenamefont
  {Ji}, \citenamefont {Liu}, \citenamefont {Liu}, \citenamefont {Zhang},\ and\
  \citenamefont {Zhao}}]{Ji:2020ect}%
  \BibitemOpen
  \bibfield  {author} {\bibinfo {author} {\bibfnamefont {X.}~\bibnamefont
  {Ji}}, \bibinfo {author} {\bibfnamefont {Y.-S.}\ \bibnamefont {Liu}},
  \bibinfo {author} {\bibfnamefont {Y.}~\bibnamefont {Liu}}, \bibinfo {author}
  {\bibfnamefont {J.-H.}\ \bibnamefont {Zhang}}, \ and\ \bibinfo {author}
  {\bibfnamefont {Y.}~\bibnamefont {Zhao}},\ }\href {\doibase
  10.1103/RevModPhys.93.035005} {\bibfield  {journal} {\bibinfo  {journal}
  {Rev. Mod. Phys.}\ }\textbf {\bibinfo {volume} {93}},\ \bibinfo {pages}
  {035005} (\bibinfo {year} {2021}{\natexlab{a}})},\ \Eprint
  {http://arxiv.org/abs/2004.03543} {arXiv:2004.03543 [hep-ph]} \BibitemShut
  {NoStop}%
\bibitem [{\citenamefont {Xiong}\ \emph {et~al.}(2014)\citenamefont {Xiong},
  \citenamefont {Ji}, \citenamefont {Zhang},\ and\ \citenamefont
  {Zhao}}]{Xiong:2013bka}%
  \BibitemOpen
  \bibfield  {author} {\bibinfo {author} {\bibfnamefont {X.}~\bibnamefont
  {Xiong}}, \bibinfo {author} {\bibfnamefont {X.}~\bibnamefont {Ji}}, \bibinfo
  {author} {\bibfnamefont {J.-H.}\ \bibnamefont {Zhang}}, \ and\ \bibinfo
  {author} {\bibfnamefont {Y.}~\bibnamefont {Zhao}},\ }\href {\doibase
  10.1103/PhysRevD.90.014051} {\bibfield  {journal} {\bibinfo  {journal} {Phys.
  Rev. D}\ }\textbf {\bibinfo {volume} {90}},\ \bibinfo {pages} {014051}
  (\bibinfo {year} {2014})},\ \Eprint {http://arxiv.org/abs/1310.7471}
  {arXiv:1310.7471 [hep-ph]} \BibitemShut {NoStop}%
\bibitem [{\citenamefont {Lin}\ \emph {et~al.}(2015)\citenamefont {Lin},
  \citenamefont {Chen}, \citenamefont {Cohen},\ and\ \citenamefont
  {Ji}}]{Lin:2014zya}%
  \BibitemOpen
  \bibfield  {author} {\bibinfo {author} {\bibfnamefont {H.-W.}\ \bibnamefont
  {Lin}}, \bibinfo {author} {\bibfnamefont {J.-W.}\ \bibnamefont {Chen}},
  \bibinfo {author} {\bibfnamefont {S.~D.}\ \bibnamefont {Cohen}}, \ and\
  \bibinfo {author} {\bibfnamefont {X.}~\bibnamefont {Ji}},\ }\href {\doibase
  10.1103/PhysRevD.91.054510} {\bibfield  {journal} {\bibinfo  {journal} {Phys.
  Rev. D}\ }\textbf {\bibinfo {volume} {91}},\ \bibinfo {pages} {054510}
  (\bibinfo {year} {2015})},\ \Eprint {http://arxiv.org/abs/1402.1462}
  {arXiv:1402.1462 [hep-ph]} \BibitemShut {NoStop}%
\bibitem [{\citenamefont {Ji}\ \emph {et~al.}(2015{\natexlab{a}})\citenamefont
  {Ji}, \citenamefont {Sun}, \citenamefont {Xiong},\ and\ \citenamefont
  {Yuan}}]{Ji:2014hxa}%
  \BibitemOpen
  \bibfield  {author} {\bibinfo {author} {\bibfnamefont {X.}~\bibnamefont
  {Ji}}, \bibinfo {author} {\bibfnamefont {P.}~\bibnamefont {Sun}}, \bibinfo
  {author} {\bibfnamefont {X.}~\bibnamefont {Xiong}}, \ and\ \bibinfo {author}
  {\bibfnamefont {F.}~\bibnamefont {Yuan}},\ }\href {\doibase
  10.1103/PhysRevD.91.074009} {\bibfield  {journal} {\bibinfo  {journal} {Phys.
  Rev. D}\ }\textbf {\bibinfo {volume} {91}},\ \bibinfo {pages} {074009}
  (\bibinfo {year} {2015}{\natexlab{a}})},\ \Eprint
  {http://arxiv.org/abs/1405.7640} {arXiv:1405.7640 [hep-ph]} \BibitemShut
  {NoStop}%
\bibitem [{\citenamefont {Ji}\ \emph {et~al.}(2015{\natexlab{b}})\citenamefont
  {Ji}, \citenamefont {Zhang},\ and\ \citenamefont {Zhao}}]{Ji:2014lra}%
  \BibitemOpen
  \bibfield  {author} {\bibinfo {author} {\bibfnamefont {X.}~\bibnamefont
  {Ji}}, \bibinfo {author} {\bibfnamefont {J.-H.}\ \bibnamefont {Zhang}}, \
  and\ \bibinfo {author} {\bibfnamefont {Y.}~\bibnamefont {Zhao}},\ }\href
  {\doibase 10.1016/j.physletb.2015.02.054} {\bibfield  {journal} {\bibinfo
  {journal} {Phys. Lett. B}\ }\textbf {\bibinfo {volume} {743}},\ \bibinfo
  {pages} {180} (\bibinfo {year} {2015}{\natexlab{b}})},\ \Eprint
  {http://arxiv.org/abs/1409.6329} {arXiv:1409.6329 [hep-ph]} \BibitemShut
  {NoStop}%
\bibitem [{\citenamefont {Sufian}\ \emph {et~al.}(2015)\citenamefont {Sufian},
  \citenamefont {Glatzmaier}, \citenamefont {Yang}, \citenamefont {Liu},\ and\
  \citenamefont {Sun}}]{Sufian:2014jma}%
  \BibitemOpen
  \bibfield  {author} {\bibinfo {author} {\bibfnamefont {R.~S.}\ \bibnamefont
  {Sufian}}, \bibinfo {author} {\bibfnamefont {M.~J.}\ \bibnamefont
  {Glatzmaier}}, \bibinfo {author} {\bibfnamefont {Y.-B.}\ \bibnamefont
  {Yang}}, \bibinfo {author} {\bibfnamefont {K.-F.}\ \bibnamefont {Liu}}, \
  and\ \bibinfo {author} {\bibfnamefont {M.}~\bibnamefont {Sun}} (\bibinfo
  {collaboration} {xQCD}),\ }\href {\doibase 10.22323/1.214.0166} {\bibfield
  {journal} {\bibinfo  {journal} {PoS}\ }\textbf {\bibinfo {volume}
  {LATTICE2014}},\ \bibinfo {pages} {166} (\bibinfo {year} {2015})},\ \Eprint
  {http://arxiv.org/abs/1412.7168} {arXiv:1412.7168 [hep-lat]} \BibitemShut
  {NoStop}%
\bibitem [{\citenamefont {Ji}\ \emph {et~al.}(2015{\natexlab{c}})\citenamefont
  {Ji}, \citenamefont {Sch\"afer}, \citenamefont {Xiong},\ and\ \citenamefont
  {Zhang}}]{Ji:2015qla}%
  \BibitemOpen
  \bibfield  {author} {\bibinfo {author} {\bibfnamefont {X.}~\bibnamefont
  {Ji}}, \bibinfo {author} {\bibfnamefont {A.}~\bibnamefont {Sch\"afer}},
  \bibinfo {author} {\bibfnamefont {X.}~\bibnamefont {Xiong}}, \ and\ \bibinfo
  {author} {\bibfnamefont {J.-H.}\ \bibnamefont {Zhang}},\ }\href {\doibase
  10.1103/PhysRevD.92.014039} {\bibfield  {journal} {\bibinfo  {journal} {Phys.
  Rev. D}\ }\textbf {\bibinfo {volume} {92}},\ \bibinfo {pages} {014039}
  (\bibinfo {year} {2015}{\natexlab{c}})},\ \Eprint
  {http://arxiv.org/abs/1506.00248} {arXiv:1506.00248 [hep-ph]} \BibitemShut
  {NoStop}%
\bibitem [{\citenamefont {Xiong}\ and\ \citenamefont
  {Zhang}(2015)}]{Xiong:2015nua}%
  \BibitemOpen
  \bibfield  {author} {\bibinfo {author} {\bibfnamefont {X.}~\bibnamefont
  {Xiong}}\ and\ \bibinfo {author} {\bibfnamefont {J.-H.}\ \bibnamefont
  {Zhang}},\ }\href {\doibase 10.1103/PhysRevD.92.054037} {\bibfield  {journal}
  {\bibinfo  {journal} {Phys. Rev. D}\ }\textbf {\bibinfo {volume} {92}},\
  \bibinfo {pages} {054037} (\bibinfo {year} {2015})},\ \Eprint
  {http://arxiv.org/abs/1509.08016} {arXiv:1509.08016 [hep-ph]} \BibitemShut
  {NoStop}%
\bibitem [{\citenamefont {Alexandrou}\ \emph {et~al.}(2016)\citenamefont
  {Alexandrou}, \citenamefont {Cichy}, \citenamefont {Constantinou},
  \citenamefont {Hadjiyiannakou}, \citenamefont {Jansen}, \citenamefont
  {Steffens},\ and\ \citenamefont {Wiese}}]{Alexandrou:2016eyt}%
  \BibitemOpen
  \bibfield  {author} {\bibinfo {author} {\bibfnamefont {C.}~\bibnamefont
  {Alexandrou}}, \bibinfo {author} {\bibfnamefont {K.}~\bibnamefont {Cichy}},
  \bibinfo {author} {\bibfnamefont {M.}~\bibnamefont {Constantinou}}, \bibinfo
  {author} {\bibfnamefont {K.}~\bibnamefont {Hadjiyiannakou}}, \bibinfo
  {author} {\bibfnamefont {K.}~\bibnamefont {Jansen}}, \bibinfo {author}
  {\bibfnamefont {F.}~\bibnamefont {Steffens}}, \ and\ \bibinfo {author}
  {\bibfnamefont {C.}~\bibnamefont {Wiese}},\ }\href {\doibase
  10.22323/1.256.0151} {\bibfield  {journal} {\bibinfo  {journal} {PoS}\
  }\textbf {\bibinfo {volume} {LATTICE2016}},\ \bibinfo {pages} {151} (\bibinfo
  {year} {2016})},\ \Eprint {http://arxiv.org/abs/1612.08728} {arXiv:1612.08728
  [hep-lat]} \BibitemShut {NoStop}%
\bibitem [{\citenamefont {Chen}\ \emph {et~al.}(2016)\citenamefont {Chen},
  \citenamefont {Cohen}, \citenamefont {Ji}, \citenamefont {Lin},\ and\
  \citenamefont {Zhang}}]{Chen:2016utp}%
  \BibitemOpen
  \bibfield  {author} {\bibinfo {author} {\bibfnamefont {J.-W.}\ \bibnamefont
  {Chen}}, \bibinfo {author} {\bibfnamefont {S.~D.}\ \bibnamefont {Cohen}},
  \bibinfo {author} {\bibfnamefont {X.}~\bibnamefont {Ji}}, \bibinfo {author}
  {\bibfnamefont {H.-W.}\ \bibnamefont {Lin}}, \ and\ \bibinfo {author}
  {\bibfnamefont {J.-H.}\ \bibnamefont {Zhang}},\ }\href {\doibase
  10.1016/j.nuclphysb.2016.07.033} {\bibfield  {journal} {\bibinfo  {journal}
  {Nucl. Phys. B}\ }\textbf {\bibinfo {volume} {911}},\ \bibinfo {pages} {246}
  (\bibinfo {year} {2016})},\ \Eprint {http://arxiv.org/abs/1603.06664}
  {arXiv:1603.06664 [hep-ph]} \BibitemShut {NoStop}%
\bibitem [{\citenamefont {Lin}(2016)}]{Lin:2016qia}%
  \BibitemOpen
  \bibfield  {author} {\bibinfo {author} {\bibfnamefont {H.-W.}\ \bibnamefont
  {Lin}},\ }\href {\doibase 10.22323/1.256.0005} {\bibfield  {journal}
  {\bibinfo  {journal} {PoS}\ }\textbf {\bibinfo {volume} {LATTICE2016}},\
  \bibinfo {pages} {005} (\bibinfo {year} {2016})},\ \Eprint
  {http://arxiv.org/abs/1612.09366} {arXiv:1612.09366 [hep-lat]} \BibitemShut
  {NoStop}%
\bibitem [{\citenamefont {Yang}\ \emph {et~al.}(2016)\citenamefont {Yang},
  \citenamefont {Sufian}, \citenamefont {Alexandru}, \citenamefont {Draper},
  \citenamefont {Glatzmaier},\ and\ \citenamefont {Liu}}]{Yang:2016nfc}%
  \BibitemOpen
  \bibfield  {author} {\bibinfo {author} {\bibfnamefont {Y.-B.}\ \bibnamefont
  {Yang}}, \bibinfo {author} {\bibfnamefont {R.~S.}\ \bibnamefont {Sufian}},
  \bibinfo {author} {\bibfnamefont {A.}~\bibnamefont {Alexandru}}, \bibinfo
  {author} {\bibfnamefont {T.}~\bibnamefont {Draper}}, \bibinfo {author}
  {\bibfnamefont {M.~J.}\ \bibnamefont {Glatzmaier}}, \ and\ \bibinfo {author}
  {\bibfnamefont {K.-F.}\ \bibnamefont {Liu}} (\bibinfo {collaboration}
  {xQCD}),\ }\href {\doibase 10.22323/1.251.0129} {\bibfield  {journal}
  {\bibinfo  {journal} {PoS}\ }\textbf {\bibinfo {volume} {LATTICE2015}},\
  \bibinfo {pages} {129} (\bibinfo {year} {2016})},\ \Eprint
  {http://arxiv.org/abs/1603.05256} {arXiv:1603.05256 [hep-ph]} \BibitemShut
  {NoStop}%
\bibitem [{\citenamefont {Yang}\ \emph {et~al.}(2017)\citenamefont {Yang},
  \citenamefont {Sufian}, \citenamefont {Alexandru}, \citenamefont {Draper},
  \citenamefont {Glatzmaier}, \citenamefont {Liu},\ and\ \citenamefont
  {Zhao}}]{Yang:2016plb}%
  \BibitemOpen
  \bibfield  {author} {\bibinfo {author} {\bibfnamefont {Y.-B.}\ \bibnamefont
  {Yang}}, \bibinfo {author} {\bibfnamefont {R.~S.}\ \bibnamefont {Sufian}},
  \bibinfo {author} {\bibfnamefont {A.}~\bibnamefont {Alexandru}}, \bibinfo
  {author} {\bibfnamefont {T.}~\bibnamefont {Draper}}, \bibinfo {author}
  {\bibfnamefont {M.~J.}\ \bibnamefont {Glatzmaier}}, \bibinfo {author}
  {\bibfnamefont {K.-F.}\ \bibnamefont {Liu}}, \ and\ \bibinfo {author}
  {\bibfnamefont {Y.}~\bibnamefont {Zhao}},\ }\href {\doibase
  10.1103/PhysRevLett.118.102001} {\bibfield  {journal} {\bibinfo  {journal}
  {Phys. Rev. Lett.}\ }\textbf {\bibinfo {volume} {118}},\ \bibinfo {pages}
  {102001} (\bibinfo {year} {2017})},\ \Eprint
  {http://arxiv.org/abs/1609.05937} {arXiv:1609.05937 [hep-ph]} \BibitemShut
  {NoStop}%
\bibitem [{\citenamefont {Alexandrou}\ \emph
  {et~al.}(2018{\natexlab{a}})\citenamefont {Alexandrou}, \citenamefont
  {Bacchio}, \citenamefont {Cichy}, \citenamefont {Constantinou}, \citenamefont
  {Hadjiyiannakou}, \citenamefont {Jansen}, \citenamefont {Koutsou},
  \citenamefont {Scapellato},\ and\ \citenamefont
  {Steffens}}]{Alexandrou:2017dzj}%
  \BibitemOpen
  \bibfield  {author} {\bibinfo {author} {\bibfnamefont {C.}~\bibnamefont
  {Alexandrou}}, \bibinfo {author} {\bibfnamefont {S.}~\bibnamefont {Bacchio}},
  \bibinfo {author} {\bibfnamefont {K.}~\bibnamefont {Cichy}}, \bibinfo
  {author} {\bibfnamefont {M.}~\bibnamefont {Constantinou}}, \bibinfo {author}
  {\bibfnamefont {K.}~\bibnamefont {Hadjiyiannakou}}, \bibinfo {author}
  {\bibfnamefont {K.}~\bibnamefont {Jansen}}, \bibinfo {author} {\bibfnamefont
  {G.}~\bibnamefont {Koutsou}}, \bibinfo {author} {\bibfnamefont
  {A.}~\bibnamefont {Scapellato}}, \ and\ \bibinfo {author} {\bibfnamefont
  {F.}~\bibnamefont {Steffens}},\ }\href {\doibase
  10.1051/epjconf/201817514008} {\bibfield  {journal} {\bibinfo  {journal} {EPJ
  Web Conf.}\ }\textbf {\bibinfo {volume} {175}},\ \bibinfo {pages} {14008}
  (\bibinfo {year} {2018}{\natexlab{a}})},\ \Eprint
  {http://arxiv.org/abs/1710.06408} {arXiv:1710.06408 [hep-lat]} \BibitemShut
  {NoStop}%
\bibitem [{\citenamefont {Alexandrou}\ \emph {et~al.}(2017)\citenamefont
  {Alexandrou}, \citenamefont {Cichy}, \citenamefont {Constantinou},
  \citenamefont {Hadjiyiannakou}, \citenamefont {Jansen}, \citenamefont
  {Panagopoulos},\ and\ \citenamefont {Steffens}}]{Alexandrou:2017huk}%
  \BibitemOpen
  \bibfield  {author} {\bibinfo {author} {\bibfnamefont {C.}~\bibnamefont
  {Alexandrou}}, \bibinfo {author} {\bibfnamefont {K.}~\bibnamefont {Cichy}},
  \bibinfo {author} {\bibfnamefont {M.}~\bibnamefont {Constantinou}}, \bibinfo
  {author} {\bibfnamefont {K.}~\bibnamefont {Hadjiyiannakou}}, \bibinfo
  {author} {\bibfnamefont {K.}~\bibnamefont {Jansen}}, \bibinfo {author}
  {\bibfnamefont {H.}~\bibnamefont {Panagopoulos}}, \ and\ \bibinfo {author}
  {\bibfnamefont {F.}~\bibnamefont {Steffens}},\ }\href {\doibase
  10.1016/j.nuclphysb.2017.08.012} {\bibfield  {journal} {\bibinfo  {journal}
  {Nucl. Phys. B}\ }\textbf {\bibinfo {volume} {923}},\ \bibinfo {pages} {394}
  (\bibinfo {year} {2017})},\ \Eprint {http://arxiv.org/abs/1706.00265}
  {arXiv:1706.00265 [hep-lat]} \BibitemShut {NoStop}%
\bibitem [{\citenamefont {Alexandrou}\ \emph
  {et~al.}(2018{\natexlab{b}})\citenamefont {Alexandrou}, \citenamefont
  {Cichy}, \citenamefont {Constantinou}, \citenamefont {Hadjiyiannakou},
  \citenamefont {Jansen}, \citenamefont {Panagopoulos}, \citenamefont
  {Scapellato},\ and\ \citenamefont {Steffens}}]{Alexandrou:2017qpu}%
  \BibitemOpen
  \bibfield  {author} {\bibinfo {author} {\bibfnamefont {C.}~\bibnamefont
  {Alexandrou}}, \bibinfo {author} {\bibfnamefont {K.}~\bibnamefont {Cichy}},
  \bibinfo {author} {\bibfnamefont {M.}~\bibnamefont {Constantinou}}, \bibinfo
  {author} {\bibfnamefont {K.}~\bibnamefont {Hadjiyiannakou}}, \bibinfo
  {author} {\bibfnamefont {K.}~\bibnamefont {Jansen}}, \bibinfo {author}
  {\bibfnamefont {H.}~\bibnamefont {Panagopoulos}}, \bibinfo {author}
  {\bibfnamefont {A.}~\bibnamefont {Scapellato}}, \ and\ \bibinfo {author}
  {\bibfnamefont {F.}~\bibnamefont {Steffens}},\ }\href {\doibase
  10.1051/epjconf/201817506021} {\bibfield  {journal} {\bibinfo  {journal} {EPJ
  Web Conf.}\ }\textbf {\bibinfo {volume} {175}},\ \bibinfo {pages} {06021}
  (\bibinfo {year} {2018}{\natexlab{b}})},\ \Eprint
  {http://arxiv.org/abs/1709.07513} {arXiv:1709.07513 [hep-lat]} \BibitemShut
  {NoStop}%
\bibitem [{\citenamefont {Chen}\ \emph {et~al.}(2019)\citenamefont {Chen},
  \citenamefont {Ishikawa}, \citenamefont {Jin}, \citenamefont {Lin},
  \citenamefont {Zhang},\ and\ \citenamefont {Zhao}}]{Chen:2017mie}%
  \BibitemOpen
  \bibfield  {author} {\bibinfo {author} {\bibfnamefont {J.-W.}\ \bibnamefont
  {Chen}}, \bibinfo {author} {\bibfnamefont {T.}~\bibnamefont {Ishikawa}},
  \bibinfo {author} {\bibfnamefont {L.}~\bibnamefont {Jin}}, \bibinfo {author}
  {\bibfnamefont {H.-W.}\ \bibnamefont {Lin}}, \bibinfo {author} {\bibfnamefont
  {J.-H.}\ \bibnamefont {Zhang}}, \ and\ \bibinfo {author} {\bibfnamefont
  {Y.}~\bibnamefont {Zhao}} (\bibinfo {collaboration} {LP3}),\ }\href {\doibase
  10.1088/1674-1137/43/10/103101} {\bibfield  {journal} {\bibinfo  {journal}
  {Chin. Phys. C}\ }\textbf {\bibinfo {volume} {43}},\ \bibinfo {pages}
  {103101} (\bibinfo {year} {2019})},\ \Eprint
  {http://arxiv.org/abs/1710.01089} {arXiv:1710.01089 [hep-lat]} \BibitemShut
  {NoStop}%
\bibitem [{\citenamefont {Chen}\ \emph
  {et~al.}(2018{\natexlab{a}})\citenamefont {Chen}, \citenamefont {Ishikawa},
  \citenamefont {Jin}, \citenamefont {Lin}, \citenamefont {Yang}, \citenamefont
  {Zhang},\ and\ \citenamefont {Zhao}}]{Chen:2017mzz}%
  \BibitemOpen
  \bibfield  {author} {\bibinfo {author} {\bibfnamefont {J.-W.}\ \bibnamefont
  {Chen}}, \bibinfo {author} {\bibfnamefont {T.}~\bibnamefont {Ishikawa}},
  \bibinfo {author} {\bibfnamefont {L.}~\bibnamefont {Jin}}, \bibinfo {author}
  {\bibfnamefont {H.-W.}\ \bibnamefont {Lin}}, \bibinfo {author} {\bibfnamefont
  {Y.-B.}\ \bibnamefont {Yang}}, \bibinfo {author} {\bibfnamefont {J.-H.}\
  \bibnamefont {Zhang}}, \ and\ \bibinfo {author} {\bibfnamefont
  {Y.}~\bibnamefont {Zhao}},\ }\href {\doibase 10.1103/PhysRevD.97.014505}
  {\bibfield  {journal} {\bibinfo  {journal} {Phys. Rev. D}\ }\textbf {\bibinfo
  {volume} {97}},\ \bibinfo {pages} {014505} (\bibinfo {year}
  {2018}{\natexlab{a}})},\ \Eprint {http://arxiv.org/abs/1706.01295}
  {arXiv:1706.01295 [hep-lat]} \BibitemShut {NoStop}%
\bibitem [{\citenamefont {Constantinou}\ and\ \citenamefont
  {Panagopoulos}(2017)}]{Constantinou:2017sej}%
  \BibitemOpen
  \bibfield  {author} {\bibinfo {author} {\bibfnamefont {M.}~\bibnamefont
  {Constantinou}}\ and\ \bibinfo {author} {\bibfnamefont {H.}~\bibnamefont
  {Panagopoulos}},\ }\href {\doibase 10.1103/PhysRevD.96.054506} {\bibfield
  {journal} {\bibinfo  {journal} {Phys. Rev. D}\ }\textbf {\bibinfo {volume}
  {96}},\ \bibinfo {pages} {054506} (\bibinfo {year} {2017})},\ \Eprint
  {http://arxiv.org/abs/1705.11193} {arXiv:1705.11193 [hep-lat]} \BibitemShut
  {NoStop}%
\bibitem [{\citenamefont {Ishikawa}\ \emph {et~al.}(2019)\citenamefont
  {Ishikawa}, \citenamefont {Jin}, \citenamefont {Lin}, \citenamefont
  {Sch\"afer}, \citenamefont {Yang}, \citenamefont {Zhang},\ and\ \citenamefont
  {Zhao}}]{Ishikawa:2017iym}%
  \BibitemOpen
  \bibfield  {author} {\bibinfo {author} {\bibfnamefont {T.}~\bibnamefont
  {Ishikawa}}, \bibinfo {author} {\bibfnamefont {L.}~\bibnamefont {Jin}},
  \bibinfo {author} {\bibfnamefont {H.-W.}\ \bibnamefont {Lin}}, \bibinfo
  {author} {\bibfnamefont {A.}~\bibnamefont {Sch\"afer}}, \bibinfo {author}
  {\bibfnamefont {Y.-B.}\ \bibnamefont {Yang}}, \bibinfo {author}
  {\bibfnamefont {J.-H.}\ \bibnamefont {Zhang}}, \ and\ \bibinfo {author}
  {\bibfnamefont {Y.}~\bibnamefont {Zhao}},\ }\href {\doibase
  10.1007/s11433-018-9375-1} {\bibfield  {journal} {\bibinfo  {journal} {Sci.
  China Phys. Mech. Astron.}\ }\textbf {\bibinfo {volume} {62}},\ \bibinfo
  {pages} {991021} (\bibinfo {year} {2019})},\ \Eprint
  {http://arxiv.org/abs/1711.07858} {arXiv:1711.07858 [hep-ph]} \BibitemShut
  {NoStop}%
\bibitem [{\citenamefont {Ji}\ \emph {et~al.}(2018)\citenamefont {Ji},
  \citenamefont {Zhang},\ and\ \citenamefont {Zhao}}]{Ji:2017oey}%
  \BibitemOpen
  \bibfield  {author} {\bibinfo {author} {\bibfnamefont {X.}~\bibnamefont
  {Ji}}, \bibinfo {author} {\bibfnamefont {J.-H.}\ \bibnamefont {Zhang}}, \
  and\ \bibinfo {author} {\bibfnamefont {Y.}~\bibnamefont {Zhao}},\ }\href
  {\doibase 10.1103/PhysRevLett.120.112001} {\bibfield  {journal} {\bibinfo
  {journal} {Phys. Rev. Lett.}\ }\textbf {\bibinfo {volume} {120}},\ \bibinfo
  {pages} {112001} (\bibinfo {year} {2018})},\ \Eprint
  {http://arxiv.org/abs/1706.08962} {arXiv:1706.08962 [hep-ph]} \BibitemShut
  {NoStop}%
\bibitem [{\citenamefont {Ji}\ \emph {et~al.}(2017)\citenamefont {Ji},
  \citenamefont {Zhang},\ and\ \citenamefont {Zhao}}]{Ji:2017rah}%
  \BibitemOpen
  \bibfield  {author} {\bibinfo {author} {\bibfnamefont {X.}~\bibnamefont
  {Ji}}, \bibinfo {author} {\bibfnamefont {J.-H.}\ \bibnamefont {Zhang}}, \
  and\ \bibinfo {author} {\bibfnamefont {Y.}~\bibnamefont {Zhao}},\ }\href
  {\doibase 10.1016/j.nuclphysb.2017.09.001} {\bibfield  {journal} {\bibinfo
  {journal} {Nucl. Phys. B}\ }\textbf {\bibinfo {volume} {924}},\ \bibinfo
  {pages} {366} (\bibinfo {year} {2017})},\ \Eprint
  {http://arxiv.org/abs/1706.07416} {arXiv:1706.07416 [hep-ph]} \BibitemShut
  {NoStop}%
\bibitem [{\citenamefont {Lin}\ \emph {et~al.}(2018{\natexlab{a}})\citenamefont
  {Lin}, \citenamefont {Chen}, \citenamefont {Ishikawa},\ and\ \citenamefont
  {Zhang}}]{Lin:2017ani}%
  \BibitemOpen
  \bibfield  {author} {\bibinfo {author} {\bibfnamefont {H.-W.}\ \bibnamefont
  {Lin}}, \bibinfo {author} {\bibfnamefont {J.-W.}\ \bibnamefont {Chen}},
  \bibinfo {author} {\bibfnamefont {T.}~\bibnamefont {Ishikawa}}, \ and\
  \bibinfo {author} {\bibfnamefont {J.-H.}\ \bibnamefont {Zhang}} (\bibinfo
  {collaboration} {LP3}),\ }\href {\doibase 10.1103/PhysRevD.98.054504}
  {\bibfield  {journal} {\bibinfo  {journal} {Phys. Rev. D}\ }\textbf {\bibinfo
  {volume} {98}},\ \bibinfo {pages} {054504} (\bibinfo {year}
  {2018}{\natexlab{a}})},\ \Eprint {http://arxiv.org/abs/1708.05301}
  {arXiv:1708.05301 [hep-lat]} \BibitemShut {NoStop}%
\bibitem [{\citenamefont {Stewart}\ and\ \citenamefont
  {Zhao}(2018)}]{Stewart:2017tvs}%
  \BibitemOpen
  \bibfield  {author} {\bibinfo {author} {\bibfnamefont {I.~W.}\ \bibnamefont
  {Stewart}}\ and\ \bibinfo {author} {\bibfnamefont {Y.}~\bibnamefont {Zhao}},\
  }\href {\doibase 10.1103/PhysRevD.97.054512} {\bibfield  {journal} {\bibinfo
  {journal} {Phys. Rev. D}\ }\textbf {\bibinfo {volume} {97}},\ \bibinfo
  {pages} {054512} (\bibinfo {year} {2018})},\ \Eprint
  {http://arxiv.org/abs/1709.04933} {arXiv:1709.04933 [hep-ph]} \BibitemShut
  {NoStop}%
\bibitem [{\citenamefont {Xiong}\ \emph {et~al.}(2017)\citenamefont {Xiong},
  \citenamefont {Luu},\ and\ \citenamefont {Mei\ss{}ner}}]{Xiong:2017jtn}%
  \BibitemOpen
  \bibfield  {author} {\bibinfo {author} {\bibfnamefont {X.}~\bibnamefont
  {Xiong}}, \bibinfo {author} {\bibfnamefont {T.}~\bibnamefont {Luu}}, \ and\
  \bibinfo {author} {\bibfnamefont {U.-G.}\ \bibnamefont {Mei\ss{}ner}},\
  }\href@noop {} {\  (\bibinfo {year} {2017})},\ \Eprint
  {http://arxiv.org/abs/1705.00246} {arXiv:1705.00246 [hep-ph]} \BibitemShut
  {NoStop}%
\bibitem [{\citenamefont {Zhang}\ \emph {et~al.}(2017)\citenamefont {Zhang},
  \citenamefont {Chen}, \citenamefont {Ji}, \citenamefont {Jin},\ and\
  \citenamefont {Lin}}]{Zhang:2017bzy}%
  \BibitemOpen
  \bibfield  {author} {\bibinfo {author} {\bibfnamefont {J.-H.}\ \bibnamefont
  {Zhang}}, \bibinfo {author} {\bibfnamefont {J.-W.}\ \bibnamefont {Chen}},
  \bibinfo {author} {\bibfnamefont {X.}~\bibnamefont {Ji}}, \bibinfo {author}
  {\bibfnamefont {L.}~\bibnamefont {Jin}}, \ and\ \bibinfo {author}
  {\bibfnamefont {H.-W.}\ \bibnamefont {Lin}},\ }\href {\doibase
  10.1103/PhysRevD.95.094514} {\bibfield  {journal} {\bibinfo  {journal} {Phys.
  Rev. D}\ }\textbf {\bibinfo {volume} {95}},\ \bibinfo {pages} {094514}
  (\bibinfo {year} {2017})},\ \Eprint {http://arxiv.org/abs/1702.00008}
  {arXiv:1702.00008 [hep-lat]} \BibitemShut {NoStop}%
\bibitem [{\citenamefont {Zhang}\ \emph
  {et~al.}(2019{\natexlab{a}})\citenamefont {Zhang}, \citenamefont {Jin},
  \citenamefont {Lin}, \citenamefont {Sch\"afer}, \citenamefont {Sun},
  \citenamefont {Yang}, \citenamefont {Zhang}, \citenamefont {Zhao},\ and\
  \citenamefont {Chen}}]{Zhang:2017zfe}%
  \BibitemOpen
  \bibfield  {author} {\bibinfo {author} {\bibfnamefont {J.-H.}\ \bibnamefont
  {Zhang}}, \bibinfo {author} {\bibfnamefont {L.}~\bibnamefont {Jin}}, \bibinfo
  {author} {\bibfnamefont {H.-W.}\ \bibnamefont {Lin}}, \bibinfo {author}
  {\bibfnamefont {A.}~\bibnamefont {Sch\"afer}}, \bibinfo {author}
  {\bibfnamefont {P.}~\bibnamefont {Sun}}, \bibinfo {author} {\bibfnamefont
  {Y.-B.}\ \bibnamefont {Yang}}, \bibinfo {author} {\bibfnamefont
  {R.}~\bibnamefont {Zhang}}, \bibinfo {author} {\bibfnamefont
  {Y.}~\bibnamefont {Zhao}}, \ and\ \bibinfo {author} {\bibfnamefont {J.-W.}\
  \bibnamefont {Chen}} (\bibinfo {collaboration} {LP3}),\ }\href {\doibase
  10.1016/j.nuclphysb.2018.12.020} {\bibfield  {journal} {\bibinfo  {journal}
  {Nucl. Phys. B}\ }\textbf {\bibinfo {volume} {939}},\ \bibinfo {pages} {429}
  (\bibinfo {year} {2019}{\natexlab{a}})},\ \Eprint
  {http://arxiv.org/abs/1712.10025} {arXiv:1712.10025 [hep-ph]} \BibitemShut
  {NoStop}%
\bibitem [{\citenamefont {Alexandrou}\ \emph
  {et~al.}(2018{\natexlab{c}})\citenamefont {Alexandrou}, \citenamefont
  {Cichy}, \citenamefont {Constantinou}, \citenamefont {Jansen}, \citenamefont
  {Scapellato},\ and\ \citenamefont {Steffens}}]{Alexandrou:2018eet}%
  \BibitemOpen
  \bibfield  {author} {\bibinfo {author} {\bibfnamefont {C.}~\bibnamefont
  {Alexandrou}}, \bibinfo {author} {\bibfnamefont {K.}~\bibnamefont {Cichy}},
  \bibinfo {author} {\bibfnamefont {M.}~\bibnamefont {Constantinou}}, \bibinfo
  {author} {\bibfnamefont {K.}~\bibnamefont {Jansen}}, \bibinfo {author}
  {\bibfnamefont {A.}~\bibnamefont {Scapellato}}, \ and\ \bibinfo {author}
  {\bibfnamefont {F.}~\bibnamefont {Steffens}},\ }\href {\doibase
  10.1103/PhysRevD.98.091503} {\bibfield  {journal} {\bibinfo  {journal} {Phys.
  Rev. D}\ }\textbf {\bibinfo {volume} {98}},\ \bibinfo {pages} {091503}
  (\bibinfo {year} {2018}{\natexlab{c}})},\ \Eprint
  {http://arxiv.org/abs/1807.00232} {arXiv:1807.00232 [hep-lat]} \BibitemShut
  {NoStop}%
\bibitem [{\citenamefont {Alexandrou}\ \emph
  {et~al.}(2018{\natexlab{d}})\citenamefont {Alexandrou}, \citenamefont
  {Cichy}, \citenamefont {Constantinou}, \citenamefont {Jansen}, \citenamefont
  {Scapellato},\ and\ \citenamefont {Steffens}}]{Alexandrou:2018pbm}%
  \BibitemOpen
  \bibfield  {author} {\bibinfo {author} {\bibfnamefont {C.}~\bibnamefont
  {Alexandrou}}, \bibinfo {author} {\bibfnamefont {K.}~\bibnamefont {Cichy}},
  \bibinfo {author} {\bibfnamefont {M.}~\bibnamefont {Constantinou}}, \bibinfo
  {author} {\bibfnamefont {K.}~\bibnamefont {Jansen}}, \bibinfo {author}
  {\bibfnamefont {A.}~\bibnamefont {Scapellato}}, \ and\ \bibinfo {author}
  {\bibfnamefont {F.}~\bibnamefont {Steffens}},\ }\href {\doibase
  10.1103/PhysRevLett.121.112001} {\bibfield  {journal} {\bibinfo  {journal}
  {Phys. Rev. Lett.}\ }\textbf {\bibinfo {volume} {121}},\ \bibinfo {pages}
  {112001} (\bibinfo {year} {2018}{\natexlab{d}})},\ \Eprint
  {http://arxiv.org/abs/1803.02685} {arXiv:1803.02685 [hep-lat]} \BibitemShut
  {NoStop}%
\bibitem [{\citenamefont {Alexandrou}\ \emph
  {et~al.}(2018{\natexlab{e}})\citenamefont {Alexandrou}, \citenamefont
  {Cichy}, \citenamefont {Constantinou}, \citenamefont {Hadjiyiannakou},
  \citenamefont {Jansen}, \citenamefont {Scapellato},\ and\ \citenamefont
  {Steffens}}]{Alexandrou:2018yuy}%
  \BibitemOpen
  \bibfield  {author} {\bibinfo {author} {\bibfnamefont {C.}~\bibnamefont
  {Alexandrou}}, \bibinfo {author} {\bibfnamefont {K.}~\bibnamefont {Cichy}},
  \bibinfo {author} {\bibfnamefont {M.}~\bibnamefont {Constantinou}}, \bibinfo
  {author} {\bibfnamefont {K.}~\bibnamefont {Hadjiyiannakou}}, \bibinfo
  {author} {\bibfnamefont {K.}~\bibnamefont {Jansen}}, \bibinfo {author}
  {\bibfnamefont {A.}~\bibnamefont {Scapellato}}, \ and\ \bibinfo {author}
  {\bibfnamefont {F.}~\bibnamefont {Steffens}},\ }\href {\doibase
  10.22323/1.334.0094} {\bibfield  {journal} {\bibinfo  {journal} {PoS}\
  }\textbf {\bibinfo {volume} {LATTICE2018}},\ \bibinfo {pages} {094} (\bibinfo
  {year} {2018}{\natexlab{e}})},\ \Eprint {http://arxiv.org/abs/1811.01588}
  {arXiv:1811.01588 [hep-lat]} \BibitemShut {NoStop}%
\bibitem [{\citenamefont {Chen}\ \emph
  {et~al.}(2018{\natexlab{b}})\citenamefont {Chen}, \citenamefont {Jin},
  \citenamefont {Lin}, \citenamefont {Liu}, \citenamefont {Yang}, \citenamefont
  {Zhang},\ and\ \citenamefont {Zhao}}]{Chen:2018xof}%
  \BibitemOpen
  \bibfield  {author} {\bibinfo {author} {\bibfnamefont {J.-W.}\ \bibnamefont
  {Chen}}, \bibinfo {author} {\bibfnamefont {L.}~\bibnamefont {Jin}}, \bibinfo
  {author} {\bibfnamefont {H.-W.}\ \bibnamefont {Lin}}, \bibinfo {author}
  {\bibfnamefont {Y.-S.}\ \bibnamefont {Liu}}, \bibinfo {author} {\bibfnamefont
  {Y.-B.}\ \bibnamefont {Yang}}, \bibinfo {author} {\bibfnamefont {J.-H.}\
  \bibnamefont {Zhang}}, \ and\ \bibinfo {author} {\bibfnamefont
  {Y.}~\bibnamefont {Zhao}},\ }\href@noop {} {\  (\bibinfo {year}
  {2018}{\natexlab{b}})},\ \Eprint {http://arxiv.org/abs/1803.04393}
  {arXiv:1803.04393 [hep-lat]} \BibitemShut {NoStop}%
\bibitem [{\citenamefont {Ebert}\ \emph {et~al.}(2019)\citenamefont {Ebert},
  \citenamefont {Stewart},\ and\ \citenamefont {Zhao}}]{Ebert:2018gzl}%
  \BibitemOpen
  \bibfield  {author} {\bibinfo {author} {\bibfnamefont {M.~A.}\ \bibnamefont
  {Ebert}}, \bibinfo {author} {\bibfnamefont {I.~W.}\ \bibnamefont {Stewart}},
  \ and\ \bibinfo {author} {\bibfnamefont {Y.}~\bibnamefont {Zhao}},\ }\href
  {\doibase 10.1103/PhysRevD.99.034505} {\bibfield  {journal} {\bibinfo
  {journal} {Phys. Rev. D}\ }\textbf {\bibinfo {volume} {99}},\ \bibinfo
  {pages} {034505} (\bibinfo {year} {2019})},\ \Eprint
  {http://arxiv.org/abs/1811.00026} {arXiv:1811.00026 [hep-ph]} \BibitemShut
  {NoStop}%
\bibitem [{\citenamefont {Fan}\ \emph {et~al.}(2018)\citenamefont {Fan},
  \citenamefont {Yang}, \citenamefont {Anthony}, \citenamefont {Lin},\ and\
  \citenamefont {Liu}}]{Fan:2018dxu}%
  \BibitemOpen
  \bibfield  {author} {\bibinfo {author} {\bibfnamefont {Z.-Y.}\ \bibnamefont
  {Fan}}, \bibinfo {author} {\bibfnamefont {Y.-B.}\ \bibnamefont {Yang}},
  \bibinfo {author} {\bibfnamefont {A.}~\bibnamefont {Anthony}}, \bibinfo
  {author} {\bibfnamefont {H.-W.}\ \bibnamefont {Lin}}, \ and\ \bibinfo
  {author} {\bibfnamefont {K.-F.}\ \bibnamefont {Liu}},\ }\href {\doibase
  10.1103/PhysRevLett.121.242001} {\bibfield  {journal} {\bibinfo  {journal}
  {Phys. Rev. Lett.}\ }\textbf {\bibinfo {volume} {121}},\ \bibinfo {pages}
  {242001} (\bibinfo {year} {2018})},\ \Eprint
  {http://arxiv.org/abs/1808.02077} {arXiv:1808.02077 [hep-lat]} \BibitemShut
  {NoStop}%
\bibitem [{\citenamefont {Izubuchi}\ \emph {et~al.}(2018)\citenamefont
  {Izubuchi}, \citenamefont {Ji}, \citenamefont {Jin}, \citenamefont
  {Stewart},\ and\ \citenamefont {Zhao}}]{Izubuchi:2018srq}%
  \BibitemOpen
  \bibfield  {author} {\bibinfo {author} {\bibfnamefont {T.}~\bibnamefont
  {Izubuchi}}, \bibinfo {author} {\bibfnamefont {X.}~\bibnamefont {Ji}},
  \bibinfo {author} {\bibfnamefont {L.}~\bibnamefont {Jin}}, \bibinfo {author}
  {\bibfnamefont {I.~W.}\ \bibnamefont {Stewart}}, \ and\ \bibinfo {author}
  {\bibfnamefont {Y.}~\bibnamefont {Zhao}},\ }\href {\doibase
  10.1103/PhysRevD.98.056004} {\bibfield  {journal} {\bibinfo  {journal} {Phys.
  Rev. D}\ }\textbf {\bibinfo {volume} {98}},\ \bibinfo {pages} {056004}
  (\bibinfo {year} {2018})},\ \Eprint {http://arxiv.org/abs/1801.03917}
  {arXiv:1801.03917 [hep-ph]} \BibitemShut {NoStop}%
\bibitem [{\citenamefont {Ji}\ \emph {et~al.}(2019)\citenamefont {Ji},
  \citenamefont {Jin}, \citenamefont {Yuan}, \citenamefont {Zhang},\ and\
  \citenamefont {Zhao}}]{Ji:2018hvs}%
  \BibitemOpen
  \bibfield  {author} {\bibinfo {author} {\bibfnamefont {X.}~\bibnamefont
  {Ji}}, \bibinfo {author} {\bibfnamefont {L.-C.}\ \bibnamefont {Jin}},
  \bibinfo {author} {\bibfnamefont {F.}~\bibnamefont {Yuan}}, \bibinfo {author}
  {\bibfnamefont {J.-H.}\ \bibnamefont {Zhang}}, \ and\ \bibinfo {author}
  {\bibfnamefont {Y.}~\bibnamefont {Zhao}},\ }\href {\doibase
  10.1103/PhysRevD.99.114006} {\bibfield  {journal} {\bibinfo  {journal} {Phys.
  Rev. D}\ }\textbf {\bibinfo {volume} {99}},\ \bibinfo {pages} {114006}
  (\bibinfo {year} {2019})},\ \Eprint {http://arxiv.org/abs/1801.05930}
  {arXiv:1801.05930 [hep-ph]} \BibitemShut {NoStop}%
\bibitem [{\citenamefont {Liu}\ \emph {et~al.}(2020)\citenamefont {Liu} \emph
  {et~al.}}]{LatticeParton:2018gjr}%
  \BibitemOpen
  \bibfield  {author} {\bibinfo {author} {\bibfnamefont {Y.-S.}\ \bibnamefont
  {Liu}} \emph {et~al.} (\bibinfo {collaboration} {Lattice Parton}),\ }\href
  {\doibase 10.1103/PhysRevD.101.034020} {\bibfield  {journal} {\bibinfo
  {journal} {Phys. Rev. D}\ }\textbf {\bibinfo {volume} {101}},\ \bibinfo
  {pages} {034020} (\bibinfo {year} {2020})},\ \Eprint
  {http://arxiv.org/abs/1807.06566} {arXiv:1807.06566 [hep-lat]} \BibitemShut
  {NoStop}%
\bibitem [{\citenamefont {Lin}\ \emph {et~al.}(2018{\natexlab{b}})\citenamefont
  {Lin}, \citenamefont {Chen}, \citenamefont {Ji}, \citenamefont {Jin},
  \citenamefont {Li}, \citenamefont {Liu}, \citenamefont {Yang}, \citenamefont
  {Zhang},\ and\ \citenamefont {Zhao}}]{Lin:2018pvv}%
  \BibitemOpen
  \bibfield  {author} {\bibinfo {author} {\bibfnamefont {H.-W.}\ \bibnamefont
  {Lin}}, \bibinfo {author} {\bibfnamefont {J.-W.}\ \bibnamefont {Chen}},
  \bibinfo {author} {\bibfnamefont {X.}~\bibnamefont {Ji}}, \bibinfo {author}
  {\bibfnamefont {L.}~\bibnamefont {Jin}}, \bibinfo {author} {\bibfnamefont
  {R.}~\bibnamefont {Li}}, \bibinfo {author} {\bibfnamefont {Y.-S.}\
  \bibnamefont {Liu}}, \bibinfo {author} {\bibfnamefont {Y.-B.}\ \bibnamefont
  {Yang}}, \bibinfo {author} {\bibfnamefont {J.-H.}\ \bibnamefont {Zhang}}, \
  and\ \bibinfo {author} {\bibfnamefont {Y.}~\bibnamefont {Zhao}},\ }\href
  {\doibase 10.1103/PhysRevLett.121.242003} {\bibfield  {journal} {\bibinfo
  {journal} {Phys. Rev. Lett.}\ }\textbf {\bibinfo {volume} {121}},\ \bibinfo
  {pages} {242003} (\bibinfo {year} {2018}{\natexlab{b}})},\ \Eprint
  {http://arxiv.org/abs/1807.07431} {arXiv:1807.07431 [hep-lat]} \BibitemShut
  {NoStop}%
\bibitem [{\citenamefont {Liu}\ \emph {et~al.}(2018)\citenamefont {Liu},
  \citenamefont {Chen}, \citenamefont {Jin}, \citenamefont {Li}, \citenamefont
  {Lin}, \citenamefont {Yang}, \citenamefont {Zhang},\ and\ \citenamefont
  {Zhao}}]{Liu:2018hxv}%
  \BibitemOpen
  \bibfield  {author} {\bibinfo {author} {\bibfnamefont {Y.-S.}\ \bibnamefont
  {Liu}}, \bibinfo {author} {\bibfnamefont {J.-W.}\ \bibnamefont {Chen}},
  \bibinfo {author} {\bibfnamefont {L.}~\bibnamefont {Jin}}, \bibinfo {author}
  {\bibfnamefont {R.}~\bibnamefont {Li}}, \bibinfo {author} {\bibfnamefont
  {H.-W.}\ \bibnamefont {Lin}}, \bibinfo {author} {\bibfnamefont {Y.-B.}\
  \bibnamefont {Yang}}, \bibinfo {author} {\bibfnamefont {J.-H.}\ \bibnamefont
  {Zhang}}, \ and\ \bibinfo {author} {\bibfnamefont {Y.}~\bibnamefont {Zhao}},\
  }\href@noop {} {\  (\bibinfo {year} {2018})},\ \Eprint
  {http://arxiv.org/abs/1810.05043} {arXiv:1810.05043 [hep-lat]} \BibitemShut
  {NoStop}%
\bibitem [{\citenamefont {Liu}\ \emph {et~al.}(2019{\natexlab{a}})\citenamefont
  {Liu}, \citenamefont {Wang}, \citenamefont {Xu}, \citenamefont {Zhang},
  \citenamefont {Zhao},\ and\ \citenamefont {Zhao}}]{Liu:2018tox}%
  \BibitemOpen
  \bibfield  {author} {\bibinfo {author} {\bibfnamefont {Y.-S.}\ \bibnamefont
  {Liu}}, \bibinfo {author} {\bibfnamefont {W.}~\bibnamefont {Wang}}, \bibinfo
  {author} {\bibfnamefont {J.}~\bibnamefont {Xu}}, \bibinfo {author}
  {\bibfnamefont {Q.-A.}\ \bibnamefont {Zhang}}, \bibinfo {author}
  {\bibfnamefont {S.}~\bibnamefont {Zhao}}, \ and\ \bibinfo {author}
  {\bibfnamefont {Y.}~\bibnamefont {Zhao}},\ }\href {\doibase
  10.1103/PhysRevD.99.094036} {\bibfield  {journal} {\bibinfo  {journal} {Phys.
  Rev. D}\ }\textbf {\bibinfo {volume} {99}},\ \bibinfo {pages} {094036}
  (\bibinfo {year} {2019}{\natexlab{a}})},\ \Eprint
  {http://arxiv.org/abs/1810.10879} {arXiv:1810.10879 [hep-ph]} \BibitemShut
  {NoStop}%
\bibitem [{\citenamefont {Zhang}\ \emph
  {et~al.}(2019{\natexlab{b}})\citenamefont {Zhang}, \citenamefont {Ji},
  \citenamefont {Sch\"afer}, \citenamefont {Wang},\ and\ \citenamefont
  {Zhao}}]{Zhang:2018diq}%
  \BibitemOpen
  \bibfield  {author} {\bibinfo {author} {\bibfnamefont {J.-H.}\ \bibnamefont
  {Zhang}}, \bibinfo {author} {\bibfnamefont {X.}~\bibnamefont {Ji}}, \bibinfo
  {author} {\bibfnamefont {A.}~\bibnamefont {Sch\"afer}}, \bibinfo {author}
  {\bibfnamefont {W.}~\bibnamefont {Wang}}, \ and\ \bibinfo {author}
  {\bibfnamefont {S.}~\bibnamefont {Zhao}},\ }\href {\doibase
  10.1103/PhysRevLett.122.142001} {\bibfield  {journal} {\bibinfo  {journal}
  {Phys. Rev. Lett.}\ }\textbf {\bibinfo {volume} {122}},\ \bibinfo {pages}
  {142001} (\bibinfo {year} {2019}{\natexlab{b}})},\ \Eprint
  {http://arxiv.org/abs/1808.10824} {arXiv:1808.10824 [hep-ph]} \BibitemShut
  {NoStop}%
\bibitem [{\citenamefont {Zhang}\ \emph
  {et~al.}(2019{\natexlab{c}})\citenamefont {Zhang}, \citenamefont {Chen},
  \citenamefont {Jin}, \citenamefont {Lin}, \citenamefont {Sch\"afer},\ and\
  \citenamefont {Zhao}}]{Zhang:2018nsy}%
  \BibitemOpen
  \bibfield  {author} {\bibinfo {author} {\bibfnamefont {J.-H.}\ \bibnamefont
  {Zhang}}, \bibinfo {author} {\bibfnamefont {J.-W.}\ \bibnamefont {Chen}},
  \bibinfo {author} {\bibfnamefont {L.}~\bibnamefont {Jin}}, \bibinfo {author}
  {\bibfnamefont {H.-W.}\ \bibnamefont {Lin}}, \bibinfo {author} {\bibfnamefont
  {A.}~\bibnamefont {Sch\"afer}}, \ and\ \bibinfo {author} {\bibfnamefont
  {Y.}~\bibnamefont {Zhao}},\ }\href {\doibase 10.1103/PhysRevD.100.034505}
  {\bibfield  {journal} {\bibinfo  {journal} {Phys. Rev. D}\ }\textbf {\bibinfo
  {volume} {100}},\ \bibinfo {pages} {034505} (\bibinfo {year}
  {2019}{\natexlab{c}})},\ \Eprint {http://arxiv.org/abs/1804.01483}
  {arXiv:1804.01483 [hep-lat]} \BibitemShut {NoStop}%
\bibitem [{\citenamefont {Zhang}\ \emph
  {et~al.}(2018{\natexlab{a}})\citenamefont {Zhang}, \citenamefont {Chen},
  \citenamefont {Jin}, \citenamefont {Lin}, \citenamefont {Liu}, \citenamefont
  {Sch\"afer}, \citenamefont {Yang},\ and\ \citenamefont
  {Zhao}}]{Zhang:2018rls}%
  \BibitemOpen
  \bibfield  {author} {\bibinfo {author} {\bibfnamefont {J.}~\bibnamefont
  {Zhang}}, \bibinfo {author} {\bibfnamefont {J.-W.}\ \bibnamefont {Chen}},
  \bibinfo {author} {\bibfnamefont {L.}~\bibnamefont {Jin}}, \bibinfo {author}
  {\bibfnamefont {H.-W.}\ \bibnamefont {Lin}}, \bibinfo {author} {\bibfnamefont
  {Y.-S.}\ \bibnamefont {Liu}}, \bibinfo {author} {\bibfnamefont
  {A.}~\bibnamefont {Sch\"afer}}, \bibinfo {author} {\bibfnamefont {Y.-B.}\
  \bibnamefont {Yang}}, \ and\ \bibinfo {author} {\bibfnamefont
  {Y.}~\bibnamefont {Zhao}},\ }\href {\doibase 10.22323/1.334.0108} {\bibfield
  {journal} {\bibinfo  {journal} {PoS}\ }\textbf {\bibinfo {volume}
  {LATTICE2018}},\ \bibinfo {pages} {108} (\bibinfo {year}
  {2018}{\natexlab{a}})}\BibitemShut {NoStop}%
\bibitem [{\citenamefont {Zhao}(2019)}]{Zhao:2018fyu}%
  \BibitemOpen
  \bibfield  {author} {\bibinfo {author} {\bibfnamefont {Y.}~\bibnamefont
  {Zhao}},\ }\href {\doibase 10.1142/S0217751X18300338} {\bibfield  {journal}
  {\bibinfo  {journal} {Int. J. Mod. Phys. A}\ }\textbf {\bibinfo {volume}
  {33}},\ \bibinfo {pages} {1830033} (\bibinfo {year} {2019})},\ \Eprint
  {http://arxiv.org/abs/1812.07192} {arXiv:1812.07192 [hep-ph]} \BibitemShut
  {NoStop}%
\bibitem [{\citenamefont {Alexandrou}\ \emph
  {et~al.}(2019{\natexlab{a}})\citenamefont {Alexandrou}, \citenamefont
  {Cichy}, \citenamefont {Constantinou}, \citenamefont {Hadjiyiannakou},
  \citenamefont {Jansen}, \citenamefont {Scapellato},\ and\ \citenamefont
  {Steffens}}]{Alexandrou:2019dax}%
  \BibitemOpen
  \bibfield  {author} {\bibinfo {author} {\bibfnamefont {C.}~\bibnamefont
  {Alexandrou}}, \bibinfo {author} {\bibfnamefont {K.}~\bibnamefont {Cichy}},
  \bibinfo {author} {\bibfnamefont {M.}~\bibnamefont {Constantinou}}, \bibinfo
  {author} {\bibfnamefont {K.}~\bibnamefont {Hadjiyiannakou}}, \bibinfo
  {author} {\bibfnamefont {K.}~\bibnamefont {Jansen}}, \bibinfo {author}
  {\bibfnamefont {A.}~\bibnamefont {Scapellato}}, \ and\ \bibinfo {author}
  {\bibfnamefont {F.}~\bibnamefont {Steffens}},\ }\href {\doibase
  10.22323/1.363.0036} {\bibfield  {journal} {\bibinfo  {journal} {PoS}\
  }\textbf {\bibinfo {volume} {LATTICE2019}},\ \bibinfo {pages} {036} (\bibinfo
  {year} {2019}{\natexlab{a}})},\ \Eprint {http://arxiv.org/abs/1910.13229}
  {arXiv:1910.13229 [hep-lat]} \BibitemShut {NoStop}%
\bibitem [{\citenamefont {Alexandrou}\ \emph
  {et~al.}(2019{\natexlab{b}})\citenamefont {Alexandrou}, \citenamefont
  {Cichy}, \citenamefont {Constantinou}, \citenamefont {Hadjiyiannakou},
  \citenamefont {Jansen}, \citenamefont {Scapellato},\ and\ \citenamefont
  {Steffens}}]{Alexandrou:2019lfo}%
  \BibitemOpen
  \bibfield  {author} {\bibinfo {author} {\bibfnamefont {C.}~\bibnamefont
  {Alexandrou}}, \bibinfo {author} {\bibfnamefont {K.}~\bibnamefont {Cichy}},
  \bibinfo {author} {\bibfnamefont {M.}~\bibnamefont {Constantinou}}, \bibinfo
  {author} {\bibfnamefont {K.}~\bibnamefont {Hadjiyiannakou}}, \bibinfo
  {author} {\bibfnamefont {K.}~\bibnamefont {Jansen}}, \bibinfo {author}
  {\bibfnamefont {A.}~\bibnamefont {Scapellato}}, \ and\ \bibinfo {author}
  {\bibfnamefont {F.}~\bibnamefont {Steffens}},\ }\href {\doibase
  10.1103/PhysRevD.99.114504} {\bibfield  {journal} {\bibinfo  {journal} {Phys.
  Rev. D}\ }\textbf {\bibinfo {volume} {99}},\ \bibinfo {pages} {114504}
  (\bibinfo {year} {2019}{\natexlab{b}})},\ \Eprint
  {http://arxiv.org/abs/1902.00587} {arXiv:1902.00587 [hep-lat]} \BibitemShut
  {NoStop}%
\bibitem [{\citenamefont {Chai}\ \emph {et~al.}(2019)\citenamefont {Chai} \emph
  {et~al.}}]{Chai:2019rer}%
  \BibitemOpen
  \bibfield  {author} {\bibinfo {author} {\bibfnamefont {Y.}~\bibnamefont
  {Chai}} \emph {et~al.},\ }\href {\doibase 10.22323/1.363.0270} {\bibfield
  {journal} {\bibinfo  {journal} {PoS}\ }\textbf {\bibinfo {volume}
  {LATTICE2019}},\ \bibinfo {pages} {270} (\bibinfo {year} {2019})},\ \Eprint
  {http://arxiv.org/abs/1907.09827} {arXiv:1907.09827 [hep-lat]} \BibitemShut
  {NoStop}%
\bibitem [{\citenamefont {Chen}\ \emph
  {et~al.}(2020{\natexlab{a}})\citenamefont {Chen}, \citenamefont {Lin},\ and\
  \citenamefont {Zhang}}]{Chen:2019lcm}%
  \BibitemOpen
  \bibfield  {author} {\bibinfo {author} {\bibfnamefont {J.-W.}\ \bibnamefont
  {Chen}}, \bibinfo {author} {\bibfnamefont {H.-W.}\ \bibnamefont {Lin}}, \
  and\ \bibinfo {author} {\bibfnamefont {J.-H.}\ \bibnamefont {Zhang}},\ }\href
  {\doibase 10.1016/j.nuclphysb.2020.114940} {\bibfield  {journal} {\bibinfo
  {journal} {Nucl. Phys. B}\ }\textbf {\bibinfo {volume} {952}},\ \bibinfo
  {pages} {114940} (\bibinfo {year} {2020}{\natexlab{a}})},\ \Eprint
  {http://arxiv.org/abs/1904.12376} {arXiv:1904.12376 [hep-lat]} \BibitemShut
  {NoStop}%
\bibitem [{\citenamefont {Constantinou}\ \emph {et~al.}(2019)\citenamefont
  {Constantinou}, \citenamefont {Panagopoulos},\ and\ \citenamefont
  {Spanoudes}}]{Constantinou:2019vyb}%
  \BibitemOpen
  \bibfield  {author} {\bibinfo {author} {\bibfnamefont {M.}~\bibnamefont
  {Constantinou}}, \bibinfo {author} {\bibfnamefont {H.}~\bibnamefont
  {Panagopoulos}}, \ and\ \bibinfo {author} {\bibfnamefont {G.}~\bibnamefont
  {Spanoudes}},\ }\href {\doibase 10.1103/PhysRevD.99.074508} {\bibfield
  {journal} {\bibinfo  {journal} {Phys. Rev. D}\ }\textbf {\bibinfo {volume}
  {99}},\ \bibinfo {pages} {074508} (\bibinfo {year} {2019})},\ \Eprint
  {http://arxiv.org/abs/1901.03862} {arXiv:1901.03862 [hep-lat]} \BibitemShut
  {NoStop}%
\bibitem [{\citenamefont {Ebert}\ \emph
  {et~al.}(2020{\natexlab{a}})\citenamefont {Ebert}, \citenamefont {Stewart},\
  and\ \citenamefont {Zhao}}]{Ebert:2019tvc}%
  \BibitemOpen
  \bibfield  {author} {\bibinfo {author} {\bibfnamefont {M.~A.}\ \bibnamefont
  {Ebert}}, \bibinfo {author} {\bibfnamefont {I.~W.}\ \bibnamefont {Stewart}},
  \ and\ \bibinfo {author} {\bibfnamefont {Y.}~\bibnamefont {Zhao}},\ }\href
  {\doibase 10.1007/JHEP03(2020)099} {\bibfield  {journal} {\bibinfo  {journal}
  {JHEP}\ }\textbf {\bibinfo {volume} {03}},\ \bibinfo {pages} {099} (\bibinfo
  {year} {2020}{\natexlab{a}})},\ \Eprint {http://arxiv.org/abs/1910.08569}
  {arXiv:1910.08569 [hep-ph]} \BibitemShut {NoStop}%
\bibitem [{\citenamefont {Liu}\ \emph {et~al.}(2019{\natexlab{b}})\citenamefont
  {Liu}, \citenamefont {Wang}, \citenamefont {Xu}, \citenamefont {Zhang},
  \citenamefont {Zhang}, \citenamefont {Zhao},\ and\ \citenamefont
  {Zhao}}]{Liu:2019urm}%
  \BibitemOpen
  \bibfield  {author} {\bibinfo {author} {\bibfnamefont {Y.-S.}\ \bibnamefont
  {Liu}}, \bibinfo {author} {\bibfnamefont {W.}~\bibnamefont {Wang}}, \bibinfo
  {author} {\bibfnamefont {J.}~\bibnamefont {Xu}}, \bibinfo {author}
  {\bibfnamefont {Q.-A.}\ \bibnamefont {Zhang}}, \bibinfo {author}
  {\bibfnamefont {J.-H.}\ \bibnamefont {Zhang}}, \bibinfo {author}
  {\bibfnamefont {S.}~\bibnamefont {Zhao}}, \ and\ \bibinfo {author}
  {\bibfnamefont {Y.}~\bibnamefont {Zhao}},\ }\href {\doibase
  10.1103/PhysRevD.100.034006} {\bibfield  {journal} {\bibinfo  {journal}
  {Phys. Rev. D}\ }\textbf {\bibinfo {volume} {100}},\ \bibinfo {pages}
  {034006} (\bibinfo {year} {2019}{\natexlab{b}})},\ \Eprint
  {http://arxiv.org/abs/1902.00307} {arXiv:1902.00307 [hep-ph]} \BibitemShut
  {NoStop}%
\bibitem [{\citenamefont {Shanahan}\ \emph
  {et~al.}(2020{\natexlab{a}})\citenamefont {Shanahan}, \citenamefont
  {Wagman},\ and\ \citenamefont {Zhao}}]{Shanahan:2019zcq}%
  \BibitemOpen
  \bibfield  {author} {\bibinfo {author} {\bibfnamefont {P.}~\bibnamefont
  {Shanahan}}, \bibinfo {author} {\bibfnamefont {M.~L.}\ \bibnamefont
  {Wagman}}, \ and\ \bibinfo {author} {\bibfnamefont {Y.}~\bibnamefont
  {Zhao}},\ }\href {\doibase 10.1103/PhysRevD.101.074505} {\bibfield  {journal}
  {\bibinfo  {journal} {Phys. Rev. D}\ }\textbf {\bibinfo {volume} {101}},\
  \bibinfo {pages} {074505} (\bibinfo {year} {2020}{\natexlab{a}})},\ \Eprint
  {http://arxiv.org/abs/1911.00800} {arXiv:1911.00800 [hep-lat]} \BibitemShut
  {NoStop}%
\bibitem [{\citenamefont {Wang}\ \emph {et~al.}(2020)\citenamefont {Wang},
  \citenamefont {Wang}, \citenamefont {Xu},\ and\ \citenamefont
  {Zhao}}]{Wang:2019msf}%
  \BibitemOpen
  \bibfield  {author} {\bibinfo {author} {\bibfnamefont {W.}~\bibnamefont
  {Wang}}, \bibinfo {author} {\bibfnamefont {Y.-M.}\ \bibnamefont {Wang}},
  \bibinfo {author} {\bibfnamefont {J.}~\bibnamefont {Xu}}, \ and\ \bibinfo
  {author} {\bibfnamefont {S.}~\bibnamefont {Zhao}},\ }\href {\doibase
  10.1103/PhysRevD.102.011502} {\bibfield  {journal} {\bibinfo  {journal}
  {Phys. Rev. D}\ }\textbf {\bibinfo {volume} {102}},\ \bibinfo {pages}
  {011502} (\bibinfo {year} {2020})},\ \Eprint
  {http://arxiv.org/abs/1908.09933} {arXiv:1908.09933 [hep-ph]} \BibitemShut
  {NoStop}%
\bibitem [{\citenamefont {Zhang}\ \emph
  {et~al.}(2020{\natexlab{a}})\citenamefont {Zhang}, \citenamefont {Fan},
  \citenamefont {Li}, \citenamefont {Lin},\ and\ \citenamefont
  {Yoon}}]{Zhang:2019qiq}%
  \BibitemOpen
  \bibfield  {author} {\bibinfo {author} {\bibfnamefont {R.}~\bibnamefont
  {Zhang}}, \bibinfo {author} {\bibfnamefont {Z.}~\bibnamefont {Fan}}, \bibinfo
  {author} {\bibfnamefont {R.}~\bibnamefont {Li}}, \bibinfo {author}
  {\bibfnamefont {H.-W.}\ \bibnamefont {Lin}}, \ and\ \bibinfo {author}
  {\bibfnamefont {B.}~\bibnamefont {Yoon}},\ }\href {\doibase
  10.1103/PhysRevD.101.034516} {\bibfield  {journal} {\bibinfo  {journal}
  {Phys. Rev. D}\ }\textbf {\bibinfo {volume} {101}},\ \bibinfo {pages}
  {034516} (\bibinfo {year} {2020}{\natexlab{a}})},\ \Eprint
  {http://arxiv.org/abs/1909.10990} {arXiv:1909.10990 [hep-lat]} \BibitemShut
  {NoStop}%
\bibitem [{\citenamefont {Alexandrou}\ \emph
  {et~al.}(2021{\natexlab{a}})\citenamefont {Alexandrou}, \citenamefont
  {Cichy}, \citenamefont {Constantinou}, \citenamefont {Green}, \citenamefont
  {Hadjiyiannakou}, \citenamefont {Jansen}, \citenamefont {Manigrasso},
  \citenamefont {Scapellato},\ and\ \citenamefont
  {Steffens}}]{Alexandrou:2020qtt}%
  \BibitemOpen
  \bibfield  {author} {\bibinfo {author} {\bibfnamefont {C.}~\bibnamefont
  {Alexandrou}}, \bibinfo {author} {\bibfnamefont {K.}~\bibnamefont {Cichy}},
  \bibinfo {author} {\bibfnamefont {M.}~\bibnamefont {Constantinou}}, \bibinfo
  {author} {\bibfnamefont {J.~R.}\ \bibnamefont {Green}}, \bibinfo {author}
  {\bibfnamefont {K.}~\bibnamefont {Hadjiyiannakou}}, \bibinfo {author}
  {\bibfnamefont {K.}~\bibnamefont {Jansen}}, \bibinfo {author} {\bibfnamefont
  {F.}~\bibnamefont {Manigrasso}}, \bibinfo {author} {\bibfnamefont
  {A.}~\bibnamefont {Scapellato}}, \ and\ \bibinfo {author} {\bibfnamefont
  {F.}~\bibnamefont {Steffens}},\ }\href {\doibase 10.1103/PhysRevD.103.094512}
  {\bibfield  {journal} {\bibinfo  {journal} {Phys. Rev. D}\ }\textbf {\bibinfo
  {volume} {103}},\ \bibinfo {pages} {094512} (\bibinfo {year}
  {2021}{\natexlab{a}})},\ \Eprint {http://arxiv.org/abs/2011.00964}
  {arXiv:2011.00964 [hep-lat]} \BibitemShut {NoStop}%
\bibitem [{\citenamefont {Alexandrou}\ \emph
  {et~al.}(2021{\natexlab{b}})\citenamefont {Alexandrou}, \citenamefont
  {Constantinou}, \citenamefont {Hadjiyiannakou}, \citenamefont {Jansen},\ and\
  \citenamefont {Manigrasso}}]{Alexandrou:2020uyt}%
  \BibitemOpen
  \bibfield  {author} {\bibinfo {author} {\bibfnamefont {C.}~\bibnamefont
  {Alexandrou}}, \bibinfo {author} {\bibfnamefont {M.}~\bibnamefont
  {Constantinou}}, \bibinfo {author} {\bibfnamefont {K.}~\bibnamefont
  {Hadjiyiannakou}}, \bibinfo {author} {\bibfnamefont {K.}~\bibnamefont
  {Jansen}}, \ and\ \bibinfo {author} {\bibfnamefont {F.}~\bibnamefont
  {Manigrasso}},\ }\href {\doibase 10.1103/PhysRevLett.126.102003} {\bibfield
  {journal} {\bibinfo  {journal} {Phys. Rev. Lett.}\ }\textbf {\bibinfo
  {volume} {126}},\ \bibinfo {pages} {102003} (\bibinfo {year}
  {2021}{\natexlab{b}})},\ \Eprint {http://arxiv.org/abs/2009.13061}
  {arXiv:2009.13061 [hep-lat]} \BibitemShut {NoStop}%
\bibitem [{\citenamefont {Alexandrou}\ \emph {et~al.}(2020)\citenamefont
  {Alexandrou}, \citenamefont {Cichy}, \citenamefont {Constantinou},
  \citenamefont {Hadjiyiannakou}, \citenamefont {Jansen}, \citenamefont
  {Scapellato},\ and\ \citenamefont {Steffens}}]{Alexandrou:2020zbe}%
  \BibitemOpen
  \bibfield  {author} {\bibinfo {author} {\bibfnamefont {C.}~\bibnamefont
  {Alexandrou}}, \bibinfo {author} {\bibfnamefont {K.}~\bibnamefont {Cichy}},
  \bibinfo {author} {\bibfnamefont {M.}~\bibnamefont {Constantinou}}, \bibinfo
  {author} {\bibfnamefont {K.}~\bibnamefont {Hadjiyiannakou}}, \bibinfo
  {author} {\bibfnamefont {K.}~\bibnamefont {Jansen}}, \bibinfo {author}
  {\bibfnamefont {A.}~\bibnamefont {Scapellato}}, \ and\ \bibinfo {author}
  {\bibfnamefont {F.}~\bibnamefont {Steffens}},\ }\href {\doibase
  10.1103/PhysRevLett.125.262001} {\bibfield  {journal} {\bibinfo  {journal}
  {Phys. Rev. Lett.}\ }\textbf {\bibinfo {volume} {125}},\ \bibinfo {pages}
  {262001} (\bibinfo {year} {2020})},\ \Eprint
  {http://arxiv.org/abs/2008.10573} {arXiv:2008.10573 [hep-lat]} \BibitemShut
  {NoStop}%
\bibitem [{\citenamefont {Bhattacharya}\ \emph
  {et~al.}(2020{\natexlab{a}})\citenamefont {Bhattacharya}, \citenamefont
  {Cichy}, \citenamefont {Constantinou}, \citenamefont {Metz}, \citenamefont
  {Scapellato},\ and\ \citenamefont {Steffens}}]{Bhattacharya:2020cen}%
  \BibitemOpen
  \bibfield  {author} {\bibinfo {author} {\bibfnamefont {S.}~\bibnamefont
  {Bhattacharya}}, \bibinfo {author} {\bibfnamefont {K.}~\bibnamefont {Cichy}},
  \bibinfo {author} {\bibfnamefont {M.}~\bibnamefont {Constantinou}}, \bibinfo
  {author} {\bibfnamefont {A.}~\bibnamefont {Metz}}, \bibinfo {author}
  {\bibfnamefont {A.}~\bibnamefont {Scapellato}}, \ and\ \bibinfo {author}
  {\bibfnamefont {F.}~\bibnamefont {Steffens}},\ }\href {\doibase
  10.1103/PhysRevD.102.111501} {\bibfield  {journal} {\bibinfo  {journal}
  {Phys. Rev. D}\ }\textbf {\bibinfo {volume} {102}},\ \bibinfo {pages}
  {111501} (\bibinfo {year} {2020}{\natexlab{a}})},\ \Eprint
  {http://arxiv.org/abs/2004.04130} {arXiv:2004.04130 [hep-lat]} \BibitemShut
  {NoStop}%
\bibitem [{\citenamefont {Bhattacharya}\ \emph
  {et~al.}(2020{\natexlab{b}})\citenamefont {Bhattacharya}, \citenamefont
  {Cichy}, \citenamefont {Constantinou}, \citenamefont {Metz}, \citenamefont
  {Scapellato},\ and\ \citenamefont {Steffens}}]{Bhattacharya:2020jfj}%
  \BibitemOpen
  \bibfield  {author} {\bibinfo {author} {\bibfnamefont {S.}~\bibnamefont
  {Bhattacharya}}, \bibinfo {author} {\bibfnamefont {K.}~\bibnamefont {Cichy}},
  \bibinfo {author} {\bibfnamefont {M.}~\bibnamefont {Constantinou}}, \bibinfo
  {author} {\bibfnamefont {A.}~\bibnamefont {Metz}}, \bibinfo {author}
  {\bibfnamefont {A.}~\bibnamefont {Scapellato}}, \ and\ \bibinfo {author}
  {\bibfnamefont {F.}~\bibnamefont {Steffens}},\ }\href {\doibase
  10.1103/PhysRevD.102.114025} {\bibfield  {journal} {\bibinfo  {journal}
  {Phys. Rev. D}\ }\textbf {\bibinfo {volume} {102}},\ \bibinfo {pages}
  {114025} (\bibinfo {year} {2020}{\natexlab{b}})},\ \Eprint
  {http://arxiv.org/abs/2006.12347} {arXiv:2006.12347 [hep-ph]} \BibitemShut
  {NoStop}%
\bibitem [{\citenamefont {Chai}\ \emph {et~al.}(2020)\citenamefont {Chai} \emph
  {et~al.}}]{Chai:2020nxw}%
  \BibitemOpen
  \bibfield  {author} {\bibinfo {author} {\bibfnamefont {Y.}~\bibnamefont
  {Chai}} \emph {et~al.},\ }\href {\doibase 10.1103/PhysRevD.102.014508}
  {\bibfield  {journal} {\bibinfo  {journal} {Phys. Rev. D}\ }\textbf {\bibinfo
  {volume} {102}},\ \bibinfo {pages} {014508} (\bibinfo {year} {2020})},\
  \Eprint {http://arxiv.org/abs/2002.12044} {arXiv:2002.12044 [hep-lat]}
  \BibitemShut {NoStop}%
\bibitem [{\citenamefont {Chen}\ \emph
  {et~al.}(2020{\natexlab{b}})\citenamefont {Chen}, \citenamefont {Wang},\ and\
  \citenamefont {Zhu}}]{Chen:2020arf}%
  \BibitemOpen
  \bibfield  {author} {\bibinfo {author} {\bibfnamefont {L.-B.}\ \bibnamefont
  {Chen}}, \bibinfo {author} {\bibfnamefont {W.}~\bibnamefont {Wang}}, \ and\
  \bibinfo {author} {\bibfnamefont {R.}~\bibnamefont {Zhu}},\ }\href {\doibase
  10.1103/PhysRevD.102.011503} {\bibfield  {journal} {\bibinfo  {journal}
  {Phys. Rev. D}\ }\textbf {\bibinfo {volume} {102}},\ \bibinfo {pages}
  {011503} (\bibinfo {year} {2020}{\natexlab{b}})},\ \Eprint
  {http://arxiv.org/abs/2005.13757} {arXiv:2005.13757 [hep-ph]} \BibitemShut
  {NoStop}%
\bibitem [{\citenamefont {Chen}\ \emph
  {et~al.}(2020{\natexlab{c}})\citenamefont {Chen}, \citenamefont {Wang},\ and\
  \citenamefont {Zhu}}]{Chen:2020iqi}%
  \BibitemOpen
  \bibfield  {author} {\bibinfo {author} {\bibfnamefont {L.-B.}\ \bibnamefont
  {Chen}}, \bibinfo {author} {\bibfnamefont {W.}~\bibnamefont {Wang}}, \ and\
  \bibinfo {author} {\bibfnamefont {R.}~\bibnamefont {Zhu}},\ }\href {\doibase
  10.1007/JHEP10(2020)079} {\bibfield  {journal} {\bibinfo  {journal} {JHEP}\
  }\textbf {\bibinfo {volume} {10}},\ \bibinfo {pages} {079} (\bibinfo {year}
  {2020}{\natexlab{c}})},\ \Eprint {http://arxiv.org/abs/2006.10917}
  {arXiv:2006.10917 [hep-ph]} \BibitemShut {NoStop}%
\bibitem [{\citenamefont {Chen}\ \emph {et~al.}(2021)\citenamefont {Chen},
  \citenamefont {Wang},\ and\ \citenamefont {Zhu}}]{Chen:2020ody}%
  \BibitemOpen
  \bibfield  {author} {\bibinfo {author} {\bibfnamefont {L.-B.}\ \bibnamefont
  {Chen}}, \bibinfo {author} {\bibfnamefont {W.}~\bibnamefont {Wang}}, \ and\
  \bibinfo {author} {\bibfnamefont {R.}~\bibnamefont {Zhu}},\ }\href {\doibase
  10.1103/PhysRevLett.126.072002} {\bibfield  {journal} {\bibinfo  {journal}
  {Phys. Rev. Lett.}\ }\textbf {\bibinfo {volume} {126}},\ \bibinfo {pages}
  {072002} (\bibinfo {year} {2021})},\ \Eprint
  {http://arxiv.org/abs/2006.14825} {arXiv:2006.14825 [hep-ph]} \BibitemShut
  {NoStop}%
\bibitem [{\citenamefont {Ebert}\ \emph
  {et~al.}(2020{\natexlab{b}})\citenamefont {Ebert}, \citenamefont {Schindler},
  \citenamefont {Stewart},\ and\ \citenamefont {Zhao}}]{Ebert:2020gxr}%
  \BibitemOpen
  \bibfield  {author} {\bibinfo {author} {\bibfnamefont {M.~A.}\ \bibnamefont
  {Ebert}}, \bibinfo {author} {\bibfnamefont {S.~T.}\ \bibnamefont
  {Schindler}}, \bibinfo {author} {\bibfnamefont {I.~W.}\ \bibnamefont
  {Stewart}}, \ and\ \bibinfo {author} {\bibfnamefont {Y.}~\bibnamefont
  {Zhao}},\ }\href {\doibase 10.1007/JHEP09(2020)099} {\bibfield  {journal}
  {\bibinfo  {journal} {JHEP}\ }\textbf {\bibinfo {volume} {09}},\ \bibinfo
  {pages} {099} (\bibinfo {year} {2020}{\natexlab{b}})},\ \Eprint
  {http://arxiv.org/abs/2004.14831} {arXiv:2004.14831 [hep-ph]} \BibitemShut
  {NoStop}%
\bibitem [{\citenamefont {Fan}\ \emph {et~al.}(2020)\citenamefont {Fan},
  \citenamefont {Gao}, \citenamefont {Li}, \citenamefont {Lin}, \citenamefont
  {Karthik}, \citenamefont {Mukherjee}, \citenamefont {Petreczky},
  \citenamefont {Syritsyn}, \citenamefont {Yang},\ and\ \citenamefont
  {Zhang}}]{Fan:2020nzz}%
  \BibitemOpen
  \bibfield  {author} {\bibinfo {author} {\bibfnamefont {Z.}~\bibnamefont
  {Fan}}, \bibinfo {author} {\bibfnamefont {X.}~\bibnamefont {Gao}}, \bibinfo
  {author} {\bibfnamefont {R.}~\bibnamefont {Li}}, \bibinfo {author}
  {\bibfnamefont {H.-W.}\ \bibnamefont {Lin}}, \bibinfo {author} {\bibfnamefont
  {N.}~\bibnamefont {Karthik}}, \bibinfo {author} {\bibfnamefont
  {S.}~\bibnamefont {Mukherjee}}, \bibinfo {author} {\bibfnamefont
  {P.}~\bibnamefont {Petreczky}}, \bibinfo {author} {\bibfnamefont
  {S.}~\bibnamefont {Syritsyn}}, \bibinfo {author} {\bibfnamefont {Y.-B.}\
  \bibnamefont {Yang}}, \ and\ \bibinfo {author} {\bibfnamefont
  {R.}~\bibnamefont {Zhang}},\ }\href {\doibase 10.1103/PhysRevD.102.074504}
  {\bibfield  {journal} {\bibinfo  {journal} {Phys. Rev. D}\ }\textbf {\bibinfo
  {volume} {102}},\ \bibinfo {pages} {074504} (\bibinfo {year} {2020})},\
  \Eprint {http://arxiv.org/abs/2005.12015} {arXiv:2005.12015 [hep-lat]}
  \BibitemShut {NoStop}%
\bibitem [{\citenamefont {Gao}\ \emph {et~al.}(2020)\citenamefont {Gao},
  \citenamefont {Jin}, \citenamefont {Kallidonis}, \citenamefont {Karthik},
  \citenamefont {Mukherjee}, \citenamefont {Petreczky}, \citenamefont
  {Shugert}, \citenamefont {Syritsyn},\ and\ \citenamefont
  {Zhao}}]{Gao:2020ito}%
  \BibitemOpen
  \bibfield  {author} {\bibinfo {author} {\bibfnamefont {X.}~\bibnamefont
  {Gao}}, \bibinfo {author} {\bibfnamefont {L.}~\bibnamefont {Jin}}, \bibinfo
  {author} {\bibfnamefont {C.}~\bibnamefont {Kallidonis}}, \bibinfo {author}
  {\bibfnamefont {N.}~\bibnamefont {Karthik}}, \bibinfo {author} {\bibfnamefont
  {S.}~\bibnamefont {Mukherjee}}, \bibinfo {author} {\bibfnamefont
  {P.}~\bibnamefont {Petreczky}}, \bibinfo {author} {\bibfnamefont
  {C.}~\bibnamefont {Shugert}}, \bibinfo {author} {\bibfnamefont
  {S.}~\bibnamefont {Syritsyn}}, \ and\ \bibinfo {author} {\bibfnamefont
  {Y.}~\bibnamefont {Zhao}},\ }\href {\doibase 10.1103/PhysRevD.102.094513}
  {\bibfield  {journal} {\bibinfo  {journal} {Phys. Rev. D}\ }\textbf {\bibinfo
  {volume} {102}},\ \bibinfo {pages} {094513} (\bibinfo {year} {2020})},\
  \Eprint {http://arxiv.org/abs/2007.06590} {arXiv:2007.06590 [hep-lat]}
  \BibitemShut {NoStop}%
\bibitem [{\citenamefont {Hua}\ \emph {et~al.}(2021)\citenamefont {Hua},
  \citenamefont {Chu}, \citenamefont {Sun}, \citenamefont {Wang}, \citenamefont
  {Xu}, \citenamefont {Yang}, \citenamefont {Zhang},\ and\ \citenamefont
  {Zhang}}]{Hua:2020gnw}%
  \BibitemOpen
  \bibfield  {author} {\bibinfo {author} {\bibfnamefont {J.}~\bibnamefont
  {Hua}}, \bibinfo {author} {\bibfnamefont {M.-H.}\ \bibnamefont {Chu}},
  \bibinfo {author} {\bibfnamefont {P.}~\bibnamefont {Sun}}, \bibinfo {author}
  {\bibfnamefont {W.}~\bibnamefont {Wang}}, \bibinfo {author} {\bibfnamefont
  {J.}~\bibnamefont {Xu}}, \bibinfo {author} {\bibfnamefont {Y.-B.}\
  \bibnamefont {Yang}}, \bibinfo {author} {\bibfnamefont {J.-H.}\ \bibnamefont
  {Zhang}}, \ and\ \bibinfo {author} {\bibfnamefont {Q.-A.}\ \bibnamefont
  {Zhang}} (\bibinfo {collaboration} {Lattice Parton}),\ }\href {\doibase
  10.1103/PhysRevLett.127.062002} {\bibfield  {journal} {\bibinfo  {journal}
  {Phys. Rev. Lett.}\ }\textbf {\bibinfo {volume} {127}},\ \bibinfo {pages}
  {062002} (\bibinfo {year} {2021})},\ \Eprint
  {http://arxiv.org/abs/2011.09788} {arXiv:2011.09788 [hep-lat]} \BibitemShut
  {NoStop}%
\bibitem [{\citenamefont {Ji}\ \emph {et~al.}(2021{\natexlab{b}})\citenamefont
  {Ji}, \citenamefont {Liu}, \citenamefont {Sch\"afer}, \citenamefont {Wang},
  \citenamefont {Yang}, \citenamefont {Zhang},\ and\ \citenamefont
  {Zhao}}]{Ji:2020brr}%
  \BibitemOpen
  \bibfield  {author} {\bibinfo {author} {\bibfnamefont {X.}~\bibnamefont
  {Ji}}, \bibinfo {author} {\bibfnamefont {Y.}~\bibnamefont {Liu}}, \bibinfo
  {author} {\bibfnamefont {A.}~\bibnamefont {Sch\"afer}}, \bibinfo {author}
  {\bibfnamefont {W.}~\bibnamefont {Wang}}, \bibinfo {author} {\bibfnamefont
  {Y.-B.}\ \bibnamefont {Yang}}, \bibinfo {author} {\bibfnamefont {J.-H.}\
  \bibnamefont {Zhang}}, \ and\ \bibinfo {author} {\bibfnamefont
  {Y.}~\bibnamefont {Zhao}},\ }\href {\doibase 10.1016/j.nuclphysb.2021.115311}
  {\bibfield  {journal} {\bibinfo  {journal} {Nucl. Phys. B}\ }\textbf
  {\bibinfo {volume} {964}},\ \bibinfo {pages} {115311} (\bibinfo {year}
  {2021}{\natexlab{b}})},\ \Eprint {http://arxiv.org/abs/2008.03886}
  {arXiv:2008.03886 [hep-ph]} \BibitemShut {NoStop}%
\bibitem [{\citenamefont {Ji}\ \emph {et~al.}(2021{\natexlab{c}})\citenamefont
  {Ji}, \citenamefont {Liu}, \citenamefont {Sch\"afer},\ and\ \citenamefont
  {Yuan}}]{Ji:2020jeb}%
  \BibitemOpen
  \bibfield  {author} {\bibinfo {author} {\bibfnamefont {X.}~\bibnamefont
  {Ji}}, \bibinfo {author} {\bibfnamefont {Y.}~\bibnamefont {Liu}}, \bibinfo
  {author} {\bibfnamefont {A.}~\bibnamefont {Sch\"afer}}, \ and\ \bibinfo
  {author} {\bibfnamefont {F.}~\bibnamefont {Yuan}},\ }\href {\doibase
  10.1103/PhysRevD.103.074005} {\bibfield  {journal} {\bibinfo  {journal}
  {Phys. Rev. D}\ }\textbf {\bibinfo {volume} {103}},\ \bibinfo {pages}
  {074005} (\bibinfo {year} {2021}{\natexlab{c}})},\ \Eprint
  {http://arxiv.org/abs/2011.13397} {arXiv:2011.13397 [hep-ph]} \BibitemShut
  {NoStop}%
\bibitem [{\citenamefont {Zhang}\ \emph
  {et~al.}(2020{\natexlab{b}})\citenamefont {Zhang} \emph
  {et~al.}}]{LatticeParton:2020uhz}%
  \BibitemOpen
  \bibfield  {author} {\bibinfo {author} {\bibfnamefont {Q.-A.}\ \bibnamefont
  {Zhang}} \emph {et~al.} (\bibinfo {collaboration} {Lattice Parton}),\ }\href
  {\doibase 10.22323/1.396.0477} {\bibfield  {journal} {\bibinfo  {journal}
  {Phys. Rev. Lett.}\ }\textbf {\bibinfo {volume} {125}},\ \bibinfo {pages}
  {192001} (\bibinfo {year} {2020}{\natexlab{b}})},\ \Eprint
  {http://arxiv.org/abs/2005.14572} {arXiv:2005.14572 [hep-lat]} \BibitemShut
  {NoStop}%
\bibitem [{\citenamefont {Lin}\ \emph {et~al.}(2020)\citenamefont {Lin},
  \citenamefont {Chen},\ and\ \citenamefont {Zhang}}]{Lin:2020fsj}%
  \BibitemOpen
  \bibfield  {author} {\bibinfo {author} {\bibfnamefont {H.-W.}\ \bibnamefont
  {Lin}}, \bibinfo {author} {\bibfnamefont {J.-W.}\ \bibnamefont {Chen}}, \
  and\ \bibinfo {author} {\bibfnamefont {R.}~\bibnamefont {Zhang}},\
  }\href@noop {} {\  (\bibinfo {year} {2020})},\ \Eprint
  {http://arxiv.org/abs/2011.14971} {arXiv:2011.14971 [hep-lat]} \BibitemShut
  {NoStop}%
\bibitem [{\citenamefont {Lin}(2021)}]{Lin:2020rxa}%
  \BibitemOpen
  \bibfield  {author} {\bibinfo {author} {\bibfnamefont {H.-W.}\ \bibnamefont
  {Lin}},\ }\href {\doibase 10.1103/PhysRevLett.127.182001} {\bibfield
  {journal} {\bibinfo  {journal} {Phys. Rev. Lett.}\ }\textbf {\bibinfo
  {volume} {127}},\ \bibinfo {pages} {182001} (\bibinfo {year} {2021})},\
  \Eprint {http://arxiv.org/abs/2008.12474} {arXiv:2008.12474 [hep-ph]}
  \BibitemShut {NoStop}%
\bibitem [{\citenamefont {Lin}\ \emph {et~al.}(2021)\citenamefont {Lin},
  \citenamefont {Chen}, \citenamefont {Fan}, \citenamefont {Zhang},\ and\
  \citenamefont {Zhang}}]{Lin:2020ssv}%
  \BibitemOpen
  \bibfield  {author} {\bibinfo {author} {\bibfnamefont {H.-W.}\ \bibnamefont
  {Lin}}, \bibinfo {author} {\bibfnamefont {J.-W.}\ \bibnamefont {Chen}},
  \bibinfo {author} {\bibfnamefont {Z.}~\bibnamefont {Fan}}, \bibinfo {author}
  {\bibfnamefont {J.-H.}\ \bibnamefont {Zhang}}, \ and\ \bibinfo {author}
  {\bibfnamefont {R.}~\bibnamefont {Zhang}},\ }\href {\doibase
  10.1103/PhysRevD.103.014516} {\bibfield  {journal} {\bibinfo  {journal}
  {Phys. Rev. D}\ }\textbf {\bibinfo {volume} {103}},\ \bibinfo {pages}
  {014516} (\bibinfo {year} {2021})},\ \Eprint
  {http://arxiv.org/abs/2003.14128} {arXiv:2003.14128 [hep-lat]} \BibitemShut
  {NoStop}%
\bibitem [{\citenamefont {Shanahan}\ \emph
  {et~al.}(2020{\natexlab{b}})\citenamefont {Shanahan}, \citenamefont
  {Wagman},\ and\ \citenamefont {Zhao}}]{Shanahan:2020zxr}%
  \BibitemOpen
  \bibfield  {author} {\bibinfo {author} {\bibfnamefont {P.}~\bibnamefont
  {Shanahan}}, \bibinfo {author} {\bibfnamefont {M.}~\bibnamefont {Wagman}}, \
  and\ \bibinfo {author} {\bibfnamefont {Y.}~\bibnamefont {Zhao}},\ }\href
  {\doibase 10.1103/PhysRevD.102.014511} {\bibfield  {journal} {\bibinfo
  {journal} {Phys. Rev. D}\ }\textbf {\bibinfo {volume} {102}},\ \bibinfo
  {pages} {014511} (\bibinfo {year} {2020}{\natexlab{b}})},\ \Eprint
  {http://arxiv.org/abs/2003.06063} {arXiv:2003.06063 [hep-lat]} \BibitemShut
  {NoStop}%
\bibitem [{\citenamefont {Shugert}\ \emph {et~al.}(2020)\citenamefont
  {Shugert}, \citenamefont {Gao}, \citenamefont {Izubichi}, \citenamefont
  {Jin}, \citenamefont {Kallidonis}, \citenamefont {Karthik}, \citenamefont
  {Mukherjee}, \citenamefont {Petreczky}, \citenamefont {Syritsyn},\ and\
  \citenamefont {Zhao}}]{Shugert:2020tgq}%
  \BibitemOpen
  \bibfield  {author} {\bibinfo {author} {\bibfnamefont {C.}~\bibnamefont
  {Shugert}}, \bibinfo {author} {\bibfnamefont {X.}~\bibnamefont {Gao}},
  \bibinfo {author} {\bibfnamefont {T.}~\bibnamefont {Izubichi}}, \bibinfo
  {author} {\bibfnamefont {L.}~\bibnamefont {Jin}}, \bibinfo {author}
  {\bibfnamefont {C.}~\bibnamefont {Kallidonis}}, \bibinfo {author}
  {\bibfnamefont {N.}~\bibnamefont {Karthik}}, \bibinfo {author} {\bibfnamefont
  {S.}~\bibnamefont {Mukherjee}}, \bibinfo {author} {\bibfnamefont
  {P.}~\bibnamefont {Petreczky}}, \bibinfo {author} {\bibfnamefont
  {S.}~\bibnamefont {Syritsyn}}, \ and\ \bibinfo {author} {\bibfnamefont
  {Y.}~\bibnamefont {Zhao}},\ }in\ \href@noop {} {\emph {\bibinfo {booktitle}
  {{37th International Symposium on Lattice Field Theory}}}}\ (\bibinfo {year}
  {2020})\ \Eprint {http://arxiv.org/abs/2001.11650} {arXiv:2001.11650
  [hep-lat]} \BibitemShut {NoStop}%
\bibitem [{\citenamefont {Vladimirov}\ and\ \citenamefont
  {Sch\"afer}(2020)}]{Vladimirov:2020ofp}%
  \BibitemOpen
  \bibfield  {author} {\bibinfo {author} {\bibfnamefont {A.~A.}\ \bibnamefont
  {Vladimirov}}\ and\ \bibinfo {author} {\bibfnamefont {A.}~\bibnamefont
  {Sch\"afer}},\ }\href {\doibase 10.1103/PhysRevD.101.074517} {\bibfield
  {journal} {\bibinfo  {journal} {Phys. Rev. D}\ }\textbf {\bibinfo {volume}
  {101}},\ \bibinfo {pages} {074517} (\bibinfo {year} {2020})},\ \Eprint
  {http://arxiv.org/abs/2002.07527} {arXiv:2002.07527 [hep-ph]} \BibitemShut
  {NoStop}%
\bibitem [{\citenamefont {Zhang}\ \emph
  {et~al.}(2021{\natexlab{a}})\citenamefont {Zhang}, \citenamefont {Lin},\ and\
  \citenamefont {Yoon}}]{Zhang:2020dkn}%
  \BibitemOpen
  \bibfield  {author} {\bibinfo {author} {\bibfnamefont {R.}~\bibnamefont
  {Zhang}}, \bibinfo {author} {\bibfnamefont {H.-W.}\ \bibnamefont {Lin}}, \
  and\ \bibinfo {author} {\bibfnamefont {B.}~\bibnamefont {Yoon}},\ }\href
  {\doibase 10.1103/PhysRevD.104.094511} {\bibfield  {journal} {\bibinfo
  {journal} {Phys. Rev. D}\ }\textbf {\bibinfo {volume} {104}},\ \bibinfo
  {pages} {094511} (\bibinfo {year} {2021}{\natexlab{a}})},\ \Eprint
  {http://arxiv.org/abs/2005.01124} {arXiv:2005.01124 [hep-lat]} \BibitemShut
  {NoStop}%
\bibitem [{\citenamefont {Zhang}\ \emph
  {et~al.}(2020{\natexlab{c}})\citenamefont {Zhang}, \citenamefont {Honkala},
  \citenamefont {Lin},\ and\ \citenamefont {Chen}}]{Zhang:2020gaj}%
  \BibitemOpen
  \bibfield  {author} {\bibinfo {author} {\bibfnamefont {R.}~\bibnamefont
  {Zhang}}, \bibinfo {author} {\bibfnamefont {C.}~\bibnamefont {Honkala}},
  \bibinfo {author} {\bibfnamefont {H.-W.}\ \bibnamefont {Lin}}, \ and\
  \bibinfo {author} {\bibfnamefont {J.-W.}\ \bibnamefont {Chen}},\ }\href
  {\doibase 10.1103/PhysRevD.102.094519} {\bibfield  {journal} {\bibinfo
  {journal} {Phys. Rev. D}\ }\textbf {\bibinfo {volume} {102}},\ \bibinfo
  {pages} {094519} (\bibinfo {year} {2020}{\natexlab{c}})},\ \Eprint
  {http://arxiv.org/abs/2005.13955} {arXiv:2005.13955 [hep-lat]} \BibitemShut
  {NoStop}%
\bibitem [{\citenamefont {Zhang}\ \emph
  {et~al.}(2021{\natexlab{b}})\citenamefont {Zhang}, \citenamefont {Li},
  \citenamefont {Huo}, \citenamefont {Sch\"afer}, \citenamefont {Sun},\ and\
  \citenamefont {Yang}}]{Zhang:2020rsx}%
  \BibitemOpen
  \bibfield  {author} {\bibinfo {author} {\bibfnamefont {K.}~\bibnamefont
  {Zhang}}, \bibinfo {author} {\bibfnamefont {Y.-Y.}\ \bibnamefont {Li}},
  \bibinfo {author} {\bibfnamefont {Y.-K.}\ \bibnamefont {Huo}}, \bibinfo
  {author} {\bibfnamefont {A.}~\bibnamefont {Sch\"afer}}, \bibinfo {author}
  {\bibfnamefont {P.}~\bibnamefont {Sun}}, \ and\ \bibinfo {author}
  {\bibfnamefont {Y.-B.}\ \bibnamefont {Yang}} (\bibinfo {collaboration}
  {\ensuremath{\chi}QCD}),\ }\href {\doibase 10.1103/PhysRevD.104.074501}
  {\bibfield  {journal} {\bibinfo  {journal} {Phys. Rev. D}\ }\textbf {\bibinfo
  {volume} {104}},\ \bibinfo {pages} {074501} (\bibinfo {year}
  {2021}{\natexlab{b}})},\ \Eprint {http://arxiv.org/abs/2012.05448}
  {arXiv:2012.05448 [hep-lat]} \BibitemShut {NoStop}%
\bibitem [{\citenamefont {Alexandrou}\ \emph {et~al.}(2022)\citenamefont
  {Alexandrou}, \citenamefont {Cichy}, \citenamefont {Constantinou},
  \citenamefont {Hadjiyiannakou}, \citenamefont {Jansen}, \citenamefont
  {Scapellato},\ and\ \citenamefont {Steffens}}]{Alexandrou:2021bbo}%
  \BibitemOpen
  \bibfield  {author} {\bibinfo {author} {\bibfnamefont {C.}~\bibnamefont
  {Alexandrou}}, \bibinfo {author} {\bibfnamefont {K.}~\bibnamefont {Cichy}},
  \bibinfo {author} {\bibfnamefont {M.}~\bibnamefont {Constantinou}}, \bibinfo
  {author} {\bibfnamefont {K.}~\bibnamefont {Hadjiyiannakou}}, \bibinfo
  {author} {\bibfnamefont {K.}~\bibnamefont {Jansen}}, \bibinfo {author}
  {\bibfnamefont {A.}~\bibnamefont {Scapellato}}, \ and\ \bibinfo {author}
  {\bibfnamefont {F.}~\bibnamefont {Steffens}},\ }\href {\doibase
  10.1103/PhysRevD.105.034501} {\bibfield  {journal} {\bibinfo  {journal}
  {Phys. Rev. D}\ }\textbf {\bibinfo {volume} {105}},\ \bibinfo {pages}
  {034501} (\bibinfo {year} {2022})},\ \Eprint
  {http://arxiv.org/abs/2108.10789} {arXiv:2108.10789 [hep-lat]} \BibitemShut
  {NoStop}%
\bibitem [{\citenamefont {Alexandrou}\ \emph
  {et~al.}(2021{\natexlab{c}})\citenamefont {Alexandrou}, \citenamefont
  {Constantinou}, \citenamefont {Hadjiyiannakou}, \citenamefont {Jansen},\ and\
  \citenamefont {Manigrasso}}]{Alexandrou:2021oih}%
  \BibitemOpen
  \bibfield  {author} {\bibinfo {author} {\bibfnamefont {C.}~\bibnamefont
  {Alexandrou}}, \bibinfo {author} {\bibfnamefont {M.}~\bibnamefont
  {Constantinou}}, \bibinfo {author} {\bibfnamefont {K.}~\bibnamefont
  {Hadjiyiannakou}}, \bibinfo {author} {\bibfnamefont {K.}~\bibnamefont
  {Jansen}}, \ and\ \bibinfo {author} {\bibfnamefont {F.}~\bibnamefont
  {Manigrasso}},\ }\href {\doibase 10.1103/PhysRevD.104.054503} {\bibfield
  {journal} {\bibinfo  {journal} {Phys. Rev. D}\ }\textbf {\bibinfo {volume}
  {104}},\ \bibinfo {pages} {054503} (\bibinfo {year} {2021}{\natexlab{c}})},\
  \Eprint {http://arxiv.org/abs/2106.16065} {arXiv:2106.16065 [hep-lat]}
  \BibitemShut {NoStop}%
\bibitem [{\citenamefont {Bhattacharya}\ \emph {et~al.}(2021)\citenamefont
  {Bhattacharya}, \citenamefont {Cichy}, \citenamefont {Constantinou},
  \citenamefont {Metz}, \citenamefont {Scapellato},\ and\ \citenamefont
  {Steffens}}]{Bhattacharya:2021moj}%
  \BibitemOpen
  \bibfield  {author} {\bibinfo {author} {\bibfnamefont {S.}~\bibnamefont
  {Bhattacharya}}, \bibinfo {author} {\bibfnamefont {K.}~\bibnamefont {Cichy}},
  \bibinfo {author} {\bibfnamefont {M.}~\bibnamefont {Constantinou}}, \bibinfo
  {author} {\bibfnamefont {A.}~\bibnamefont {Metz}}, \bibinfo {author}
  {\bibfnamefont {A.}~\bibnamefont {Scapellato}}, \ and\ \bibinfo {author}
  {\bibfnamefont {F.}~\bibnamefont {Steffens}},\ }\href {\doibase
  10.1103/PhysRevD.104.114510} {\bibfield  {journal} {\bibinfo  {journal}
  {Phys. Rev. D}\ }\textbf {\bibinfo {volume} {104}},\ \bibinfo {pages}
  {114510} (\bibinfo {year} {2021})},\ \Eprint
  {http://arxiv.org/abs/2107.02574} {arXiv:2107.02574 [hep-lat]} \BibitemShut
  {NoStop}%
\bibitem [{\citenamefont {Bhattacharya}\ \emph
  {et~al.}(2022{\natexlab{a}})\citenamefont {Bhattacharya}, \citenamefont
  {Cichy}, \citenamefont {Constantinou}, \citenamefont {Metz}, \citenamefont
  {Scapellato},\ and\ \citenamefont {Steffens}}]{Bhattacharya:2021rua}%
  \BibitemOpen
  \bibfield  {author} {\bibinfo {author} {\bibfnamefont {S.}~\bibnamefont
  {Bhattacharya}}, \bibinfo {author} {\bibfnamefont {K.}~\bibnamefont {Cichy}},
  \bibinfo {author} {\bibfnamefont {M.}~\bibnamefont {Constantinou}}, \bibinfo
  {author} {\bibfnamefont {A.}~\bibnamefont {Metz}}, \bibinfo {author}
  {\bibfnamefont {A.}~\bibnamefont {Scapellato}}, \ and\ \bibinfo {author}
  {\bibfnamefont {F.}~\bibnamefont {Steffens}},\ }\href {\doibase
  10.21468/SciPostPhysProc.8.057} {\bibfield  {journal} {\bibinfo  {journal}
  {SciPost Phys. Proc.}\ }\textbf {\bibinfo {volume} {8}},\ \bibinfo {pages}
  {057} (\bibinfo {year} {2022}{\natexlab{a}})},\ \Eprint
  {http://arxiv.org/abs/2107.12818} {arXiv:2107.12818 [hep-lat]} \BibitemShut
  {NoStop}%
\bibitem [{\citenamefont {Constantinou}\ \emph {et~al.}(2022)\citenamefont
  {Constantinou}, \citenamefont {Bhattacharya}, \citenamefont {Cichy},
  \citenamefont {Metz}, \citenamefont {Scapellato},\ and\ \citenamefont
  {Steffens}}]{Constantinou:2021nbn}%
  \BibitemOpen
  \bibfield  {author} {\bibinfo {author} {\bibfnamefont {M.}~\bibnamefont
  {Constantinou}}, \bibinfo {author} {\bibfnamefont {S.}~\bibnamefont
  {Bhattacharya}}, \bibinfo {author} {\bibfnamefont {K.}~\bibnamefont {Cichy}},
  \bibinfo {author} {\bibfnamefont {A.}~\bibnamefont {Metz}}, \bibinfo {author}
  {\bibfnamefont {A.}~\bibnamefont {Scapellato}}, \ and\ \bibinfo {author}
  {\bibfnamefont {F.}~\bibnamefont {Steffens}},\ }\href {\doibase
  10.22323/1.396.0391} {\bibfield  {journal} {\bibinfo  {journal} {PoS}\
  }\textbf {\bibinfo {volume} {LATTICE2021}},\ \bibinfo {pages} {391} (\bibinfo
  {year} {2022})},\ \Eprint {http://arxiv.org/abs/2111.01056} {arXiv:2111.01056
  [hep-lat]} \BibitemShut {NoStop}%
\bibitem [{\citenamefont {Dodson}\ \emph {et~al.}(2022)\citenamefont {Dodson},
  \citenamefont {Bhattacharya}, \citenamefont {Cichy}, \citenamefont
  {Constantinou}, \citenamefont {Metz}, \citenamefont {Scapellato},\ and\
  \citenamefont {Steffens}}]{Dodson:2021rdq}%
  \BibitemOpen
  \bibfield  {author} {\bibinfo {author} {\bibfnamefont {J.}~\bibnamefont
  {Dodson}}, \bibinfo {author} {\bibfnamefont {S.}~\bibnamefont
  {Bhattacharya}}, \bibinfo {author} {\bibfnamefont {K.}~\bibnamefont {Cichy}},
  \bibinfo {author} {\bibfnamefont {M.}~\bibnamefont {Constantinou}}, \bibinfo
  {author} {\bibfnamefont {A.}~\bibnamefont {Metz}}, \bibinfo {author}
  {\bibfnamefont {A.}~\bibnamefont {Scapellato}}, \ and\ \bibinfo {author}
  {\bibfnamefont {F.}~\bibnamefont {Steffens}},\ }\href {\doibase
  10.22323/1.396.0054} {\bibfield  {journal} {\bibinfo  {journal} {PoS}\
  }\textbf {\bibinfo {volume} {LATTICE2021}},\ \bibinfo {pages} {054} (\bibinfo
  {year} {2022})},\ \Eprint {http://arxiv.org/abs/2112.05538} {arXiv:2112.05538
  [hep-lat]} \BibitemShut {NoStop}%
\bibitem [{\citenamefont {Gao}\ \emph {et~al.}(2021{\natexlab{a}})\citenamefont
  {Gao}, \citenamefont {Lee}, \citenamefont {Mukherjee}, \citenamefont
  {Shugert},\ and\ \citenamefont {Zhao}}]{Gao:2021hxl}%
  \BibitemOpen
  \bibfield  {author} {\bibinfo {author} {\bibfnamefont {X.}~\bibnamefont
  {Gao}}, \bibinfo {author} {\bibfnamefont {K.}~\bibnamefont {Lee}}, \bibinfo
  {author} {\bibfnamefont {S.}~\bibnamefont {Mukherjee}}, \bibinfo {author}
  {\bibfnamefont {C.}~\bibnamefont {Shugert}}, \ and\ \bibinfo {author}
  {\bibfnamefont {Y.}~\bibnamefont {Zhao}},\ }\href {\doibase
  10.1103/PhysRevD.103.094504} {\bibfield  {journal} {\bibinfo  {journal}
  {Phys. Rev. D}\ }\textbf {\bibinfo {volume} {103}},\ \bibinfo {pages}
  {094504} (\bibinfo {year} {2021}{\natexlab{a}})},\ \Eprint
  {http://arxiv.org/abs/2102.01101} {arXiv:2102.01101 [hep-ph]} \BibitemShut
  {NoStop}%
\bibitem [{\citenamefont {Gao}\ \emph {et~al.}(2022{\natexlab{a}})\citenamefont
  {Gao}, \citenamefont {Hanlon}, \citenamefont {Mukherjee}, \citenamefont
  {Petreczky}, \citenamefont {Scior}, \citenamefont {Syritsyn},\ and\
  \citenamefont {Zhao}}]{Gao:2021dbh}%
  \BibitemOpen
  \bibfield  {author} {\bibinfo {author} {\bibfnamefont {X.}~\bibnamefont
  {Gao}}, \bibinfo {author} {\bibfnamefont {A.~D.}\ \bibnamefont {Hanlon}},
  \bibinfo {author} {\bibfnamefont {S.}~\bibnamefont {Mukherjee}}, \bibinfo
  {author} {\bibfnamefont {P.}~\bibnamefont {Petreczky}}, \bibinfo {author}
  {\bibfnamefont {P.}~\bibnamefont {Scior}}, \bibinfo {author} {\bibfnamefont
  {S.}~\bibnamefont {Syritsyn}}, \ and\ \bibinfo {author} {\bibfnamefont
  {Y.}~\bibnamefont {Zhao}},\ }\href {\doibase 10.1103/PhysRevLett.128.142003}
  {\bibfield  {journal} {\bibinfo  {journal} {Phys. Rev. Lett.}\ }\textbf
  {\bibinfo {volume} {128}},\ \bibinfo {pages} {142003} (\bibinfo {year}
  {2022}{\natexlab{a}})},\ \Eprint {http://arxiv.org/abs/2112.02208}
  {arXiv:2112.02208 [hep-lat]} \BibitemShut {NoStop}%
\bibitem [{\citenamefont {Huo}\ \emph {et~al.}(2021)\citenamefont {Huo} \emph
  {et~al.}}]{LatticePartonLPC:2021gpi}%
  \BibitemOpen
  \bibfield  {author} {\bibinfo {author} {\bibfnamefont {Y.-K.}\ \bibnamefont
  {Huo}} \emph {et~al.} (\bibinfo {collaboration} {Lattice Parton (LPC)}),\
  }\href {\doibase 10.1016/j.nuclphysb.2021.115443} {\bibfield  {journal}
  {\bibinfo  {journal} {Nucl. Phys. B}\ }\textbf {\bibinfo {volume} {969}},\
  \bibinfo {pages} {115443} (\bibinfo {year} {2021})},\ \Eprint
  {http://arxiv.org/abs/2103.02965} {arXiv:2103.02965 [hep-lat]} \BibitemShut
  {NoStop}%
\bibitem [{\citenamefont {Li}\ \emph {et~al.}(2022)\citenamefont {Li} \emph
  {et~al.}}]{Li:2021wvl}%
  \BibitemOpen
  \bibfield  {author} {\bibinfo {author} {\bibfnamefont {Y.}~\bibnamefont {Li}}
  \emph {et~al.},\ }\href {\doibase 10.1103/PhysRevLett.128.062002} {\bibfield
  {journal} {\bibinfo  {journal} {Phys. Rev. Lett.}\ }\textbf {\bibinfo
  {volume} {128}},\ \bibinfo {pages} {062002} (\bibinfo {year} {2022})},\
  \Eprint {http://arxiv.org/abs/2106.13027} {arXiv:2106.13027 [hep-lat]}
  \BibitemShut {NoStop}%
\bibitem [{\citenamefont {Lin}(2022{\natexlab{a}})}]{Lin:2021brq}%
  \BibitemOpen
  \bibfield  {author} {\bibinfo {author} {\bibfnamefont {H.-W.}\ \bibnamefont
  {Lin}},\ }\href {\doibase 10.1016/j.physletb.2021.136821} {\bibfield
  {journal} {\bibinfo  {journal} {Phys. Lett. B}\ }\textbf {\bibinfo {volume}
  {824}},\ \bibinfo {pages} {136821} (\bibinfo {year} {2022}{\natexlab{a}})},\
  \Eprint {http://arxiv.org/abs/2112.07519} {arXiv:2112.07519 [hep-lat]}
  \BibitemShut {NoStop}%
\bibitem [{\citenamefont {Lin}(2022{\natexlab{b}})}]{Lin:2021ukf}%
  \BibitemOpen
  \bibfield  {author} {\bibinfo {author} {\bibfnamefont {H.-W.}\ \bibnamefont
  {Lin}},\ }\href {\doibase 10.22323/1.396.0276} {\bibfield  {journal}
  {\bibinfo  {journal} {PoS}\ }\textbf {\bibinfo {volume} {LATTICE2021}},\
  \bibinfo {pages} {276} (\bibinfo {year} {2022}{\natexlab{b}})},\ \Eprint
  {http://arxiv.org/abs/2110.06779} {arXiv:2110.06779 [hep-lat]} \BibitemShut
  {NoStop}%
\bibitem [{\citenamefont {Scapellato}\ \emph
  {et~al.}(2022{\natexlab{a}})\citenamefont {Scapellato}, \citenamefont
  {Alexandrou}, \citenamefont {Cichy}, \citenamefont {Constantinou},
  \citenamefont {Hadjiyiannakou}, \citenamefont {Jansen},\ and\ \citenamefont
  {Steffens}}]{Scapellato:2021uke}%
  \BibitemOpen
  \bibfield  {author} {\bibinfo {author} {\bibfnamefont {A.}~\bibnamefont
  {Scapellato}}, \bibinfo {author} {\bibfnamefont {C.}~\bibnamefont
  {Alexandrou}}, \bibinfo {author} {\bibfnamefont {K.}~\bibnamefont {Cichy}},
  \bibinfo {author} {\bibfnamefont {M.}~\bibnamefont {Constantinou}}, \bibinfo
  {author} {\bibfnamefont {K.}~\bibnamefont {Hadjiyiannakou}}, \bibinfo
  {author} {\bibfnamefont {K.}~\bibnamefont {Jansen}}, \ and\ \bibinfo {author}
  {\bibfnamefont {F.}~\bibnamefont {Steffens}},\ }\href {\doibase
  10.22323/1.396.0129} {\bibfield  {journal} {\bibinfo  {journal} {PoS}\
  }\textbf {\bibinfo {volume} {LATTICE2021}},\ \bibinfo {pages} {129} (\bibinfo
  {year} {2022}{\natexlab{a}})},\ \Eprint {http://arxiv.org/abs/2111.03226}
  {arXiv:2111.03226 [hep-lat]} \BibitemShut {NoStop}%
\bibitem [{\citenamefont {Schlemmer}\ \emph {et~al.}(2021)\citenamefont
  {Schlemmer}, \citenamefont {Vladimirov}, \citenamefont {Zimmermann},
  \citenamefont {Engelhardt},\ and\ \citenamefont
  {Sch\"afer}}]{Schlemmer:2021aij}%
  \BibitemOpen
  \bibfield  {author} {\bibinfo {author} {\bibfnamefont {M.}~\bibnamefont
  {Schlemmer}}, \bibinfo {author} {\bibfnamefont {A.}~\bibnamefont
  {Vladimirov}}, \bibinfo {author} {\bibfnamefont {C.}~\bibnamefont
  {Zimmermann}}, \bibinfo {author} {\bibfnamefont {M.}~\bibnamefont
  {Engelhardt}}, \ and\ \bibinfo {author} {\bibfnamefont {A.}~\bibnamefont
  {Sch\"afer}},\ }\href {\doibase 10.1007/JHEP08(2021)004} {\bibfield
  {journal} {\bibinfo  {journal} {JHEP}\ }\textbf {\bibinfo {volume} {08}},\
  \bibinfo {pages} {004} (\bibinfo {year} {2021})},\ \Eprint
  {http://arxiv.org/abs/2103.16991} {arXiv:2103.16991 [hep-lat]} \BibitemShut
  {NoStop}%
\bibitem [{\citenamefont {Shanahan}\ \emph {et~al.}(2021)\citenamefont
  {Shanahan}, \citenamefont {Wagman},\ and\ \citenamefont
  {Zhao}}]{Shanahan:2021tst}%
  \BibitemOpen
  \bibfield  {author} {\bibinfo {author} {\bibfnamefont {P.}~\bibnamefont
  {Shanahan}}, \bibinfo {author} {\bibfnamefont {M.}~\bibnamefont {Wagman}}, \
  and\ \bibinfo {author} {\bibfnamefont {Y.}~\bibnamefont {Zhao}},\ }\href
  {\doibase 10.1103/PhysRevD.104.114502} {\bibfield  {journal} {\bibinfo
  {journal} {Phys. Rev. D}\ }\textbf {\bibinfo {volume} {104}},\ \bibinfo
  {pages} {114502} (\bibinfo {year} {2021})},\ \Eprint
  {http://arxiv.org/abs/2107.11930} {arXiv:2107.11930 [hep-lat]} \BibitemShut
  {NoStop}%
\bibitem [{\citenamefont {Bhattacharya}\ \emph
  {et~al.}(2022{\natexlab{b}})\citenamefont {Bhattacharya}, \citenamefont
  {Cichy}, \citenamefont {Constantinou}, \citenamefont {Dodson}, \citenamefont
  {Gao}, \citenamefont {Metz}, \citenamefont {Mukherjee}, \citenamefont
  {Scapellato}, \citenamefont {Steffens},\ and\ \citenamefont
  {Zhao}}]{Bhattacharya:2022aob}%
  \BibitemOpen
  \bibfield  {author} {\bibinfo {author} {\bibfnamefont {S.}~\bibnamefont
  {Bhattacharya}}, \bibinfo {author} {\bibfnamefont {K.}~\bibnamefont {Cichy}},
  \bibinfo {author} {\bibfnamefont {M.}~\bibnamefont {Constantinou}}, \bibinfo
  {author} {\bibfnamefont {J.}~\bibnamefont {Dodson}}, \bibinfo {author}
  {\bibfnamefont {X.}~\bibnamefont {Gao}}, \bibinfo {author} {\bibfnamefont
  {A.}~\bibnamefont {Metz}}, \bibinfo {author} {\bibfnamefont {S.}~\bibnamefont
  {Mukherjee}}, \bibinfo {author} {\bibfnamefont {A.}~\bibnamefont
  {Scapellato}}, \bibinfo {author} {\bibfnamefont {F.}~\bibnamefont
  {Steffens}}, \ and\ \bibinfo {author} {\bibfnamefont {Y.}~\bibnamefont
  {Zhao}},\ }\href {\doibase 10.1103/PhysRevD.106.114512} {\bibfield  {journal}
  {\bibinfo  {journal} {Phys. Rev. D}\ }\textbf {\bibinfo {volume} {106}},\
  \bibinfo {pages} {114512} (\bibinfo {year} {2022}{\natexlab{b}})},\ \Eprint
  {http://arxiv.org/abs/2209.05373} {arXiv:2209.05373 [hep-lat]} \BibitemShut
  {NoStop}%
\bibitem [{\citenamefont {Constantinou}\ \emph {et~al.}(2023)\citenamefont
  {Constantinou}, \citenamefont {Bhattacharya}, \citenamefont {Cichy},
  \citenamefont {Dodson}, \citenamefont {Gao}, \citenamefont {Metz},
  \citenamefont {Mukherjee}, \citenamefont {Scapellato}, \citenamefont
  {Steffens},\ and\ \citenamefont {Zhao}}]{Constantinou:2022fqt}%
  \BibitemOpen
  \bibfield  {author} {\bibinfo {author} {\bibfnamefont {M.}~\bibnamefont
  {Constantinou}}, \bibinfo {author} {\bibfnamefont {S.}~\bibnamefont
  {Bhattacharya}}, \bibinfo {author} {\bibfnamefont {K.}~\bibnamefont {Cichy}},
  \bibinfo {author} {\bibfnamefont {J.}~\bibnamefont {Dodson}}, \bibinfo
  {author} {\bibfnamefont {X.}~\bibnamefont {Gao}}, \bibinfo {author}
  {\bibfnamefont {A.}~\bibnamefont {Metz}}, \bibinfo {author} {\bibfnamefont
  {S.}~\bibnamefont {Mukherjee}}, \bibinfo {author} {\bibfnamefont
  {A.}~\bibnamefont {Scapellato}}, \bibinfo {author} {\bibfnamefont
  {F.}~\bibnamefont {Steffens}}, \ and\ \bibinfo {author} {\bibfnamefont
  {Y.}~\bibnamefont {Zhao}},\ }\href {\doibase 10.22323/1.430.0096} {\bibfield
  {journal} {\bibinfo  {journal} {PoS}\ }\textbf {\bibinfo {volume}
  {LATTICE2022}},\ \bibinfo {pages} {096} (\bibinfo {year} {2023})},\ \Eprint
  {http://arxiv.org/abs/2212.09818} {arXiv:2212.09818 [hep-lat]} \BibitemShut
  {NoStop}%
\bibitem [{\citenamefont {Deng}\ \emph {et~al.}(2022)\citenamefont {Deng},
  \citenamefont {Wang},\ and\ \citenamefont {Zeng}}]{Deng:2022gzi}%
  \BibitemOpen
  \bibfield  {author} {\bibinfo {author} {\bibfnamefont {Z.-F.}\ \bibnamefont
  {Deng}}, \bibinfo {author} {\bibfnamefont {W.}~\bibnamefont {Wang}}, \ and\
  \bibinfo {author} {\bibfnamefont {J.}~\bibnamefont {Zeng}},\ }\href {\doibase
  10.1007/JHEP09(2022)046} {\bibfield  {journal} {\bibinfo  {journal} {JHEP}\
  }\textbf {\bibinfo {volume} {09}},\ \bibinfo {pages} {046} (\bibinfo {year}
  {2022})},\ \Eprint {http://arxiv.org/abs/2207.07280} {arXiv:2207.07280
  [hep-th]} \BibitemShut {NoStop}%
\bibitem [{\citenamefont {Ebert}\ \emph {et~al.}(2022)\citenamefont {Ebert},
  \citenamefont {Schindler}, \citenamefont {Stewart},\ and\ \citenamefont
  {Zhao}}]{Ebert:2022fmh}%
  \BibitemOpen
  \bibfield  {author} {\bibinfo {author} {\bibfnamefont {M.~A.}\ \bibnamefont
  {Ebert}}, \bibinfo {author} {\bibfnamefont {S.~T.}\ \bibnamefont
  {Schindler}}, \bibinfo {author} {\bibfnamefont {I.~W.}\ \bibnamefont
  {Stewart}}, \ and\ \bibinfo {author} {\bibfnamefont {Y.}~\bibnamefont
  {Zhao}},\ }\href {\doibase 10.1007/JHEP04(2022)178} {\bibfield  {journal}
  {\bibinfo  {journal} {JHEP}\ }\textbf {\bibinfo {volume} {04}},\ \bibinfo
  {pages} {178} (\bibinfo {year} {2022})},\ \Eprint
  {http://arxiv.org/abs/2201.08401} {arXiv:2201.08401 [hep-ph]} \BibitemShut
  {NoStop}%
\bibitem [{\citenamefont {Gao}\ \emph {et~al.}(2022{\natexlab{b}})\citenamefont
  {Gao}, \citenamefont {Hanlon}, \citenamefont {Karthik}, \citenamefont
  {Mukherjee}, \citenamefont {Petreczky}, \citenamefont {Scior}, \citenamefont
  {Shi}, \citenamefont {Syritsyn}, \citenamefont {Zhao},\ and\ \citenamefont
  {Zhou}}]{Gao:2022iex}%
  \BibitemOpen
  \bibfield  {author} {\bibinfo {author} {\bibfnamefont {X.}~\bibnamefont
  {Gao}}, \bibinfo {author} {\bibfnamefont {A.~D.}\ \bibnamefont {Hanlon}},
  \bibinfo {author} {\bibfnamefont {N.}~\bibnamefont {Karthik}}, \bibinfo
  {author} {\bibfnamefont {S.}~\bibnamefont {Mukherjee}}, \bibinfo {author}
  {\bibfnamefont {P.}~\bibnamefont {Petreczky}}, \bibinfo {author}
  {\bibfnamefont {P.}~\bibnamefont {Scior}}, \bibinfo {author} {\bibfnamefont
  {S.}~\bibnamefont {Shi}}, \bibinfo {author} {\bibfnamefont {S.}~\bibnamefont
  {Syritsyn}}, \bibinfo {author} {\bibfnamefont {Y.}~\bibnamefont {Zhao}}, \
  and\ \bibinfo {author} {\bibfnamefont {K.}~\bibnamefont {Zhou}},\ }\href
  {\doibase 10.1103/PhysRevD.106.114510} {\bibfield  {journal} {\bibinfo
  {journal} {Phys. Rev. D}\ }\textbf {\bibinfo {volume} {106}},\ \bibinfo
  {pages} {114510} (\bibinfo {year} {2022}{\natexlab{b}})},\ \Eprint
  {http://arxiv.org/abs/2208.02297} {arXiv:2208.02297 [hep-lat]} \BibitemShut
  {NoStop}%
\bibitem [{\citenamefont {Gao}\ \emph {et~al.}(2023)\citenamefont {Gao},
  \citenamefont {Hanlon}, \citenamefont {Holligan}, \citenamefont {Karthik},
  \citenamefont {Mukherjee}, \citenamefont {Petreczky}, \citenamefont
  {Syritsyn},\ and\ \citenamefont {Zhao}}]{Gao:2022uhg}%
  \BibitemOpen
  \bibfield  {author} {\bibinfo {author} {\bibfnamefont {X.}~\bibnamefont
  {Gao}}, \bibinfo {author} {\bibfnamefont {A.~D.}\ \bibnamefont {Hanlon}},
  \bibinfo {author} {\bibfnamefont {J.}~\bibnamefont {Holligan}}, \bibinfo
  {author} {\bibfnamefont {N.}~\bibnamefont {Karthik}}, \bibinfo {author}
  {\bibfnamefont {S.}~\bibnamefont {Mukherjee}}, \bibinfo {author}
  {\bibfnamefont {P.}~\bibnamefont {Petreczky}}, \bibinfo {author}
  {\bibfnamefont {S.}~\bibnamefont {Syritsyn}}, \ and\ \bibinfo {author}
  {\bibfnamefont {Y.}~\bibnamefont {Zhao}},\ }\href {\doibase
  10.1103/PhysRevD.107.074509} {\bibfield  {journal} {\bibinfo  {journal}
  {Phys. Rev. D}\ }\textbf {\bibinfo {volume} {107}},\ \bibinfo {pages}
  {074509} (\bibinfo {year} {2023})},\ \Eprint
  {http://arxiv.org/abs/2212.12569} {arXiv:2212.12569 [hep-lat]} \BibitemShut
  {NoStop}%
\bibitem [{\citenamefont {Yao}\ \emph {et~al.}(2023{\natexlab{a}})\citenamefont
  {Yao} \emph {et~al.}}]{LatticeParton:2022xsd}%
  \BibitemOpen
  \bibfield  {author} {\bibinfo {author} {\bibfnamefont {F.}~\bibnamefont
  {Yao}} \emph {et~al.} (\bibinfo {collaboration} {Lattice Parton}),\ }\href
  {\doibase 10.1103/PhysRevLett.131.261901} {\bibfield  {journal} {\bibinfo
  {journal} {Phys. Rev. Lett.}\ }\textbf {\bibinfo {volume} {131}},\ \bibinfo
  {pages} {261901} (\bibinfo {year} {2023}{\natexlab{a}})},\ \Eprint
  {http://arxiv.org/abs/2208.08008} {arXiv:2208.08008 [hep-lat]} \BibitemShut
  {NoStop}%
\bibitem [{\citenamefont {Hua}\ \emph {et~al.}(2022)\citenamefont {Hua} \emph
  {et~al.}}]{LatticeParton:2022zqc}%
  \BibitemOpen
  \bibfield  {author} {\bibinfo {author} {\bibfnamefont {J.}~\bibnamefont
  {Hua}} \emph {et~al.} (\bibinfo {collaboration} {Lattice Parton}),\ }\href
  {\doibase 10.1103/PhysRevLett.129.132001} {\bibfield  {journal} {\bibinfo
  {journal} {Phys. Rev. Lett.}\ }\textbf {\bibinfo {volume} {129}},\ \bibinfo
  {pages} {132001} (\bibinfo {year} {2022})},\ \Eprint
  {http://arxiv.org/abs/2201.09173} {arXiv:2201.09173 [hep-lat]} \BibitemShut
  {NoStop}%
\bibitem [{\citenamefont {He}\ \emph {et~al.}(2024)\citenamefont {He},
  \citenamefont {Chu}, \citenamefont {Hua}, \citenamefont {Ji}, \citenamefont
  {Sch\"afer}, \citenamefont {Su}, \citenamefont {Wang}, \citenamefont {Yang},
  \citenamefont {Zhang},\ and\ \citenamefont
  {Zhang}}]{LatticePartonCollaborationLPC:2022myp}%
  \BibitemOpen
  \bibfield  {author} {\bibinfo {author} {\bibfnamefont {J.-C.}\ \bibnamefont
  {He}}, \bibinfo {author} {\bibfnamefont {M.-H.}\ \bibnamefont {Chu}},
  \bibinfo {author} {\bibfnamefont {J.}~\bibnamefont {Hua}}, \bibinfo {author}
  {\bibfnamefont {X.}~\bibnamefont {Ji}}, \bibinfo {author} {\bibfnamefont
  {A.}~\bibnamefont {Sch\"afer}}, \bibinfo {author} {\bibfnamefont
  {Y.}~\bibnamefont {Su}}, \bibinfo {author} {\bibfnamefont {W.}~\bibnamefont
  {Wang}}, \bibinfo {author} {\bibfnamefont {Y.-B.}\ \bibnamefont {Yang}},
  \bibinfo {author} {\bibfnamefont {J.-H.}\ \bibnamefont {Zhang}}, \ and\
  \bibinfo {author} {\bibfnamefont {Q.-A.}\ \bibnamefont {Zhang}} (\bibinfo
  {collaboration} {Lattice Parton Collaboration (LPC)}),\ }\href {\doibase
  10.1103/PhysRevD.109.114513} {\bibfield  {journal} {\bibinfo  {journal}
  {Phys. Rev. D}\ }\textbf {\bibinfo {volume} {109}},\ \bibinfo {pages}
  {114513} (\bibinfo {year} {2024})},\ \Eprint
  {http://arxiv.org/abs/2211.02340} {arXiv:2211.02340 [hep-lat]} \BibitemShut
  {NoStop}%
\bibitem [{\citenamefont {Chu}\ \emph {et~al.}(2022)\citenamefont {Chu} \emph
  {et~al.}}]{LatticePartonLPC:2022eev}%
  \BibitemOpen
  \bibfield  {author} {\bibinfo {author} {\bibfnamefont {M.-H.}\ \bibnamefont
  {Chu}} \emph {et~al.} (\bibinfo {collaboration} {Lattice Parton (LPC)}),\
  }\href {\doibase 10.1103/PhysRevD.106.034509} {\bibfield  {journal} {\bibinfo
   {journal} {Phys. Rev. D}\ }\textbf {\bibinfo {volume} {106}},\ \bibinfo
  {pages} {034509} (\bibinfo {year} {2022})},\ \Eprint
  {http://arxiv.org/abs/2204.00200} {arXiv:2204.00200 [hep-lat]} \BibitemShut
  {NoStop}%
\bibitem [{\citenamefont {Scapellato}\ \emph
  {et~al.}(2022{\natexlab{b}})\citenamefont {Scapellato}, \citenamefont
  {Alexandrou}, \citenamefont {Cichy}, \citenamefont {Constantinou},
  \citenamefont {Hadjiyiannakou}, \citenamefont {Jansen},\ and\ \citenamefont
  {Steffens}}]{Scapellato:2022mai}%
  \BibitemOpen
  \bibfield  {author} {\bibinfo {author} {\bibfnamefont {A.}~\bibnamefont
  {Scapellato}}, \bibinfo {author} {\bibfnamefont {C.}~\bibnamefont
  {Alexandrou}}, \bibinfo {author} {\bibfnamefont {K.}~\bibnamefont {Cichy}},
  \bibinfo {author} {\bibfnamefont {M.}~\bibnamefont {Constantinou}}, \bibinfo
  {author} {\bibfnamefont {K.}~\bibnamefont {Hadjiyiannakou}}, \bibinfo
  {author} {\bibfnamefont {K.}~\bibnamefont {Jansen}}, \ and\ \bibinfo {author}
  {\bibfnamefont {F.}~\bibnamefont {Steffens}},\ }\href {\doibase
  10.31349/SuplRevMexFis.3.0308104} {\bibfield  {journal} {\bibinfo  {journal}
  {Rev. Mex. Fis. Suppl.}\ }\textbf {\bibinfo {volume} {3}},\ \bibinfo {pages}
  {0308104} (\bibinfo {year} {2022}{\natexlab{b}})},\ \Eprint
  {http://arxiv.org/abs/2201.06519} {arXiv:2201.06519 [hep-lat]} \BibitemShut
  {NoStop}%
\bibitem [{\citenamefont {Schindler}\ \emph {et~al.}(2022)\citenamefont
  {Schindler}, \citenamefont {Stewart},\ and\ \citenamefont
  {Zhao}}]{Schindler:2022eva}%
  \BibitemOpen
  \bibfield  {author} {\bibinfo {author} {\bibfnamefont {S.~T.}\ \bibnamefont
  {Schindler}}, \bibinfo {author} {\bibfnamefont {I.~W.}\ \bibnamefont
  {Stewart}}, \ and\ \bibinfo {author} {\bibfnamefont {Y.}~\bibnamefont
  {Zhao}},\ }\href {\doibase 10.1007/JHEP08(2022)084} {\bibfield  {journal}
  {\bibinfo  {journal} {JHEP}\ }\textbf {\bibinfo {volume} {08}},\ \bibinfo
  {pages} {084} (\bibinfo {year} {2022})},\ \Eprint
  {http://arxiv.org/abs/2205.12369} {arXiv:2205.12369 [hep-ph]} \BibitemShut
  {NoStop}%
\bibitem [{\citenamefont {Zhang}\ \emph {et~al.}(2022)\citenamefont {Zhang},
  \citenamefont {Ji}, \citenamefont {Yang}, \citenamefont {Yao},\ and\
  \citenamefont {Zhang}}]{Zhang:2022xuw}%
  \BibitemOpen
  \bibfield  {author} {\bibinfo {author} {\bibfnamefont {K.}~\bibnamefont
  {Zhang}}, \bibinfo {author} {\bibfnamefont {X.}~\bibnamefont {Ji}}, \bibinfo
  {author} {\bibfnamefont {Y.-B.}\ \bibnamefont {Yang}}, \bibinfo {author}
  {\bibfnamefont {F.}~\bibnamefont {Yao}}, \ and\ \bibinfo {author}
  {\bibfnamefont {J.-H.}\ \bibnamefont {Zhang}} (\bibinfo {collaboration}
  {Lattice Parton (LPC)}),\ }\href {\doibase 10.1103/PhysRevLett.129.082002}
  {\bibfield  {journal} {\bibinfo  {journal} {Phys. Rev. Lett.}\ }\textbf
  {\bibinfo {volume} {129}},\ \bibinfo {pages} {082002} (\bibinfo {year}
  {2022})},\ \Eprint {http://arxiv.org/abs/2205.13402} {arXiv:2205.13402
  [hep-lat]} \BibitemShut {NoStop}%
\bibitem [{\citenamefont {Alexandrou}\ \emph {et~al.}(2023)\citenamefont
  {Alexandrou} \emph {et~al.}}]{Alexandrou:2023ucc}%
  \BibitemOpen
  \bibfield  {author} {\bibinfo {author} {\bibfnamefont {C.}~\bibnamefont
  {Alexandrou}} \emph {et~al.},\ }\href {\doibase 10.1103/PhysRevD.108.114503}
  {\bibfield  {journal} {\bibinfo  {journal} {Phys. Rev. D}\ }\textbf {\bibinfo
  {volume} {108}},\ \bibinfo {pages} {114503} (\bibinfo {year} {2023})},\
  \Eprint {http://arxiv.org/abs/2305.11824} {arXiv:2305.11824 [hep-lat]}
  \BibitemShut {NoStop}%
\bibitem [{\citenamefont {Avkhadiev}\ \emph {et~al.}(2023)\citenamefont
  {Avkhadiev}, \citenamefont {Shanahan}, \citenamefont {Wagman},\ and\
  \citenamefont {Zhao}}]{Avkhadiev:2023poz}%
  \BibitemOpen
  \bibfield  {author} {\bibinfo {author} {\bibfnamefont {A.}~\bibnamefont
  {Avkhadiev}}, \bibinfo {author} {\bibfnamefont {P.~E.}\ \bibnamefont
  {Shanahan}}, \bibinfo {author} {\bibfnamefont {M.~L.}\ \bibnamefont
  {Wagman}}, \ and\ \bibinfo {author} {\bibfnamefont {Y.}~\bibnamefont
  {Zhao}},\ }\href {\doibase 10.1103/PhysRevD.108.114505} {\bibfield  {journal}
  {\bibinfo  {journal} {Phys. Rev. D}\ }\textbf {\bibinfo {volume} {108}},\
  \bibinfo {pages} {114505} (\bibinfo {year} {2023})},\ \Eprint
  {http://arxiv.org/abs/2307.12359} {arXiv:2307.12359 [hep-lat]} \BibitemShut
  {NoStop}%
\bibitem [{\citenamefont {Bhattacharya}\ \emph
  {et~al.}(2024{\natexlab{a}})\citenamefont {Bhattacharya} \emph
  {et~al.}}]{Bhattacharya:2023jsc}%
  \BibitemOpen
  \bibfield  {author} {\bibinfo {author} {\bibfnamefont {S.}~\bibnamefont
  {Bhattacharya}} \emph {et~al.},\ }\href {\doibase
  10.1103/PhysRevD.109.034508} {\bibfield  {journal} {\bibinfo  {journal}
  {Phys. Rev. D}\ }\textbf {\bibinfo {volume} {109}},\ \bibinfo {pages}
  {034508} (\bibinfo {year} {2024}{\natexlab{a}})},\ \Eprint
  {http://arxiv.org/abs/2310.13114} {arXiv:2310.13114 [hep-lat]} \BibitemShut
  {NoStop}%
\bibitem [{\citenamefont {Bhattacharya}\ \emph
  {et~al.}(2023{\natexlab{a}})\citenamefont {Bhattacharya}, \citenamefont
  {Cichy}, \citenamefont {Constantinou}, \citenamefont {Dodson}, \citenamefont
  {Metz}, \citenamefont {Scapellato},\ and\ \citenamefont
  {Steffens}}]{Bhattacharya:2023nmv}%
  \BibitemOpen
  \bibfield  {author} {\bibinfo {author} {\bibfnamefont {S.}~\bibnamefont
  {Bhattacharya}}, \bibinfo {author} {\bibfnamefont {K.}~\bibnamefont {Cichy}},
  \bibinfo {author} {\bibfnamefont {M.}~\bibnamefont {Constantinou}}, \bibinfo
  {author} {\bibfnamefont {J.}~\bibnamefont {Dodson}}, \bibinfo {author}
  {\bibfnamefont {A.}~\bibnamefont {Metz}}, \bibinfo {author} {\bibfnamefont
  {A.}~\bibnamefont {Scapellato}}, \ and\ \bibinfo {author} {\bibfnamefont
  {F.}~\bibnamefont {Steffens}},\ }\href {\doibase 10.1103/PhysRevD.108.054501}
  {\bibfield  {journal} {\bibinfo  {journal} {Phys. Rev. D}\ }\textbf {\bibinfo
  {volume} {108}},\ \bibinfo {pages} {054501} (\bibinfo {year}
  {2023}{\natexlab{a}})},\ \Eprint {http://arxiv.org/abs/2306.05533}
  {arXiv:2306.05533 [hep-lat]} \BibitemShut {NoStop}%
\bibitem [{\citenamefont {Bhattacharya}\ \emph
  {et~al.}(2023{\natexlab{b}})\citenamefont {Bhattacharya}, \citenamefont
  {Cichy}, \citenamefont {Constantinou}, \citenamefont {Dodson}, \citenamefont
  {Gao}, \citenamefont {Metz}, \citenamefont {Mukherjee}, \citenamefont
  {Scapellato}, \citenamefont {Steffens},\ and\ \citenamefont
  {Zhao}}]{Bhattacharya:2023tik}%
  \BibitemOpen
  \bibfield  {author} {\bibinfo {author} {\bibfnamefont {S.}~\bibnamefont
  {Bhattacharya}}, \bibinfo {author} {\bibfnamefont {K.}~\bibnamefont {Cichy}},
  \bibinfo {author} {\bibfnamefont {M.}~\bibnamefont {Constantinou}}, \bibinfo
  {author} {\bibfnamefont {J.}~\bibnamefont {Dodson}}, \bibinfo {author}
  {\bibfnamefont {X.}~\bibnamefont {Gao}}, \bibinfo {author} {\bibfnamefont
  {A.}~\bibnamefont {Metz}}, \bibinfo {author} {\bibfnamefont {S.}~\bibnamefont
  {Mukherjee}}, \bibinfo {author} {\bibfnamefont {A.}~\bibnamefont
  {Scapellato}}, \bibinfo {author} {\bibfnamefont {F.}~\bibnamefont
  {Steffens}}, \ and\ \bibinfo {author} {\bibfnamefont {Y.}~\bibnamefont
  {Zhao}},\ }\href {\doibase 10.22323/1.430.0095} {\bibfield  {journal}
  {\bibinfo  {journal} {PoS}\ }\textbf {\bibinfo {volume} {LATTICE2022}},\
  \bibinfo {pages} {095} (\bibinfo {year} {2023}{\natexlab{b}})},\ \Eprint
  {http://arxiv.org/abs/2301.03400} {arXiv:2301.03400 [hep-lat]} \BibitemShut
  {NoStop}%
\bibitem [{\citenamefont {Cichy}\ \emph {et~al.}(2023)\citenamefont {Cichy}
  \emph {et~al.}}]{Cichy:2023dgk}%
  \BibitemOpen
  \bibfield  {author} {\bibinfo {author} {\bibfnamefont {K.}~\bibnamefont
  {Cichy}} \emph {et~al.},\ }\href {\doibase 10.5506/APhysPolBSupp.16.7-A6}
  {\bibfield  {journal} {\bibinfo  {journal} {Acta Phys. Polon. Supp.}\
  }\textbf {\bibinfo {volume} {16}},\ \bibinfo {pages} {7} (\bibinfo {year}
  {2023})},\ \Eprint {http://arxiv.org/abs/2304.14970} {arXiv:2304.14970
  [hep-lat]} \BibitemShut {NoStop}%
\bibitem [{\citenamefont {Deng}\ \emph {et~al.}(2023)\citenamefont {Deng},
  \citenamefont {Han}, \citenamefont {Wang}, \citenamefont {Zeng},\ and\
  \citenamefont {Zhang}}]{Deng:2023csv}%
  \BibitemOpen
  \bibfield  {author} {\bibinfo {author} {\bibfnamefont {Z.-F.}\ \bibnamefont
  {Deng}}, \bibinfo {author} {\bibfnamefont {C.}~\bibnamefont {Han}}, \bibinfo
  {author} {\bibfnamefont {W.}~\bibnamefont {Wang}}, \bibinfo {author}
  {\bibfnamefont {J.}~\bibnamefont {Zeng}}, \ and\ \bibinfo {author}
  {\bibfnamefont {J.-L.}\ \bibnamefont {Zhang}},\ }\href {\doibase
  10.1007/JHEP07(2023)191} {\bibfield  {journal} {\bibinfo  {journal} {JHEP}\
  }\textbf {\bibinfo {volume} {07}},\ \bibinfo {pages} {191} (\bibinfo {year}
  {2023})},\ \Eprint {http://arxiv.org/abs/2304.09004} {arXiv:2304.09004
  [hep-ph]} \BibitemShut {NoStop}%
\bibitem [{\citenamefont {Gao}\ \emph {et~al.}(2024{\natexlab{a}})\citenamefont
  {Gao}, \citenamefont {Hanlon}, \citenamefont {Mukherjee}, \citenamefont
  {Petreczky}, \citenamefont {Shi}, \citenamefont {Syritsyn},\ and\
  \citenamefont {Zhao}}]{Gao:2023ktu}%
  \BibitemOpen
  \bibfield  {author} {\bibinfo {author} {\bibfnamefont {X.}~\bibnamefont
  {Gao}}, \bibinfo {author} {\bibfnamefont {A.~D.}\ \bibnamefont {Hanlon}},
  \bibinfo {author} {\bibfnamefont {S.}~\bibnamefont {Mukherjee}}, \bibinfo
  {author} {\bibfnamefont {P.}~\bibnamefont {Petreczky}}, \bibinfo {author}
  {\bibfnamefont {Q.}~\bibnamefont {Shi}}, \bibinfo {author} {\bibfnamefont
  {S.}~\bibnamefont {Syritsyn}}, \ and\ \bibinfo {author} {\bibfnamefont
  {Y.}~\bibnamefont {Zhao}},\ }\href {\doibase 10.1103/PhysRevD.109.054506}
  {\bibfield  {journal} {\bibinfo  {journal} {Phys. Rev. D}\ }\textbf {\bibinfo
  {volume} {109}},\ \bibinfo {pages} {054506} (\bibinfo {year}
  {2024}{\natexlab{a}})},\ \Eprint {http://arxiv.org/abs/2310.19047}
  {arXiv:2310.19047 [hep-lat]} \BibitemShut {NoStop}%
\bibitem [{\citenamefont {Gao}\ \emph {et~al.}(2024{\natexlab{b}})\citenamefont
  {Gao}, \citenamefont {Liu},\ and\ \citenamefont {Zhao}}]{Gao:2023lny}%
  \BibitemOpen
  \bibfield  {author} {\bibinfo {author} {\bibfnamefont {X.}~\bibnamefont
  {Gao}}, \bibinfo {author} {\bibfnamefont {W.-Y.}\ \bibnamefont {Liu}}, \ and\
  \bibinfo {author} {\bibfnamefont {Y.}~\bibnamefont {Zhao}},\ }\href {\doibase
  10.1103/PhysRevD.109.094506} {\bibfield  {journal} {\bibinfo  {journal}
  {Phys. Rev. D}\ }\textbf {\bibinfo {volume} {109}},\ \bibinfo {pages}
  {094506} (\bibinfo {year} {2024}{\natexlab{b}})},\ \Eprint
  {http://arxiv.org/abs/2306.14960} {arXiv:2306.14960 [hep-ph]} \BibitemShut
  {NoStop}%
\bibitem [{\citenamefont {Holligan}\ and\ \citenamefont
  {Lin}(2024{\natexlab{a}})}]{Holligan:2023jqh}%
  \BibitemOpen
  \bibfield  {author} {\bibinfo {author} {\bibfnamefont {J.}~\bibnamefont
  {Holligan}}\ and\ \bibinfo {author} {\bibfnamefont {H.-W.}\ \bibnamefont
  {Lin}},\ }\href {\doibase 10.1103/PhysRevD.110.034503} {\bibfield  {journal}
  {\bibinfo  {journal} {Phys. Rev. D}\ }\textbf {\bibinfo {volume} {110}},\
  \bibinfo {pages} {034503} (\bibinfo {year} {2024}{\natexlab{a}})},\ \Eprint
  {http://arxiv.org/abs/2312.10829} {arXiv:2312.10829 [hep-lat]} \BibitemShut
  {NoStop}%
\bibitem [{\citenamefont {Holligan}\ \emph {et~al.}(2023)\citenamefont
  {Holligan}, \citenamefont {Ji}, \citenamefont {Lin}, \citenamefont {Su},\
  and\ \citenamefont {Zhang}}]{Holligan:2023rex}%
  \BibitemOpen
  \bibfield  {author} {\bibinfo {author} {\bibfnamefont {J.}~\bibnamefont
  {Holligan}}, \bibinfo {author} {\bibfnamefont {X.}~\bibnamefont {Ji}},
  \bibinfo {author} {\bibfnamefont {H.-W.}\ \bibnamefont {Lin}}, \bibinfo
  {author} {\bibfnamefont {Y.}~\bibnamefont {Su}}, \ and\ \bibinfo {author}
  {\bibfnamefont {R.}~\bibnamefont {Zhang}},\ }\href {\doibase
  10.1016/j.nuclphysb.2023.116282} {\bibfield  {journal} {\bibinfo  {journal}
  {Nucl. Phys. B}\ }\textbf {\bibinfo {volume} {993}},\ \bibinfo {pages}
  {116282} (\bibinfo {year} {2023})},\ \Eprint
  {http://arxiv.org/abs/2301.10372} {arXiv:2301.10372 [hep-lat]} \BibitemShut
  {NoStop}%
\bibitem [{\citenamefont {Ji}\ \emph {et~al.}(2023)\citenamefont {Ji},
  \citenamefont {Liu},\ and\ \citenamefont {Su}}]{Ji:2023pba}%
  \BibitemOpen
  \bibfield  {author} {\bibinfo {author} {\bibfnamefont {X.}~\bibnamefont
  {Ji}}, \bibinfo {author} {\bibfnamefont {Y.}~\bibnamefont {Liu}}, \ and\
  \bibinfo {author} {\bibfnamefont {Y.}~\bibnamefont {Su}},\ }\href {\doibase
  10.1007/JHEP08(2023)037} {\bibfield  {journal} {\bibinfo  {journal} {JHEP}\
  }\textbf {\bibinfo {volume} {08}},\ \bibinfo {pages} {037} (\bibinfo {year}
  {2023})},\ \Eprint {http://arxiv.org/abs/2305.04416} {arXiv:2305.04416
  [hep-ph]} \BibitemShut {NoStop}%
\bibitem [{\citenamefont {Chu}\ \emph {et~al.}(2024)\citenamefont {Chu} \emph
  {et~al.}}]{LatticeParton:2023xdl}%
  \BibitemOpen
  \bibfield  {author} {\bibinfo {author} {\bibfnamefont {M.-H.}\ \bibnamefont
  {Chu}} \emph {et~al.} (\bibinfo {collaboration} {Lattice Parton}),\ }\href
  {\doibase 10.1103/PhysRevD.109.L091503} {\bibfield  {journal} {\bibinfo
  {journal} {Phys. Rev. D}\ }\textbf {\bibinfo {volume} {109}},\ \bibinfo
  {pages} {L091503} (\bibinfo {year} {2024})},\ \Eprint
  {http://arxiv.org/abs/2302.09961} {arXiv:2302.09961 [hep-lat]} \BibitemShut
  {NoStop}%
\bibitem [{\citenamefont {Chu}\ \emph {et~al.}(2023)\citenamefont {Chu} \emph
  {et~al.}}]{LatticePartonLPC:2023pdv}%
  \BibitemOpen
  \bibfield  {author} {\bibinfo {author} {\bibfnamefont {M.-H.}\ \bibnamefont
  {Chu}} \emph {et~al.} (\bibinfo {collaboration} {Lattice Parton (LPC)}),\
  }\href {\doibase 10.1007/JHEP08(2023)172} {\bibfield  {journal} {\bibinfo
  {journal} {JHEP}\ }\textbf {\bibinfo {volume} {08}},\ \bibinfo {pages} {172}
  (\bibinfo {year} {2023})},\ \Eprint {http://arxiv.org/abs/2306.06488}
  {arXiv:2306.06488 [hep-lat]} \BibitemShut {NoStop}%
\bibitem [{\citenamefont {Lin}(2023)}]{Lin:2023gxz}%
  \BibitemOpen
  \bibfield  {author} {\bibinfo {author} {\bibfnamefont {H.-W.}\ \bibnamefont
  {Lin}},\ }\href {\doibase 10.1016/j.physletb.2023.138181} {\bibfield
  {journal} {\bibinfo  {journal} {Phys. Lett. B}\ }\textbf {\bibinfo {volume}
  {846}},\ \bibinfo {pages} {138181} (\bibinfo {year} {2023})},\ \Eprint
  {http://arxiv.org/abs/2310.10579} {arXiv:2310.10579 [hep-lat]} \BibitemShut
  {NoStop}%
\bibitem [{\citenamefont {Zhao}(2024)}]{Zhao:2023ptv}%
  \BibitemOpen
  \bibfield  {author} {\bibinfo {author} {\bibfnamefont {Y.}~\bibnamefont
  {Zhao}},\ }\href {\doibase 10.1103/PhysRevLett.133.241904} {\bibfield
  {journal} {\bibinfo  {journal} {Phys. Rev. Lett.}\ }\textbf {\bibinfo
  {volume} {133}},\ \bibinfo {pages} {241904} (\bibinfo {year} {2024})},\
  \Eprint {http://arxiv.org/abs/2311.01391} {arXiv:2311.01391 [hep-ph]}
  \BibitemShut {NoStop}%
\bibitem [{\citenamefont {Liu}\ and\ \citenamefont {Su}(2024)}]{Liu:2023onm}%
  \BibitemOpen
  \bibfield  {author} {\bibinfo {author} {\bibfnamefont {Y.}~\bibnamefont
  {Liu}}\ and\ \bibinfo {author} {\bibfnamefont {Y.}~\bibnamefont {Su}},\
  }\href {\doibase 10.1007/JHEP02(2024)204} {\bibfield  {journal} {\bibinfo
  {journal} {JHEP}\ }\textbf {\bibinfo {volume} {2024}},\ \bibinfo {pages}
  {204} (\bibinfo {year} {2024})},\ \Eprint {http://arxiv.org/abs/2311.06907}
  {arXiv:2311.06907 [hep-ph]} \BibitemShut {NoStop}%
\bibitem [{\citenamefont {Avkhadiev}\ \emph {et~al.}(2024)\citenamefont
  {Avkhadiev}, \citenamefont {Shanahan}, \citenamefont {Wagman},\ and\
  \citenamefont {Zhao}}]{Avkhadiev:2024mgd}%
  \BibitemOpen
  \bibfield  {author} {\bibinfo {author} {\bibfnamefont {A.}~\bibnamefont
  {Avkhadiev}}, \bibinfo {author} {\bibfnamefont {P.~E.}\ \bibnamefont
  {Shanahan}}, \bibinfo {author} {\bibfnamefont {M.~L.}\ \bibnamefont
  {Wagman}}, \ and\ \bibinfo {author} {\bibfnamefont {Y.}~\bibnamefont
  {Zhao}},\ }\href {\doibase 10.1103/PhysRevLett.132.231901} {\bibfield
  {journal} {\bibinfo  {journal} {Phys. Rev. Lett.}\ }\textbf {\bibinfo
  {volume} {132}},\ \bibinfo {pages} {231901} (\bibinfo {year} {2024})},\
  \Eprint {http://arxiv.org/abs/2402.06725} {arXiv:2402.06725 [hep-lat]}
  \BibitemShut {NoStop}%
\bibitem [{\citenamefont {Baker}\ \emph {et~al.}(2024)\citenamefont {Baker},
  \citenamefont {Bollweg}, \citenamefont {Boyle}, \citenamefont {Clo\"et},
  \citenamefont {Gao}, \citenamefont {Mukherjee}, \citenamefont {Petreczky},
  \citenamefont {Zhang},\ and\ \citenamefont {Zhao}}]{Baker:2024zcd}%
  \BibitemOpen
  \bibfield  {author} {\bibinfo {author} {\bibfnamefont {E.}~\bibnamefont
  {Baker}}, \bibinfo {author} {\bibfnamefont {D.}~\bibnamefont {Bollweg}},
  \bibinfo {author} {\bibfnamefont {P.}~\bibnamefont {Boyle}}, \bibinfo
  {author} {\bibfnamefont {I.}~\bibnamefont {Clo\"et}}, \bibinfo {author}
  {\bibfnamefont {X.}~\bibnamefont {Gao}}, \bibinfo {author} {\bibfnamefont
  {S.}~\bibnamefont {Mukherjee}}, \bibinfo {author} {\bibfnamefont
  {P.}~\bibnamefont {Petreczky}}, \bibinfo {author} {\bibfnamefont
  {R.}~\bibnamefont {Zhang}}, \ and\ \bibinfo {author} {\bibfnamefont
  {Y.}~\bibnamefont {Zhao}},\ }\href {\doibase 10.1007/JHEP07(2024)211}
  {\bibfield  {journal} {\bibinfo  {journal} {JHEP}\ }\textbf {\bibinfo
  {volume} {07}},\ \bibinfo {pages} {211} (\bibinfo {year} {2024})},\ \Eprint
  {http://arxiv.org/abs/2405.20120} {arXiv:2405.20120 [hep-lat]} \BibitemShut
  {NoStop}%
\bibitem [{\citenamefont {Bollweg}\ \emph {et~al.}(2024)\citenamefont
  {Bollweg}, \citenamefont {Gao}, \citenamefont {Mukherjee},\ and\
  \citenamefont {Zhao}}]{Bollweg:2024zet}%
  \BibitemOpen
  \bibfield  {author} {\bibinfo {author} {\bibfnamefont {D.}~\bibnamefont
  {Bollweg}}, \bibinfo {author} {\bibfnamefont {X.}~\bibnamefont {Gao}},
  \bibinfo {author} {\bibfnamefont {S.}~\bibnamefont {Mukherjee}}, \ and\
  \bibinfo {author} {\bibfnamefont {Y.}~\bibnamefont {Zhao}},\ }\href {\doibase
  10.1016/j.physletb.2024.138617} {\bibfield  {journal} {\bibinfo  {journal}
  {Phys. Lett. B}\ }\textbf {\bibinfo {volume} {852}},\ \bibinfo {pages}
  {138617} (\bibinfo {year} {2024})},\ \Eprint
  {http://arxiv.org/abs/2403.00664} {arXiv:2403.00664 [hep-lat]} \BibitemShut
  {NoStop}%
\bibitem [{\citenamefont {Chen}\ \emph {et~al.}(2025)\citenamefont {Chen},
  \citenamefont {Geng}, \citenamefont {Liu}, \citenamefont {Sun}, \citenamefont
  {Yang}, \citenamefont {Yao}, \citenamefont {Zhang},\ and\ \citenamefont
  {Zhang}}]{Chen:2024rgi}%
  \BibitemOpen
  \bibfield  {author} {\bibinfo {author} {\bibfnamefont {C.}~\bibnamefont
  {Chen}}, \bibinfo {author} {\bibfnamefont {Y.}~\bibnamefont {Geng}}, \bibinfo
  {author} {\bibfnamefont {L.}~\bibnamefont {Liu}}, \bibinfo {author}
  {\bibfnamefont {P.}~\bibnamefont {Sun}}, \bibinfo {author} {\bibfnamefont
  {Y.-B.}\ \bibnamefont {Yang}}, \bibinfo {author} {\bibfnamefont
  {F.}~\bibnamefont {Yao}}, \bibinfo {author} {\bibfnamefont {J.-H.}\
  \bibnamefont {Zhang}}, \ and\ \bibinfo {author} {\bibfnamefont
  {K.}~\bibnamefont {Zhang}} (\bibinfo {collaboration} {CLQCD, Lattice
  Parton}),\ }\href {\doibase 10.1103/PhysRevD.111.074506} {\bibfield
  {journal} {\bibinfo  {journal} {Phys. Rev. D}\ }\textbf {\bibinfo {volume}
  {111}},\ \bibinfo {pages} {074506} (\bibinfo {year} {2025})},\ \Eprint
  {http://arxiv.org/abs/2408.12819} {arXiv:2408.12819 [hep-lat]} \BibitemShut
  {NoStop}%
\bibitem [{\citenamefont {Cloet}\ \emph {et~al.}(2024)\citenamefont {Cloet},
  \citenamefont {Gao}, \citenamefont {Mukherjee}, \citenamefont {Syritsyn},
  \citenamefont {Karthik}, \citenamefont {Petreczky}, \citenamefont {Zhang},\
  and\ \citenamefont {Zhao}}]{Cloet:2024vbv}%
  \BibitemOpen
  \bibfield  {author} {\bibinfo {author} {\bibfnamefont {I.}~\bibnamefont
  {Cloet}}, \bibinfo {author} {\bibfnamefont {X.}~\bibnamefont {Gao}}, \bibinfo
  {author} {\bibfnamefont {S.}~\bibnamefont {Mukherjee}}, \bibinfo {author}
  {\bibfnamefont {S.}~\bibnamefont {Syritsyn}}, \bibinfo {author}
  {\bibfnamefont {N.}~\bibnamefont {Karthik}}, \bibinfo {author} {\bibfnamefont
  {P.}~\bibnamefont {Petreczky}}, \bibinfo {author} {\bibfnamefont
  {R.}~\bibnamefont {Zhang}}, \ and\ \bibinfo {author} {\bibfnamefont
  {Y.}~\bibnamefont {Zhao}},\ }\href {\doibase 10.1103/PhysRevD.110.114502}
  {\bibfield  {journal} {\bibinfo  {journal} {Phys. Rev. D}\ }\textbf {\bibinfo
  {volume} {110}},\ \bibinfo {pages} {114502} (\bibinfo {year} {2024})},\
  \Eprint {http://arxiv.org/abs/2407.00206} {arXiv:2407.00206 [hep-lat]}
  \BibitemShut {NoStop}%
\bibitem [{\citenamefont {Ding}\ \emph {et~al.}(2025)\citenamefont {Ding},
  \citenamefont {Gao}, \citenamefont {Mukherjee}, \citenamefont {Petreczky},
  \citenamefont {Shi}, \citenamefont {Syritsyn},\ and\ \citenamefont
  {Zhao}}]{Ding:2024saz}%
  \BibitemOpen
  \bibfield  {author} {\bibinfo {author} {\bibfnamefont {H.-T.}\ \bibnamefont
  {Ding}}, \bibinfo {author} {\bibfnamefont {X.}~\bibnamefont {Gao}}, \bibinfo
  {author} {\bibfnamefont {S.}~\bibnamefont {Mukherjee}}, \bibinfo {author}
  {\bibfnamefont {P.}~\bibnamefont {Petreczky}}, \bibinfo {author}
  {\bibfnamefont {Q.}~\bibnamefont {Shi}}, \bibinfo {author} {\bibfnamefont
  {S.}~\bibnamefont {Syritsyn}}, \ and\ \bibinfo {author} {\bibfnamefont
  {Y.}~\bibnamefont {Zhao}},\ }\href {\doibase 10.1007/JHEP02(2025)056}
  {\bibfield  {journal} {\bibinfo  {journal} {JHEP}\ }\textbf {\bibinfo
  {volume} {02}},\ \bibinfo {pages} {056} (\bibinfo {year} {2025})},\ \Eprint
  {http://arxiv.org/abs/2407.03516} {arXiv:2407.03516 [hep-lat]} \BibitemShut
  {NoStop}%
\bibitem [{\citenamefont {Gao}\ \emph {et~al.}(2024{\natexlab{c}})\citenamefont
  {Gao}, \citenamefont {He}, \citenamefont {Zhang},\ and\ \citenamefont
  {Zhao}}]{Gao:2024fbh}%
  \BibitemOpen
  \bibfield  {author} {\bibinfo {author} {\bibfnamefont {X.}~\bibnamefont
  {Gao}}, \bibinfo {author} {\bibfnamefont {J.}~\bibnamefont {He}}, \bibinfo
  {author} {\bibfnamefont {R.}~\bibnamefont {Zhang}}, \ and\ \bibinfo {author}
  {\bibfnamefont {Y.}~\bibnamefont {Zhao}},\ }\href {\doibase
  10.1088/0256-307X/41/12/121201} {\bibfield  {journal} {\bibinfo  {journal}
  {Chin. Phys. Lett.}\ }\textbf {\bibinfo {volume} {41}},\ \bibinfo {pages}
  {121201} (\bibinfo {year} {2024}{\natexlab{c}})},\ \Eprint
  {http://arxiv.org/abs/2408.05910} {arXiv:2408.05910 [hep-lat]} \BibitemShut
  {NoStop}%
\bibitem [{\citenamefont {Good}\ \emph {et~al.}(2025)\citenamefont {Good},
  \citenamefont {Hasan},\ and\ \citenamefont {Lin}}]{Good:2024iur}%
  \BibitemOpen
  \bibfield  {author} {\bibinfo {author} {\bibfnamefont {W.}~\bibnamefont
  {Good}}, \bibinfo {author} {\bibfnamefont {K.}~\bibnamefont {Hasan}}, \ and\
  \bibinfo {author} {\bibfnamefont {H.-W.}\ \bibnamefont {Lin}},\ }\href
  {\doibase 10.1088/1361-6471/ada815} {\bibfield  {journal} {\bibinfo
  {journal} {J. Phys. G}\ }\textbf {\bibinfo {volume} {52}},\ \bibinfo {pages}
  {035105} (\bibinfo {year} {2025})},\ \Eprint
  {http://arxiv.org/abs/2409.02750} {arXiv:2409.02750 [hep-lat]} \BibitemShut
  {NoStop}%
\bibitem [{\citenamefont {Han}\ \emph {et~al.}(2024{\natexlab{a}})\citenamefont
  {Han}, \citenamefont {Hua}, \citenamefont {Ji}, \citenamefont {L\"u},
  \citenamefont {Wang}, \citenamefont {Xu}, \citenamefont {Zhang},\ and\
  \citenamefont {Zhao}}]{Han:2024min}%
  \BibitemOpen
  \bibfield  {author} {\bibinfo {author} {\bibfnamefont {X.-Y.}\ \bibnamefont
  {Han}}, \bibinfo {author} {\bibfnamefont {J.}~\bibnamefont {Hua}}, \bibinfo
  {author} {\bibfnamefont {X.}~\bibnamefont {Ji}}, \bibinfo {author}
  {\bibfnamefont {C.-D.}\ \bibnamefont {L\"u}}, \bibinfo {author}
  {\bibfnamefont {W.}~\bibnamefont {Wang}}, \bibinfo {author} {\bibfnamefont
  {J.}~\bibnamefont {Xu}}, \bibinfo {author} {\bibfnamefont {Q.-A.}\
  \bibnamefont {Zhang}}, \ and\ \bibinfo {author} {\bibfnamefont
  {S.}~\bibnamefont {Zhao}},\ }\href@noop {} {\  (\bibinfo {year}
  {2024}{\natexlab{a}})},\ \Eprint {http://arxiv.org/abs/2403.17492}
  {arXiv:2403.17492 [hep-ph]} \BibitemShut {NoStop}%
\bibitem [{\citenamefont {Holligan}\ and\ \citenamefont
  {Lin}(2024{\natexlab{b}})}]{Holligan:2024umc}%
  \BibitemOpen
  \bibfield  {author} {\bibinfo {author} {\bibfnamefont {J.}~\bibnamefont
  {Holligan}}\ and\ \bibinfo {author} {\bibfnamefont {H.-W.}\ \bibnamefont
  {Lin}},\ }\href {\doibase 10.1088/1361-6471/ad3162} {\bibfield  {journal}
  {\bibinfo  {journal} {J. Phys. G}\ }\textbf {\bibinfo {volume} {51}},\
  \bibinfo {pages} {065101} (\bibinfo {year} {2024}{\natexlab{b}})},\ \Eprint
  {http://arxiv.org/abs/2404.14525} {arXiv:2404.14525 [hep-lat]} \BibitemShut
  {NoStop}%
\bibitem [{\citenamefont {Holligan}\ and\ \citenamefont
  {Lin}(2024{\natexlab{c}})}]{Holligan:2024wpv}%
  \BibitemOpen
  \bibfield  {author} {\bibinfo {author} {\bibfnamefont {J.}~\bibnamefont
  {Holligan}}\ and\ \bibinfo {author} {\bibfnamefont {H.-W.}\ \bibnamefont
  {Lin}},\ }\href {\doibase 10.1016/j.physletb.2024.138731} {\bibfield
  {journal} {\bibinfo  {journal} {Phys. Lett. B}\ }\textbf {\bibinfo {volume}
  {854}},\ \bibinfo {pages} {138731} (\bibinfo {year} {2024}{\natexlab{c}})},\
  \Eprint {http://arxiv.org/abs/2405.18238} {arXiv:2405.18238 [hep-lat]}
  \BibitemShut {NoStop}%
\bibitem [{\citenamefont {Ji}\ \emph {et~al.}(2025{\natexlab{a}})\citenamefont
  {Ji}, \citenamefont {Liu}, \citenamefont {Su},\ and\ \citenamefont
  {Zhang}}]{Ji:2024hit}%
  \BibitemOpen
  \bibfield  {author} {\bibinfo {author} {\bibfnamefont {X.}~\bibnamefont
  {Ji}}, \bibinfo {author} {\bibfnamefont {Y.}~\bibnamefont {Liu}}, \bibinfo
  {author} {\bibfnamefont {Y.}~\bibnamefont {Su}}, \ and\ \bibinfo {author}
  {\bibfnamefont {R.}~\bibnamefont {Zhang}},\ }\href {\doibase
  10.1007/JHEP03(2025)045} {\bibfield  {journal} {\bibinfo  {journal} {JHEP}\
  }\textbf {\bibinfo {volume} {03}},\ \bibinfo {pages} {045} (\bibinfo {year}
  {2025}{\natexlab{a}})},\ \Eprint {http://arxiv.org/abs/2410.12910}
  {arXiv:2410.12910 [hep-ph]} \BibitemShut {NoStop}%
\bibitem [{\citenamefont {Walter}\ \emph {et~al.}(2025)\citenamefont {Walter}
  \emph {et~al.}}]{LatticeParton:2024mxp}%
  \BibitemOpen
  \bibfield  {author} {\bibinfo {author} {\bibfnamefont {L.}~\bibnamefont
  {Walter}} \emph {et~al.} (\bibinfo {collaboration} {Lattice Parton, LPC}),\
  }\href {\doibase 10.1103/PhysRevD.111.094507} {\bibfield  {journal} {\bibinfo
   {journal} {Phys. Rev. D}\ }\textbf {\bibinfo {volume} {111}},\ \bibinfo
  {pages} {094507} (\bibinfo {year} {2025})},\ \Eprint
  {http://arxiv.org/abs/2412.19988} {arXiv:2412.19988 [hep-lat]} \BibitemShut
  {NoStop}%
\bibitem [{\citenamefont {Chu}\ \emph {et~al.}(2025)\citenamefont {Chu} \emph
  {et~al.}}]{LatticeParton:2024vck}%
  \BibitemOpen
  \bibfield  {author} {\bibinfo {author} {\bibfnamefont {M.-H.}\ \bibnamefont
  {Chu}} \emph {et~al.} (\bibinfo {collaboration} {Lattice Parton}),\ }\href
  {\doibase 10.1103/PhysRevD.111.034510} {\bibfield  {journal} {\bibinfo
  {journal} {Phys. Rev. D}\ }\textbf {\bibinfo {volume} {111}},\ \bibinfo
  {pages} {034510} (\bibinfo {year} {2025})},\ \Eprint
  {http://arxiv.org/abs/2411.12554} {arXiv:2411.12554 [hep-lat]} \BibitemShut
  {NoStop}%
\bibitem [{\citenamefont {Han}\ \emph {et~al.}(2025{\natexlab{a}})\citenamefont
  {Han} \emph {et~al.}}]{LatticeParton:2024zko}%
  \BibitemOpen
  \bibfield  {author} {\bibinfo {author} {\bibfnamefont {X.-Y.}\ \bibnamefont
  {Han}} \emph {et~al.} (\bibinfo {collaboration} {Lattice Parton}),\ }\href
  {\doibase 10.1103/PhysRevD.111.034503} {\bibfield  {journal} {\bibinfo
  {journal} {Phys. Rev. D}\ }\textbf {\bibinfo {volume} {111}},\ \bibinfo
  {pages} {034503} (\bibinfo {year} {2025}{\natexlab{a}})},\ \Eprint
  {http://arxiv.org/abs/2410.18654} {arXiv:2410.18654 [hep-lat]} \BibitemShut
  {NoStop}%
\bibitem [{\citenamefont {Miller}\ \emph {et~al.}(2024)\citenamefont {Miller},
  \citenamefont {Bhattacharya}, \citenamefont {Cichy}, \citenamefont
  {Constantinou}, \citenamefont {Gao}, \citenamefont {Metz}, \citenamefont
  {Mukherjee}, \citenamefont {Petreczky}, \citenamefont {Steffens},\ and\
  \citenamefont {Zhao}}]{Miller:2024yfw}%
  \BibitemOpen
  \bibfield  {author} {\bibinfo {author} {\bibfnamefont {J.}~\bibnamefont
  {Miller}}, \bibinfo {author} {\bibfnamefont {S.}~\bibnamefont
  {Bhattacharya}}, \bibinfo {author} {\bibfnamefont {K.}~\bibnamefont {Cichy}},
  \bibinfo {author} {\bibfnamefont {M.}~\bibnamefont {Constantinou}}, \bibinfo
  {author} {\bibfnamefont {X.}~\bibnamefont {Gao}}, \bibinfo {author}
  {\bibfnamefont {A.}~\bibnamefont {Metz}}, \bibinfo {author} {\bibfnamefont
  {S.}~\bibnamefont {Mukherjee}}, \bibinfo {author} {\bibfnamefont
  {P.}~\bibnamefont {Petreczky}}, \bibinfo {author} {\bibfnamefont
  {F.}~\bibnamefont {Steffens}}, \ and\ \bibinfo {author} {\bibfnamefont
  {Y.}~\bibnamefont {Zhao}},\ }\href {\doibase 10.22323/1.453.0310} {\bibfield
  {journal} {\bibinfo  {journal} {PoS}\ }\textbf {\bibinfo {volume}
  {LATTICE2023}},\ \bibinfo {pages} {310} (\bibinfo {year} {2024})},\ \Eprint
  {http://arxiv.org/abs/2403.05282} {arXiv:2403.05282 [hep-lat]} \BibitemShut
  {NoStop}%
\bibitem [{\citenamefont {Mukherjee}\ \emph {et~al.}(2025)\citenamefont
  {Mukherjee}, \citenamefont {Bollweg}, \citenamefont {Gao},\ and\
  \citenamefont {Zhao}}]{Mukherjee:2024xie}%
  \BibitemOpen
  \bibfield  {author} {\bibinfo {author} {\bibfnamefont {S.}~\bibnamefont
  {Mukherjee}}, \bibinfo {author} {\bibfnamefont {D.}~\bibnamefont {Bollweg}},
  \bibinfo {author} {\bibfnamefont {X.}~\bibnamefont {Gao}}, \ and\ \bibinfo
  {author} {\bibfnamefont {Y.}~\bibnamefont {Zhao}},\ }\href {\doibase
  10.22323/1.469.0238} {\bibfield  {journal} {\bibinfo  {journal} {PoS}\
  }\textbf {\bibinfo {volume} {DIS2024}},\ \bibinfo {pages} {238} (\bibinfo
  {year} {2025})},\ \Eprint {http://arxiv.org/abs/2407.10739} {arXiv:2407.10739
  [hep-lat]} \BibitemShut {NoStop}%
\bibitem [{\citenamefont {Spanoudes}\ \emph {et~al.}(2024)\citenamefont
  {Spanoudes}, \citenamefont {Constantinou},\ and\ \citenamefont
  {Panagopoulos}}]{Spanoudes:2024kpb}%
  \BibitemOpen
  \bibfield  {author} {\bibinfo {author} {\bibfnamefont {G.}~\bibnamefont
  {Spanoudes}}, \bibinfo {author} {\bibfnamefont {M.}~\bibnamefont
  {Constantinou}}, \ and\ \bibinfo {author} {\bibfnamefont {H.}~\bibnamefont
  {Panagopoulos}},\ }\href {\doibase 10.1103/PhysRevD.109.114501} {\bibfield
  {journal} {\bibinfo  {journal} {Phys. Rev. D}\ }\textbf {\bibinfo {volume}
  {109}},\ \bibinfo {pages} {114501} (\bibinfo {year} {2024})},\ \Eprint
  {http://arxiv.org/abs/2401.01182} {arXiv:2401.01182 [hep-lat]} \BibitemShut
  {NoStop}%
\bibitem [{\citenamefont {Zhang}\ \emph {et~al.}(2024)\citenamefont {Zhang},
  \citenamefont {Huo}, \citenamefont {Ji}, \citenamefont {Schaefer},
  \citenamefont {Shi}, \citenamefont {Sun}, \citenamefont {Wang}, \citenamefont
  {Yang},\ and\ \citenamefont {Zhang}}]{Zhang:2024omt}%
  \BibitemOpen
  \bibfield  {author} {\bibinfo {author} {\bibfnamefont {K.}~\bibnamefont
  {Zhang}}, \bibinfo {author} {\bibfnamefont {Y.-K.}\ \bibnamefont {Huo}},
  \bibinfo {author} {\bibfnamefont {X.}~\bibnamefont {Ji}}, \bibinfo {author}
  {\bibfnamefont {A.}~\bibnamefont {Schaefer}}, \bibinfo {author}
  {\bibfnamefont {C.-J.}\ \bibnamefont {Shi}}, \bibinfo {author} {\bibfnamefont
  {P.}~\bibnamefont {Sun}}, \bibinfo {author} {\bibfnamefont {W.}~\bibnamefont
  {Wang}}, \bibinfo {author} {\bibfnamefont {Y.-B.}\ \bibnamefont {Yang}}, \
  and\ \bibinfo {author} {\bibfnamefont {J.-H.}\ \bibnamefont {Zhang}}
  (\bibinfo {collaboration} {Lattice Parton}),\ }\href {\doibase
  10.1103/PhysRevD.110.074505} {\bibfield  {journal} {\bibinfo  {journal}
  {Phys. Rev. D}\ }\textbf {\bibinfo {volume} {110}},\ \bibinfo {pages}
  {074505} (\bibinfo {year} {2024})},\ \Eprint
  {http://arxiv.org/abs/2405.14097} {arXiv:2405.14097 [hep-lat]} \BibitemShut
  {NoStop}%
\bibitem [{\citenamefont {Bollweg}\ \emph
  {et~al.}(2025{\natexlab{a}})\citenamefont {Bollweg}, \citenamefont {Gao},
  \citenamefont {Mukherjee},\ and\ \citenamefont {Zhao}}]{Bollweg:2025ecn}%
  \BibitemOpen
  \bibfield  {author} {\bibinfo {author} {\bibfnamefont {D.}~\bibnamefont
  {Bollweg}}, \bibinfo {author} {\bibfnamefont {X.}~\bibnamefont {Gao}},
  \bibinfo {author} {\bibfnamefont {S.}~\bibnamefont {Mukherjee}}, \ and\
  \bibinfo {author} {\bibfnamefont {Y.}~\bibnamefont {Zhao}},\ }\href@noop {}
  {\  (\bibinfo {year} {2025}{\natexlab{a}})},\ \Eprint
  {http://arxiv.org/abs/2505.18430} {arXiv:2505.18430 [hep-lat]} \BibitemShut
  {NoStop}%
\bibitem [{\citenamefont {Bollweg}\ \emph
  {et~al.}(2025{\natexlab{b}})\citenamefont {Bollweg}, \citenamefont {Gao},
  \citenamefont {He}, \citenamefont {Mukherjee},\ and\ \citenamefont
  {Zhao}}]{Bollweg:2025iol}%
  \BibitemOpen
  \bibfield  {author} {\bibinfo {author} {\bibfnamefont {D.}~\bibnamefont
  {Bollweg}}, \bibinfo {author} {\bibfnamefont {X.}~\bibnamefont {Gao}},
  \bibinfo {author} {\bibfnamefont {J.}~\bibnamefont {He}}, \bibinfo {author}
  {\bibfnamefont {S.}~\bibnamefont {Mukherjee}}, \ and\ \bibinfo {author}
  {\bibfnamefont {Y.}~\bibnamefont {Zhao}},\ }\href@noop {} {\  (\bibinfo
  {year} {2025}{\natexlab{b}})},\ \Eprint {http://arxiv.org/abs/2504.04625}
  {arXiv:2504.04625 [hep-lat]} \BibitemShut {NoStop}%
\bibitem [{\citenamefont {Han}\ \emph {et~al.}(2025{\natexlab{b}})\citenamefont
  {Han} \emph {et~al.}}]{Han:2025odf}%
  \BibitemOpen
  \bibfield  {author} {\bibinfo {author} {\bibfnamefont {X.-Y.}\ \bibnamefont
  {Han}} \emph {et~al.},\ }in\ \href@noop {} {\emph {\bibinfo {booktitle}
  {{16th Conference on Quark Confinement and the Hadron Spectrum}}}}\ (\bibinfo
  {year} {2025})\ \Eprint {http://arxiv.org/abs/2504.02453} {arXiv:2504.02453
  [hep-lat]} \BibitemShut {NoStop}%
\bibitem [{\citenamefont {Holligan}\ \emph {et~al.}(2025)\citenamefont
  {Holligan}, \citenamefont {Lin}, \citenamefont {Zhang},\ and\ \citenamefont
  {Zhao}}]{Holligan:2025ydm}%
  \BibitemOpen
  \bibfield  {author} {\bibinfo {author} {\bibfnamefont {J.}~\bibnamefont
  {Holligan}}, \bibinfo {author} {\bibfnamefont {H.-W.}\ \bibnamefont {Lin}},
  \bibinfo {author} {\bibfnamefont {R.}~\bibnamefont {Zhang}}, \ and\ \bibinfo
  {author} {\bibfnamefont {Y.}~\bibnamefont {Zhao}},\ }\href@noop {} {\
  (\bibinfo {year} {2025})},\ \Eprint {http://arxiv.org/abs/2501.19225}
  {arXiv:2501.19225 [hep-ph]} \BibitemShut {NoStop}%
\bibitem [{\citenamefont {Ma}\ \emph {et~al.}(2025)\citenamefont {Ma} \emph
  {et~al.}}]{LPC:2025spt}%
  \BibitemOpen
  \bibfield  {author} {\bibinfo {author} {\bibfnamefont {L.}~\bibnamefont {Ma}}
  \emph {et~al.} (\bibinfo {collaboration} {LPC}),\ }\href@noop {} {\
  (\bibinfo {year} {2025})},\ \Eprint {http://arxiv.org/abs/2502.11807}
  {arXiv:2502.11807 [hep-lat]} \BibitemShut {NoStop}%
\bibitem [{\citenamefont {Zhang}\ \emph
  {et~al.}(2025{\natexlab{a}})\citenamefont {Zhang}, \citenamefont {He},
  \citenamefont {Chu}, \citenamefont {Hua}, \citenamefont {Ji}, \citenamefont
  {Sch\"afer}, \citenamefont {Su}, \citenamefont {Wang}, \citenamefont {Yang},\
  and\ \citenamefont {Zhang}}]{Zhang:2025hvf}%
  \BibitemOpen
  \bibfield  {author} {\bibinfo {author} {\bibfnamefont {Q.-A.}\ \bibnamefont
  {Zhang}}, \bibinfo {author} {\bibfnamefont {J.-C.}\ \bibnamefont {He}},
  \bibinfo {author} {\bibfnamefont {M.-H.}\ \bibnamefont {Chu}}, \bibinfo
  {author} {\bibfnamefont {J.}~\bibnamefont {Hua}}, \bibinfo {author}
  {\bibfnamefont {X.}~\bibnamefont {Ji}}, \bibinfo {author} {\bibfnamefont
  {A.}~\bibnamefont {Sch\"afer}}, \bibinfo {author} {\bibfnamefont
  {Y.}~\bibnamefont {Su}}, \bibinfo {author} {\bibfnamefont {W.}~\bibnamefont
  {Wang}}, \bibinfo {author} {\bibfnamefont {Y.-B.}\ \bibnamefont {Yang}}, \
  and\ \bibinfo {author} {\bibfnamefont {J.-H.}\ \bibnamefont {Zhang}},\ }\href
  {\doibase 10.22323/1.465.0052} {\bibfield  {journal} {\bibinfo  {journal}
  {PoS}\ }\textbf {\bibinfo {volume} {QNP2024}},\ \bibinfo {pages} {052}
  (\bibinfo {year} {2025}{\natexlab{a}})}\BibitemShut {NoStop}%
\bibitem [{\citenamefont {Ji}\ \emph {et~al.}(2025{\natexlab{b}})\citenamefont
  {Ji}, \citenamefont {Yao},\ and\ \citenamefont {Zhang}}]{Ji:2025mvk}%
  \BibitemOpen
  \bibfield  {author} {\bibinfo {author} {\bibfnamefont {Y.}~\bibnamefont
  {Ji}}, \bibinfo {author} {\bibfnamefont {F.}~\bibnamefont {Yao}}, \ and\
  \bibinfo {author} {\bibfnamefont {J.-H.}\ \bibnamefont {Zhang}},\ }\href@noop
  {} {\  (\bibinfo {year} {2025}{\natexlab{b}})},\ \Eprint
  {http://arxiv.org/abs/2504.09367} {arXiv:2504.09367 [hep-ph]} \BibitemShut
  {NoStop}%
\bibitem [{\citenamefont {Ji}\ and\ \citenamefont {Zhang}(2015)}]{Ji:2015jwa}%
  \BibitemOpen
  \bibfield  {author} {\bibinfo {author} {\bibfnamefont {X.}~\bibnamefont
  {Ji}}\ and\ \bibinfo {author} {\bibfnamefont {J.-H.}\ \bibnamefont {Zhang}},\
  }\href {\doibase 10.1103/PhysRevD.92.034006} {\bibfield  {journal} {\bibinfo
  {journal} {Phys. Rev. D}\ }\textbf {\bibinfo {volume} {92}},\ \bibinfo
  {pages} {034006} (\bibinfo {year} {2015})},\ \Eprint
  {http://arxiv.org/abs/1505.07699} {arXiv:1505.07699 [hep-ph]} \BibitemShut
  {NoStop}%
\bibitem [{\citenamefont {Chen}\ \emph {et~al.}(2017)\citenamefont {Chen},
  \citenamefont {Ji},\ and\ \citenamefont {Zhang}}]{Chen:2016fxx}%
  \BibitemOpen
  \bibfield  {author} {\bibinfo {author} {\bibfnamefont {J.-W.}\ \bibnamefont
  {Chen}}, \bibinfo {author} {\bibfnamefont {X.}~\bibnamefont {Ji}}, \ and\
  \bibinfo {author} {\bibfnamefont {J.-H.}\ \bibnamefont {Zhang}},\ }\href
  {\doibase 10.1016/j.nuclphysb.2016.12.004} {\bibfield  {journal} {\bibinfo
  {journal} {Nucl. Phys. B}\ }\textbf {\bibinfo {volume} {915}},\ \bibinfo
  {pages} {1} (\bibinfo {year} {2017})},\ \Eprint
  {http://arxiv.org/abs/1609.08102} {arXiv:1609.08102 [hep-ph]} \BibitemShut
  {NoStop}%
\bibitem [{\citenamefont {Zhang}\ \emph
  {et~al.}(2018{\natexlab{b}})\citenamefont {Zhang}, \citenamefont {Chen},\
  and\ \citenamefont {Monahan}}]{Zhang:2018ggy}%
  \BibitemOpen
  \bibfield  {author} {\bibinfo {author} {\bibfnamefont {J.-H.}\ \bibnamefont
  {Zhang}}, \bibinfo {author} {\bibfnamefont {J.-W.}\ \bibnamefont {Chen}}, \
  and\ \bibinfo {author} {\bibfnamefont {C.}~\bibnamefont {Monahan}},\ }\href
  {\doibase 10.1103/PhysRevD.97.074508} {\bibfield  {journal} {\bibinfo
  {journal} {Phys. Rev. D}\ }\textbf {\bibinfo {volume} {97}},\ \bibinfo
  {pages} {074508} (\bibinfo {year} {2018}{\natexlab{b}})},\ \Eprint
  {http://arxiv.org/abs/1801.03023} {arXiv:1801.03023 [hep-ph]} \BibitemShut
  {NoStop}%
\bibitem [{\citenamefont {Braun}\ \emph {et~al.}(2019)\citenamefont {Braun},
  \citenamefont {Vladimirov},\ and\ \citenamefont {Zhang}}]{Braun:2018brg}%
  \BibitemOpen
  \bibfield  {author} {\bibinfo {author} {\bibfnamefont {V.~M.}\ \bibnamefont
  {Braun}}, \bibinfo {author} {\bibfnamefont {A.}~\bibnamefont {Vladimirov}}, \
  and\ \bibinfo {author} {\bibfnamefont {J.-H.}\ \bibnamefont {Zhang}},\ }\href
  {\doibase 10.1103/PhysRevD.99.014013} {\bibfield  {journal} {\bibinfo
  {journal} {Phys. Rev. D}\ }\textbf {\bibinfo {volume} {99}},\ \bibinfo
  {pages} {014013} (\bibinfo {year} {2019})},\ \Eprint
  {http://arxiv.org/abs/1810.00048} {arXiv:1810.00048 [hep-ph]} \BibitemShut
  {NoStop}%
\bibitem [{\citenamefont {Wang}\ \emph {et~al.}(2019)\citenamefont {Wang},
  \citenamefont {Zhang}, \citenamefont {Zhao},\ and\ \citenamefont
  {Zhu}}]{Wang:2019tgg}%
  \BibitemOpen
  \bibfield  {author} {\bibinfo {author} {\bibfnamefont {W.}~\bibnamefont
  {Wang}}, \bibinfo {author} {\bibfnamefont {J.-H.}\ \bibnamefont {Zhang}},
  \bibinfo {author} {\bibfnamefont {S.}~\bibnamefont {Zhao}}, \ and\ \bibinfo
  {author} {\bibfnamefont {R.}~\bibnamefont {Zhu}},\ }\href {\doibase
  10.1103/PhysRevD.100.074509} {\bibfield  {journal} {\bibinfo  {journal}
  {Phys. Rev. D}\ }\textbf {\bibinfo {volume} {100}},\ \bibinfo {pages}
  {074509} (\bibinfo {year} {2019})},\ \Eprint
  {http://arxiv.org/abs/1904.00978} {arXiv:1904.00978 [hep-ph]} \BibitemShut
  {NoStop}%
\bibitem [{\citenamefont {Ji}\ \emph {et~al.}(2021{\natexlab{d}})\citenamefont
  {Ji}, \citenamefont {Zhang}, \citenamefont {Zhao},\ and\ \citenamefont
  {Zhu}}]{Ji:2021uvr}%
  \BibitemOpen
  \bibfield  {author} {\bibinfo {author} {\bibfnamefont {Y.}~\bibnamefont
  {Ji}}, \bibinfo {author} {\bibfnamefont {J.-H.}\ \bibnamefont {Zhang}},
  \bibinfo {author} {\bibfnamefont {S.}~\bibnamefont {Zhao}}, \ and\ \bibinfo
  {author} {\bibfnamefont {R.}~\bibnamefont {Zhu}},\ }\href {\doibase
  10.1103/PhysRevD.104.094510} {\bibfield  {journal} {\bibinfo  {journal}
  {Phys. Rev. D}\ }\textbf {\bibinfo {volume} {104}},\ \bibinfo {pages}
  {094510} (\bibinfo {year} {2021}{\natexlab{d}})},\ \Eprint
  {http://arxiv.org/abs/2104.13345} {arXiv:2104.13345 [hep-ph]} \BibitemShut
  {NoStop}%
\bibitem [{\citenamefont {Su}\ \emph {et~al.}(2023)\citenamefont {Su},
  \citenamefont {Holligan}, \citenamefont {Ji}, \citenamefont {Yao},
  \citenamefont {Zhang},\ and\ \citenamefont {Zhang}}]{Su:2022fiu}%
  \BibitemOpen
  \bibfield  {author} {\bibinfo {author} {\bibfnamefont {Y.}~\bibnamefont
  {Su}}, \bibinfo {author} {\bibfnamefont {J.}~\bibnamefont {Holligan}},
  \bibinfo {author} {\bibfnamefont {X.}~\bibnamefont {Ji}}, \bibinfo {author}
  {\bibfnamefont {F.}~\bibnamefont {Yao}}, \bibinfo {author} {\bibfnamefont
  {J.-H.}\ \bibnamefont {Zhang}}, \ and\ \bibinfo {author} {\bibfnamefont
  {R.}~\bibnamefont {Zhang}},\ }\href {\doibase
  10.1016/j.nuclphysb.2023.116201} {\bibfield  {journal} {\bibinfo  {journal}
  {Nucl. Phys. B}\ }\textbf {\bibinfo {volume} {991}},\ \bibinfo {pages}
  {116201} (\bibinfo {year} {2023})},\ \Eprint
  {http://arxiv.org/abs/2209.01236} {arXiv:2209.01236 [hep-ph]} \BibitemShut
  {NoStop}%
\bibitem [{\citenamefont {Zhu}\ \emph {et~al.}(2023)\citenamefont {Zhu},
  \citenamefont {Ji}, \citenamefont {Zhang},\ and\ \citenamefont
  {Zhao}}]{Zhu:2022bja}%
  \BibitemOpen
  \bibfield  {author} {\bibinfo {author} {\bibfnamefont {R.}~\bibnamefont
  {Zhu}}, \bibinfo {author} {\bibfnamefont {Y.}~\bibnamefont {Ji}}, \bibinfo
  {author} {\bibfnamefont {J.-H.}\ \bibnamefont {Zhang}}, \ and\ \bibinfo
  {author} {\bibfnamefont {S.}~\bibnamefont {Zhao}},\ }\href {\doibase
  10.1007/JHEP02(2023)114} {\bibfield  {journal} {\bibinfo  {journal} {JHEP}\
  }\textbf {\bibinfo {volume} {02}},\ \bibinfo {pages} {114} (\bibinfo {year}
  {2023})},\ \Eprint {http://arxiv.org/abs/2209.05443} {arXiv:2209.05443
  [hep-ph]} \BibitemShut {NoStop}%
\bibitem [{\citenamefont {Yao}\ \emph {et~al.}(2023{\natexlab{b}})\citenamefont
  {Yao}, \citenamefont {Ji},\ and\ \citenamefont {Zhang}}]{Yao:2022vtp}%
  \BibitemOpen
  \bibfield  {author} {\bibinfo {author} {\bibfnamefont {F.}~\bibnamefont
  {Yao}}, \bibinfo {author} {\bibfnamefont {Y.}~\bibnamefont {Ji}}, \ and\
  \bibinfo {author} {\bibfnamefont {J.-H.}\ \bibnamefont {Zhang}},\ }\href
  {\doibase 10.1007/JHEP11(2023)021} {\bibfield  {journal} {\bibinfo  {journal}
  {JHEP}\ }\textbf {\bibinfo {volume} {11}},\ \bibinfo {pages} {021} (\bibinfo
  {year} {2023}{\natexlab{b}})},\ \Eprint {http://arxiv.org/abs/2212.14415}
  {arXiv:2212.14415 [hep-ph]} \BibitemShut {NoStop}%
\bibitem [{\citenamefont {Pang}\ \emph
  {et~al.}(2024{\natexlab{a}})\citenamefont {Pang}, \citenamefont {Yao},\ and\
  \citenamefont {Zhang}}]{Pang:2024sdl}%
  \BibitemOpen
  \bibfield  {author} {\bibinfo {author} {\bibfnamefont {Z.}~\bibnamefont
  {Pang}}, \bibinfo {author} {\bibfnamefont {F.}~\bibnamefont {Yao}}, \ and\
  \bibinfo {author} {\bibfnamefont {J.-H.}\ \bibnamefont {Zhang}},\ }\href
  {\doibase 10.1007/JHEP07(2024)222} {\bibfield  {journal} {\bibinfo  {journal}
  {JHEP}\ }\textbf {\bibinfo {volume} {07}},\ \bibinfo {pages} {222} (\bibinfo
  {year} {2024}{\natexlab{a}})},\ \Eprint {http://arxiv.org/abs/2404.00693}
  {arXiv:2404.00693 [hep-ph]} \BibitemShut {NoStop}%
\bibitem [{\citenamefont {Han}\ \emph {et~al.}(2024{\natexlab{b}})\citenamefont
  {Han}, \citenamefont {Wang}, \citenamefont {Zhang},\ and\ \citenamefont
  {Zhang}}]{Han:2024cht}%
  \BibitemOpen
  \bibfield  {author} {\bibinfo {author} {\bibfnamefont {C.}~\bibnamefont
  {Han}}, \bibinfo {author} {\bibfnamefont {W.}~\bibnamefont {Wang}}, \bibinfo
  {author} {\bibfnamefont {J.-L.}\ \bibnamefont {Zhang}}, \ and\ \bibinfo
  {author} {\bibfnamefont {J.-H.}\ \bibnamefont {Zhang}},\ }\href {\doibase
  10.1103/PhysRevD.110.094038} {\bibfield  {journal} {\bibinfo  {journal}
  {Phys. Rev. D}\ }\textbf {\bibinfo {volume} {110}},\ \bibinfo {pages}
  {094038} (\bibinfo {year} {2024}{\natexlab{b}})},\ \Eprint
  {http://arxiv.org/abs/2408.13486} {arXiv:2408.13486 [hep-ph]} \BibitemShut
  {NoStop}%
\bibitem [{\citenamefont {Ji}(2024)}]{Ji:2024oka}%
  \BibitemOpen
  \bibfield  {author} {\bibinfo {author} {\bibfnamefont {X.}~\bibnamefont
  {Ji}},\ }\href {\doibase 10.1016/j.nuclphysb.2024.116670} {\bibfield
  {journal} {\bibinfo  {journal} {Nucl. Phys. B}\ }\textbf {\bibinfo {volume}
  {1007}},\ \bibinfo {pages} {116670} (\bibinfo {year} {2024})},\ \Eprint
  {http://arxiv.org/abs/2408.03378} {arXiv:2408.03378 [hep-ph]} \BibitemShut
  {NoStop}%
\bibitem [{\citenamefont {Orginos}\ \emph {et~al.}(2017)\citenamefont
  {Orginos}, \citenamefont {Radyushkin}, \citenamefont {Karpie},\ and\
  \citenamefont {Zafeiropoulos}}]{Orginos:2017kos}%
  \BibitemOpen
  \bibfield  {author} {\bibinfo {author} {\bibfnamefont {K.}~\bibnamefont
  {Orginos}}, \bibinfo {author} {\bibfnamefont {A.}~\bibnamefont {Radyushkin}},
  \bibinfo {author} {\bibfnamefont {J.}~\bibnamefont {Karpie}}, \ and\ \bibinfo
  {author} {\bibfnamefont {S.}~\bibnamefont {Zafeiropoulos}},\ }\href {\doibase
  10.1103/PhysRevD.96.094503} {\bibfield  {journal} {\bibinfo  {journal} {Phys.
  Rev. D}\ }\textbf {\bibinfo {volume} {96}},\ \bibinfo {pages} {094503}
  (\bibinfo {year} {2017})},\ \Eprint {http://arxiv.org/abs/1706.05373}
  {arXiv:1706.05373 [hep-ph]} \BibitemShut {NoStop}%
\bibitem [{\citenamefont {Karpie}\ \emph
  {et~al.}(2018{\natexlab{a}})\citenamefont {Karpie}, \citenamefont {Orginos},
  \citenamefont {Radyushkin},\ and\ \citenamefont
  {Zafeiropoulos}}]{Karpie:2017bzm}%
  \BibitemOpen
  \bibfield  {author} {\bibinfo {author} {\bibfnamefont {J.}~\bibnamefont
  {Karpie}}, \bibinfo {author} {\bibfnamefont {K.}~\bibnamefont {Orginos}},
  \bibinfo {author} {\bibfnamefont {A.}~\bibnamefont {Radyushkin}}, \ and\
  \bibinfo {author} {\bibfnamefont {S.}~\bibnamefont {Zafeiropoulos}},\ }\href
  {\doibase 10.1051/epjconf/201817506032} {\bibfield  {journal} {\bibinfo
  {journal} {EPJ Web Conf.}\ }\textbf {\bibinfo {volume} {175}},\ \bibinfo
  {pages} {06032} (\bibinfo {year} {2018}{\natexlab{a}})},\ \Eprint
  {http://arxiv.org/abs/1710.08288} {arXiv:1710.08288 [hep-lat]} \BibitemShut
  {NoStop}%
\bibitem [{\citenamefont
  {Radyushkin}(2018{\natexlab{a}})}]{Radyushkin:2017lvu}%
  \BibitemOpen
  \bibfield  {author} {\bibinfo {author} {\bibfnamefont {A.~V.}\ \bibnamefont
  {Radyushkin}},\ }\href {\doibase 10.1016/j.physletb.2018.04.023} {\bibfield
  {journal} {\bibinfo  {journal} {Phys. Lett. B}\ }\textbf {\bibinfo {volume}
  {781}},\ \bibinfo {pages} {433} (\bibinfo {year} {2018}{\natexlab{a}})},\
  \Eprint {http://arxiv.org/abs/1710.08813} {arXiv:1710.08813 [hep-ph]}
  \BibitemShut {NoStop}%
\bibitem [{\citenamefont
  {Radyushkin}(2017{\natexlab{b}})}]{Radyushkin:2017sfi}%
  \BibitemOpen
  \bibfield  {author} {\bibinfo {author} {\bibfnamefont {A.}~\bibnamefont
  {Radyushkin}},\ }\href {\doibase 10.22323/1.308.0021} {\bibfield  {journal}
  {\bibinfo  {journal} {PoS}\ }\textbf {\bibinfo {volume} {QCDEV2017}},\
  \bibinfo {pages} {021} (\bibinfo {year} {2017}{\natexlab{b}})},\ \Eprint
  {http://arxiv.org/abs/1711.06031} {arXiv:1711.06031 [hep-ph]} \BibitemShut
  {NoStop}%
\bibitem [{\citenamefont {Karpie}\ \emph
  {et~al.}(2018{\natexlab{b}})\citenamefont {Karpie}, \citenamefont {Orginos},\
  and\ \citenamefont {Zafeiropoulos}}]{Karpie:2018zaz}%
  \BibitemOpen
  \bibfield  {author} {\bibinfo {author} {\bibfnamefont {J.}~\bibnamefont
  {Karpie}}, \bibinfo {author} {\bibfnamefont {K.}~\bibnamefont {Orginos}}, \
  and\ \bibinfo {author} {\bibfnamefont {S.}~\bibnamefont {Zafeiropoulos}},\
  }\href {\doibase 10.1007/JHEP11(2018)178} {\bibfield  {journal} {\bibinfo
  {journal} {JHEP}\ }\textbf {\bibinfo {volume} {11}},\ \bibinfo {pages} {178}
  (\bibinfo {year} {2018}{\natexlab{b}})},\ \Eprint
  {http://arxiv.org/abs/1807.10933} {arXiv:1807.10933 [hep-lat]} \BibitemShut
  {NoStop}%
\bibitem [{\citenamefont
  {Radyushkin}(2018{\natexlab{b}})}]{Radyushkin:2018cvn}%
  \BibitemOpen
  \bibfield  {author} {\bibinfo {author} {\bibfnamefont {A.}~\bibnamefont
  {Radyushkin}},\ }\href {\doibase 10.1103/PhysRevD.98.014019} {\bibfield
  {journal} {\bibinfo  {journal} {Phys. Rev. D}\ }\textbf {\bibinfo {volume}
  {98}},\ \bibinfo {pages} {014019} (\bibinfo {year} {2018}{\natexlab{b}})},\
  \Eprint {http://arxiv.org/abs/1801.02427} {arXiv:1801.02427 [hep-ph]}
  \BibitemShut {NoStop}%
\bibitem [{\citenamefont
  {Radyushkin}(2019{\natexlab{a}})}]{Radyushkin:2018nbf}%
  \BibitemOpen
  \bibfield  {author} {\bibinfo {author} {\bibfnamefont {A.~V.}\ \bibnamefont
  {Radyushkin}},\ }\href {\doibase 10.1016/j.physletb.2018.11.047} {\bibfield
  {journal} {\bibinfo  {journal} {Phys. Lett. B}\ }\textbf {\bibinfo {volume}
  {788}},\ \bibinfo {pages} {380} (\bibinfo {year} {2019}{\natexlab{a}})},\
  \Eprint {http://arxiv.org/abs/1807.07509} {arXiv:1807.07509 [hep-ph]}
  \BibitemShut {NoStop}%
\bibitem [{\citenamefont {Balitsky}\ \emph {et~al.}(2020)\citenamefont
  {Balitsky}, \citenamefont {Morris},\ and\ \citenamefont
  {Radyushkin}}]{Balitsky:2019krf}%
  \BibitemOpen
  \bibfield  {author} {\bibinfo {author} {\bibfnamefont {I.}~\bibnamefont
  {Balitsky}}, \bibinfo {author} {\bibfnamefont {W.}~\bibnamefont {Morris}}, \
  and\ \bibinfo {author} {\bibfnamefont {A.}~\bibnamefont {Radyushkin}},\
  }\href {\doibase 10.1016/j.physletb.2020.135621} {\bibfield  {journal}
  {\bibinfo  {journal} {Phys. Lett. B}\ }\textbf {\bibinfo {volume} {808}},\
  \bibinfo {pages} {135621} (\bibinfo {year} {2020})},\ \Eprint
  {http://arxiv.org/abs/1910.13963} {arXiv:1910.13963 [hep-ph]} \BibitemShut
  {NoStop}%
\bibitem [{\citenamefont {Jo\'o}\ \emph
  {et~al.}(2019{\natexlab{a}})\citenamefont {Jo\'o}, \citenamefont {Karpie},
  \citenamefont {Orginos}, \citenamefont {Radyushkin}, \citenamefont
  {Richards}, \citenamefont {Sufian},\ and\ \citenamefont
  {Zafeiropoulos}}]{Joo:2019bzr}%
  \BibitemOpen
  \bibfield  {author} {\bibinfo {author} {\bibfnamefont {B.}~\bibnamefont
  {Jo\'o}}, \bibinfo {author} {\bibfnamefont {J.}~\bibnamefont {Karpie}},
  \bibinfo {author} {\bibfnamefont {K.}~\bibnamefont {Orginos}}, \bibinfo
  {author} {\bibfnamefont {A.~V.}\ \bibnamefont {Radyushkin}}, \bibinfo
  {author} {\bibfnamefont {D.~G.}\ \bibnamefont {Richards}}, \bibinfo {author}
  {\bibfnamefont {R.~S.}\ \bibnamefont {Sufian}}, \ and\ \bibinfo {author}
  {\bibfnamefont {S.}~\bibnamefont {Zafeiropoulos}},\ }\href {\doibase
  10.1103/PhysRevD.100.114512} {\bibfield  {journal} {\bibinfo  {journal}
  {Phys. Rev. D}\ }\textbf {\bibinfo {volume} {100}},\ \bibinfo {pages}
  {114512} (\bibinfo {year} {2019}{\natexlab{a}})},\ \Eprint
  {http://arxiv.org/abs/1909.08517} {arXiv:1909.08517 [hep-lat]} \BibitemShut
  {NoStop}%
\bibitem [{\citenamefont {Jo\'o}\ \emph
  {et~al.}(2019{\natexlab{b}})\citenamefont {Jo\'o}, \citenamefont {Karpie},
  \citenamefont {Orginos}, \citenamefont {Radyushkin}, \citenamefont
  {Richards},\ and\ \citenamefont {Zafeiropoulos}}]{Joo:2019jct}%
  \BibitemOpen
  \bibfield  {author} {\bibinfo {author} {\bibfnamefont {B.}~\bibnamefont
  {Jo\'o}}, \bibinfo {author} {\bibfnamefont {J.}~\bibnamefont {Karpie}},
  \bibinfo {author} {\bibfnamefont {K.}~\bibnamefont {Orginos}}, \bibinfo
  {author} {\bibfnamefont {A.}~\bibnamefont {Radyushkin}}, \bibinfo {author}
  {\bibfnamefont {D.}~\bibnamefont {Richards}}, \ and\ \bibinfo {author}
  {\bibfnamefont {S.}~\bibnamefont {Zafeiropoulos}},\ }\href {\doibase
  10.1007/JHEP12(2019)081} {\bibfield  {journal} {\bibinfo  {journal} {JHEP}\
  }\textbf {\bibinfo {volume} {12}},\ \bibinfo {pages} {081} (\bibinfo {year}
  {2019}{\natexlab{b}})},\ \Eprint {http://arxiv.org/abs/1908.09771}
  {arXiv:1908.09771 [hep-lat]} \BibitemShut {NoStop}%
\bibitem [{\citenamefont {Karpie}\ \emph {et~al.}(2019)\citenamefont {Karpie},
  \citenamefont {Orginos}, \citenamefont {Rothkopf},\ and\ \citenamefont
  {Zafeiropoulos}}]{Karpie:2019eiq}%
  \BibitemOpen
  \bibfield  {author} {\bibinfo {author} {\bibfnamefont {J.}~\bibnamefont
  {Karpie}}, \bibinfo {author} {\bibfnamefont {K.}~\bibnamefont {Orginos}},
  \bibinfo {author} {\bibfnamefont {A.}~\bibnamefont {Rothkopf}}, \ and\
  \bibinfo {author} {\bibfnamefont {S.}~\bibnamefont {Zafeiropoulos}},\ }\href
  {\doibase 10.1007/JHEP04(2019)057} {\bibfield  {journal} {\bibinfo  {journal}
  {JHEP}\ }\textbf {\bibinfo {volume} {04}},\ \bibinfo {pages} {057} (\bibinfo
  {year} {2019})},\ \Eprint {http://arxiv.org/abs/1901.05408} {arXiv:1901.05408
  [hep-lat]} \BibitemShut {NoStop}%
\bibitem [{\citenamefont {Radyushkin}(2020)}]{Radyushkin:2019mye}%
  \BibitemOpen
  \bibfield  {author} {\bibinfo {author} {\bibfnamefont {A.~V.}\ \bibnamefont
  {Radyushkin}},\ }\href {\doibase 10.1142/S0217751X20300021} {\bibfield
  {journal} {\bibinfo  {journal} {Int. J. Mod. Phys. A}\ }\textbf {\bibinfo
  {volume} {35}},\ \bibinfo {pages} {2030002} (\bibinfo {year} {2020})},\
  \Eprint {http://arxiv.org/abs/1912.04244} {arXiv:1912.04244 [hep-ph]}
  \BibitemShut {NoStop}%
\bibitem [{\citenamefont
  {Radyushkin}(2019{\natexlab{b}})}]{Radyushkin:2019owq}%
  \BibitemOpen
  \bibfield  {author} {\bibinfo {author} {\bibfnamefont {A.~V.}\ \bibnamefont
  {Radyushkin}},\ }\href {\doibase 10.1103/PhysRevD.100.116011} {\bibfield
  {journal} {\bibinfo  {journal} {Phys. Rev. D}\ }\textbf {\bibinfo {volume}
  {100}},\ \bibinfo {pages} {116011} (\bibinfo {year} {2019}{\natexlab{b}})},\
  \Eprint {http://arxiv.org/abs/1909.08474} {arXiv:1909.08474 [hep-ph]}
  \BibitemShut {NoStop}%
\bibitem [{\citenamefont {Sufian}\ \emph {et~al.}(2019)\citenamefont {Sufian},
  \citenamefont {Karpie}, \citenamefont {Egerer}, \citenamefont {Orginos},
  \citenamefont {Qiu},\ and\ \citenamefont {Richards}}]{Sufian:2019bol}%
  \BibitemOpen
  \bibfield  {author} {\bibinfo {author} {\bibfnamefont {R.~S.}\ \bibnamefont
  {Sufian}}, \bibinfo {author} {\bibfnamefont {J.}~\bibnamefont {Karpie}},
  \bibinfo {author} {\bibfnamefont {C.}~\bibnamefont {Egerer}}, \bibinfo
  {author} {\bibfnamefont {K.}~\bibnamefont {Orginos}}, \bibinfo {author}
  {\bibfnamefont {J.-W.}\ \bibnamefont {Qiu}}, \ and\ \bibinfo {author}
  {\bibfnamefont {D.~G.}\ \bibnamefont {Richards}},\ }\href {\doibase
  10.1103/PhysRevD.99.074507} {\bibfield  {journal} {\bibinfo  {journal} {Phys.
  Rev. D}\ }\textbf {\bibinfo {volume} {99}},\ \bibinfo {pages} {074507}
  (\bibinfo {year} {2019})},\ \Eprint {http://arxiv.org/abs/1901.03921}
  {arXiv:1901.03921 [hep-lat]} \BibitemShut {NoStop}%
\bibitem [{\citenamefont {Bhat}\ \emph {et~al.}(2021)\citenamefont {Bhat},
  \citenamefont {Cichy}, \citenamefont {Constantinou},\ and\ \citenamefont
  {Scapellato}}]{Bhat:2020ktg}%
  \BibitemOpen
  \bibfield  {author} {\bibinfo {author} {\bibfnamefont {M.}~\bibnamefont
  {Bhat}}, \bibinfo {author} {\bibfnamefont {K.}~\bibnamefont {Cichy}},
  \bibinfo {author} {\bibfnamefont {M.}~\bibnamefont {Constantinou}}, \ and\
  \bibinfo {author} {\bibfnamefont {A.}~\bibnamefont {Scapellato}},\ }\href
  {\doibase 10.1103/PhysRevD.103.034510} {\bibfield  {journal} {\bibinfo
  {journal} {Phys. Rev. D}\ }\textbf {\bibinfo {volume} {103}},\ \bibinfo
  {pages} {034510} (\bibinfo {year} {2021})},\ \Eprint
  {http://arxiv.org/abs/2005.02102} {arXiv:2005.02102 [hep-lat]} \BibitemShut
  {NoStop}%
\bibitem [{\citenamefont {Bringewatt}\ \emph {et~al.}(2021)\citenamefont
  {Bringewatt}, \citenamefont {Sato}, \citenamefont {Melnitchouk},
  \citenamefont {Qiu}, \citenamefont {Steffens},\ and\ \citenamefont
  {Constantinou}}]{Bringewatt:2020ixn}%
  \BibitemOpen
  \bibfield  {author} {\bibinfo {author} {\bibfnamefont {J.}~\bibnamefont
  {Bringewatt}}, \bibinfo {author} {\bibfnamefont {N.}~\bibnamefont {Sato}},
  \bibinfo {author} {\bibfnamefont {W.}~\bibnamefont {Melnitchouk}}, \bibinfo
  {author} {\bibfnamefont {J.-W.}\ \bibnamefont {Qiu}}, \bibinfo {author}
  {\bibfnamefont {F.}~\bibnamefont {Steffens}}, \ and\ \bibinfo {author}
  {\bibfnamefont {M.}~\bibnamefont {Constantinou}},\ }\href {\doibase
  10.1103/PhysRevD.103.016003} {\bibfield  {journal} {\bibinfo  {journal}
  {Phys. Rev. D}\ }\textbf {\bibinfo {volume} {103}},\ \bibinfo {pages}
  {016003} (\bibinfo {year} {2021})},\ \Eprint
  {http://arxiv.org/abs/2010.00548} {arXiv:2010.00548 [hep-ph]} \BibitemShut
  {NoStop}%
\bibitem [{\citenamefont {Del~Debbio}\ \emph {et~al.}(2021)\citenamefont
  {Del~Debbio}, \citenamefont {Giani}, \citenamefont {Karpie}, \citenamefont
  {Orginos}, \citenamefont {Radyushkin},\ and\ \citenamefont
  {Zafeiropoulos}}]{DelDebbio:2020rgv}%
  \BibitemOpen
  \bibfield  {author} {\bibinfo {author} {\bibfnamefont {L.}~\bibnamefont
  {Del~Debbio}}, \bibinfo {author} {\bibfnamefont {T.}~\bibnamefont {Giani}},
  \bibinfo {author} {\bibfnamefont {J.}~\bibnamefont {Karpie}}, \bibinfo
  {author} {\bibfnamefont {K.}~\bibnamefont {Orginos}}, \bibinfo {author}
  {\bibfnamefont {A.}~\bibnamefont {Radyushkin}}, \ and\ \bibinfo {author}
  {\bibfnamefont {S.}~\bibnamefont {Zafeiropoulos}},\ }\href {\doibase
  10.1007/JHEP02(2021)138} {\bibfield  {journal} {\bibinfo  {journal} {JHEP}\
  }\textbf {\bibinfo {volume} {02}},\ \bibinfo {pages} {138} (\bibinfo {year}
  {2021})},\ \Eprint {http://arxiv.org/abs/2010.03996} {arXiv:2010.03996
  [hep-ph]} \BibitemShut {NoStop}%
\bibitem [{\citenamefont {Fan}\ \emph {et~al.}(2021)\citenamefont {Fan},
  \citenamefont {Zhang},\ and\ \citenamefont {Lin}}]{Fan:2020cpa}%
  \BibitemOpen
  \bibfield  {author} {\bibinfo {author} {\bibfnamefont {Z.}~\bibnamefont
  {Fan}}, \bibinfo {author} {\bibfnamefont {R.}~\bibnamefont {Zhang}}, \ and\
  \bibinfo {author} {\bibfnamefont {H.-W.}\ \bibnamefont {Lin}},\ }\href
  {\doibase 10.1142/S0217751X21500809} {\bibfield  {journal} {\bibinfo
  {journal} {Int. J. Mod. Phys. A}\ }\textbf {\bibinfo {volume} {36}},\
  \bibinfo {pages} {2150080} (\bibinfo {year} {2021})},\ \Eprint
  {http://arxiv.org/abs/2007.16113} {arXiv:2007.16113 [hep-lat]} \BibitemShut
  {NoStop}%
\bibitem [{\citenamefont {Jo\'o}\ \emph {et~al.}(2020)\citenamefont {Jo\'o},
  \citenamefont {Karpie}, \citenamefont {Orginos}, \citenamefont {Radyushkin},
  \citenamefont {Richards},\ and\ \citenamefont {Zafeiropoulos}}]{Joo:2020spy}%
  \BibitemOpen
  \bibfield  {author} {\bibinfo {author} {\bibfnamefont {B.}~\bibnamefont
  {Jo\'o}}, \bibinfo {author} {\bibfnamefont {J.}~\bibnamefont {Karpie}},
  \bibinfo {author} {\bibfnamefont {K.}~\bibnamefont {Orginos}}, \bibinfo
  {author} {\bibfnamefont {A.~V.}\ \bibnamefont {Radyushkin}}, \bibinfo
  {author} {\bibfnamefont {D.~G.}\ \bibnamefont {Richards}}, \ and\ \bibinfo
  {author} {\bibfnamefont {S.}~\bibnamefont {Zafeiropoulos}},\ }\href {\doibase
  10.1103/PhysRevLett.125.232003} {\bibfield  {journal} {\bibinfo  {journal}
  {Phys. Rev. Lett.}\ }\textbf {\bibinfo {volume} {125}},\ \bibinfo {pages}
  {232003} (\bibinfo {year} {2020})},\ \Eprint
  {http://arxiv.org/abs/2004.01687} {arXiv:2004.01687 [hep-lat]} \BibitemShut
  {NoStop}%
\bibitem [{\citenamefont {Li}\ \emph {et~al.}(2021)\citenamefont {Li},
  \citenamefont {Ma},\ and\ \citenamefont {Qiu}}]{Li:2020xml}%
  \BibitemOpen
  \bibfield  {author} {\bibinfo {author} {\bibfnamefont {Z.-Y.}\ \bibnamefont
  {Li}}, \bibinfo {author} {\bibfnamefont {Y.-Q.}\ \bibnamefont {Ma}}, \ and\
  \bibinfo {author} {\bibfnamefont {J.-W.}\ \bibnamefont {Qiu}},\ }\href
  {\doibase 10.1103/PhysRevLett.126.072001} {\bibfield  {journal} {\bibinfo
  {journal} {Phys. Rev. Lett.}\ }\textbf {\bibinfo {volume} {126}},\ \bibinfo
  {pages} {072001} (\bibinfo {year} {2021})},\ \Eprint
  {http://arxiv.org/abs/2006.12370} {arXiv:2006.12370 [hep-ph]} \BibitemShut
  {NoStop}%
\bibitem [{\citenamefont {Sufian}\ \emph {et~al.}(2020)\citenamefont {Sufian},
  \citenamefont {Egerer}, \citenamefont {Karpie}, \citenamefont {Edwards},
  \citenamefont {Jo\'o}, \citenamefont {Ma}, \citenamefont {Orginos},
  \citenamefont {Qiu},\ and\ \citenamefont {Richards}}]{Sufian:2020vzb}%
  \BibitemOpen
  \bibfield  {author} {\bibinfo {author} {\bibfnamefont {R.~S.}\ \bibnamefont
  {Sufian}}, \bibinfo {author} {\bibfnamefont {C.}~\bibnamefont {Egerer}},
  \bibinfo {author} {\bibfnamefont {J.}~\bibnamefont {Karpie}}, \bibinfo
  {author} {\bibfnamefont {R.~G.}\ \bibnamefont {Edwards}}, \bibinfo {author}
  {\bibfnamefont {B.}~\bibnamefont {Jo\'o}}, \bibinfo {author} {\bibfnamefont
  {Y.-Q.}\ \bibnamefont {Ma}}, \bibinfo {author} {\bibfnamefont
  {K.}~\bibnamefont {Orginos}}, \bibinfo {author} {\bibfnamefont {J.-W.}\
  \bibnamefont {Qiu}}, \ and\ \bibinfo {author} {\bibfnamefont {D.~G.}\
  \bibnamefont {Richards}},\ }\href {\doibase 10.1103/PhysRevD.102.054508}
  {\bibfield  {journal} {\bibinfo  {journal} {Phys. Rev. D}\ }\textbf {\bibinfo
  {volume} {102}},\ \bibinfo {pages} {054508} (\bibinfo {year} {2020})},\
  \Eprint {http://arxiv.org/abs/2001.04960} {arXiv:2001.04960 [hep-lat]}
  \BibitemShut {NoStop}%
\bibitem [{\citenamefont {Zhao}\ and\ \citenamefont
  {Radyushkin}(2021)}]{Zhao:2020bsx}%
  \BibitemOpen
  \bibfield  {author} {\bibinfo {author} {\bibfnamefont {S.}~\bibnamefont
  {Zhao}}\ and\ \bibinfo {author} {\bibfnamefont {A.~V.}\ \bibnamefont
  {Radyushkin}},\ }\href {\doibase 10.1103/PhysRevD.103.054022} {\bibfield
  {journal} {\bibinfo  {journal} {Phys. Rev. D}\ }\textbf {\bibinfo {volume}
  {103}},\ \bibinfo {pages} {054022} (\bibinfo {year} {2021})},\ \Eprint
  {http://arxiv.org/abs/2006.05663} {arXiv:2006.05663 [hep-ph]} \BibitemShut
  {NoStop}%
\bibitem [{\citenamefont {Balitsky}\ \emph
  {et~al.}(2022{\natexlab{a}})\citenamefont {Balitsky}, \citenamefont
  {Morris},\ and\ \citenamefont {Radyushkin}}]{Balitsky:2021bds}%
  \BibitemOpen
  \bibfield  {author} {\bibinfo {author} {\bibfnamefont {I.}~\bibnamefont
  {Balitsky}}, \bibinfo {author} {\bibfnamefont {W.}~\bibnamefont {Morris}}, \
  and\ \bibinfo {author} {\bibfnamefont {A.}~\bibnamefont {Radyushkin}},\
  }\href {\doibase 10.21468/SciPostPhysProc.8.161} {\bibfield  {journal}
  {\bibinfo  {journal} {SciPost Phys. Proc.}\ }\textbf {\bibinfo {volume}
  {8}},\ \bibinfo {pages} {161} (\bibinfo {year} {2022}{\natexlab{a}})},\
  \Eprint {http://arxiv.org/abs/2106.01916} {arXiv:2106.01916 [hep-ph]}
  \BibitemShut {NoStop}%
\bibitem [{\citenamefont {Balitsky}\ \emph
  {et~al.}(2022{\natexlab{b}})\citenamefont {Balitsky}, \citenamefont
  {Morris},\ and\ \citenamefont {Radyushkin}}]{Balitsky:2021cwr}%
  \BibitemOpen
  \bibfield  {author} {\bibinfo {author} {\bibfnamefont {I.}~\bibnamefont
  {Balitsky}}, \bibinfo {author} {\bibfnamefont {W.}~\bibnamefont {Morris}}, \
  and\ \bibinfo {author} {\bibfnamefont {A.}~\bibnamefont {Radyushkin}},\
  }\href {\doibase 10.1007/JHEP02(2022)193} {\bibfield  {journal} {\bibinfo
  {journal} {JHEP}\ }\textbf {\bibinfo {volume} {02}},\ \bibinfo {pages} {193}
  (\bibinfo {year} {2022}{\natexlab{b}})},\ \Eprint
  {http://arxiv.org/abs/2112.02011} {arXiv:2112.02011 [hep-ph]} \BibitemShut
  {NoStop}%
\bibitem [{\citenamefont {Balitsky}\ \emph
  {et~al.}(2022{\natexlab{c}})\citenamefont {Balitsky}, \citenamefont
  {Morris},\ and\ \citenamefont {Radyushkin}}]{Balitsky:2021qsr}%
  \BibitemOpen
  \bibfield  {author} {\bibinfo {author} {\bibfnamefont {I.}~\bibnamefont
  {Balitsky}}, \bibinfo {author} {\bibfnamefont {W.}~\bibnamefont {Morris}}, \
  and\ \bibinfo {author} {\bibfnamefont {A.}~\bibnamefont {Radyushkin}},\
  }\href {\doibase 10.1103/PhysRevD.105.014008} {\bibfield  {journal} {\bibinfo
   {journal} {Phys. Rev. D}\ }\textbf {\bibinfo {volume} {105}},\ \bibinfo
  {pages} {014008} (\bibinfo {year} {2022}{\natexlab{c}})},\ \Eprint
  {http://arxiv.org/abs/2111.06797} {arXiv:2111.06797 [hep-ph]} \BibitemShut
  {NoStop}%
\bibitem [{\citenamefont {Egerer}\ \emph {et~al.}(2021)\citenamefont {Egerer},
  \citenamefont {Edwards}, \citenamefont {Kallidonis}, \citenamefont {Orginos},
  \citenamefont {Radyushkin}, \citenamefont {Richards}, \citenamefont
  {Romero},\ and\ \citenamefont {Zafeiropoulos}}]{Egerer:2021ymv}%
  \BibitemOpen
  \bibfield  {author} {\bibinfo {author} {\bibfnamefont {C.}~\bibnamefont
  {Egerer}}, \bibinfo {author} {\bibfnamefont {R.~G.}\ \bibnamefont {Edwards}},
  \bibinfo {author} {\bibfnamefont {C.}~\bibnamefont {Kallidonis}}, \bibinfo
  {author} {\bibfnamefont {K.}~\bibnamefont {Orginos}}, \bibinfo {author}
  {\bibfnamefont {A.~V.}\ \bibnamefont {Radyushkin}}, \bibinfo {author}
  {\bibfnamefont {D.~G.}\ \bibnamefont {Richards}}, \bibinfo {author}
  {\bibfnamefont {E.}~\bibnamefont {Romero}}, \ and\ \bibinfo {author}
  {\bibfnamefont {S.}~\bibnamefont {Zafeiropoulos}} (\bibinfo {collaboration}
  {HadStruc}),\ }\href {\doibase 10.1007/JHEP11(2021)148} {\bibfield  {journal}
  {\bibinfo  {journal} {JHEP}\ }\textbf {\bibinfo {volume} {11}},\ \bibinfo
  {pages} {148} (\bibinfo {year} {2021})},\ \Eprint
  {http://arxiv.org/abs/2107.05199} {arXiv:2107.05199 [hep-lat]} \BibitemShut
  {NoStop}%
\bibitem [{\citenamefont {Fan}\ and\ \citenamefont {Lin}(2021)}]{Fan:2021bcr}%
  \BibitemOpen
  \bibfield  {author} {\bibinfo {author} {\bibfnamefont {Z.}~\bibnamefont
  {Fan}}\ and\ \bibinfo {author} {\bibfnamefont {H.-W.}\ \bibnamefont {Lin}},\
  }\href {\doibase 10.1016/j.physletb.2021.136778} {\bibfield  {journal}
  {\bibinfo  {journal} {Phys. Lett. B}\ }\textbf {\bibinfo {volume} {823}},\
  \bibinfo {pages} {136778} (\bibinfo {year} {2021})},\ \Eprint
  {http://arxiv.org/abs/2104.06372} {arXiv:2104.06372 [hep-lat]} \BibitemShut
  {NoStop}%
\bibitem [{\citenamefont {Egerer}\ \emph
  {et~al.}(2022{\natexlab{a}})\citenamefont {Egerer} \emph
  {et~al.}}]{HadStruc:2021qdf}%
  \BibitemOpen
  \bibfield  {author} {\bibinfo {author} {\bibfnamefont {C.}~\bibnamefont
  {Egerer}} \emph {et~al.} (\bibinfo {collaboration} {HadStruc}),\ }\href
  {\doibase 10.1103/PhysRevD.105.034507} {\bibfield  {journal} {\bibinfo
  {journal} {Phys. Rev. D}\ }\textbf {\bibinfo {volume} {105}},\ \bibinfo
  {pages} {034507} (\bibinfo {year} {2022}{\natexlab{a}})},\ \Eprint
  {http://arxiv.org/abs/2111.01808} {arXiv:2111.01808 [hep-lat]} \BibitemShut
  {NoStop}%
\bibitem [{\citenamefont {Khan}\ \emph {et~al.}(2021)\citenamefont {Khan} \emph
  {et~al.}}]{HadStruc:2021wmh}%
  \BibitemOpen
  \bibfield  {author} {\bibinfo {author} {\bibfnamefont {T.}~\bibnamefont
  {Khan}} \emph {et~al.} (\bibinfo {collaboration} {HadStruc}),\ }\href
  {\doibase 10.1103/PhysRevD.104.094516} {\bibfield  {journal} {\bibinfo
  {journal} {Phys. Rev. D}\ }\textbf {\bibinfo {volume} {104}},\ \bibinfo
  {pages} {094516} (\bibinfo {year} {2021})},\ \Eprint
  {http://arxiv.org/abs/2107.08960} {arXiv:2107.08960 [hep-lat]} \BibitemShut
  {NoStop}%
\bibitem [{\citenamefont {Karpie}\ \emph {et~al.}(2021)\citenamefont {Karpie},
  \citenamefont {Orginos}, \citenamefont {Radyushkin},\ and\ \citenamefont
  {Zafeiropoulos}}]{Karpie:2021pap}%
  \BibitemOpen
  \bibfield  {author} {\bibinfo {author} {\bibfnamefont {J.}~\bibnamefont
  {Karpie}}, \bibinfo {author} {\bibfnamefont {K.}~\bibnamefont {Orginos}},
  \bibinfo {author} {\bibfnamefont {A.}~\bibnamefont {Radyushkin}}, \ and\
  \bibinfo {author} {\bibfnamefont {S.}~\bibnamefont {Zafeiropoulos}} (\bibinfo
  {collaboration} {HadStruc}),\ }\href {\doibase 10.1007/JHEP11(2021)024}
  {\bibfield  {journal} {\bibinfo  {journal} {JHEP}\ }\textbf {\bibinfo
  {volume} {11}},\ \bibinfo {pages} {024} (\bibinfo {year} {2021})},\ \Eprint
  {http://arxiv.org/abs/2105.13313} {arXiv:2105.13313 [hep-lat]} \BibitemShut
  {NoStop}%
\bibitem [{\citenamefont {Salas-Chavira}\ \emph {et~al.}(2022)\citenamefont
  {Salas-Chavira}, \citenamefont {Fan},\ and\ \citenamefont
  {Lin}}]{Salas-Chavira:2021wui}%
  \BibitemOpen
  \bibfield  {author} {\bibinfo {author} {\bibfnamefont {A.}~\bibnamefont
  {Salas-Chavira}}, \bibinfo {author} {\bibfnamefont {Z.}~\bibnamefont {Fan}},
  \ and\ \bibinfo {author} {\bibfnamefont {H.-W.}\ \bibnamefont {Lin}},\ }\href
  {\doibase 10.1103/PhysRevD.106.094510} {\bibfield  {journal} {\bibinfo
  {journal} {Phys. Rev. D}\ }\textbf {\bibinfo {volume} {106}},\ \bibinfo
  {pages} {094510} (\bibinfo {year} {2022})},\ \Eprint
  {http://arxiv.org/abs/2112.03124} {arXiv:2112.03124 [hep-lat]} \BibitemShut
  {NoStop}%
\bibitem [{\citenamefont {Bhat}\ \emph {et~al.}(2023)\citenamefont {Bhat},
  \citenamefont {Chomicki}, \citenamefont {Cichy}, \citenamefont
  {Constantinou}, \citenamefont {Green},\ and\ \citenamefont
  {Scapellato}}]{Bhat:2022mjv}%
  \BibitemOpen
  \bibfield  {author} {\bibinfo {author} {\bibfnamefont {M.}~\bibnamefont
  {Bhat}}, \bibinfo {author} {\bibfnamefont {W.}~\bibnamefont {Chomicki}},
  \bibinfo {author} {\bibfnamefont {K.}~\bibnamefont {Cichy}}, \bibinfo
  {author} {\bibfnamefont {M.}~\bibnamefont {Constantinou}}, \bibinfo {author}
  {\bibfnamefont {J.~R.}\ \bibnamefont {Green}}, \ and\ \bibinfo {author}
  {\bibfnamefont {A.}~\bibnamefont {Scapellato}},\ }\href {\doibase
  10.22323/1.430.0094} {\bibfield  {journal} {\bibinfo  {journal} {PoS}\
  }\textbf {\bibinfo {volume} {LATTICE2022}},\ \bibinfo {pages} {094} (\bibinfo
  {year} {2023})},\ \Eprint {http://arxiv.org/abs/2212.06201} {arXiv:2212.06201
  [hep-lat]} \BibitemShut {NoStop}%
\bibitem [{\citenamefont {Bhat}\ \emph {et~al.}(2022)\citenamefont {Bhat},
  \citenamefont {Chomicki}, \citenamefont {Cichy}, \citenamefont
  {Constantinou}, \citenamefont {Green},\ and\ \citenamefont
  {Scapellato}}]{Bhat:2022zrw}%
  \BibitemOpen
  \bibfield  {author} {\bibinfo {author} {\bibfnamefont {M.}~\bibnamefont
  {Bhat}}, \bibinfo {author} {\bibfnamefont {W.}~\bibnamefont {Chomicki}},
  \bibinfo {author} {\bibfnamefont {K.}~\bibnamefont {Cichy}}, \bibinfo
  {author} {\bibfnamefont {M.}~\bibnamefont {Constantinou}}, \bibinfo {author}
  {\bibfnamefont {J.~R.}\ \bibnamefont {Green}}, \ and\ \bibinfo {author}
  {\bibfnamefont {A.}~\bibnamefont {Scapellato}},\ }\href {\doibase
  10.1103/PhysRevD.106.054504} {\bibfield  {journal} {\bibinfo  {journal}
  {Phys. Rev. D}\ }\textbf {\bibinfo {volume} {106}},\ \bibinfo {pages}
  {054504} (\bibinfo {year} {2022})},\ \Eprint
  {http://arxiv.org/abs/2205.07585} {arXiv:2205.07585 [hep-lat]} \BibitemShut
  {NoStop}%
\bibitem [{\citenamefont {Delmar}\ \emph
  {et~al.}(2023{\natexlab{a}})\citenamefont {Delmar}, \citenamefont
  {Alexandrou}, \citenamefont {Cichy}, \citenamefont {Constantinou},\ and\
  \citenamefont {Hadjiyiannakou}}]{Delmar:2022plq}%
  \BibitemOpen
  \bibfield  {author} {\bibinfo {author} {\bibfnamefont {J.}~\bibnamefont
  {Delmar}}, \bibinfo {author} {\bibfnamefont {C.}~\bibnamefont {Alexandrou}},
  \bibinfo {author} {\bibfnamefont {K.}~\bibnamefont {Cichy}}, \bibinfo
  {author} {\bibfnamefont {M.}~\bibnamefont {Constantinou}}, \ and\ \bibinfo
  {author} {\bibfnamefont {K.}~\bibnamefont {Hadjiyiannakou}},\ }\href
  {\doibase 10.22323/1.430.0099} {\bibfield  {journal} {\bibinfo  {journal}
  {PoS}\ }\textbf {\bibinfo {volume} {LATTICE2022}},\ \bibinfo {pages} {099}
  (\bibinfo {year} {2023}{\natexlab{a}})},\ \Eprint
  {http://arxiv.org/abs/2212.11399} {arXiv:2212.11399 [hep-lat]} \BibitemShut
  {NoStop}%
\bibitem [{\citenamefont {Fan}\ \emph {et~al.}(2023)\citenamefont {Fan},
  \citenamefont {Good},\ and\ \citenamefont {Lin}}]{Fan:2022kcb}%
  \BibitemOpen
  \bibfield  {author} {\bibinfo {author} {\bibfnamefont {Z.}~\bibnamefont
  {Fan}}, \bibinfo {author} {\bibfnamefont {W.}~\bibnamefont {Good}}, \ and\
  \bibinfo {author} {\bibfnamefont {H.-W.}\ \bibnamefont {Lin}},\ }\href
  {\doibase 10.1103/PhysRevD.108.014508} {\bibfield  {journal} {\bibinfo
  {journal} {Phys. Rev. D}\ }\textbf {\bibinfo {volume} {108}},\ \bibinfo
  {pages} {014508} (\bibinfo {year} {2023})},\ \Eprint
  {http://arxiv.org/abs/2210.09985} {arXiv:2210.09985 [hep-lat]} \BibitemShut
  {NoStop}%
\bibitem [{\citenamefont {Gao}\ \emph {et~al.}(2022{\natexlab{c}})\citenamefont
  {Gao}, \citenamefont {Hanlon}, \citenamefont {Karthik}, \citenamefont
  {Mukherjee}, \citenamefont {Petreczky}, \citenamefont {Scior}, \citenamefont
  {Syritsyn},\ and\ \citenamefont {Zhao}}]{Gao:2022vyh}%
  \BibitemOpen
  \bibfield  {author} {\bibinfo {author} {\bibfnamefont {X.}~\bibnamefont
  {Gao}}, \bibinfo {author} {\bibfnamefont {A.~D.}\ \bibnamefont {Hanlon}},
  \bibinfo {author} {\bibfnamefont {N.}~\bibnamefont {Karthik}}, \bibinfo
  {author} {\bibfnamefont {S.}~\bibnamefont {Mukherjee}}, \bibinfo {author}
  {\bibfnamefont {P.}~\bibnamefont {Petreczky}}, \bibinfo {author}
  {\bibfnamefont {P.}~\bibnamefont {Scior}}, \bibinfo {author} {\bibfnamefont
  {S.}~\bibnamefont {Syritsyn}}, \ and\ \bibinfo {author} {\bibfnamefont
  {Y.}~\bibnamefont {Zhao}},\ }\href {\doibase 10.1103/PhysRevD.106.074505}
  {\bibfield  {journal} {\bibinfo  {journal} {Phys. Rev. D}\ }\textbf {\bibinfo
  {volume} {106}},\ \bibinfo {pages} {074505} (\bibinfo {year}
  {2022}{\natexlab{c}})},\ \Eprint {http://arxiv.org/abs/2206.04084}
  {arXiv:2206.04084 [hep-lat]} \BibitemShut {NoStop}%
\bibitem [{\citenamefont {Edwards}\ \emph {et~al.}(2023)\citenamefont {Edwards}
  \emph {et~al.}}]{HadStruc:2022nay}%
  \BibitemOpen
  \bibfield  {author} {\bibinfo {author} {\bibfnamefont {R.~G.}\ \bibnamefont
  {Edwards}} \emph {et~al.} (\bibinfo {collaboration} {HadStruc}),\ }\href
  {\doibase 10.1007/JHEP03(2023)086} {\bibfield  {journal} {\bibinfo  {journal}
  {JHEP}\ }\textbf {\bibinfo {volume} {03}},\ \bibinfo {pages} {086} (\bibinfo
  {year} {2023})},\ \Eprint {http://arxiv.org/abs/2211.04434} {arXiv:2211.04434
  [hep-lat]} \BibitemShut {NoStop}%
\bibitem [{\citenamefont {Egerer}\ \emph
  {et~al.}(2022{\natexlab{b}})\citenamefont {Egerer} \emph
  {et~al.}}]{HadStruc:2022yaw}%
  \BibitemOpen
  \bibfield  {author} {\bibinfo {author} {\bibfnamefont {C.}~\bibnamefont
  {Egerer}} \emph {et~al.} (\bibinfo {collaboration} {HadStruc}),\ }\href
  {\doibase 10.1103/PhysRevD.106.094511} {\bibfield  {journal} {\bibinfo
  {journal} {Phys. Rev. D}\ }\textbf {\bibinfo {volume} {106}},\ \bibinfo
  {pages} {094511} (\bibinfo {year} {2022}{\natexlab{b}})},\ \Eprint
  {http://arxiv.org/abs/2207.08733} {arXiv:2207.08733 [hep-lat]} \BibitemShut
  {NoStop}%
\bibitem [{\citenamefont {Barry}\ \emph {et~al.}(2022)\citenamefont {Barry}
  \emph {et~al.}}]{JeffersonLabAngularMomentumJAM:2022aix}%
  \BibitemOpen
  \bibfield  {author} {\bibinfo {author} {\bibfnamefont {P.~C.}\ \bibnamefont
  {Barry}} \emph {et~al.} (\bibinfo {collaboration} {Jefferson Lab Angular
  Momentum (JAM), HadStruc}),\ }\href {\doibase 10.1103/PhysRevD.105.114051}
  {\bibfield  {journal} {\bibinfo  {journal} {Phys. Rev. D}\ }\textbf {\bibinfo
  {volume} {105}},\ \bibinfo {pages} {114051} (\bibinfo {year} {2022})},\
  \Eprint {http://arxiv.org/abs/2204.00543} {arXiv:2204.00543 [hep-ph]}
  \BibitemShut {NoStop}%
\bibitem [{\citenamefont {Bhattacharya}\ \emph
  {et~al.}(2023{\natexlab{c}})\citenamefont {Bhattacharya}, \citenamefont
  {Cichy}, \citenamefont {Constantinou}, \citenamefont {Gao}, \citenamefont
  {Metz}, \citenamefont {Miller}, \citenamefont {Mukherjee}, \citenamefont
  {Petreczky}, \citenamefont {Steffens},\ and\ \citenamefont
  {Zhao}}]{Bhattacharya:2023ays}%
  \BibitemOpen
  \bibfield  {author} {\bibinfo {author} {\bibfnamefont {S.}~\bibnamefont
  {Bhattacharya}}, \bibinfo {author} {\bibfnamefont {K.}~\bibnamefont {Cichy}},
  \bibinfo {author} {\bibfnamefont {M.}~\bibnamefont {Constantinou}}, \bibinfo
  {author} {\bibfnamefont {X.}~\bibnamefont {Gao}}, \bibinfo {author}
  {\bibfnamefont {A.}~\bibnamefont {Metz}}, \bibinfo {author} {\bibfnamefont
  {J.}~\bibnamefont {Miller}}, \bibinfo {author} {\bibfnamefont
  {S.}~\bibnamefont {Mukherjee}}, \bibinfo {author} {\bibfnamefont
  {P.}~\bibnamefont {Petreczky}}, \bibinfo {author} {\bibfnamefont
  {F.}~\bibnamefont {Steffens}}, \ and\ \bibinfo {author} {\bibfnamefont
  {Y.}~\bibnamefont {Zhao}},\ }\href {\doibase 10.1103/PhysRevD.108.014507}
  {\bibfield  {journal} {\bibinfo  {journal} {Phys. Rev. D}\ }\textbf {\bibinfo
  {volume} {108}},\ \bibinfo {pages} {014507} (\bibinfo {year}
  {2023}{\natexlab{c}})},\ \Eprint {http://arxiv.org/abs/2305.11117}
  {arXiv:2305.11117 [hep-lat]} \BibitemShut {NoStop}%
\bibitem [{\citenamefont {Delmar}\ \emph
  {et~al.}(2023{\natexlab{b}})\citenamefont {Delmar}, \citenamefont
  {Alexandrou}, \citenamefont {Cichy}, \citenamefont {Constantinou},\ and\
  \citenamefont {Hadjiyiannakou}}]{Delmar:2023agv}%
  \BibitemOpen
  \bibfield  {author} {\bibinfo {author} {\bibfnamefont {J.}~\bibnamefont
  {Delmar}}, \bibinfo {author} {\bibfnamefont {C.}~\bibnamefont {Alexandrou}},
  \bibinfo {author} {\bibfnamefont {K.}~\bibnamefont {Cichy}}, \bibinfo
  {author} {\bibfnamefont {M.}~\bibnamefont {Constantinou}}, \ and\ \bibinfo
  {author} {\bibfnamefont {K.}~\bibnamefont {Hadjiyiannakou}},\ }\href
  {\doibase 10.1103/PhysRevD.108.094515} {\bibfield  {journal} {\bibinfo
  {journal} {Phys. Rev. D}\ }\textbf {\bibinfo {volume} {108}},\ \bibinfo
  {pages} {094515} (\bibinfo {year} {2023}{\natexlab{b}})},\ \Eprint
  {http://arxiv.org/abs/2310.01389} {arXiv:2310.01389 [hep-lat]} \BibitemShut
  {NoStop}%
\bibitem [{\citenamefont {Dutrieux}\ \emph
  {et~al.}(2024{\natexlab{a}})\citenamefont {Dutrieux}, \citenamefont {Karpie},
  \citenamefont {Monahan}, \citenamefont {Orginos},\ and\ \citenamefont
  {Zafeiropoulos}}]{Dutrieux:2023zpy}%
  \BibitemOpen
  \bibfield  {author} {\bibinfo {author} {\bibfnamefont {H.}~\bibnamefont
  {Dutrieux}}, \bibinfo {author} {\bibfnamefont {J.}~\bibnamefont {Karpie}},
  \bibinfo {author} {\bibfnamefont {C.}~\bibnamefont {Monahan}}, \bibinfo
  {author} {\bibfnamefont {K.}~\bibnamefont {Orginos}}, \ and\ \bibinfo
  {author} {\bibfnamefont {S.}~\bibnamefont {Zafeiropoulos}} (\bibinfo
  {collaboration} {HadStruc}),\ }\href {\doibase 10.1007/JHEP04(2024)061}
  {\bibfield  {journal} {\bibinfo  {journal} {JHEP}\ }\textbf {\bibinfo
  {volume} {04}},\ \bibinfo {pages} {061} (\bibinfo {year}
  {2024}{\natexlab{a}})},\ \Eprint {http://arxiv.org/abs/2310.19926}
  {arXiv:2310.19926 [hep-lat]} \BibitemShut {NoStop}%
\bibitem [{\citenamefont {Good}\ \emph {et~al.}(2023)\citenamefont {Good},
  \citenamefont {Fan},\ and\ \citenamefont {Lin}}]{Good:2023gai}%
  \BibitemOpen
  \bibfield  {author} {\bibinfo {author} {\bibfnamefont {W.}~\bibnamefont
  {Good}}, \bibinfo {author} {\bibfnamefont {Z.}~\bibnamefont {Fan}}, \ and\
  \bibinfo {author} {\bibfnamefont {H.-W.}\ \bibnamefont {Lin}},\ }in\
  \href@noop {} {\emph {\bibinfo {booktitle} {{30th International Workshop on
  Deep-Inelastic Scattering and Related Subjects}}}}\ (\bibinfo {year} {2023})\
  \Eprint {http://arxiv.org/abs/2307.06916} {arXiv:2307.06916 [hep-lat]}
  \BibitemShut {NoStop}%
\bibitem [{\citenamefont {Karpie}\ \emph {et~al.}(2024)\citenamefont {Karpie},
  \citenamefont {Whitehill}, \citenamefont {Melnitchouk}, \citenamefont
  {Monahan}, \citenamefont {Orginos}, \citenamefont {Qiu}, \citenamefont
  {Richards}, \citenamefont {Sato},\ and\ \citenamefont
  {Zafeiropoulos}}]{Karpie:2023nyg}%
  \BibitemOpen
  \bibfield  {author} {\bibinfo {author} {\bibfnamefont {J.}~\bibnamefont
  {Karpie}}, \bibinfo {author} {\bibfnamefont {R.~M.}\ \bibnamefont
  {Whitehill}}, \bibinfo {author} {\bibfnamefont {W.}~\bibnamefont
  {Melnitchouk}}, \bibinfo {author} {\bibfnamefont {C.}~\bibnamefont
  {Monahan}}, \bibinfo {author} {\bibfnamefont {K.}~\bibnamefont {Orginos}},
  \bibinfo {author} {\bibfnamefont {J.~W.}\ \bibnamefont {Qiu}}, \bibinfo
  {author} {\bibfnamefont {D.~G.}\ \bibnamefont {Richards}}, \bibinfo {author}
  {\bibfnamefont {N.}~\bibnamefont {Sato}}, \ and\ \bibinfo {author}
  {\bibfnamefont {S.}~\bibnamefont {Zafeiropoulos}} (\bibinfo {collaboration}
  {Jefferson Lab Angular Momentum, HadStruc}),\ }\href {\doibase
  10.1103/PhysRevD.109.036031} {\bibfield  {journal} {\bibinfo  {journal}
  {Phys. Rev. D}\ }\textbf {\bibinfo {volume} {109}},\ \bibinfo {pages}
  {036031} (\bibinfo {year} {2024})},\ \Eprint
  {http://arxiv.org/abs/2310.18179} {arXiv:2310.18179 [hep-ph]} \BibitemShut
  {NoStop}%
\bibitem [{\citenamefont {Nurminen}\ \emph {et~al.}(2024)\citenamefont
  {Nurminen}, \citenamefont {Bhattacharya}, \citenamefont {Chomicki},
  \citenamefont {Cichy}, \citenamefont {Constantinou}, \citenamefont {Metz},\
  and\ \citenamefont {Steffens}}]{Nurminen:2023qok}%
  \BibitemOpen
  \bibfield  {author} {\bibinfo {author} {\bibfnamefont {N.}~\bibnamefont
  {Nurminen}}, \bibinfo {author} {\bibfnamefont {S.}~\bibnamefont
  {Bhattacharya}}, \bibinfo {author} {\bibfnamefont {W.}~\bibnamefont
  {Chomicki}}, \bibinfo {author} {\bibfnamefont {K.}~\bibnamefont {Cichy}},
  \bibinfo {author} {\bibfnamefont {M.}~\bibnamefont {Constantinou}}, \bibinfo
  {author} {\bibfnamefont {A.}~\bibnamefont {Metz}}, \ and\ \bibinfo {author}
  {\bibfnamefont {F.}~\bibnamefont {Steffens}},\ }\href {\doibase
  10.22323/1.453.0318} {\bibfield  {journal} {\bibinfo  {journal} {PoS}\
  }\textbf {\bibinfo {volume} {LATTICE2023}},\ \bibinfo {pages} {318} (\bibinfo
  {year} {2024})},\ \Eprint {http://arxiv.org/abs/2311.18502} {arXiv:2311.18502
  [hep-lat]} \BibitemShut {NoStop}%
\bibitem [{\citenamefont {Radyushkin}(2024)}]{Radyushkin:2023ref}%
  \BibitemOpen
  \bibfield  {author} {\bibinfo {author} {\bibfnamefont {A.~V.}\ \bibnamefont
  {Radyushkin}},\ }\href {\doibase 10.1103/PhysRevD.109.014514} {\bibfield
  {journal} {\bibinfo  {journal} {Phys. Rev. D}\ }\textbf {\bibinfo {volume}
  {109}},\ \bibinfo {pages} {014514} (\bibinfo {year} {2024})},\ \Eprint
  {http://arxiv.org/abs/2311.06007} {arXiv:2311.06007 [hep-ph]} \BibitemShut
  {NoStop}%
\bibitem [{\citenamefont {Bhattacharya}\ \emph
  {et~al.}(2024{\natexlab{b}})\citenamefont {Bhattacharya}, \citenamefont
  {Cichy}, \citenamefont {Constantinou}, \citenamefont {Metz}, \citenamefont
  {Nurminen},\ and\ \citenamefont {Steffens}}]{Bhattacharya:2024qpp}%
  \BibitemOpen
  \bibfield  {author} {\bibinfo {author} {\bibfnamefont {S.}~\bibnamefont
  {Bhattacharya}}, \bibinfo {author} {\bibfnamefont {K.}~\bibnamefont {Cichy}},
  \bibinfo {author} {\bibfnamefont {M.}~\bibnamefont {Constantinou}}, \bibinfo
  {author} {\bibfnamefont {A.}~\bibnamefont {Metz}}, \bibinfo {author}
  {\bibfnamefont {N.}~\bibnamefont {Nurminen}}, \ and\ \bibinfo {author}
  {\bibfnamefont {F.}~\bibnamefont {Steffens}},\ }\href {\doibase
  10.1103/PhysRevD.110.054502} {\bibfield  {journal} {\bibinfo  {journal}
  {Phys. Rev. D}\ }\textbf {\bibinfo {volume} {110}},\ \bibinfo {pages}
  {054502} (\bibinfo {year} {2024}{\natexlab{b}})},\ \Eprint
  {http://arxiv.org/abs/2405.04414} {arXiv:2405.04414 [hep-lat]} \BibitemShut
  {NoStop}%
\bibitem [{\citenamefont {Bhattacharya}\ \emph {et~al.}(2025)\citenamefont
  {Bhattacharya}, \citenamefont {Cichy}, \citenamefont {Constantinou},
  \citenamefont {Gao}, \citenamefont {Metz}, \citenamefont {Miller},
  \citenamefont {Mukherjee}, \citenamefont {Petreczky}, \citenamefont
  {Steffens},\ and\ \citenamefont {Zhao}}]{Bhattacharya:2024wtg}%
  \BibitemOpen
  \bibfield  {author} {\bibinfo {author} {\bibfnamefont {S.}~\bibnamefont
  {Bhattacharya}}, \bibinfo {author} {\bibfnamefont {K.}~\bibnamefont {Cichy}},
  \bibinfo {author} {\bibfnamefont {M.}~\bibnamefont {Constantinou}}, \bibinfo
  {author} {\bibfnamefont {X.}~\bibnamefont {Gao}}, \bibinfo {author}
  {\bibfnamefont {A.}~\bibnamefont {Metz}}, \bibinfo {author} {\bibfnamefont
  {J.}~\bibnamefont {Miller}}, \bibinfo {author} {\bibfnamefont
  {S.}~\bibnamefont {Mukherjee}}, \bibinfo {author} {\bibfnamefont
  {P.}~\bibnamefont {Petreczky}}, \bibinfo {author} {\bibfnamefont
  {F.}~\bibnamefont {Steffens}}, \ and\ \bibinfo {author} {\bibfnamefont
  {Y.}~\bibnamefont {Zhao}},\ }\href {\doibase 10.1007/JHEP01(2025)146}
  {\bibfield  {journal} {\bibinfo  {journal} {JHEP}\ }\textbf {\bibinfo
  {volume} {01}},\ \bibinfo {pages} {146} (\bibinfo {year} {2025})},\ \Eprint
  {http://arxiv.org/abs/2410.03539} {arXiv:2410.03539 [hep-lat]} \BibitemShut
  {NoStop}%
\bibitem [{\citenamefont {Blossier}\ \emph {et~al.}(2024)\citenamefont
  {Blossier}, \citenamefont {Mangin-Brinet}, \citenamefont {Morgado~Ch\'avez},\
  and\ \citenamefont {San~Jos\'e}}]{Blossier:2024wyx}%
  \BibitemOpen
  \bibfield  {author} {\bibinfo {author} {\bibfnamefont {B.}~\bibnamefont
  {Blossier}}, \bibinfo {author} {\bibfnamefont {M.}~\bibnamefont
  {Mangin-Brinet}}, \bibinfo {author} {\bibfnamefont {J.~M.}\ \bibnamefont
  {Morgado~Ch\'avez}}, \ and\ \bibinfo {author} {\bibfnamefont
  {T.}~\bibnamefont {San~Jos\'e}},\ }\href {\doibase 10.1007/JHEP09(2024)079}
  {\bibfield  {journal} {\bibinfo  {journal} {JHEP}\ }\textbf {\bibinfo
  {volume} {09}},\ \bibinfo {pages} {079} (\bibinfo {year} {2024})},\ \Eprint
  {http://arxiv.org/abs/2406.04668} {arXiv:2406.04668 [hep-lat]} \BibitemShut
  {NoStop}%
\bibitem [{\citenamefont {Cheng}\ \emph {et~al.}(2024)\citenamefont {Cheng},
  \citenamefont {Huang}, \citenamefont {Li}, \citenamefont {Li},\ and\
  \citenamefont {Ma}}]{Cheng:2024wyu}%
  \BibitemOpen
  \bibfield  {author} {\bibinfo {author} {\bibfnamefont {C.}~\bibnamefont
  {Cheng}}, \bibinfo {author} {\bibfnamefont {L.-H.}\ \bibnamefont {Huang}},
  \bibinfo {author} {\bibfnamefont {X.}~\bibnamefont {Li}}, \bibinfo {author}
  {\bibfnamefont {Z.-Y.}\ \bibnamefont {Li}}, \ and\ \bibinfo {author}
  {\bibfnamefont {Y.-Q.}\ \bibnamefont {Ma}},\ }\href@noop {} {\  (\bibinfo
  {year} {2024})},\ \Eprint {http://arxiv.org/abs/2410.05141} {arXiv:2410.05141
  [hep-ph]} \BibitemShut {NoStop}%
\bibitem [{\citenamefont {Cichy}\ \emph {et~al.}(2024)\citenamefont {Cichy},
  \citenamefont {Constantinou}, \citenamefont {Sznajder},\ and\ \citenamefont
  {Wagner}}]{Cichy:2024afd}%
  \BibitemOpen
  \bibfield  {author} {\bibinfo {author} {\bibfnamefont {K.}~\bibnamefont
  {Cichy}}, \bibinfo {author} {\bibfnamefont {M.}~\bibnamefont {Constantinou}},
  \bibinfo {author} {\bibfnamefont {P.}~\bibnamefont {Sznajder}}, \ and\
  \bibinfo {author} {\bibfnamefont {J.}~\bibnamefont {Wagner}},\ }\href
  {\doibase 10.1103/PhysRevD.110.114025} {\bibfield  {journal} {\bibinfo
  {journal} {Phys. Rev. D}\ }\textbf {\bibinfo {volume} {110}},\ \bibinfo
  {pages} {114025} (\bibinfo {year} {2024})},\ \Eprint
  {http://arxiv.org/abs/2409.17955} {arXiv:2409.17955 [hep-ph]} \BibitemShut
  {NoStop}%
\bibitem [{\citenamefont {Dutrieux}\ \emph
  {et~al.}(2025{\natexlab{b}})\citenamefont {Dutrieux}, \citenamefont {Karpie},
  \citenamefont {Orginos},\ and\ \citenamefont
  {Zafeiropoulos}}]{Dutrieux:2024rem}%
  \BibitemOpen
  \bibfield  {author} {\bibinfo {author} {\bibfnamefont {H.}~\bibnamefont
  {Dutrieux}}, \bibinfo {author} {\bibfnamefont {J.}~\bibnamefont {Karpie}},
  \bibinfo {author} {\bibfnamefont {K.}~\bibnamefont {Orginos}}, \ and\
  \bibinfo {author} {\bibfnamefont {S.}~\bibnamefont {Zafeiropoulos}},\ }\href
  {\doibase 10.1103/PhysRevD.111.034515} {\bibfield  {journal} {\bibinfo
  {journal} {Phys. Rev. D}\ }\textbf {\bibinfo {volume} {111}},\ \bibinfo
  {pages} {034515} (\bibinfo {year} {2025}{\natexlab{b}})},\ \Eprint
  {http://arxiv.org/abs/2412.05227} {arXiv:2412.05227 [hep-lat]} \BibitemShut
  {NoStop}%
\bibitem [{\citenamefont {Dutrieux}\ \emph
  {et~al.}(2024{\natexlab{b}})\citenamefont {Dutrieux}, \citenamefont
  {Edwards}, \citenamefont {Egerer}, \citenamefont {Karpie}, \citenamefont
  {Monahan}, \citenamefont {Orginos}, \citenamefont {Radyushkin}, \citenamefont
  {Richards}, \citenamefont {Romero},\ and\ \citenamefont
  {Zafeiropoulos}}]{HadStruc:2024rix}%
  \BibitemOpen
  \bibfield  {author} {\bibinfo {author} {\bibfnamefont {H.}~\bibnamefont
  {Dutrieux}}, \bibinfo {author} {\bibfnamefont {R.~G.}\ \bibnamefont
  {Edwards}}, \bibinfo {author} {\bibfnamefont {C.}~\bibnamefont {Egerer}},
  \bibinfo {author} {\bibfnamefont {J.}~\bibnamefont {Karpie}}, \bibinfo
  {author} {\bibfnamefont {C.}~\bibnamefont {Monahan}}, \bibinfo {author}
  {\bibfnamefont {K.}~\bibnamefont {Orginos}}, \bibinfo {author} {\bibfnamefont
  {A.}~\bibnamefont {Radyushkin}}, \bibinfo {author} {\bibfnamefont
  {D.}~\bibnamefont {Richards}}, \bibinfo {author} {\bibfnamefont
  {E.}~\bibnamefont {Romero}}, \ and\ \bibinfo {author} {\bibfnamefont
  {S.}~\bibnamefont {Zafeiropoulos}} (\bibinfo {collaboration} {HadStruc}),\
  }\href {\doibase 10.1007/JHEP08(2024)162} {\bibfield  {journal} {\bibinfo
  {journal} {JHEP}\ }\textbf {\bibinfo {volume} {08}},\ \bibinfo {pages} {162}
  (\bibinfo {year} {2024}{\natexlab{b}})},\ \Eprint
  {http://arxiv.org/abs/2405.10304} {arXiv:2405.10304 [hep-lat]} \BibitemShut
  {NoStop}%
\bibitem [{\citenamefont {Karpie}\ \emph {et~al.}(2025)\citenamefont {Karpie},
  \citenamefont {Monahan},\ and\ \citenamefont {Radyushkin}}]{Karpie:2024bof}%
  \BibitemOpen
  \bibfield  {author} {\bibinfo {author} {\bibfnamefont {J.}~\bibnamefont
  {Karpie}}, \bibinfo {author} {\bibfnamefont {C.}~\bibnamefont {Monahan}}, \
  and\ \bibinfo {author} {\bibfnamefont {A.}~\bibnamefont {Radyushkin}},\
  }\href {\doibase 10.1103/PhysRevD.111.054503} {\bibfield  {journal} {\bibinfo
   {journal} {Phys. Rev. D}\ }\textbf {\bibinfo {volume} {111}},\ \bibinfo
  {pages} {054503} (\bibinfo {year} {2025})},\ \Eprint
  {http://arxiv.org/abs/2407.16577} {arXiv:2407.16577 [hep-lat]} \BibitemShut
  {NoStop}%
\bibitem [{\citenamefont {Kovner}\ \emph {et~al.}(2024)\citenamefont {Kovner},
  \citenamefont {Karpie}, \citenamefont {Orginos}, \citenamefont {Radyushkin},\
  and\ \citenamefont {Zafeiropoulos}}]{Kovner:2024pwl}%
  \BibitemOpen
  \bibfield  {author} {\bibinfo {author} {\bibfnamefont {D.}~\bibnamefont
  {Kovner}}, \bibinfo {author} {\bibfnamefont {J.}~\bibnamefont {Karpie}},
  \bibinfo {author} {\bibfnamefont {K.}~\bibnamefont {Orginos}}, \bibinfo
  {author} {\bibfnamefont {A.}~\bibnamefont {Radyushkin}}, \ and\ \bibinfo
  {author} {\bibfnamefont {S.}~\bibnamefont {Zafeiropoulos}} (\bibinfo
  {collaboration} {HadStruc}),\ }\href {\doibase 10.22323/1.453.0300}
  {\bibfield  {journal} {\bibinfo  {journal} {PoS}\ }\textbf {\bibinfo {volume}
  {LATTICE2023}},\ \bibinfo {pages} {300} (\bibinfo {year} {2024})},\ \Eprint
  {http://arxiv.org/abs/2401.06858} {arXiv:2401.06858 [hep-lat]} \BibitemShut
  {NoStop}%
\bibitem [{\citenamefont {San Jos\'e~P\'erez}\ \emph
  {et~al.}(2025{\natexlab{a}})\citenamefont {San Jos\'e~P\'erez}, \citenamefont
  {Blossier}, \citenamefont {Mangin-Brinet},\ and\ \citenamefont
  {Morgado~Ch\'avez}}]{SanJosePerez:2024axh}%
  \BibitemOpen
  \bibfield  {author} {\bibinfo {author} {\bibfnamefont {M.~T.}\ \bibnamefont
  {San Jos\'e~P\'erez}}, \bibinfo {author} {\bibfnamefont {B.}~\bibnamefont
  {Blossier}}, \bibinfo {author} {\bibfnamefont {M.}~\bibnamefont
  {Mangin-Brinet}}, \ and\ \bibinfo {author} {\bibfnamefont {J.~M.}\
  \bibnamefont {Morgado~Ch\'avez}},\ }\href {\doibase 10.22323/1.469.0042}
  {\bibfield  {journal} {\bibinfo  {journal} {PoS}\ }\textbf {\bibinfo {volume}
  {DIS2024}},\ \bibinfo {pages} {042} (\bibinfo {year} {2025}{\natexlab{a}})},\
  \Eprint {http://arxiv.org/abs/2406.12361} {arXiv:2406.12361 [hep-lat]}
  \BibitemShut {NoStop}%
\bibitem [{\citenamefont {San Jos\'e~P\'erez}\ \emph
  {et~al.}(2025{\natexlab{b}})\citenamefont {San Jos\'e~P\'erez}, \citenamefont
  {Blossier}, \citenamefont {Mangin-Brinet},\ and\ \citenamefont
  {Morgado~Ch\'avez}}]{SanJosePerez:2024fad}%
  \BibitemOpen
  \bibfield  {author} {\bibinfo {author} {\bibfnamefont {M.~T.}\ \bibnamefont
  {San Jos\'e~P\'erez}}, \bibinfo {author} {\bibfnamefont {B.}~\bibnamefont
  {Blossier}}, \bibinfo {author} {\bibfnamefont {M.}~\bibnamefont
  {Mangin-Brinet}}, \ and\ \bibinfo {author} {\bibfnamefont {J.~M.}\
  \bibnamefont {Morgado~Ch\'avez}},\ }\href {\doibase 10.22323/1.466.0313}
  {\bibfield  {journal} {\bibinfo  {journal} {PoS}\ }\textbf {\bibinfo {volume}
  {LATTICE2024}},\ \bibinfo {pages} {313} (\bibinfo {year}
  {2025}{\natexlab{b}})},\ \Eprint {http://arxiv.org/abs/2409.12084}
  {arXiv:2409.12084 [hep-lat]} \BibitemShut {NoStop}%
\bibitem [{\citenamefont {Pang}\ \emph
  {et~al.}(2024{\natexlab{b}})\citenamefont {Pang}, \citenamefont {Zhang},\
  and\ \citenamefont {Zhao}}]{Pang:2024kza}%
  \BibitemOpen
  \bibfield  {author} {\bibinfo {author} {\bibfnamefont {Z.}~\bibnamefont
  {Pang}}, \bibinfo {author} {\bibfnamefont {J.-H.}\ \bibnamefont {Zhang}}, \
  and\ \bibinfo {author} {\bibfnamefont {D.-J.}\ \bibnamefont {Zhao}},\
  }\href@noop {} {\  (\bibinfo {year} {2024}{\natexlab{b}})},\ \Eprint
  {http://arxiv.org/abs/2412.19862} {arXiv:2412.19862 [hep-lat]} \BibitemShut
  {NoStop}%
\bibitem [{\citenamefont {Nowak}\ \emph {et~al.}(1996)\citenamefont {Nowak},
  \citenamefont {Rho},\ and\ \citenamefont {Zahed}}]{Nowak:1996aj}%
  \BibitemOpen
  \bibfield  {author} {\bibinfo {author} {\bibfnamefont {M.~A.}\ \bibnamefont
  {Nowak}}, \bibinfo {author} {\bibfnamefont {M.}~\bibnamefont {Rho}}, \ and\
  \bibinfo {author} {\bibfnamefont {I.}~\bibnamefont {Zahed}},\ }\href@noop {}
  {\emph {\bibinfo {title} {{Chiral nuclear dynamics}}}}\ (\bibinfo {year}
  {1996})\BibitemShut {NoStop}%
\bibitem [{\citenamefont {Sch\"afer}\ and\ \citenamefont
  {Shuryak}(1998)}]{Schafer:1996wv}%
  \BibitemOpen
  \bibfield  {author} {\bibinfo {author} {\bibfnamefont {T.}~\bibnamefont
  {Sch\"afer}}\ and\ \bibinfo {author} {\bibfnamefont {E.~V.}\ \bibnamefont
  {Shuryak}},\ }\href {\doibase 10.1103/RevModPhys.70.323} {\bibfield
  {journal} {\bibinfo  {journal} {Rev. Mod. Phys.}\ }\textbf {\bibinfo {volume}
  {70}},\ \bibinfo {pages} {323} (\bibinfo {year} {1998})},\ \Eprint
  {http://arxiv.org/abs/hep-ph/9610451} {arXiv:hep-ph/9610451} \BibitemShut
  {NoStop}%
\bibitem [{\citenamefont {Ji}(2020)}]{Ji:2020byp}%
  \BibitemOpen
  \bibfield  {author} {\bibinfo {author} {\bibfnamefont {X.}~\bibnamefont
  {Ji}},\ }\href@noop {} {\  (\bibinfo {year} {2020})},\ \Eprint
  {http://arxiv.org/abs/2007.06613} {arXiv:2007.06613 [hep-ph]} \BibitemShut
  {NoStop}%
\bibitem [{\citenamefont {Ji}(2025)}]{Ji:2022ezo}%
  \BibitemOpen
  \bibfield  {author} {\bibinfo {author} {\bibfnamefont {X.}~\bibnamefont
  {Ji}},\ }\href {\doibase 10.34133/research.0695} {\bibfield  {journal}
  {\bibinfo  {journal} {Research}\ }\textbf {\bibinfo {volume} {8}},\ \bibinfo
  {pages} {0695} (\bibinfo {year} {2025})},\ \Eprint
  {http://arxiv.org/abs/2209.09332} {arXiv:2209.09332 [hep-lat]} \BibitemShut
  {NoStop}%
\bibitem [{\citenamefont {Rietsch}(1976)}]{rietsch1976maximum}%
  \BibitemOpen
  \bibfield  {author} {\bibinfo {author} {\bibfnamefont {E.}~\bibnamefont
  {Rietsch}},\ }\href
  {https://journal.geophysicsjournal.com/JofG/article/view/87} {\bibfield
  {journal} {\bibinfo  {journal} {Journal of Geophysics}\ }\textbf {\bibinfo
  {volume} {42}},\ \bibinfo {pages} {489} (\bibinfo {year} {1976})}\BibitemShut
  {NoStop}%
\bibitem [{\citenamefont {Burnier}\ and\ \citenamefont
  {Rothkopf}(2013)}]{burnier2013bayesian}%
  \BibitemOpen
  \bibfield  {author} {\bibinfo {author} {\bibfnamefont {Y.}~\bibnamefont
  {Burnier}}\ and\ \bibinfo {author} {\bibfnamefont {A.}~\bibnamefont
  {Rothkopf}},\ }\href {\doibase 10.1103/PhysRevLett.111.182003} {\bibfield
  {journal} {\bibinfo  {journal} {Physical Review Letters}\ }\textbf {\bibinfo
  {volume} {111}},\ \bibinfo {pages} {182003} (\bibinfo {year}
  {2013})}\BibitemShut {NoStop}%
\bibitem [{\citenamefont {Liang}\ \emph {et~al.}(2020)\citenamefont {Liang},
  \citenamefont {Draper}, \citenamefont {Liu}, \citenamefont {Rothkopf},\ and\
  \citenamefont {Yang}}]{liang2020towards}%
  \BibitemOpen
  \bibfield  {author} {\bibinfo {author} {\bibfnamefont {J.}~\bibnamefont
  {Liang}}, \bibinfo {author} {\bibfnamefont {T.}~\bibnamefont {Draper}},
  \bibinfo {author} {\bibfnamefont {K.-F.}\ \bibnamefont {Liu}}, \bibinfo
  {author} {\bibfnamefont {A.}~\bibnamefont {Rothkopf}}, \ and\ \bibinfo
  {author} {\bibfnamefont {Y.-B.}\ \bibnamefont {Yang}},\ }\href {\doibase
  10.1103/PhysRevD.101.114503} {\bibfield  {journal} {\bibinfo  {journal}
  {Physical Review D}\ }\textbf {\bibinfo {volume} {101}},\ \bibinfo {pages}
  {114503} (\bibinfo {year} {2020})}\BibitemShut {NoStop}%
\bibitem [{\citenamefont {Backus}\ and\ \citenamefont
  {Gilbert}(1968)}]{backus1968resolving}%
  \BibitemOpen
  \bibfield  {author} {\bibinfo {author} {\bibfnamefont {G.}~\bibnamefont
  {Backus}}\ and\ \bibinfo {author} {\bibfnamefont {F.}~\bibnamefont
  {Gilbert}},\ }\href {\doibase 10.1111/j.1365-246X.1968.tb00216.x} {\bibfield
  {journal} {\bibinfo  {journal} {Geophysical Journal International}\ }\textbf
  {\bibinfo {volume} {16}},\ \bibinfo {pages} {169} (\bibinfo {year}
  {1968})}\BibitemShut {NoStop}%
\bibitem [{\citenamefont {Debbio}\ \emph {et~al.}(2025)\citenamefont {Debbio},
  \citenamefont {Lupo}, \citenamefont {Panero},\ and\ \citenamefont
  {Tantalo}}]{DelDebbio2025Bayesian}%
  \BibitemOpen
  \bibfield  {author} {\bibinfo {author} {\bibfnamefont {L.~D.}\ \bibnamefont
  {Debbio}}, \bibinfo {author} {\bibfnamefont {A.}~\bibnamefont {Lupo}},
  \bibinfo {author} {\bibfnamefont {M.}~\bibnamefont {Panero}}, \ and\ \bibinfo
  {author} {\bibfnamefont {N.}~\bibnamefont {Tantalo}},\ }\href {\doibase
  10.1140/epjc/s10052-025-13885-9} {\bibfield  {journal} {\bibinfo  {journal}
  {The European Physical Journal C}\ }\textbf {\bibinfo {volume} {85}},\
  \bibinfo {pages} {13885} (\bibinfo {year} {2025})}\BibitemShut {NoStop}%
\bibitem [{\citenamefont {Dutrieux}\ \emph
  {et~al.}(2025{\natexlab{c}})\citenamefont {Dutrieux}, \citenamefont {Karpie},
  \citenamefont {Monahan}, \citenamefont {Orginos}, \citenamefont {Radyushkin},
  \citenamefont {Richards},\ and\ \citenamefont
  {Zafeiropoulos}}]{Dutrieux:2025axb}%
  \BibitemOpen
  \bibfield  {author} {\bibinfo {author} {\bibfnamefont {H.}~\bibnamefont
  {Dutrieux}}, \bibinfo {author} {\bibfnamefont {J.}~\bibnamefont {Karpie}},
  \bibinfo {author} {\bibfnamefont {C.~J.}\ \bibnamefont {Monahan}}, \bibinfo
  {author} {\bibfnamefont {K.}~\bibnamefont {Orginos}}, \bibinfo {author}
  {\bibfnamefont {A.}~\bibnamefont {Radyushkin}}, \bibinfo {author}
  {\bibfnamefont {D.}~\bibnamefont {Richards}}, \ and\ \bibinfo {author}
  {\bibfnamefont {S.}~\bibnamefont {Zafeiropoulos}},\ }\href@noop {} {\
  (\bibinfo {year} {2025}{\natexlab{c}})},\ \Eprint
  {http://arxiv.org/abs/2506.24037} {arXiv:2506.24037 [hep-lat]} \BibitemShut
  {NoStop}%
\bibitem [{\citenamefont {Zhang}\ \emph {et~al.}(2023)\citenamefont {Zhang},
  \citenamefont {Holligan}, \citenamefont {Ji},\ and\ \citenamefont
  {Su}}]{Zhang:2023bxs}%
  \BibitemOpen
  \bibfield  {author} {\bibinfo {author} {\bibfnamefont {R.}~\bibnamefont
  {Zhang}}, \bibinfo {author} {\bibfnamefont {J.}~\bibnamefont {Holligan}},
  \bibinfo {author} {\bibfnamefont {X.}~\bibnamefont {Ji}}, \ and\ \bibinfo
  {author} {\bibfnamefont {Y.}~\bibnamefont {Su}},\ }\href {\doibase
  10.1016/j.physletb.2023.138081} {\bibfield  {journal} {\bibinfo  {journal}
  {Phys. Lett. B}\ }\textbf {\bibinfo {volume} {844}},\ \bibinfo {pages}
  {138081} (\bibinfo {year} {2023})},\ \Eprint
  {http://arxiv.org/abs/2305.05212} {arXiv:2305.05212 [hep-lat]} \BibitemShut
  {NoStop}%
\bibitem [{\citenamefont {Ji}\ \emph {et~al.}(2026)\citenamefont {Ji},
  \citenamefont {Liu},\ and\ \citenamefont {Su}}]{Ji:2026vir}%
  \BibitemOpen
  \bibfield  {author} {\bibinfo {author} {\bibfnamefont {X.}~\bibnamefont
  {Ji}}, \bibinfo {author} {\bibfnamefont {Y.}~\bibnamefont {Liu}}, \ and\
  \bibinfo {author} {\bibfnamefont {Y.}~\bibnamefont {Su}},\ }\href@noop {} {\
  (\bibinfo {year} {2026})},\ \Eprint {http://arxiv.org/abs/2601.12189}
  {arXiv:2601.12189 [hep-lat]} \BibitemShut {NoStop}%
\bibitem [{\citenamefont {Mondal}\ \emph
  {et~al.}(2020{\natexlab{a}})\citenamefont {Mondal}, \citenamefont {Gupta},
  \citenamefont {Park}, \citenamefont {Yoon}, \citenamefont {Bhattacharya},\
  and\ \citenamefont {Lin}}]{Mondal:2020cmt}%
  \BibitemOpen
  \bibfield  {author} {\bibinfo {author} {\bibfnamefont {S.}~\bibnamefont
  {Mondal}}, \bibinfo {author} {\bibfnamefont {R.}~\bibnamefont {Gupta}},
  \bibinfo {author} {\bibfnamefont {S.}~\bibnamefont {Park}}, \bibinfo {author}
  {\bibfnamefont {B.}~\bibnamefont {Yoon}}, \bibinfo {author} {\bibfnamefont
  {T.}~\bibnamefont {Bhattacharya}}, \ and\ \bibinfo {author} {\bibfnamefont
  {H.-W.}\ \bibnamefont {Lin}},\ }\href {\doibase 10.1103/PhysRevD.102.054512}
  {\bibfield  {journal} {\bibinfo  {journal} {Phys. Rev. D}\ }\textbf {\bibinfo
  {volume} {102}},\ \bibinfo {pages} {054512} (\bibinfo {year}
  {2020}{\natexlab{a}})},\ \Eprint {http://arxiv.org/abs/2005.13779}
  {arXiv:2005.13779 [hep-lat]} \BibitemShut {NoStop}%
\bibitem [{\citenamefont {Mondal}\ \emph
  {et~al.}(2020{\natexlab{b}})\citenamefont {Mondal}, \citenamefont {Gupta},
  \citenamefont {Park}, \citenamefont {Yoon}, \citenamefont {Bhattacharya},
  \citenamefont {Jo\'o},\ and\ \citenamefont {Winter}}]{Mondal:2020ela}%
  \BibitemOpen
  \bibfield  {author} {\bibinfo {author} {\bibfnamefont {S.}~\bibnamefont
  {Mondal}}, \bibinfo {author} {\bibfnamefont {R.}~\bibnamefont {Gupta}},
  \bibinfo {author} {\bibfnamefont {S.}~\bibnamefont {Park}}, \bibinfo {author}
  {\bibfnamefont {B.}~\bibnamefont {Yoon}}, \bibinfo {author} {\bibfnamefont
  {T.}~\bibnamefont {Bhattacharya}}, \bibinfo {author} {\bibfnamefont
  {B.}~\bibnamefont {Jo\'o}}, \ and\ \bibinfo {author} {\bibfnamefont
  {F.}~\bibnamefont {Winter}} (\bibinfo {collaboration} {Nucleon Matrix
  Elements (NME)}),\ }\href {\doibase 10.1007/JHEP04(2021)044} {\bibfield
  {journal} {\bibinfo  {journal} {JHEP}\ }\textbf {\bibinfo {volume} {21}},\
  \bibinfo {pages} {004} (\bibinfo {year} {2020}{\natexlab{b}})},\ \Eprint
  {http://arxiv.org/abs/2011.12787} {arXiv:2011.12787 [hep-lat]} \BibitemShut
  {NoStop}%
\bibitem [{\citenamefont {Mondal}\ \emph {et~al.}(2021)\citenamefont {Mondal},
  \citenamefont {Bhattacharya}, \citenamefont {Gupta}, \citenamefont {Jo\'o},
  \citenamefont {Lin}, \citenamefont {Park}, \citenamefont {Winter},\ and\
  \citenamefont {Yoon}}]{Mondal:2021oot}%
  \BibitemOpen
  \bibfield  {author} {\bibinfo {author} {\bibfnamefont {S.}~\bibnamefont
  {Mondal}}, \bibinfo {author} {\bibfnamefont {T.}~\bibnamefont
  {Bhattacharya}}, \bibinfo {author} {\bibfnamefont {R.}~\bibnamefont {Gupta}},
  \bibinfo {author} {\bibfnamefont {B.}~\bibnamefont {Jo\'o}}, \bibinfo
  {author} {\bibfnamefont {H.-W.}\ \bibnamefont {Lin}}, \bibinfo {author}
  {\bibfnamefont {S.}~\bibnamefont {Park}}, \bibinfo {author} {\bibfnamefont
  {F.}~\bibnamefont {Winter}}, \ and\ \bibinfo {author} {\bibfnamefont
  {B.}~\bibnamefont {Yoon}},\ }\href {\doibase 10.22323/1.396.0513} {\bibfield
  {journal} {\bibinfo  {journal} {PoS}\ }\textbf {\bibinfo {volume}
  {LATTICE2021}},\ \bibinfo {pages} {513} (\bibinfo {year} {2021})},\ \Eprint
  {http://arxiv.org/abs/2201.00067} {arXiv:2201.00067 [hep-lat]} \BibitemShut
  {NoStop}%
\bibitem [{\citenamefont {Chen}\ and\ \citenamefont {Ji}(2001)}]{Chen:2001eg}%
  \BibitemOpen
  \bibfield  {author} {\bibinfo {author} {\bibfnamefont {J.-W.}\ \bibnamefont
  {Chen}}\ and\ \bibinfo {author} {\bibfnamefont {X.-d.}\ \bibnamefont {Ji}},\
  }\href {\doibase 10.1016/S0370-2693(01)01337-5} {\bibfield  {journal}
  {\bibinfo  {journal} {Phys. Lett. B}\ }\textbf {\bibinfo {volume} {523}},\
  \bibinfo {pages} {107} (\bibinfo {year} {2001})},\ \Eprint
  {http://arxiv.org/abs/hep-ph/0105197} {arXiv:hep-ph/0105197} \BibitemShut
  {NoStop}%
\bibitem [{\citenamefont {Arndt}\ and\ \citenamefont
  {Savage}(2002)}]{Arndt:2001ye}%
  \BibitemOpen
  \bibfield  {author} {\bibinfo {author} {\bibfnamefont {D.}~\bibnamefont
  {Arndt}}\ and\ \bibinfo {author} {\bibfnamefont {M.~J.}\ \bibnamefont
  {Savage}},\ }\href {\doibase 10.1016/S0375-9474(01)01223-4} {\bibfield
  {journal} {\bibinfo  {journal} {Nucl. Phys. A}\ }\textbf {\bibinfo {volume}
  {697}},\ \bibinfo {pages} {429} (\bibinfo {year} {2002})},\ \Eprint
  {http://arxiv.org/abs/nucl-th/0105045} {arXiv:nucl-th/0105045} \BibitemShut
  {NoStop}%
\bibitem [{\citenamefont {Chang}\ \emph {et~al.}(2018)\citenamefont {Chang}
  \emph {et~al.}}]{Chang:2018uxx}%
  \BibitemOpen
  \bibfield  {author} {\bibinfo {author} {\bibfnamefont {C.~C.}\ \bibnamefont
  {Chang}} \emph {et~al.},\ }\href {\doibase 10.1038/s41586-018-0161-8}
  {\bibfield  {journal} {\bibinfo  {journal} {Nature}\ }\textbf {\bibinfo
  {volume} {558}},\ \bibinfo {pages} {91} (\bibinfo {year} {2018})},\ \Eprint
  {http://arxiv.org/abs/1805.12130} {arXiv:1805.12130 [hep-lat]} \BibitemShut
  {NoStop}%
\bibitem [{\citenamefont {Hadamard}(1923)}]{Hadamard:1923}%
  \BibitemOpen
  \bibfield  {author} {\bibinfo {author} {\bibfnamefont {J.}~\bibnamefont
  {Hadamard}},\ }\href@noop {} {\emph {\bibinfo {title} {{Lectures on the
  Cauchy Problem in Linear Partial Differential Equations}}}}\ (\bibinfo
  {publisher} {Yale University Press},\ \bibinfo {address} {New Haven},\
  \bibinfo {year} {1923})\BibitemShut {NoStop}%
\bibitem [{\citenamefont {Kirsch}(2011)}]{Kirsch:2011}%
  \BibitemOpen
  \bibfield  {author} {\bibinfo {author} {\bibfnamefont {A.}~\bibnamefont
  {Kirsch}},\ }\href {\doibase 10.1007/978-1-4419-8474-6} {\emph {\bibinfo
  {title} {An Introduction to the Mathematical Theory of Inverse Problems}}},\
  \bibinfo {edition} {2nd}\ ed.,\ \bibinfo {series} {Applied Mathematical
  Sciences}, Vol.\ \bibinfo {volume} {120}\ (\bibinfo  {publisher} {Springer},\
  \bibinfo {address} {New York, NY},\ \bibinfo {year} {2011})\BibitemShut
  {NoStop}%
\bibitem [{\citenamefont {Khan}\ \emph {et~al.}(2023)\citenamefont {Khan},
  \citenamefont {Liu},\ and\ \citenamefont {Sufian}}]{Khan:2022vot}%
  \BibitemOpen
  \bibfield  {author} {\bibinfo {author} {\bibfnamefont {T.}~\bibnamefont
  {Khan}}, \bibinfo {author} {\bibfnamefont {T.}~\bibnamefont {Liu}}, \ and\
  \bibinfo {author} {\bibfnamefont {R.~S.}\ \bibnamefont {Sufian}},\ }\href
  {\doibase 10.1103/PhysRevD.108.074502} {\bibfield  {journal} {\bibinfo
  {journal} {Phys. Rev. D}\ }\textbf {\bibinfo {volume} {108}},\ \bibinfo
  {pages} {074502} (\bibinfo {year} {2023})},\ \Eprint
  {http://arxiv.org/abs/2211.15587} {arXiv:2211.15587 [hep-lat]} \BibitemShut
  {NoStop}%
\bibitem [{\citenamefont {Chowdhury}\ \emph {et~al.}(2025)\citenamefont
  {Chowdhury}, \citenamefont {Izubuchi}, \citenamefont {Kamruzzaman},
  \citenamefont {Karthik}, \citenamefont {Khan}, \citenamefont {Liu},
  \citenamefont {Paul}, \citenamefont {Schoenleber},\ and\ \citenamefont
  {Sufian}}]{Chowdhury:2024ymm}%
  \BibitemOpen
  \bibfield  {author} {\bibinfo {author} {\bibfnamefont {T.~A.}\ \bibnamefont
  {Chowdhury}}, \bibinfo {author} {\bibfnamefont {T.}~\bibnamefont {Izubuchi}},
  \bibinfo {author} {\bibfnamefont {M.}~\bibnamefont {Kamruzzaman}}, \bibinfo
  {author} {\bibfnamefont {N.}~\bibnamefont {Karthik}}, \bibinfo {author}
  {\bibfnamefont {T.}~\bibnamefont {Khan}}, \bibinfo {author} {\bibfnamefont
  {T.}~\bibnamefont {Liu}}, \bibinfo {author} {\bibfnamefont {A.}~\bibnamefont
  {Paul}}, \bibinfo {author} {\bibfnamefont {J.}~\bibnamefont {Schoenleber}}, \
  and\ \bibinfo {author} {\bibfnamefont {R.~S.}\ \bibnamefont {Sufian}},\
  }\href {\doibase 10.1103/PhysRevD.111.074509} {\bibfield  {journal} {\bibinfo
   {journal} {Phys. Rev. D}\ }\textbf {\bibinfo {volume} {111}},\ \bibinfo
  {pages} {074509} (\bibinfo {year} {2025})},\ \Eprint
  {http://arxiv.org/abs/2409.17234} {arXiv:2409.17234 [hep-lat]} \BibitemShut
  {NoStop}%
\bibitem [{\citenamefont {Della~Morte}\ \emph {et~al.}(2017)\citenamefont
  {Della~Morte}, \citenamefont {Francis}, \citenamefont {G{\"u}lpers},
  \citenamefont {Herdo{\'\i}za}, \citenamefont {von Hippel}, \citenamefont
  {Horch}, \citenamefont {J{\"a}ger}, \citenamefont {Meyer}, \citenamefont
  {Nyffeler},\ and\ \citenamefont {Wittig}}]{DellaMorte:2017dyu}%
  \BibitemOpen
  \bibfield  {author} {\bibinfo {author} {\bibfnamefont {M.}~\bibnamefont
  {Della~Morte}}, \bibinfo {author} {\bibfnamefont {A.}~\bibnamefont
  {Francis}}, \bibinfo {author} {\bibfnamefont {V.}~\bibnamefont
  {G{\"u}lpers}}, \bibinfo {author} {\bibfnamefont {G.}~\bibnamefont
  {Herdo{\'\i}za}}, \bibinfo {author} {\bibfnamefont {G.}~\bibnamefont {von
  Hippel}}, \bibinfo {author} {\bibfnamefont {H.}~\bibnamefont {Horch}},
  \bibinfo {author} {\bibfnamefont {B.}~\bibnamefont {J{\"a}ger}}, \bibinfo
  {author} {\bibfnamefont {H.~B.}\ \bibnamefont {Meyer}}, \bibinfo {author}
  {\bibfnamefont {A.}~\bibnamefont {Nyffeler}}, \ and\ \bibinfo {author}
  {\bibfnamefont {H.}~\bibnamefont {Wittig}},\ }\href {\doibase
  10.1007/JHEP10(2017)020} {\bibfield  {journal} {\bibinfo  {journal} {JHEP}\
  }\textbf {\bibinfo {volume} {10}},\ \bibinfo {pages} {020} (\bibinfo {year}
  {2017})},\ \Eprint {http://arxiv.org/abs/1705.01775} {arXiv:1705.01775
  [hep-lat]} \BibitemShut {NoStop}%
\bibitem [{\citenamefont {Borsanyi}\ \emph {et~al.}(2017)\citenamefont
  {Borsanyi}, \citenamefont {Fodor}, \citenamefont {Kawanai}, \citenamefont
  {Krieg}, \citenamefont {Lellouch}, \citenamefont {Malak}, \citenamefont
  {Miura}, \citenamefont {Szabo}, \citenamefont {Torrero},\ and\ \citenamefont
  {Toth}}]{Borsanyi:2016lpl}%
  \BibitemOpen
  \bibfield  {author} {\bibinfo {author} {\bibfnamefont {S.}~\bibnamefont
  {Borsanyi}}, \bibinfo {author} {\bibfnamefont {Z.}~\bibnamefont {Fodor}},
  \bibinfo {author} {\bibfnamefont {T.}~\bibnamefont {Kawanai}}, \bibinfo
  {author} {\bibfnamefont {S.}~\bibnamefont {Krieg}}, \bibinfo {author}
  {\bibfnamefont {L.}~\bibnamefont {Lellouch}}, \bibinfo {author}
  {\bibfnamefont {R.}~\bibnamefont {Malak}}, \bibinfo {author} {\bibfnamefont
  {K.}~\bibnamefont {Miura}}, \bibinfo {author} {\bibfnamefont {K.~K.}\
  \bibnamefont {Szabo}}, \bibinfo {author} {\bibfnamefont {C.}~\bibnamefont
  {Torrero}}, \ and\ \bibinfo {author} {\bibfnamefont {B.}~\bibnamefont
  {Toth}},\ }\href {\doibase 10.1103/PhysRevD.96.074507} {\bibfield  {journal}
  {\bibinfo  {journal} {Phys. Rev. D}\ }\textbf {\bibinfo {volume} {96}},\
  \bibinfo {pages} {074507} (\bibinfo {year} {2017})},\ \Eprint
  {http://arxiv.org/abs/1612.02364} {arXiv:1612.02364 [hep-lat]} \BibitemShut
  {NoStop}%
\bibitem [{\citenamefont {Borsanyi}\ \emph {et~al.}(2018)\citenamefont
  {Borsanyi} \emph {et~al.}}]{Budapest-Marseille-Wuppertal:2017okr}%
  \BibitemOpen
  \bibfield  {author} {\bibinfo {author} {\bibfnamefont {S.}~\bibnamefont
  {Borsanyi}} \emph {et~al.} (\bibinfo {collaboration}
  {Budapest-Marseille-Wuppertal}),\ }\href {\doibase
  10.1103/PhysRevLett.121.022002} {\bibfield  {journal} {\bibinfo  {journal}
  {Phys. Rev. Lett.}\ }\textbf {\bibinfo {volume} {121}},\ \bibinfo {pages}
  {022002} (\bibinfo {year} {2018})},\ \Eprint
  {http://arxiv.org/abs/1711.04980} {arXiv:1711.04980 [hep-lat]} \BibitemShut
  {NoStop}%
\bibitem [{\citenamefont {Blum}\ \emph {et~al.}(2018)\citenamefont {Blum},
  \citenamefont {Boyle}, \citenamefont {G{\"u}lpers}, \citenamefont {Izubuchi},
  \citenamefont {Jin}, \citenamefont {Jung}, \citenamefont {J{\"u}ttner},
  \citenamefont {Lehner}, \citenamefont {Portelli},\ and\ \citenamefont
  {Tsang}}]{RBC:2018dos}%
  \BibitemOpen
  \bibfield  {author} {\bibinfo {author} {\bibfnamefont {T.}~\bibnamefont
  {Blum}}, \bibinfo {author} {\bibfnamefont {P.~A.}\ \bibnamefont {Boyle}},
  \bibinfo {author} {\bibfnamefont {V.}~\bibnamefont {G{\"u}lpers}}, \bibinfo
  {author} {\bibfnamefont {T.}~\bibnamefont {Izubuchi}}, \bibinfo {author}
  {\bibfnamefont {L.}~\bibnamefont {Jin}}, \bibinfo {author} {\bibfnamefont
  {C.}~\bibnamefont {Jung}}, \bibinfo {author} {\bibfnamefont {A.}~\bibnamefont
  {J{\"u}ttner}}, \bibinfo {author} {\bibfnamefont {C.}~\bibnamefont {Lehner}},
  \bibinfo {author} {\bibfnamefont {A.}~\bibnamefont {Portelli}}, \ and\
  \bibinfo {author} {\bibfnamefont {J.~T.}\ \bibnamefont {Tsang}} (\bibinfo
  {collaboration} {RBC, UKQCD}),\ }\href {\doibase
  10.1103/PhysRevLett.121.022003} {\bibfield  {journal} {\bibinfo  {journal}
  {Phys. Rev. Lett.}\ }\textbf {\bibinfo {volume} {121}},\ \bibinfo {pages}
  {022003} (\bibinfo {year} {2018})},\ \Eprint
  {http://arxiv.org/abs/1801.07224} {arXiv:1801.07224 [hep-lat]} \BibitemShut
  {NoStop}%
\bibitem [{\citenamefont {G{\'e}rardin}\ \emph {et~al.}(2019)\citenamefont
  {G{\'e}rardin}, \citenamefont {C{\`e}}, \citenamefont {von Hippel},
  \citenamefont {H{\"o}rz}, \citenamefont {Meyer}, \citenamefont {Mohler},
  \citenamefont {Ottnad}, \citenamefont {Wilhelm},\ and\ \citenamefont
  {Wittig}}]{Gerardin:2019rua}%
  \BibitemOpen
  \bibfield  {author} {\bibinfo {author} {\bibfnamefont {A.}~\bibnamefont
  {G{\'e}rardin}}, \bibinfo {author} {\bibfnamefont {M.}~\bibnamefont
  {C{\`e}}}, \bibinfo {author} {\bibfnamefont {G.}~\bibnamefont {von Hippel}},
  \bibinfo {author} {\bibfnamefont {B.}~\bibnamefont {H{\"o}rz}}, \bibinfo
  {author} {\bibfnamefont {H.~B.}\ \bibnamefont {Meyer}}, \bibinfo {author}
  {\bibfnamefont {D.}~\bibnamefont {Mohler}}, \bibinfo {author} {\bibfnamefont
  {K.}~\bibnamefont {Ottnad}}, \bibinfo {author} {\bibfnamefont
  {J.}~\bibnamefont {Wilhelm}}, \ and\ \bibinfo {author} {\bibfnamefont
  {H.}~\bibnamefont {Wittig}},\ }\href {\doibase 10.1103/PhysRevD.100.014510}
  {\bibfield  {journal} {\bibinfo  {journal} {Phys. Rev. D}\ }\textbf {\bibinfo
  {volume} {100}},\ \bibinfo {pages} {014510} (\bibinfo {year} {2019})},\
  \Eprint {http://arxiv.org/abs/1904.03120} {arXiv:1904.03120 [hep-lat]}
  \BibitemShut {NoStop}%
\bibitem [{\citenamefont {Djukanovic}\ \emph {et~al.}(2025)\citenamefont
  {Djukanovic}, \citenamefont {von Hippel}, \citenamefont {Kuberski},
  \citenamefont {Meyer}, \citenamefont {Miller}, \citenamefont {Ottnad},
  \citenamefont {Parrino}, \citenamefont {Risch},\ and\ \citenamefont
  {Wittig}}]{Djukanovic:2024cmq}%
  \BibitemOpen
  \bibfield  {author} {\bibinfo {author} {\bibfnamefont {D.}~\bibnamefont
  {Djukanovic}}, \bibinfo {author} {\bibfnamefont {G.}~\bibnamefont {von
  Hippel}}, \bibinfo {author} {\bibfnamefont {S.}~\bibnamefont {Kuberski}},
  \bibinfo {author} {\bibfnamefont {H.~B.}\ \bibnamefont {Meyer}}, \bibinfo
  {author} {\bibfnamefont {N.}~\bibnamefont {Miller}}, \bibinfo {author}
  {\bibfnamefont {K.}~\bibnamefont {Ottnad}}, \bibinfo {author} {\bibfnamefont
  {J.}~\bibnamefont {Parrino}}, \bibinfo {author} {\bibfnamefont
  {A.}~\bibnamefont {Risch}}, \ and\ \bibinfo {author} {\bibfnamefont
  {H.}~\bibnamefont {Wittig}},\ }\href {\doibase 10.1007/JHEP04(2025)098}
  {\bibfield  {journal} {\bibinfo  {journal} {JHEP}\ }\textbf {\bibinfo
  {volume} {04}},\ \bibinfo {pages} {098} (\bibinfo {year} {2025})},\ \Eprint
  {http://arxiv.org/abs/2411.07969} {arXiv:2411.07969 [hep-lat]} \BibitemShut
  {NoStop}%
\bibitem [{\citenamefont {Jay}\ and\ \citenamefont {Neil}(2021)}]{Jay:2020jkz}%
  \BibitemOpen
  \bibfield  {author} {\bibinfo {author} {\bibfnamefont {W.~I.}\ \bibnamefont
  {Jay}}\ and\ \bibinfo {author} {\bibfnamefont {E.~T.}\ \bibnamefont {Neil}},\
  }\href {\doibase 10.1103/PhysRevD.103.114502} {\bibfield  {journal} {\bibinfo
   {journal} {Phys. Rev. D}\ }\textbf {\bibinfo {volume} {103}},\ \bibinfo
  {pages} {114502} (\bibinfo {year} {2021})},\ \Eprint
  {http://arxiv.org/abs/2008.01069} {arXiv:2008.01069 [stat.ME]} \BibitemShut
  {NoStop}%
\bibitem [{\citenamefont {Bazavov}\ \emph
  {et~al.}(2025{\natexlab{a}})\citenamefont {Bazavov} \emph
  {et~al.}}]{MILC:2024ryz}%
  \BibitemOpen
  \bibfield  {author} {\bibinfo {author} {\bibfnamefont {A.}~\bibnamefont
  {Bazavov}} \emph {et~al.} (\bibinfo {collaboration} {MILC, Fermilab Lattice,
  HPQCD}),\ }\href {\doibase 10.1103/PhysRevD.111.094508} {\bibfield  {journal}
  {\bibinfo  {journal} {Phys. Rev. D}\ }\textbf {\bibinfo {volume} {111}},\
  \bibinfo {pages} {094508} (\bibinfo {year} {2025}{\natexlab{a}})},\ \Eprint
  {http://arxiv.org/abs/2411.09656} {arXiv:2411.09656 [hep-lat]} \BibitemShut
  {NoStop}%
\bibitem [{\citenamefont {Bazavov}\ \emph
  {et~al.}(2025{\natexlab{b}})\citenamefont {Bazavov} \emph
  {et~al.}}]{FermilabLatticeHPQCD:2024ppc}%
  \BibitemOpen
  \bibfield  {author} {\bibinfo {author} {\bibfnamefont {A.}~\bibnamefont
  {Bazavov}} \emph {et~al.} (\bibinfo {collaboration} {Fermilab Lattice,
  HPQCD,, MILC}),\ }\href {\doibase 10.1103/d583-yhfs} {\bibfield  {journal}
  {\bibinfo  {journal} {Phys. Rev. Lett.}\ }\textbf {\bibinfo {volume} {135}},\
  \bibinfo {pages} {011901} (\bibinfo {year} {2025}{\natexlab{b}})},\ \Eprint
  {http://arxiv.org/abs/2412.18491} {arXiv:2412.18491 [hep-lat]} \BibitemShut
  {NoStop}%
\bibitem [{\citenamefont {Gao}\ \emph {et~al.}(2021{\natexlab{b}})\citenamefont
  {Gao}, \citenamefont {Karthik}, \citenamefont {Mukherjee}, \citenamefont
  {Petreczky}, \citenamefont {Syritsyn},\ and\ \citenamefont
  {Zhao}}]{Gao:2021xsm}%
  \BibitemOpen
  \bibfield  {author} {\bibinfo {author} {\bibfnamefont {X.}~\bibnamefont
  {Gao}}, \bibinfo {author} {\bibfnamefont {N.}~\bibnamefont {Karthik}},
  \bibinfo {author} {\bibfnamefont {S.}~\bibnamefont {Mukherjee}}, \bibinfo
  {author} {\bibfnamefont {P.}~\bibnamefont {Petreczky}}, \bibinfo {author}
  {\bibfnamefont {S.}~\bibnamefont {Syritsyn}}, \ and\ \bibinfo {author}
  {\bibfnamefont {Y.}~\bibnamefont {Zhao}},\ }\href {\doibase
  10.1103/PhysRevD.104.114515} {\bibfield  {journal} {\bibinfo  {journal}
  {Phys. Rev. D}\ }\textbf {\bibinfo {volume} {104}},\ \bibinfo {pages}
  {114515} (\bibinfo {year} {2021}{\natexlab{b}})},\ \Eprint
  {http://arxiv.org/abs/2102.06047} {arXiv:2102.06047 [hep-lat]} \BibitemShut
  {NoStop}%
\bibitem [{\citenamefont {Zhang}\ \emph
  {et~al.}(2025{\natexlab{b}})\citenamefont {Zhang}, \citenamefont {Grebe},
  \citenamefont {Hackett}, \citenamefont {Wagman},\ and\ \citenamefont
  {Zhao}}]{Zhang:2025hyo}%
  \BibitemOpen
  \bibfield  {author} {\bibinfo {author} {\bibfnamefont {R.}~\bibnamefont
  {Zhang}}, \bibinfo {author} {\bibfnamefont {A.~V.}\ \bibnamefont {Grebe}},
  \bibinfo {author} {\bibfnamefont {D.~C.}\ \bibnamefont {Hackett}}, \bibinfo
  {author} {\bibfnamefont {M.~L.}\ \bibnamefont {Wagman}}, \ and\ \bibinfo
  {author} {\bibfnamefont {Y.}~\bibnamefont {Zhao}},\ }\href@noop {} {\
  (\bibinfo {year} {2025}{\natexlab{b}})},\ \Eprint
  {http://arxiv.org/abs/2501.00729} {arXiv:2501.00729 [hep-lat]} \BibitemShut
  {NoStop}%
\bibitem [{\citenamefont {Wagman}(2025)}]{Wagman:2024rid}%
  \BibitemOpen
  \bibfield  {author} {\bibinfo {author} {\bibfnamefont {M.~L.}\ \bibnamefont
  {Wagman}},\ }\href {\doibase 10.1103/pcvc-734h} {\bibfield  {journal}
  {\bibinfo  {journal} {Phys. Rev. Lett.}\ }\textbf {\bibinfo {volume} {134}},\
  \bibinfo {pages} {241901} (\bibinfo {year} {2025})},\ \Eprint
  {http://arxiv.org/abs/2406.20009} {arXiv:2406.20009 [hep-lat]} \BibitemShut
  {NoStop}%
\bibitem [{\citenamefont {Hackett}\ and\ \citenamefont
  {Wagman}(2025)}]{Hackett:2024nbe}%
  \BibitemOpen
  \bibfield  {author} {\bibinfo {author} {\bibfnamefont {D.~C.}\ \bibnamefont
  {Hackett}}\ and\ \bibinfo {author} {\bibfnamefont {M.~L.}\ \bibnamefont
  {Wagman}},\ }\href {\doibase 10.1103/fp74-q35q} {\bibfield  {journal}
  {\bibinfo  {journal} {Phys. Rev. D}\ }\textbf {\bibinfo {volume} {112}},\
  \bibinfo {pages} {014514} (\bibinfo {year} {2025})},\ \Eprint
  {http://arxiv.org/abs/2412.04444} {arXiv:2412.04444 [hep-lat]} \BibitemShut
  {NoStop}%
\bibitem [{\citenamefont {Young}(2012)}]{Young:2012kg}%
  \BibitemOpen
  \bibfield  {author} {\bibinfo {author} {\bibfnamefont {P.}~\bibnamefont
  {Young}},\ }\href@noop {} {\  (\bibinfo {year} {2012})},\ \Eprint
  {http://arxiv.org/abs/1210.3781} {arXiv:1210.3781 [physics.data-an]}
  \BibitemShut {NoStop}%
\end{thebibliography}%
\bibliographystyle{apsrev4-1}

\end{document}